\documentclass[final,1p]{elsarticle}
\usepackage{lineno}
\usepackage{hyperref}
\usepackage[hyperref]{xcolor}
\hypersetup{
    colorlinks=true,
}
\usepackage{bm}%
\usepackage{amsfonts}
\usepackage{amssymb}
\usepackage{amsmath}
\usepackage{bbold}
\journal{Progress in Particle and Nuclear Physics}
\biboptions{sort&compress}

\newcommand{\si}{\sigma}
\newcommand{\al}{\alpha}

\newcommand{\az}{\varphi}
\newcommand{\ro}{\rho}

\newcommand{\be}{\beta}

\newcommand{\oeq}{\begin{equation}}
\newcommand{\ceq}{\end{equation}}
\newcommand{\oeqn}{\begin{eqnarray}}
\newcommand{\ceqn}{\end{eqnarray}}

\renewcommand{\>}{\rangle}
\newcommand{\<}{\langle}
\renewcommand{\(}{\left(}
\renewcommand{\)}{\right)}
\renewcommand{\[}{\left[}
\renewcommand{\]}{\right]}

\newcommand{\stf}{\,\,\,}
\newcommand{\sdf}{\,\,}
\newcommand{\stb}{\!\!\!}

\newcommand{\kfi}{|\phi \>}

\newcommand{\bfi}{\<\phi |}

\newcommand{\oQ}{\hat{Q}}

\newcommand{\oP}{\hat{P}}

\newcommand{\oH}{\hat{H}}
\newcommand{\oK}{\hat{K}}
\newcommand{\ox}{\hat{x}}

\newcommand{\oV}{\hat{V}}
\newcommand{\oU}{\hat{U}}
\newcommand{\oD}{\hat{D}}

\newcommand{\ovp}{\hat{\bf p}}

\newcommand{\of}{\hat{f}}

\newcommand{\oad}{\hat{a}^\dagger}
\newcommand{\oa}{\hat{a}}
\newcommand{\oA}{\hat{A}}

\newcommand{\oF}{\hat{F}}
\newcommand{\oN}{\hat{N}}

\newcommand{\ov}{\hat{v}}

\newcommand{\opsi}{\hat{\psi}}
\newcommand{\opsid}{\hat{\psi}^\dagger}

\newcommand{\del}{\delta\!}
\newcommand{\dt}{\frac{\partial}{\partial t}}

\renewcommand{\d}{{\mbox d}}

\newcommand{\hb}{\hbar}

\newcommand{\vr}{{\bf r}}
\newcommand{\vx}{{\bf x}}
\newcommand{\vy}{{\bf y}}
\newcommand{\vz}{{\bf z}}

\newcommand{\vR}{{\bf R}}

\newcommand{\vp}{{\bf p}}

\newcommand{\mD}{{\mathcal{D}}}
\newcommand{\mP}{{\mathcal{P}}}

\newcommand{\Tr}{\mbox{Tr}}
\newcommand{\tr}{\mbox{tr}}

\newcommand{\bra}[1]{\langle #1 \vert}
\newcommand{\ket}[1]{\vert #1 \rangle}

\newcommand{\com}[1]{\left[ #1 \right]}

\newcommand{\anni}{\hat{a}}

\newcommand{\hba}{\hat{\beta}}

\newcommand{\ofR}{(\mathbf{r})}

\newcommand{\Smat}[4]{\left( \begin{array}{cc} #1 & #2 \\ #3 & #4 \end{array} \right)}
\newcommand{\Svec}[2]{\left( \begin{array}{c} #1 \\ #2 \end{array} \right)}
\newcommand{\Hbogo}{\mathcal{H}}
\newcommand{\Rbogo}{\mathcal{R}}

\begin{document}
\hypersetup{urlcolor=blue,citecolor=blue,linkcolor=blue, hypertexnames=false}

\begin{frontmatter}

\title{Heavy-ion collisions and fission dynamics with the time-dependent Hartree-Fock theory and its extensions}

%% Group authors per affiliation:
\author{C. Simenel\corref{cor1}}
\address{Department of Theoretical Physics and Department of Nuclear Physics, Research School of Physics and Engineering,
The Australian National University, Canberra ACT  2601, Australia}
\cortext[cor1]{Corresponding author}
\ead{cedric.simenel@anu.edu.au}

\author{A.S. Umar}
\address{Department of Physics and Astronomy, Vanderbilt University, Nashville, TN 37235, USA}

\begin{abstract}
Microscopic methods and tools to describe nuclear dynamics have considerably been improved in the past few years. 
They are based on the time-dependent Hartree-Fock (TDHF) theory and its extensions to include pairing correlations and
quantum fluctuations. 
The TDHF theory is the lowest level of approximation of a range of methods to solve the quantum many-body problem, 
showing its universality to describe many-fermion dynamics at the mean-field level. 
The range of applications of TDHF to describe realistic systems allowing for detailed comparisons with experiment 
has considerably increased. 
For instance, TDHF is now commonly used to investigate fusion, multi-nucleon transfer and quasi-fission reactions. 
Thanks to the inclusion of pairing correlations, it has also recently led to breakthroughs in our description of the saddle to scission evolution, and, in particular, 
the non-adiabatic effects near scission. 
Beyond mean-field approaches such as the time-dependent random-phase approximation (TDRPA) and stochastic mean-field methods 
have reached the point where they can be used for realistic applications. 
We review recent progresses in both techniques and applications to heavy-ion collision and fission. 
\end{abstract}

\begin{keyword}
\texttt{Time-Dependent Hartree-Fock, Heavy-Ion Fusion, Fission, 
Nucleus-Nucleus Potential.}
\end{keyword}

\end{frontmatter}

\section{Introduction: nuclear quantum many-body dynamics}

The quantum many-body problem is vital to many areas of physics.
Indeed, the description of complex quantum objects of interacting particles is of interests for many physical systems, from quarks and gluons in a nucleon to macromolecules, such as fullerenes, to Bose-Einstein condensates. 
Consequently, major developments in the description of quantum many-body systems are often of interest to many different fields. 
For instance, the BCS theory introduced to describe superconductivity~\cite{bardeen1957} 
is also widely used in nuclear physics to incorporate the effect of  pairing correlations. 
Similarly, the tools used to study low-energy fusion with multi-channel tunneling~\cite{dasgupta1998} 
are also used to describe dissociative adsorption of molecules on a surface~\cite{hagino2012}.

The similarity between dynamical processes in quantum many-body systems, whether their constituents are nucleons, electrons, or atoms, is quite striking.
It is possible to make such systems vibrate, rotate, fuse, transfer particles, fission and break up. 
This is of course true for nuclear systems, where 
each of these processes can be used to learn about specific aspects of quantum many-body dynamics.
For instance,  the study of nuclear vibrations tells us how single-particles can produce collective motion,
 what is its interplay with the underlying shell structure, what are the source of non-linearities leading to anharmonicities, 
 and how collective modes get damped and decay. 
These concepts and many others can also be studied via heavy-ion collisions:
\begin{itemize}
\item Heavy-ion fusion could be a tool to understand thermalization of a many-body system initially out of equilibrium. 
 Fusion is also well suited to investigate the quantum tunneling of complex systems 
 over orders of magnitudes in terms of barrier transmission probabilities.
 This allows to study the coupling between relative motion (the main collective degree of freedom used to characterize fusion) 
  with other internal degrees of freedom (vibrations, rotations, single-particle excitations...).  
 These internal degrees of freedom could also be responsible for dissipation and decoherence processes, 
 whose descriptions remain problematic with fully quantal treatments. 
\item Transfer reactions between heavy-ions are another example of mechanism strongly driven by quantum dynamical processes.
Such reactions produce entangled fragments in coherent superpositions of proton and neutron numbers. 
The measurement of the properties of one fragment (e.g., particle number or kinetic energy) 
induces a projection of the quantum state of the other fragment. 
Transfer reactions are also ideal to investigate clustering and superfluidity (via, e.g., the excitation of pairing vibrations).
In addition, they are thought to be a doorway to dissipation in heavy-ion collisions. 
Interesting questions regarding the indistinguishibility in the transfer of identical particles could also be raised. 
\item At higher energies, deep-inelastic collisions (DIC) can be used to investigate quantum fluctuations, 
via, e.g., the measurement of fragment mass and charge distributions. 
In addition, dissipation and fluctuations are correlated in a dynamical way. 
The finite contact times between the fragments give access to equilibration times
of initial asymmetries between the collision partners (e.g., mass and isospin asymmetries). 
DIC and quasi-fission reactions can also be used to test the persistence of the quantum shell corrections 
(responsible, e.g., for the formation of magic fragments) with respect to the excitation energy of the fragments.
\item Finally, nuclear fission is probably the mechanism which has the most to teach us about the diversity of quantum many-body dynamical aspects. 
It is then not surprising that, in turns, it is the most complicated reaction to describe. 
Not only it is a large amplitude collective motion which involves the overcoming of one or several barriers (where quantum tunneling could happen), 
it is also strongly affected by shell effects, superfluidity, fluctuations and dissipation. 
Naively, it could be seen as the reverse process of fusion. 
However, the mechanisms which produce  fission fragments are often much more difficult to describe than in the fusion case. 
It has to do essentially with the adiabatic (slow motion in a space of collective degrees of freedom where the other internal degrees of freedoms are assumed to be at equilibrium) to non-adiabatic transition occurring between the fission saddle point and the scission point where the entangled fragments are formed. 
\end{itemize}

\begin{table}[h]
\centering
\caption{Examples of many-body systems and their environments.}
\label{tab1}       
\begin{tabular}{l|lll}
\hline
& Nuclei & Atoms & Molecules  \\\hline
Size & $\sim10^{-14}$~m & $\sim10^{-10}$~m& $\sim10^{-9}$~m \\
Time scale & $10^{-21}$~s=1~zs& $10^{-18}$~s=1~as & $10^{-15}$~s=1~fs \\
Environment & none& EM& EM+gas\\\hline
Equation of& \textit{Quantum} & \textit{Quantum} & \textit{Classical} \\
motion& & \textit{(Classical)} & \textit{(Quantum)} \\\hline
\end{tabular}
\end{table}

In addition to the diversity of phenomena accessible in heavy-ion collisions, 
what makes them special to test predictions of quantum many-body theories is their almost complete isolation from external environments. 
It is well known that the coupling of a system to an external environment induces a decoherence process 
which is responsible for the quantum to classical transition \cite{zurek1991,joos2003}. 
Atoms and molecules have different sizes and native time scales than atomic nuclei and, 
in turn, do not interact with the same environments (see Tab.~\ref{tab1}).
In particular, atomic and molecular dynamics are easily affected by the coupling to electromagnetic field 
(and  to collisions with  the particles of the surrounding gas for the largest systems). 
The typical time scale for the collision of atomic nuclei, however, 
is of the order of few zeptoseconds (1~zs$=10^{-21}$~s), which is usually too short to emit or absorb photons. 
The quantum coherence built up during the collision is then preserved until after the outcome of the collision has been decided. 
In other words, no major decoherence processes are expected to occur and affect the quantum nature of the reaction. 
Heavy-ion collisions are then ideal to study fundamental concepts of quantum physics, 
such as the existence of collective motion in many-particle systems~\cite{bohr1975}, 
the interplay between quantum tunneling and dissipation~\cite{caldeira1981}, 
coupled reaction channels effects \cite{hagino2012}, and entanglement properties~\cite{lamehi-rachti1976}.\par

Nuclear physicists can put together up to about 500 interacting nucleons 
 over a very short time (few zeptoseconds) thanks to actinide collisions. 
(Of course, many more nucleons may interact in astrophysical systems such as neutron stars.) 
Describing such reactions and predicting their outcomes  is of course a great challenge.
As discussed above, several reactions could occur, which are sometimes in competition with each other for a given collision energy. 
Ideally, the same theoretical model should be able to describe all types of nuclear dynamics.
To some extent, this is possible at the mean-field level, where the particles evolve independently in the mean-field generated by the ensemble of particles. 
For fermions, this is described by the time-dependent Hartree-Fock (TDHF) theory proposed by Dirac in 1930 \cite{dirac1930}.

In his original work, Dirac introduced the TDHF equation to describe the electrons in atoms. 
The first applications of the TDHF formalism to nuclear systems was due to Bonche and co-workers in 1976~\cite{bonche1976}. 
Since then, several groups have developed and used TDHF codes to study nuclear dynamics, 
with a strong revival of the approach in the past ten years as it became clear 
that the TDHF theory was a tool of choice to describe low-energy reactions to be studied with the upcoming radioactive beam facilities. 
Indeed, applications to heavy-ion collisions at energies close to the Coulomb barrier with stable nuclei have led to realistic description of nuclear dynamics, 
motivating new TDHF studies focussing on reactions with exotic nuclei at low-energy.  
Although the applicability of the TDHF approach at low-energy has been  demonstrated, 
it is not clear how high in energy one can go before the mean-field approximation breaks down. 
TDHF calculations of heavy-ion collisions at more than twice the barrier energy have given realistic results \cite{simenel2011} 
as long as one focusses on expectation values of one-body observables for which TDHF is optimised \cite{balian1981}. 
Collisions up to 40~MeV/nucleon have also been simulated with TDHF to investigate charge equilibration \cite{iwata2010} 
and reactions leading to supranormal density \cite{stone2017}.
However, all applications presented in this review are much lower in energy. 
Nevertheless, it would be interesting in the future to investigate the energy range for applicability of the TDHF theory to heavy-ion collisions. 

Earlier investigations of nuclear dynamics within the  TDHF framework have been summarized in previous reviews~\cite{negele1982,simenel2012,nakatsukasa2016}.
Nevertheless, several new techniques have been developed  and new applications have been made in the past few years.
These new techniques include the extraction of microscopic nucleus-nucleus potentials in fusion reactions, 
the development of projection techniques to better characterize the fragment properties,
the inclusion of time-dependent pairing correlations, 
beyond mean-field quantum fluctuations,
optimization of collective coordinates, improvement of boundary conditions, etc.
These led to new applications such as to better understand fusion with exotic nuclei, multi-nucleon transfer reactions, 
dissipative processes in DIC, the role of shell effects in quasi-fission, non-adiabatic effects in fission, etc. 
The present review will focus on these recent works.

\section{Formalism}

\subsection{Introduction}
\begin{figure}[h]
\begin{center}
\includegraphics[width=10cm]{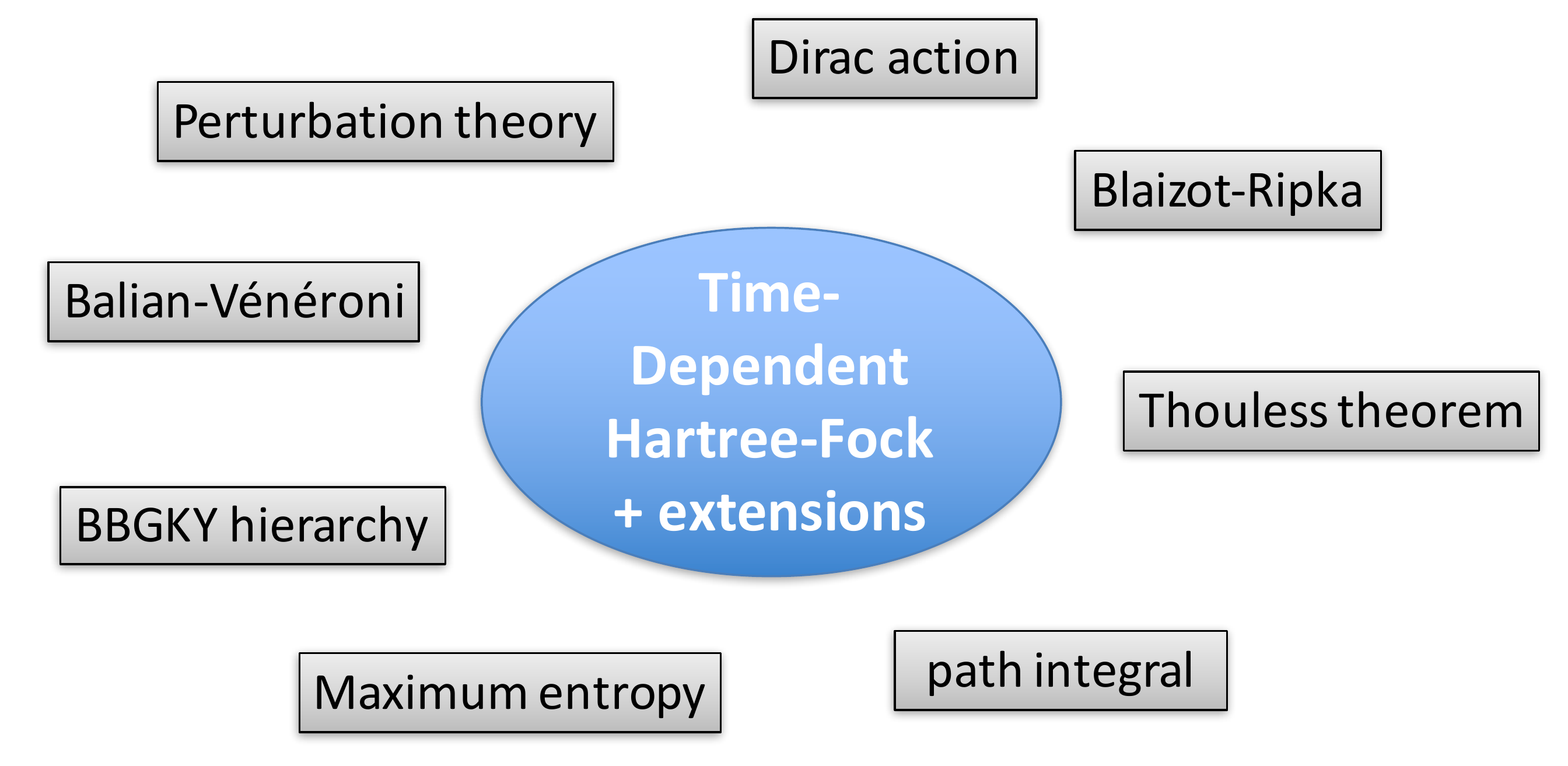}
\end{center}
\caption{Various approaches leading to the TDHF equation and to its extensions. 
\label{fig:TDHFderiv}}
\end{figure}

Various formalisms have been developed in the past to describe the dynamics of quantum many-body systems (see Fig.~\ref{fig:TDHFderiv}).
Some of these approaches are based on variational principles, while others rely on perturbation theory.
Interestingly, at the lowest level of approximation they usually lead to the time-dependent Hartree-Fock self-consistent mean-field theory~\cite{dirac1930}.
The fact that the TDHF equation can be derived in so many ways illustrates its universality. 
It is also interesting to see what are the underlying approximations 
used in deriving the TDHF equation in these various approaches as they inform us on the domain of applicability of the theory. 
For instance, the Thouless theorem~\cite{thouless1960} can be used to extract  the mean-field contribution to the dynamics (see, e.g., \cite{lacroix2014}),
while the approach of Blaizot and Ripka is  optimized to scattering amplitudes \cite{blaizot1981}. 
The next section describes how the TDHF equation is obtained in different approaches. 

In the following, we give only a brief description of standard approaches to microscopic dynamics based on the mean-field approximation. 
For more details, see earlier reviews \cite{negele1982,simenel2012,lacroix2014,nakatsukasa2016}.

For non-relativistic systems, the evolution of the state $|\Psi(t)\>$ is given by the Schr\"odinger equation
\oeq
i\hb\frac{\partial}{\partial t}|\Psi(t)\>=\oH|\Psi(t)\>.
\label{eq:Schrodinger}
\ceq
In the case of a two-body interaction $\ov(1,2)$, the Hamiltonian can be written as
\oeq
\oH= \sum_{i=1}^{A}\frac{\ovp(i)^2}{2m}+ \sum_{i>j=1}^{A}\ov(i,j) \equiv 
 \sum_{ij} \sdf t_{ij}\sdf \oad_i \oa_j + \frac{1}{4} \sum_{ijkl} \sdf  \bar{v}_{ijkl} \sdf \oad_i \oad_j \oa_l \oa_k ,
 \label{eq:oH}
\ceq
where $A$ is the total number of particles and $\bar{v}$ denotes antisymmetrized matrix elements.
The many-body state $|\Psi\>$ contains all the information on the system, including all types of correlations between the nucleons.
We often require only a small subset of this information. 
For instance, we can often content ourself with the knowledge of one-body observables of the form 
$
\oF = \sum_{i=1}^{A} \of(i) ,
$ such as particle number operators, kinetic energy, momentum, multipole moment among others.
If our focus is on the expectation value of such an operator, then its time evolution is given by the Ehrenfest theorem
\oeq
\dt  \<\oF\>_\psi   = \frac{i}{\hb} \< \, [\oH,\oF ] \, \>_\psi. \label{eq:Ehrenfest}
\ceq
The above equation is exact and valid for any (correlated or not) state. 
The expectation value of~$\oF$,
\oeq
\<\oF\> = \sum_{\al\be} f_{\al\be}\rho_{\be\al} = \tr (\rho f),
\label{eq:oF}
\ceq
only requires the knowledge of the one-body density matrix $\rho$ with matrix elements
\oeq
\rho_{\al\be} = \<\Psi|\oad_\be\oa_\al|\Psi\>,
\label{eq:rho}
\ceq
where $\oa_\al$ ($\oad_\al$) annihilates (creates) a particle in the single-particle state $|\al\>$. 
It is therefore not surprising that the one-body density matrix plays a central role in the various approximation schemes discussed below.

\subsection{Quantum many-body perturbation theory \label{sec:perturb}}

Quantum many-body perturbation theory offers an elegant way to derive the Hartree-Fock equation (see, e.g., \cite{fetter2003}). 
In particular, the origin of the self-consistency of the theory is relatively clear in this approach: 
It is due to the fact that the propagation of a particle is affected by its interaction with other particles, which are themselves interacting particles. 
An important quantity is then  $\<\Psi_0|\opsi_H(x's'q')\opsid_H(xsq)|\Psi_0\>$ in the Heisenberg representation\footnote{In the case of a time-independent Hamiltonian, we have $\oF_H(t)=e^{i\oH t/\hb}\oF e^{-i\oH t/\hb}$.}.
It is associated with the amplitude of probability for a nucleon with spin $s$ and isospin $q$ created at 
 space-time position $x=(\vx,t)$ on top of the interacting ground-state $\Psi_0$ to  propagate and be found 
 at  $x'=(\vx',t'>t)$ with $s'$ and $q'$ where it is annihilated\footnote{Of course, nucleons are indistinguishable fermions and there is no guarantee that the nucleon which is annihilated is the same as the one which is created. In fact, the question of ``which one'' is created or annihilated does not make sense.}. 
 Similarly, the quantity  $\<\Psi_0|\opsid_H(x's'q')\opsi_H(xsq)|\Psi_0\>$, where the annihilation occurs before the creation, plays also an essential role. 
 
Both terms are combined in the single-particle Green's function or Feynman propagator
\oeqn
iG(x,x')&=&\<\Psi_0|T[\opsi_H(x)\opsid_H(x')]|\Psi_0\>\nonumber \\
&=&\<\Psi_0|\opsi_H(x)\opsid_H(x')|\Psi_0\>\Theta(t-t')-\<\Psi_0|\opsid_H(x')\opsi_H(x)|\Psi_0\>\Theta(t'-t),
\ceqn
where $T$ is the time ordering operator, and where spin and isospin indices have been dropped for simplicity\footnote{See, e.g., Ref. \cite{fetter2003} for a derivation including spin.}. 
The minus sign in front of the last term is due to the fermionic nature of the system. 
The Green's function is central in quantum field theory as it can be used to compute important quantities 
such as the one-body density matrix (in the interacting ground-state):
$\rho(\vr,\vr';t)=\<\Psi_0|\opsid(\vr't)\opsi(\vr t)|\Psi_0\> = -iG(\vr t,\vr't^+)$.

The perturbation theory is based on the separation of the Hamiltonian (which we assume to be time independent) into 
$\oH=\oH_0+\oV$ where $\oH_0$ is the non interacting Hamiltonian (kinetic energy plus an eventual external one-body potential)
and $\oV$ is the interaction [see eq.~(\ref{eq:oH})]. 
It is then more convenient to use the interaction representation where the operators evolve according to the non-interacting Hamiltonian only: 
$\oF_I(t)=e^{i\oH_0t/\hb}\oF e^{-i\oH_0t/\hb}$.
The Green's function can then be written as 
\oeq
iG(x,y)=\sum_{n=0}^\infty \(\frac{-i}{\hb}\)^n\frac{1}{n!}\int_{-\infty}^{+\infty} dt_1 \cdots dt_n 
\frac{\<\Phi_0| T\[\oV_I(t_1)\cdots \oV_I(t_n)\,\opsi_I(x)\opsid_I(y)\] |\Phi_0\>}{\<\Phi_0|\oU(-\infty,+\infty)|\Phi_0\>}
\ceq
where $\Phi_0$ is the non-interacting ground-state of $\oH_0$ and $\oU$ the evolution operator (with interaction). 

Feynman diagrams (see Appendix) provide a powerful way to calculating the terms in the expansion of $G(x,y)$. 
In particular, all non-connected diagrams in the numerator can be factorised and  cancelled by the denominator, leading to 
\oeq
iG(x,y)=\sum_{n=0}^\infty \(\frac{-i}{\hb}\)^n\frac{1}{n!}\int_{-\infty}^{+\infty} dt_1 \cdots dt_n 
\,\,\,\<\Phi_0| T\[\oV(t_1)\cdots \oV(t_n)\,\opsi(x)\opsid(y)\] |\Phi_0\>_{connect.}
\ceq
where only connected diagrams remain. Here and in the following, the subscript $I$ of the interaction representation is dropped to simplify the notation. 
The above equation can be expressed as 
\begin{figure}[h]
\begin{center}
\includegraphics[width=9cm]{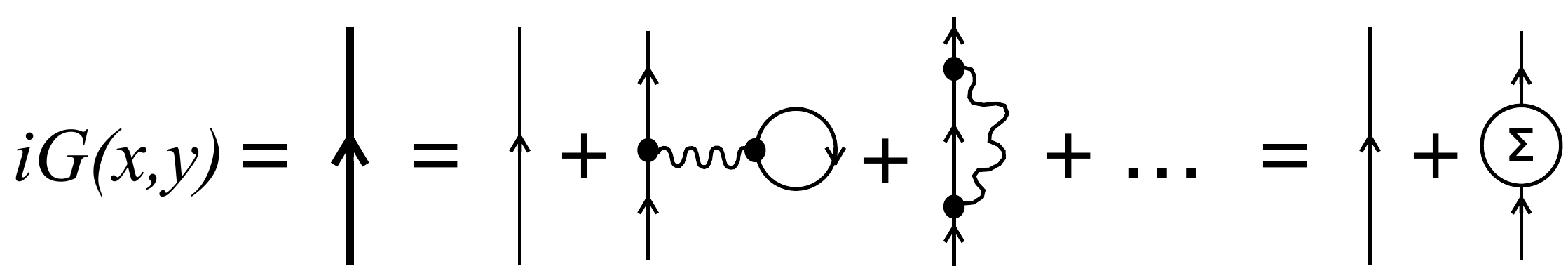}
\end{center}
\end{figure}\\
where the thin lines refer to the non-interacting system with the propagator $iG^0(x,y)=\<\Phi_0|T[\opsi(x)\opsid(y)]|\Phi_0\>$,  
the curly lines correspond to interaction terms $\mathcal{V}(x,x')$, and $\Sigma$ is the self energy of the interacting system.
The first and second diagrams with an interaction line are the ``direct'' and ``exchange'' contributions to the first order term of the expansion, respectively.  
Using Feynman rules (see Appendix), the direct and exchange terms are respectively 
$$ \frac{i}{\hb}\int d^4z d^4z' (-1) G^0(x,z)\mathcal{V}(z,z')G^0(z,y)G^0(z',z') $$
and
$$ \frac{i}{\hb}\int d^4z d^4z' G^0(x,z)\mathcal{V}(z,z')G^0(z,z')G^0(z',y).$$
We see that each diagram involves single-particle non-interacting Green's function $G^0$, which can easily be calculated, as well as matrix elements of the interaction. 

There is of course an infinite number of diagrams to calculate which makes the perturbation theory useless 
unless the interaction is small enough to neglect diagrams above a given order in the expansion. 
One could for instance neglect diagrams above the first order. 
We see that, in this case, the particle propagating between $x$ and $y$ space-time positions 
interacts with another particle which itself propagates without interaction as its propagator is $G^0$. 
The resulting theory is therefore not self-consistent. 

It is possible to construct a self-consistent theory by rewriting the self energy $\Sigma$ in terms of ``proper self energy'' insertions, i.e., 
diagrams with an incoming and an outgoing particle and which cannot be cut into two pieces by cutting a single-particle line. 
The proper self energy $\Sigma^\star$ is the sum of these diagrams:
\begin{figure}[h]
\begin{center}
\includegraphics[width=9cm]{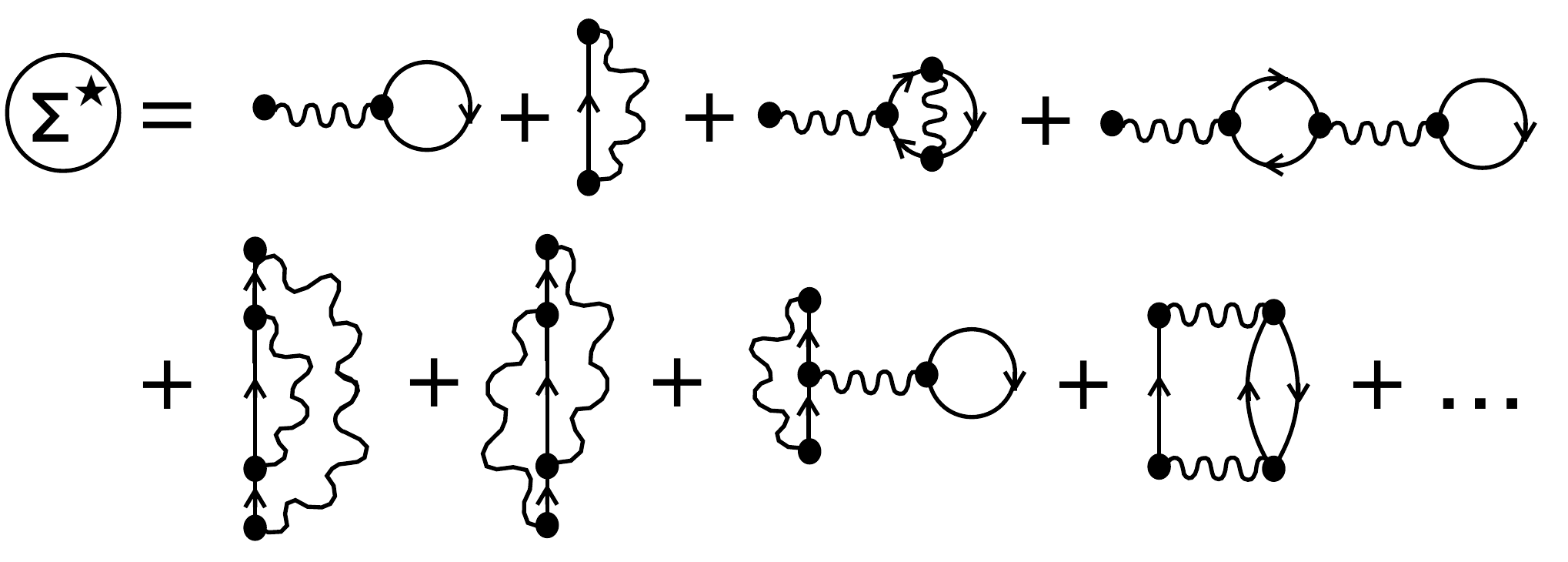}
\end{center}
\end{figure}\\
The self energy can then be written as
\begin{figure}[h]
\begin{center}
\includegraphics[width=6cm]{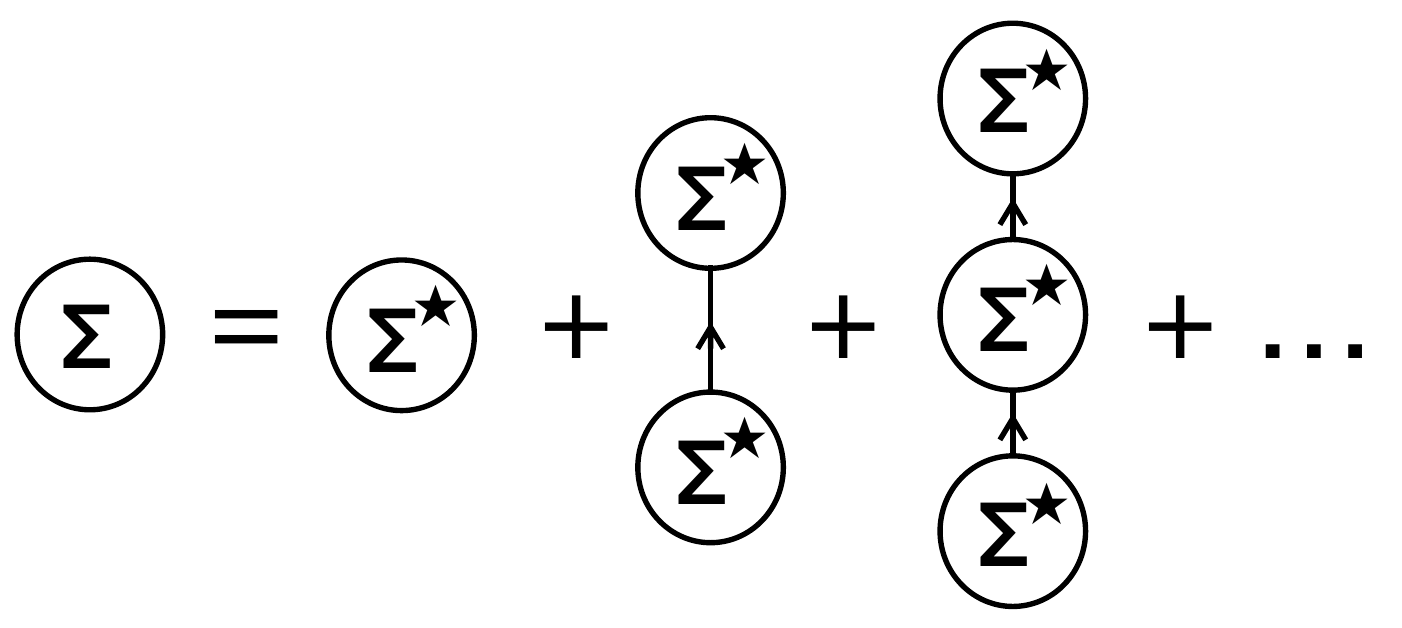}
\end{center}
\end{figure}\\
leading to Dyson's equation for $G(x,y)$:
\begin{figure}[h]
\begin{center}
\includegraphics[width=8cm]{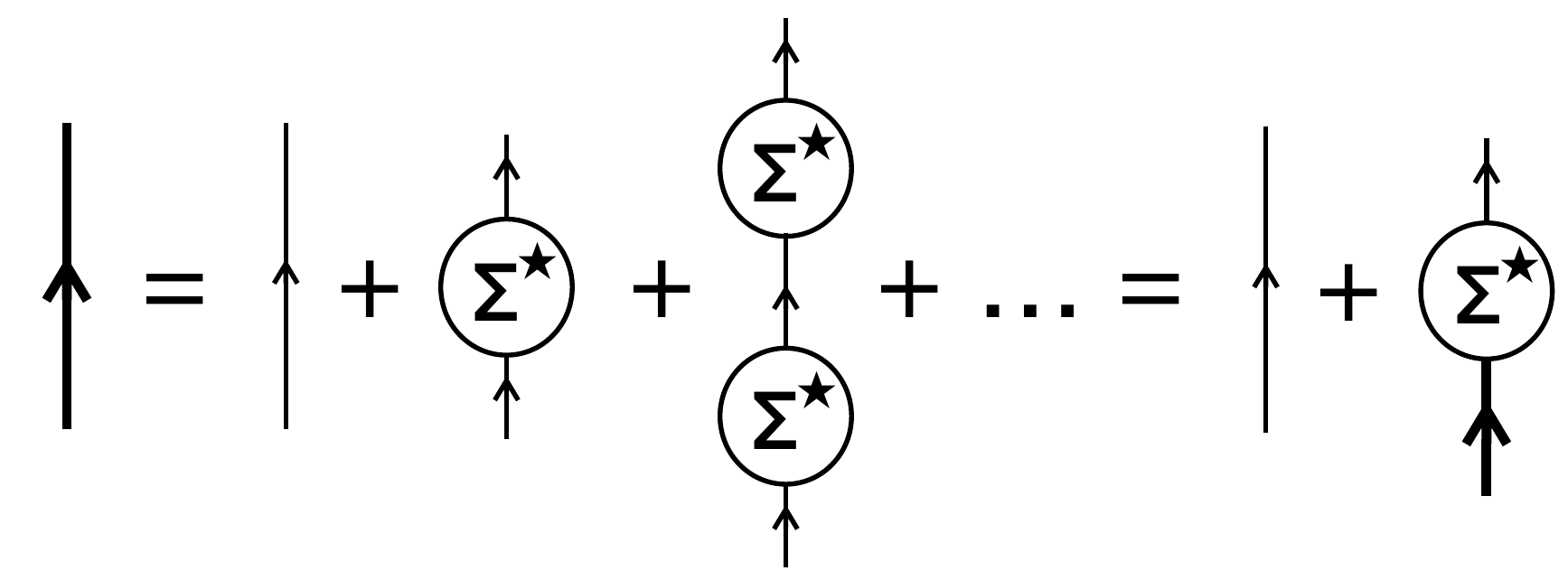}
\end{center}
\end{figure}\\
or equivalently
\oeq 
G(x,y) = G^0(x,y) + \int d^4zd^4z' \,\, G^0(x,z) \Sigma^\star(z,z') G(z',y).
\ceq
We see that the interacting single-particle Green's function $G$ is now present on both sides of the equation.

So far no approximations have been made and the problem remains impractical as the proper self energy $\Sigma^\star$ contains an infinite number of diagrams. 
A possible approximation is to keep only the first order diagrams in $\Sigma^\star$:

\begin{figure}[h]
\begin{center}
\includegraphics[width=6cm]{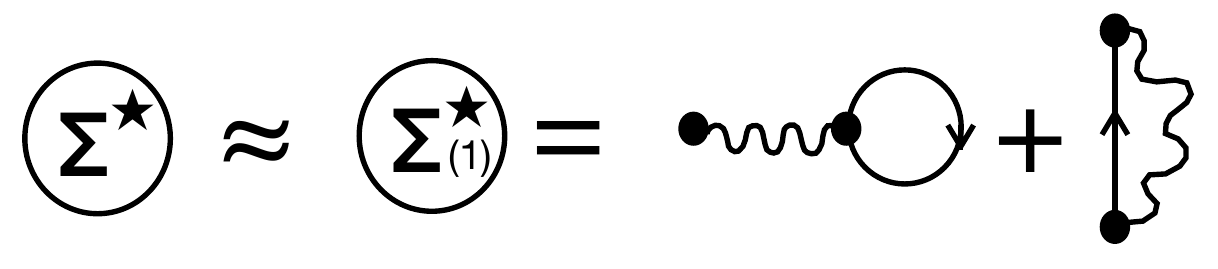}
\end{center}
\end{figure}
However, this again leads to a lack of self consistency as the particle line is the non-interacting one. 

In the self-consistent Hatree-Fock approximation, the above diagrams are replaced by 
\begin{figure}[h]
\begin{center}
\includegraphics[width=6cm]{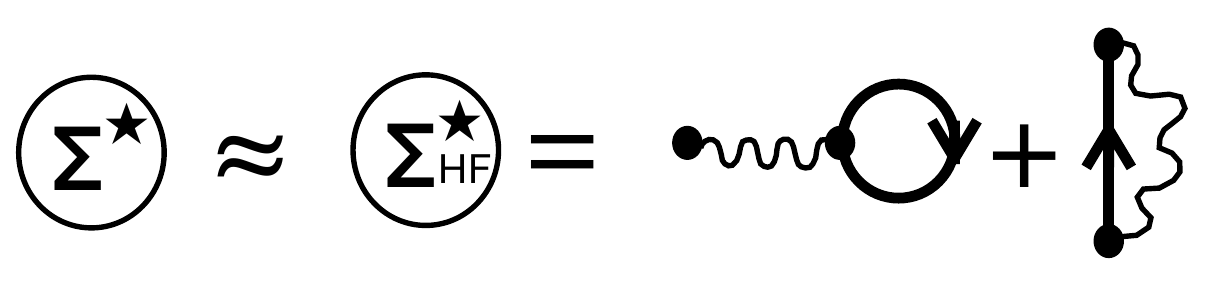}
\end{center}
\end{figure}\\
where the interaction occurs with an interacting particle (i.e., with the propagator $G$ instead of $G^0$). 
For simplicity, we consider an interaction of the form  $\mathcal{V}(x,x')=V(\vx-\vx')\delta(t-t')$.
Using Feynman rules, we get 
\begin{align}
\Sigma^\star_{HF}(x,x')=\frac{-i\delta(t-t')}{\hb}\,\,\Big[&\delta(\vx-\vx')\int d^3x" G(\vx"t,\vx"t^+)V(\vx-\vx")\nonumber \\
&- V(\vx-\vx')G(\vx t,\vx' t^+)\Big]. \label{eq:SigmaHF}
\end{align}
The first term in the right hand side is the direct (Hartree) contribution and the last term is the exchange (Fock) one. 
The Dyson's equation for $G(x,y)$ in the self-consistent Hartree-Fock approximation becomes
\begin{figure}[h]
\begin{center}
\includegraphics[width=6cm]{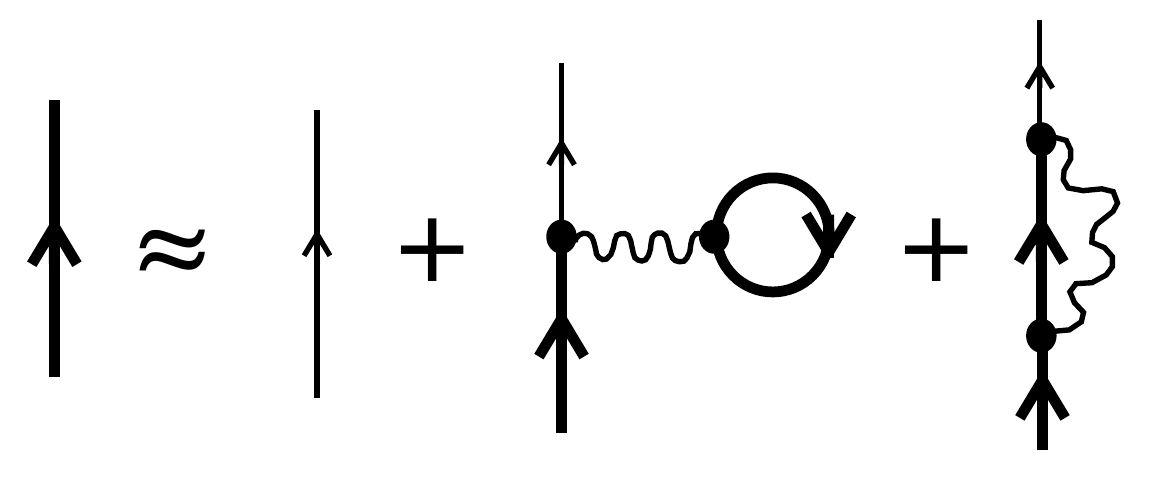}
\end{center}
\end{figure}\\
that is,
\oeq
G(x,y) \approx G^0(x,y) + \int d^4zd^4z' \,\, G^0(x,z) \Sigma^\star_{HF}(z,z') G(z',y).
\ceq
Rewriting the proper self energy as $\Sigma^\star_{HF}(x,y)=\delta(t_x-t_y)\Sigma^\star_{HF}(\vx,\vy)$ 
and using Fourier transform in the time domain defined as 
\oeq
G(x,y)=\frac{1}{2\pi}\int_{-\infty}^{+\infty}d\omega \,e^{-i\omega(t_x-t_y)}G(\vx,\vy,\omega)\label{eq:GFT}
\ceq
we get
\oeq
G(\vx,\vy,\omega)=G^0(\vx,\vy,\omega)+\int d^3zd^3z'\, \, G^0(\vx,\vz,\omega)\, \Sigma^\star_{HF}(\vz,\vz')\, G(\vz',\vy,\omega).
\label{eq:DysonHF}
\ceq

It is convenient to use the single-particle states $\az_i^0$ with single-particle energy $e_i^0$  
associated with eigenstates of the non-interacting Hamiltonian $\oH_0$. 
Using $\opsi(x)=\sum_i\az_i^0(\vx) \oa_i(t)$ and its Hermitian conjugate, the non-interacting single-particle Green's function can be written as
\oeqn
iG^0(x,x')&=&\<\Phi_0|T\Big[\sum_{ij}\az_i^0(\vx)\,{\az_j^0}^*(\vx')\, \oa_i(t)\,  {\oad_j}(t')\Big]|\Phi_0\>\nonumber\\
&=&\<\Phi_0|T\Big[\sum_{ij}\az_i^0(\vx)\,{\az_j^0}^*(\vx')\, e^{i\oH_0t/\hb}\oa_i\,e^{-i\oH_0(t-t')/\hb}\,  {\oad_j}e^{-i\oH_0t'/\hb}\Big]|\Phi_0\>\nonumber\\
&=&\sum_{i}\az_i^0(\vx)\,{\az_i^0}^*(\vx')\, e^{-ie_i^0(t-t')/\hb}\big[\Theta(t-t')\Theta(e_i^0-\epsilon^0_F)-\Theta(t'-t)\Theta(\epsilon^0_F-e_i^0)\big]\nonumber
\ceqn
where we have used $\<\Phi_0|\oa_i\oad_j|\Phi_0\>=\delta_{ij}\Theta(e_i^0-\epsilon_F^0)$ 
and $\<\Phi_0|\oad_j\oa_i|\Phi_0\>=\delta_{ij}\Theta(\epsilon_F^0-e_i^0)$
with the Fermi energy $\epsilon_F^0$. 
This leads to the time-Fourier transform
\oeqn
G^0(\vx,\vy,\omega) &=& \sum_{i}\az_i^0(\vx)\,{\az_i^0}^*(\vy)\,\[\frac{\Theta(e_i^0-\epsilon_F^0)}{\omega-\frac{e_i^0}{\hb}+i\eta}
+\frac{\Theta(\epsilon_F^0-e_i^0)}{\omega-\frac{e_i^0}{\hb}-i\eta}\] \nonumber\\
&=&\sum_{i}\az_i^0(\vx)\,{\az_i^0}^*(\vy)\,\frac{1}{\omega-\frac{e_i^0}{\hb}+\mbox{sign}(e_i^0-\epsilon_F^0)i\eta}.
\ceqn

By analogy with $G^0(\vx,\vy,\omega)$ and using the fact that $\Sigma^\star_{HF}$ does not depend on $\omega$, 
we can search for solutions of the Dyson's equation of the form
\oeq
G(\vx,\vy,\omega)=\sum_{i}\az_i(\vx)\,{\az_i}^*(\vy)\,\frac{1}{\omega-\frac{e_i}{\hb}+\mbox{sign}(e_i-\epsilon_F)i\eta}, \label{eq:Gform}
\ceq
where $\az_i(\vx)$ is a single-particle eigenstate of the correlated system with single-particle energy $e_i$ and Fermi energy $\epsilon_F$. 
Note that, by construction, the many-body ground-state of the interacting system is approximated by a Slater determinant, 
with single-particle states fully occupied below $\epsilon_F$ and empty above. 
However, unlike the expression for $G^0$ where $\az_i^0$ and $e_i^0$ are assumed to be known, 
the single-particle states $\az_i$ and energies $e_i$ are unknown in the interacting system. 

To find an equation for $\az_i(\vx)$ and $e_i$, we first apply $\hb\omega-H_0$ to the left of Eq.~(\ref{eq:DysonHF}).
Noting that 
$$\(\hb\omega-H_0\)G^0(\vx,\vy,\omega)= \hb \sum_i \az_i^0(\vx)\az_i^0(\vy)^*=\hb\delta(\vx-\vy),$$
we get
\oeq
(\hb\omega-H_0)\,G(\vx,\vy,\omega)=\hb\delta(\vx-\vy)+  \int d^3z\,\, \hb\,\Sigma^\star_{HF}(\vx,\vz)\,G(\vz,\vy,\omega).
\ceq
Using expression (\ref{eq:Gform}) for $G$, multiplying by $\az_j(\vy)$ and integrating over $\vy$ gives 
\oeq
H_0\az_i(\vx)+\int d^3y \,\,\hb\, \Sigma^\star_{HF}(\vx,\vy)\,\az_i(\vy) = e_i \az_i(\vx). \label{eq:H0sp}
\ceq

For a system of fermions without external potential, the non-interacting Hamiltonian is just the kinetic energy, i.e., $H_0=\frac{-\hb^2\nabla^2}{2m}$. 
The self-energy in Eq.~(\ref{eq:H0sp}) acts as a mean-field potential induced by the interactions between particles. 
We then recognise a single-particle Schr\"odinger equation $h\az_i=e_i\az_i$ with the single-particle Hamiltonian $h=\frac{\vp^2}{2m}+\hb\Sigma_{HF}^\star$. 
Using the expressions~(\ref{eq:GFT}) and~(\ref{eq:Gform}) for $G$, we can show that the one-body density matrix $\rho(\vx,\vy)=-iG(\vx t,\vy t^+)$ in the HF ground-state
is expressed as $\rho(\vx,\vy)=\sum_{i=1}^A\az_i(\vx)\az_i(\vy)$ with the sum running over the occupied single-particle states with $e_i<\epsilon_F$.
We then get the HF equation in its usual form
\oeq
\[h,\rho\]=0.
\ceq

The expression for the mean-field potential can be obtained from Eq.~(\ref{eq:SigmaHF}):
\oeqn
\hb\Sigma^\star_{HF}(\vx,\vx')&=&-\delta(\vx-\vx')\int d^3x" \,\,iG(\vx"t,\vx"t^+)V(\vx-\vx")\,+\, V(\vx-\vx')\,iG(\vx t,\vx' t^+)\nonumber\\
&=&\delta(\vx-\vx')\int d^3x" \,\,\rho(\vx",\vx")V(\vx-\vx")\,- \, V(\vx-\vx')\,\rho(\vx,\vx').
\ceqn
The above equation leads to the usual expression for the mean-field
\oeqn
\hb\Sigma^\star_{HF}(\vx,\vy)&=&\int d^3z\d^3z' \,\, \rho(\vz',\vz) \, V(\vx-\vz)\big[\delta(\vx-\vy)\,\delta(\vz-\vz')\,- \,\delta(\vx-\vz') \delta(\vz-\vy)\big]\nonumber\\
&=&\int d^3z\d^3z' \,\, \rho(\vz',\vz) \, \<1:\vx,2:\vz|\oV(1,2)\big[|1:\vy,2:\vz'\>-|1:\vz',2:\vy\>\big]\nonumber\\
&=&\<\vx| \,\tr_2\[\hat{\overline{V}}(1,2)\hat{\rho}(2)\]\,|\vy\>
\ceqn
where $\tr_2$ denotes the partial trace over the degrees of freedom of the particle ``2'', and $\overline{V}$ is antisymmetrised.
We see that the mean-field is self-consistent as it depends on $\rho$. 

This derivation from the quantum many-body perturbation theory gives a natural explanation for the self-consistency of the HF mean-field. 
We see that it assumes an independent particle form of the many-body state, i.e., written as a Slater determinant of occupied single-particle wave-functions.
It is interesting to see that the approach includes diagrams to all orders in the self energy $\Sigma$.
However, the proper self energy $\Sigma^\star$, which is used to build $\Sigma$, contains only a subset of diagrams. 
The validity of this approximation is not {\it a priori} straightforward. 
In fact, like the truncation of the BBGKY hierarchy which we see next, one drawback of this approach 
is that it is based on the approximation that some terms in the exact expansion can be neglected. 
This is in contrast with variational approaches which provide an optimal description within a sub-space of states 
(e.g., approximating the state by a single Slater determinant).

\subsection{Truncation of the BBGKY hierarchy \label{sec:BBGKY}}

A one-body operator can be written as a linear combination of $\oad_i\oa_j$. Therefore, it is sufficient to 
follow the expectation values of  $\< \oad_i\oa_j \>$, which are nothing but the matrix elements of the one-body density matrix $\ro$.
Using Erhenfest's theorem in Eq.~(\ref{eq:Ehrenfest}), we get
\oeq
\dt \sdf \<\oad_i\oa_j\>_\psi = \dt \sdf \ro_{ji} 
= \frac{i}{\hb} \< \, [\oH,\oad_i\, \oa_j ] \,\>_\psi.
\label{eq:erhenfest}
\ceq 
Using the expression~(\ref{eq:oH}) for the Hamiltonian, we can rewrite the above equation as
\oeq
i\hb\frac{\partial}{\partial t}\rho(t) = \[h[\rho],\rho\] + \tr_2 \[v^{(res)}_{(1,2)}, C_{(1,2)}\], 
\label{eq:1body}
\ceq
where $C_{12}$ is the two-body correlation matrix with elements
\oeq
{C_{{ijkl}}} =  \< \oad_l \, \oad_k \, \oa_i\, \oa_j\>_\psi  - {\ro_{{jl}}} \sdf {\ro_{{ik}}} + {\ro_{{il}}} \sdf {\ro_{{jk}}} \;.
\ceq
The partial trace is taken over the degrees of freedom of the particle ``2'',
\oeq
\tr_2 \{v_{(1,2)} C_{(1,2)} \}_{\al\be} = \sum_{\gamma\delta\epsilon} v_{\al\delta\gamma\epsilon}C_{\gamma\epsilon\be\delta}.
\ceq
The Hartree-Fock single-particle Hamiltonian $h[\rho]$ is a matrix with elements
\oeq
h_{\al\be}[\rho] = \frac{\delta \<\Psi|\oH|\Psi\>}{\delta\rho_{\be\al}}.
\ceq
In the energy density functional (EDF) formalism\footnote{As shown in section~\ref{sec:equivalence}, the EDF and Hamiltonian approaches are equivalent, in the sense that there is no need to assume an explicit expression of the Hamiltonian [as in Eq.~(\ref{eq:oH})] to get the TDHF equation.}, the above equation is replaced by
\oeq
h_{\al\be}[\rho] = \frac{\delta E[\rho]}{\delta\rho_{\be\al}},
\ceq
where the functional $E[\rho]$ is directly fitted for a set of observables (nuclear matter equation of state, masses, radii, spin-orbit splitting, etc.).

Equation (\ref{eq:1body}) contains two unknown quantities, namely the one-body density matrix $\rho$ and the two-body correlation matrix $C$. 
According to the Bogoliubov-Born-Green-Kirkwood-Yvon (BBGKY) hierarchy \cite{bogoliubov1946,born1946,kirkwood1946}, the evolution of $C$ depends on the three-body correlation matrix and so on. 
The mean-field approximation is  obtained by truncating the BBGKY hierarchy to the lowest order, i.e., 
assuming $C=0$ at all time. We end up with the time-dependent Hartree-Fock (TDHF) equation proposed by Dirac \cite{dirac1930}
\oeq
i\hb\frac{\partial}{\partial t}\rho(t) = \[h[\rho],\rho\].
\label{eq:tdhf1}
\ceq
As the two-body correlations have been neglected, the state of the system reduces to an independent particle-state at all times. 
Thus, the TDHF equation describes the evolution of independent particles in the mean-field generated by all the particles. 

\subsection{Beyond TDHF correlations}

The above approach, where the TDHF equation is obtained by truncating the BBGKY hierarchy at lowest order, 
tells us that the TDHF approach neglects two-body correlations at all time. 
For the mean-field approximation to be valid, it is therefore necessary for the two-body correlations to remain a small perturbation during the dynamics. 
These correlations include, e.g., nucleon-nucleon collisions, pairing and quantum fluctuations of collective observables:
\begin{itemize}
\item {\it Nucleon-nucleon collisions} are expected to be small at low energy due to the Pauli exclusion principle.
Indeed, due to conservation of energy when two nucleons collide, one nucleon has an increased energy while the other one would go to a lower energy state which is likely to be already occupied. 
As a result of this Pauli blocking, the collision is forbidden and the mean-free path of the nucleon is increased to the order of the size of the nucleus.
The above argument is in fact simplistic as it assumes that the nucleons have a well defined energy.
Unlike in the static HF case, where $h$ and $\rho$ commute and thus have the same eigenstates, 
in TDHF an occupied single-particle state (eigenstate of $\rho$) does not have a well defined single-particle energy.
As a result, collisions (neglected in TDHF) could happen even at low energy. 
Ultimately, the theoretical approach has to be validated by detailed comparison with experimental data.
\item {\it Pairing correlations}, however,
which are known to play an important role in structure, could affect some reaction mechanisms such as multi-nucleon transfer. 
The inclusion of pairing dynamics at the BCS or  Bogoliubov levels has been the subject of recent works \cite{avez2008,ebata2010,scamps2013a,stetcu2011} which will be discussed in section~\ref{sec:pairing}.
\item {\it Quantum fluctuations} of observables are sometimes poorly described in TDHF. 
This is the case, for instance, with the fluctuation of the particle number in deep-inelastic collisions \cite{dasso1979}. 
Beyond TDHF methods which rely on existing codes have thus been developed, 
based on the time-dependent RPA \cite{balian1984} approach or on the stochastic mean-field method \cite{ayik2008}.
Recent applications will be discussed in section~\ref{sec:beyond}.
\end{itemize}

\subsection{Variational principle with the Dirac action \label{sec:Dirac}}

The Dirac action is defined as
\oeq
S \equiv S_{t_0,t_1}[\Psi] = \int_{t_0}^{t_1} \stb d t \stf \<\Psi (t)| \( i\hb \frac{d }{{d} t} - \oH \) |\Psi(t)\>.
\ceq
If the many-body state $|\Psi(t)\>$ is free to explore the entire Hilbert space, 
then requesting stationarity of the action leads to the exact Schr\"odinger equation. 

If, however, the variational space is restricted to independent particle states $|\phi\>$ corresponding to Slater determinants of the occupied single-particle states $|\az_i\>$, then the action can be expressed as 
\oeqn
S = \int_{t_0}^{t_1} \stb  d t \stf \(i\hb \sum_{i=1}^N \sdf \int \stb  d x \stf \az_i^*(x,t) \sdf \frac{ d}{ d t} {\az}_i(x,t)
-E[\ro(t)]\) \label{eq:dirac_action} 
\ceqn
where $E[\ro] = \bfi \oH \kfi$.
For simplicity, the equations are written in one-dimension and spin is omitted. 
The stationarity of the action imposes
\oeq
\frac{\delta S}{\delta  \az^*_\al(x,t)} = i\hb \sdf \frac{ d}{ d t} \az_\al(x,t) - 
 \int_{t_0}^{t_1} \stb  d t' \stf \frac{\delta E\[\ro(t')\]}{\delta  \az^*_\al(x,t)}=0.
\ceq
The variation of $E$ can be expressed as
\oeq 
 \frac{\delta E\[\ro(t')\]}{\delta  \az^*_\al(x,t)}
 =  \int \stb  d y \,  d y' \stf \frac{\delta E\[\ro(t')\]}{\delta  \ro(y,y';t')}
\sdf  \frac{\delta  \ro(y,y';t')}{\delta  \az^*_\al(x,t)} 
=  \int \stb  d y \stf h(x,y;t)
\sdf   \az_\al(y,t')  \sdf \delta(t-t'),
\ceq
where we have defined the single-particle Hartree-Fock Hamiltonian $h$ as
\oeq
h(x,y;t) = \frac{\delta E\[\ro(t)\]}{\del \ro(y,x;t)}.
\label{eq:hEHF}
\ceq
The TDHF equation for the occupied states finally read
\oeq
 i\hb \sdf \frac{ d }{ d t} \az_\al(x,t) = \int \stb  d y \stf h(x,y;t) \sdf \az_\al(y,t)\label{eq:TDHF_az}\;,
\ceq
or, equivalently,
\oeq
i\hb \frac{\partial}{\partial t} \az_i(t) = h[\rho(t)] \az_i(t) \stf\stf \mbox{ for } 1\le i\le A.
\label{eq:tdhf2}
\ceq
This set of non-linear Schr\"odinger like equations is fully equivalent to the TDHF equation~(\ref{eq:tdhf1}).

In practice, TDHF codes usually solve the mean-field dynamics in the canonical basis using Eq.~(\ref{eq:tdhf2}) \cite{kim1997,umar2005a,nakatsukasa2005,maruhn2014}.
They provide a time-evolution of each single-particle wave-function $\az_i(t)$ 
which can be used to compute expectation values of one body observables, 
as well as single-particle properties such as the transfer from specific single-particle states \cite{umar2008a}. 

We see from the above derivation that a restriction to independent particle states at all times, 
which is equivalent to neglecting the effect of  two-body correlations on the dynamics, leads to the TDHF formalism.
Interestingly, neither the BBGKY truncation nor the Dirac variational principle impose any restriction on the type of observables which could in principle be computed within the TDHF framework.  

\subsection{Balian-V\'en\'eroni variational principle \label{sec:BV}}

The Balian-V\'en\'eroni action differs from the Dirac one as it considers both the state of the system, represented by its density matrix operator $\oD(t)$ (in the Schr\"odinger picture),  and the observable $\oA(t)$ (in the Heisenberg picture) to be variational quantities.
Their action is written as \cite{balian1981}
\oeq
J = \Tr\[ \oA(t_1)\oD(t_1) \]
- \int_{t_0}^{t_1} \stb dt \,\Tr\!\[ \oA(t) \(\frac{d\oD(t)}{dt} + i[\oH(t),\oD(t)]\) \],
\label{eq:JA}
\ceq
where $\hb=1$. The boundary conditions are given by 
\oeq
\oD(t_0)=\oD_0,
\label{eq:BCD}
\ceq
where  $\oD_0$ is the initial state,  and
\oeq
\oA(t_1) = \oA_1,
\label{eq:BCA}
\ceq
where $\oA_1$ is the operator at the final time. 
We see that, unlike the previous methods, the choice of the observable to compute plays a crucial role in this approach.

The TDHF equation is obtained by constraining the variation  $\delta_A$ to leave $\oA$ in the space of one-body operators,
which is achieved with
\oeq
\delta_A\oA(t)\equiv\oad_\al\oa_\be
\ceq
and $\delta\oA(t_1)=0$ due to the boundary condition in Eq.~(\ref{eq:BCA}).
Imposing the stationarity of the action with respect to variations of $A$, i.e., $\delta_AJ=0$, we get 
\oeq
\Tr\[\oad_\al\oa_\be\(\frac{\partial\oD}{\partial t}+i[\oH,\oD]\)\]=0.
\ceq
As before, we also restrict the variational space for $\oD$ to independent particle states, i.e., $\oD(t)=|\phi(t)\>\,\<\phi(t)|$ where $|\phi(t)\>$ is a Slater determinant. 

Using $\ro_{\al\be}=\bfi \oad_\be\oa_\al \kfi$, we get
\oeq
i\frac{\partial\ro_{\be\al}(t)}{\partial t}=\<\phi(t)|\[\oad_\al\oa_\be,\oH\]|\phi(t)\>.
\label{eq:idrodt}
\ceq
This is Eq.~(\ref{eq:erhenfest}) with the additional restriction that $|\phi\>$ is a Slater determinant. 
As a result, the two-body correlation matrix $C$ is zero and we get the TDHF equation~(\ref{eq:tdhf1}).

The Balian-V\'en\'eroni variational principle tells us that, in addition to being an approach to independent particles, 
the TDHF theory is optimized to expectation values of one-body observables. 
In principle, the TDHF approach is not adequate to compute two-body operators or fluctuations of one-body operators.   
For the latter, the TDRPA equation has been derived from the same variational principle with a larger variational space for the observable \cite{balian1984} (see section~\ref{sec:TDRPA}). 

\subsection{Path integral formulation}

In his many-path formulation of quantum mechanics~\cite{feynman1948}, 
Feynman showed that the amplitude of probability for a particle to go from a position $q_i$ at time $t_i$ 
to a position $q_f$ at time $t_f$ can be expressed as
$$
\<q_f|\oU(t_f,t_i)|q_i\> = \int\mD[q]\, e^{iS[q]/\hb},
$$
where $S[q]=\int_{t_i}^{t_f} dt \, L(q,\dot{q})$ is the action 
and $L=\frac{m}{2}\dot{q}^2-V(q)$ is the Lagrangian of the particle. 
The stationary phase approximation (SPA), valid for classical systems where $S\gg\hb$, 
leads to $ \int\mD[q]\, e^{iS[q]/\hb} \sim e^{iS[q_c]/\hb}$ with the classical path obeying the variational principle $\delta S[q]|_{q=q_c}=0$. 

Starting from the Hamiltonian $\oH=\sum_{i=1}^A \oK_i+\frac{1}{2}\sum_{i,j=1}^A V(\ox_i-\ox_j)$, Levit showed that
the amplitude of probability to go from the many-body state $|i\>$ to $|f\>$ could be written as a similar path integral \cite{levit1980a}, 
$$
\<f|\oU(t_f,t_i)|i\>= \int\mD[\sigma]\, e^{iS_{eff}[\sigma]},
$$
where $\sigma(x,t)$ is the scalar auxiliary field arising from the use of Hubbard-Stratonovich transformation and the effective action is given by ($\hb=1$)
$$
S_{eff}[\sigma]= \frac{1}{2}\int_{t_i}^{t_f}dt \, \int dx\,dx'\, \sigma(x,t)V(x-x')\sigma(x',t') - i \ln \<f|\hat{U}_\sigma|i\>.
$$
Here, $\hat{U}_\sigma(t_f-t_i)$ is the evolution operator in the interaction representation with Hamiltonian
$$
\oH_\sigma = \sum_{i=1}^A \[\oK_i-\frac{V(0)}{2}+\int dx \, V(\ox_i-x) \sigma(x,t)\].
$$
Imposing $|i\>$ and $|f\>$ to be Slater determinants and using the SPA, i.e.,  requiring $\delta S_{eff}[\sigma]=0$,  lead to the mean-field equation
$$
i\frac{\partial\varphi_i(x)}{\partial t} = \( \frac{p^2}{2m} -\frac{V(0)}{2} + \int dx' \,\rho(x',t) V(x-x')\) \, \varphi_i(x).
$$
This is the time-dependent Hartree equation. 
Going to the next order beyond the SPA leads to the Fock term, gets rid of the self-interaction energy term $\frac{V(0)}{2}$, and also contains some RPA corrections \cite{levit1980b}. 

We see that the SPA applied to the path integral formulation leads to classical physics for one-particle system 
while it produces a mean-field approach for many-particle systems. 
We then understand the classical behavior of the TDHF equation in terms of collective coordinates as well as its lack of quantum fluctuations. 
In particular, this is a possible explanation for the inability of TDHF to incorporate quantum many-body tunneling of strongly interacting systems like nuclei. 

\subsection{Equivalence between Hamiltonian and energy density functional approaches \label{sec:equivalence}}

We conclude the formalism section by showing the equivalence between Hamiltonian and energy density functional approaches.
It is indeed interesting to note that the derivation of the TDHF equation from the variational principle with the Dirac action 
(see section~\ref{sec:Dirac}) does not require an explicit expression for the Hamiltonian $\oH$. 
This is because the Dirac action can be directly defined with the energy density functional $E[\rho]=\<\phi|\oH|\phi\>$ in it [see Eq.~(\ref{eq:dirac_action})]. 
On the contrary, other derivations of the TDHF equation from the truncation of the BBGKY hierarchy (section~\ref{sec:BBGKY}) 
or from the Balian-V\'en\'eroni variational principle (section~\ref{sec:BV}) require an explicit expression of the Hamiltonian, as in Eq.~(\ref{eq:oH}), in order to compute the quantity $\<[\oad_i\oa_j,\oH]\>$ appearing in Eqs.~(\ref{eq:erhenfest}) and~(\ref{eq:idrodt}). 
Here, we show that such an explicit expression of $\oH$ is, in fact, not necessary. 

Our goal is to show that 
\oeq\<\phi|\[\oad_\beta\oa_\al\,,\,\oH\]|\phi\>=\[h,\rho\]_{\al\beta}\label{eq:goal}\ceq
for any Hamiltonian $\oH$ and Slater $|\phi\>$, with 
\oeq h_{\al\be}=\frac{\delta \<\phi|\oH|\phi\>}{\delta \rho_{\be\al}}=\frac{\delta E[\rho]}{\delta \rho_{\be\al}}.\ceq
Indeed, using Eq.~(\ref{eq:goal}) in Eqs.~(\ref{eq:erhenfest}) or~(\ref{eq:idrodt}) would give the TDHF equation~(\ref{eq:tdhf1}).

We can rewrite the left hand side of Eq.~(\ref{eq:goal}) using the basis $\{|\nu\>\}$ of eigenstates of $\oH$ as 
\oeqn
\<\phi|\[\oad_\beta\oa_\al\,,\,\oH\]|\phi\>&=&\<\phi|\[\oad_\beta\oa_\al\,,\,\sum_\nu|\nu\>\<\nu|\oH\]|\phi\> \nonumber\\
&=&\sum_\nu E_\nu\<\phi|\[\oad_\be\oa_\al\,,\,|\nu\>\<\nu|\]|\phi\>,
\ceqn
where $E_\nu$ is the eigenenergy of the state $|\nu\>$. 
Let us write  $|\nu\>$ in the $n-$particle $n-$hole basis built from the Slater $|\phi\>$ as 
\oeq
|\nu\> = C^\nu_0|\phi\> + \sum_{ph} C^\nu_{ph} \oad_p\oa_h|\phi\>+ \sum_{pp'hh'} C^\nu_{pp'hh'}\oad_p\oad_{p'}\oa_h\oa_{h'}+\cdots,
\label{eq:nu}
\ceq
with $C^\nu_{ph}=\<\phi|\oad_h\oa_p|\nu\>$. (Note that $C^\nu_{hp}=0$.) This leads to 
\oeq
\<\phi|\[\oad_\be\oa_\al,\oH\]|\phi\> = \sum_\nu E_\nu \(C^\nu_{\al\be} {C_0^\nu}^*-C_0^\nu {C^\nu_{\be\al}}^*\).
\label{eq:commoad}
\ceq

We now look for an expression for the term  $[h,\rho]$ as a function of the same quantities. 
Using $\rho_{\al\be}=\<\phi|\oad_\be\oa_\al|\phi\>\equiv \<\oad_\be\oa_\al\>$, we can write 
\oeqn
\[h,\rho\]_{\al\be} &=& \sum_\gamma \(\frac{\delta E[\rho]}{\delta\rho_{\gamma\alpha}}\rho_{\gamma\be} - \rho_{\al\gamma} \frac{\delta E[\rho]}{\delta\rho_{\beta\gamma}}\) \nonumber \\
&=& \sum_\gamma \(\frac{\delta \<\oH\>}{\delta\<\oad_\al\oa_\gamma\>}\<\oad_\be\oa_\gamma\> - \<\oad_\gamma\oa_\al\>\frac{\delta \<\oH\>}{\delta\<\oad_\gamma\oa_\beta\>}\) \label{eq:comm}.
\ceqn
Notice that, in the canonical basis, $\gamma$ has to be a hole state.  
Using, again, the eigenbasis $\{|\nu\>\}$, we have $\<\oH\>=\sum_\nu E_\nu \<\phi|\nu\>\<\nu|\phi\>$. 
Using this and Eq.~(\ref{eq:nu}), we see that the functional derivatives in Eq.~(\ref{eq:comm}) include terms like
\oeqn
\frac{\delta ( \<\phi|\nu\>\<\nu|\phi\>)}{\delta\<\oad_\al\oa_\gamma\>} &=& \sum_{ph} C^\nu_{ph} \frac{\delta  \<\oad_p\oa_h\>}{\delta\<\oad_\al\oa_\gamma\>} \<\nu|\phi\> + \sum_{ph} {C^\nu_{ph}}^* \<\phi|\nu\> \frac{\delta  \<\oad_h\oa_p\>}{\delta\<\oad_\al\oa_\gamma\>} \nonumber\\
&=& C^\nu_{\al\gamma}\<\nu|\phi\> + {C^\nu_{\gamma\al}}^* \<\phi|\nu\>, 
\ceqn
where $C_{\al{p}}=C_{h\al}=0$, i.e., only $C_{ph}$ terms are non zero. 
Similarly, for a Slater determinant we have $\<\oad_\al\oa_p\>=\<\oad_p\oa_\al\>=0$, i.e., the only non-zero terms are  $\rho_{hh'}=\delta_{hh'}$. 
Consequently, terms like $C^\nu_{\gamma\al}\rho_{\gamma\be}$ and $C^\nu_{\gamma\al}\rho_{\be\gamma}$ 
are equal to  zero because $\gamma$ cannot be a particle and a hole at the same time. 
Equation~(\ref{eq:comm}) then becomes
\oeqn
\[h,\rho\]_{\al\be}&=&\sum_\nu E_\nu \sum_\gamma \(C_{\al{\gamma}}^\nu \<\nu|\phi\>\rho_{\gamma\be}-\rho_{\al{\gamma}} {C_{\be{\gamma}}^\nu}^*\<\phi|\nu\>\)\nonumber\\
&=& \sum_\nu E_\nu \(C_{\al{\be}}^\nu {C_0^\nu}^*- {C_{\be{\al}}^\nu}^*C_0^\nu\),
\ceqn
which, according to Eq.~(\ref{eq:commoad}), is equal to $\<\phi|\[\oad_\be\oa_\al,\oH\]|\phi\>$. 
Equation~(\ref{eq:goal}) is then verified, implying that  derivations of the TDHF equation from Hamiltonian approaches such as the truncation of the BBGKY hierarchy or the Balian-V\'en\'eroni variational principle do not depend on the specific expression of the Hamiltonian.

\section{New Techniques and their applications}

A number of new techniques have been recently developed to utilize HF and TDHF calculations
to extract information that could be directly compared with experimental observations. These include the
microscopic calculation of fusion barriers and cross-sections, methods to use TDHF to study scission dynamics
with and without pairing, projection methods to compute fluctuation of various observables, beyond mean-field
methods based on TDHF, improvements in effective interactions and boundary conditions. In this section, we examine
the theoretical underpinnings of these new developments.

\subsection{Nucleus-nucleus potentials and fusion reactions}
In the absence of a practical many-body approach to study sub-barrier fusion, the problem is transformed to that of
a nucleus-nucleus interaction barrier or a series of barriers to incorporate other degrees of freedom such as static
deformations. Historically, 
phenomenological ion-ion potentials such as the Bass potential~\cite{bass1974,bass1980},
the proximity potential~\cite{blocki1977,randrup1978a,seiwert1984,birkelund1983}, or potentials obtained via
the double-folding method~\cite{satchler1979,bertsch1977} have been some of the popular choices.
Some of these potentials have been fitted to experimental fusion barrier heights and have been quite successful in describing scattering data
with stable nuclei.

The reduction of the many-body dynamics to a two-body problem is  a 
serious approximation. It is needed to 
describe quantities which are outside the range of applicability of TDHF, such as 
sub-barrier fusion, or the coherence between reaction channels associated with 
different barriers (where TDHF reduces the problem to an average barrier).
Microscopic approaches can then be used to provide inputs to such two-body calculations,
in particular the nucleus-nucleus potential, the coordinate dependent mass 
(which differs from the asymptotic reduced mass when the densities overlap), the moment
of inertia and the friction parameter. In principle, one needs all of these
to compute fusion cross-sections. The potential will be discussed
at length. By construction, 
potentials extracted from TDHF dynamics are real as they incorporate (at the mean-field 
level) all reaction channels. 
The coordinate dependent mass is discussed in section~\ref{sec:cross-sections} where 
the method to calculate cross-sections are introduced. In principle, 
this is all we need to compute fusion cross-sections if one makes
the isocentrifugal approximation, i.e., neglecting non-central effects, 
and if dissipation is small. The isocentrifugal approximation is very common 
(e.g., in most coupled-channel calculations) and will be discussed in section~\ref{sec:cross-sections}. 
If dissipation is neglected (see section~\ref{sec:dissipation}) the reduction of the dynamics
to a two-body problem leads to a Schr\"odinger equation, as in the coupled-channel framework.

Traditionally, the nucleus-nucleus potential is used to calculate fusion cross-sections with a quantum mechanical barrier 
penetration model~\cite{bass1980,takigawa1984,balantekin1998}, or the coupled-channels approach~\cite{landowne1984,hagino1999,esbensen2004},
for the transmission coefficients.
The coupled-channels method includes various excitations of the
target and/or projectile, as well as
a simplified description of neutron transfer (see, e.g.,~\cite{karpov2015}), and can be consistently applied at energies above
and below the barrier~\cite{balantekin1998,hagino1999} (see~\cite{hagino2012} for a review).

One possible way to construct a nucleus-nucleus potential from microscopic densities is via the double-folding model. 
In this approach, static densities of the two 
nuclei, separated by a relative coordinate $\mathbf{R}$, are folded with a phenomenological effective nucleon-nucleon interaction
to compute the interaction barrier:
\begin{equation}
V_F(R)=\int {\mbox d}\mathbf{r}_1 {\mbox d}\mathbf{r}_2 \;\; \rho_1(\mathbf{r}_1)\rho_2(\mathbf{r}_2) v(\mathbf{r}_{12})\, ,
\label{vf}
\end{equation}
where $R$ is the norm of $\mathbf{R}$ and $\mathbf{r}_{12}=\mathbf{R}-(\mathbf{r}_1- \mathbf{r}_2)$.
The double-folding potential allows the use of microscopic approaches, such as the Hartree-Fock-Bogoliubov (HFB) or the
relativistic mean-field model (RMF), to compute the nuclear densities in Eq.~(\ref{vf}).
Double-folding potentials have also been used in conjunction with the S\~{a}o Paulo barrier penetration model
to simulate the energy dependence of barriers~\cite{gasques2004,chamon2002} for astrophysical fusion cross-sections.
Note that the interaction used in the structure part and the one used in the calculation of the potential often differ, 
introducing a lack of consistency in the method. 
In this section, we discuss microscopic approaches to compute nucleus-nucleus potentials which overcome this limitation
as both  potentials and densities are computed from the same Skyrme EDF.

One common physical assumption used in many of these
calculations is the frozen density (or ``sudden'') 
approximation. In this approximation the nuclear densities
are unchanged during the computation of the ion-ion potential as a function
of the internuclear distance. 
This leads to a {\it bare} nucleus-nucleus potential. 
Methods to compute bare potentials from HF densities will be discussed in sections~\ref{sec:FHF} and~\ref{sec:DCFHF}.
Of course, the possibility for the nuclei to change their shape dynamically must be accounted for one way or another.
Indeed, in principle the bare potential provides a realistic description of the interaction between the nuclei at high energy only
(where the time scale of the reaction is too short for the nuclei to polarise), 
and at distances large enough so that the overlap between the densities of the fragments is not unrealistic 
(typically with a density in the neck between the fragments that does not exceed the saturation density). 

Dynamical change of shapes of the nuclei is accounted for in coupled-channels calculations via the couplings to low-lying collective states. 
This means that  coupled-channels calculations require
properties of low-lying collective states in the projectile and target nuclei
such as excitation energies and reduced transition probabilities B(EL)  which are usually taken
from experiment, but not always available (in particular for exotic nuclei). 
RPA strength functions (e.g., computed from TDHF response in the small amplitude limit) can also be used to provide these inputs to 
coupled-channels calculations, as illustrated in section~\ref{sec:couplings}.
Although these couplings may accurately represent the early stages of the collision
process, the situation may become more complicated when the nuclei overlap 
due to a modification of the collective excitations~\cite{ichikawa2015b}.
An alternative approach to coupled-channels method is to incorporate dynamical effects 
directly in the potential by calculating the latter from TDHF densities (see section~\ref{sec:DCTDHF}).

Issues with the two-body approach to heavy-ion collisions become much more prevalent when studying fusion at deep sub-barrier energies.
In particular, the large overlap of the frozen nuclear densities lead to an unphysical
attraction in the inner barrier region.
To remedy this drawback, various modifications of the nucleus-nucleus potential have been suggested. These include
the addition of a repulsive core resulting in a \textit{shallow potential}
pocket~\cite{misicu2006} or a two-step model for fusion in which the effects of
neck formation are approximately included~\cite{ichikawa2007}.
Microscopic approaches discussed here provide a physical insight into the mechanisms at play when the nuclei overlap.
For instance, they can be used to account for the repulsion in the bare potential 
induced by the Pauli exclusion principle between nucleons of different collision partners 
(see section~\ref{sec:DCFHF}). 
This Pauli repulsion is of course naturally included in TDHF which describes the dynamics of a fully antisymmetrised many-body state. 

In recent years, a number of new developments have been introduced to employ HF or TDHF calculations
to compute nucleus-nucleus potentials and fusion cross-sections in a more microscopic and self-consistent manner. In this section,
we review these new methods and present some selected applications.

\subsubsection{Frozen Hartree-Fock Approach \label{sec:FHF}}

\begin{figure}[!tb]
\centering
\includegraphics[width=7cm]{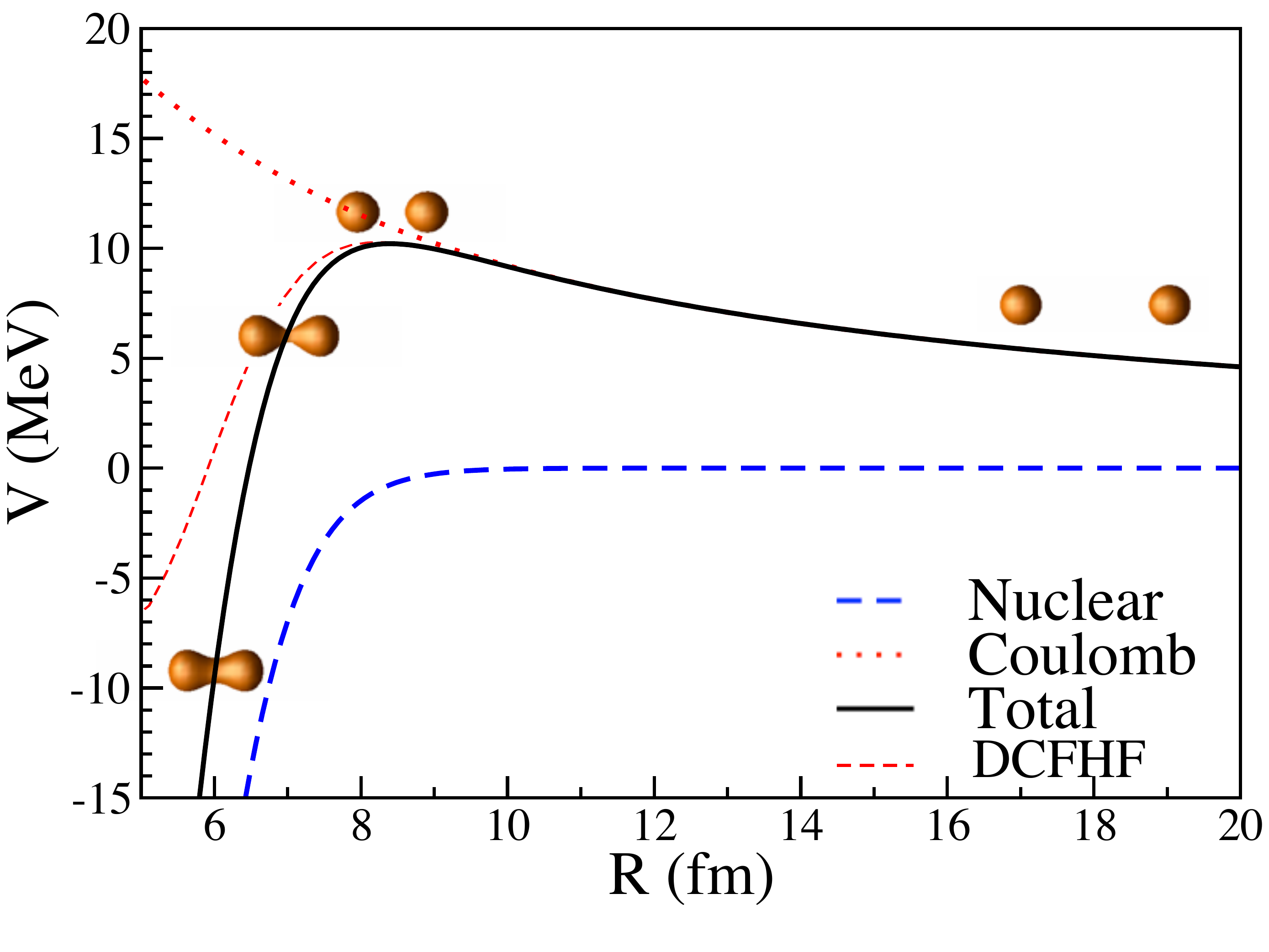}
\caption{Nuclear (long-dashed line), Coulomb (dotted line) and total (solid line) contributions to the Frozen Hartree-Fock (FHF) potential calculated for the $^{16}$O+$^{16}$O system with the SLy4d parametrization of the Skyrme EDF. The density-constrained FHF (DCFHF) potential including Pauli repulsion is shown with a short dashed line. 
Isodensities at half the saturation density $\rho_0/2=0.08$~fm$^{-3}$ from a TDHF calculation at centre of mass energy  $E_{cm}=12$~MeV are also shown.} 
\label{fig:Frozen_O+O}   
\end{figure}

A procedure for computing nucleus-nucleus potentials microscopically using frozen static HF densities
 has been developed.
Construction of such microscopic bare potentials could provide
input to coupled-channels calculations since frozen densities are
used and no excitations have been accounted for.
The approach is based on energy density functional (EDF) $E[\rho]$
written as an integral of an energy density $\mathcal{H}[\rho(\mathbf{r})]$ \cite{weizsacker1935,bethe1936}, i.e.,
\begin{equation}
E[\rho]=\int{\mbox d} \mathbf{r} \,\, \mathcal{H}[\rho(\mathbf{r})]\,.
\end{equation}
One can adopt the idea of Brueckner \textit{et al.}\,\cite{brueckner1968}
to derive the bare potential from an EDF.
The bare potential is obtained by requiring frozen ground-state densities $\rho_{i}$
of each nucleus ($i=1,2$), which we compute
using the HF mean-field approximation. %CS\,\cite{hartree1928,fock1930}.

Typically, the Skyrme EDF\,\cite{skyrme1956} is used both in HF calculations and to compute the bare potential.
This ensures that all of the structure information contained in the EDF is part of the bare potential, such
as the effects of neutron skin for neutron-rich nuclei~\cite{reinhard2016a}.
Neglecting the Pauli exclusion principle between nucleons in different nuclei
leads to the usual frozen Hartree-Fock (FHF)
potential\,\cite{denisov2002,washiyama2008,simenel2008,simenel2012}
\begin{equation}
V_{FHF}(\hat{\bf R})=\int {\mbox d}\mathbf{r} \;\; \mathcal{H}[\rho_1(\mathbf{r})+\rho_2(\mathbf{r}-\hat{\bf R})] - E[\rho_1] -E[\rho_2]\;,
\label{eq:frozen}
\end{equation}
where $\hat{\bf R}$ is the distance vector between the centres of mass of the nuclei.

The first term in the r.h.s. of  Eq.~(\ref{eq:frozen}) is calculated by placing the two static HF solutions
at a distance $R$ from each other in a TDHF code without a boost and computing the
energy of the combined system, including the Coulomb contribution.
An example of FHF potential for the $^{16}$O$+^{16}$O reaction is shown in Fig.~\ref{fig:Frozen_O+O}. 

\begin{figure}[!tb]
\centering
\includegraphics[width=7cm]{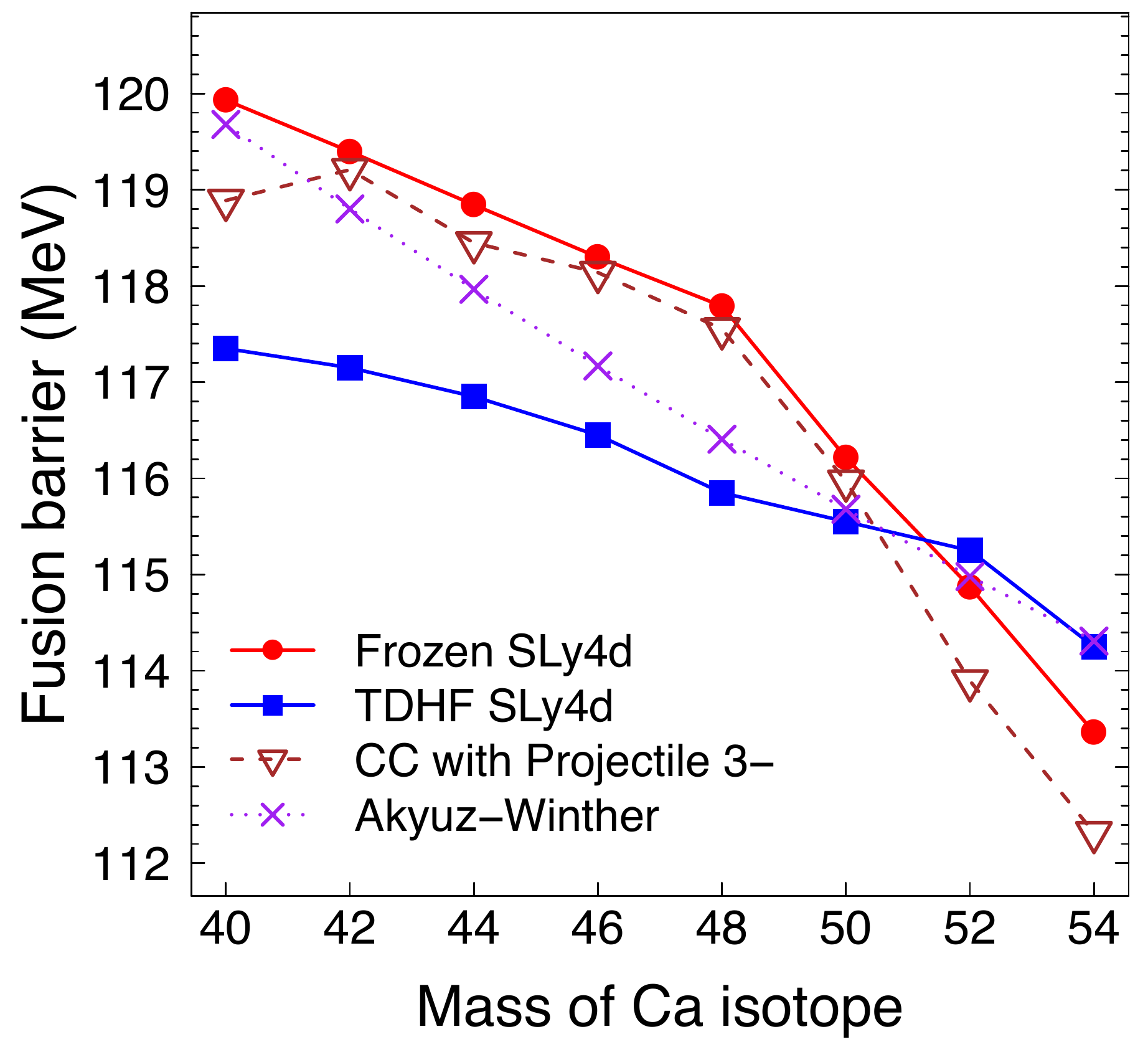}
\caption{Fusion barriers in $^A$Ca$+^{116}$Sn obtained from the Aky\"uz-Winter potential~\cite{akyuz1981} (crosses), FHF (circles), coupled-channel calculations (triangles) and  TDHF (squares). } 
\label{fig:CaSn}   
\end{figure}

Figure \ref{fig:CaSn} shows the change of the  bare potential barrier using FHF (circles) 
with the mass of calcium projectiles incident on a $^{116}$Sn target~\cite{vophuoc2016}.
A clear change of slope is observed near $^{48}$Ca which is interpreted as an effect of the neutron skin in neutron rich calcium isotopes which increases the barrier radius, and then reduces the Coulomb repulsion as well as the barrier height. 
This effect is not observed in the phenomenological Aky\"uz-Winter potential (crosses)~\cite{akyuz1981}. 
This shows that microscopic potentials like FHF are more promising for reactions with exotic beams. 

\subsubsection{Couplings to collective vibrations \label{sec:couplings}}

Computing fusion cross-sections via a one-barrier penetration model with bare FHF potentials
 usually underestimates fusion cross-sections at near and below barrier energies.
 This is because couplings to low-lying states, which are known to enhance fusion at these energies\footnote{
 Couplings between relative motion and internal excitations leads to 
off-diagonal terms in the Hamiltonian. After diagonalisation, the new
eigenchannels correspond to mixture of ground and excited states. 
The new eigenchannel with lowest energy is pushed down in energy with 
respect to the uncoupled ground-state channel. See
 Ref~\cite{hagino2012} for a review.}, are not accounted for. 
 FHF potentials have then been used in coupled-channels calculations in order to take into account these couplings in the calculations of cross-sections~\cite{simenel2013b,vophuoc2016,bourgin2016}. 
 
%The FHF potential, assumed to be central, can then directly be used to compute
%fusion cross-sections\,\cite{simenel2013b,bourgin2016,vophuoc2016}.

%CS Furthermore, with the development of new RIB facilities fusion studies of neutron-rich nuclei
%will be possible. For these nuclei spectroscopic information may
%not be available as input to the coupled-channel codes.

In addition to the nucleus-nucleus potential, coupled-channels calculations require the energy and transition probability 
of the collective states. 
Such information on low-lying rotational and vibrational states may not be readily available for exotic nuclei.
Consequently, it is also essential to develop methods to calculate collective
excitation energies and transition probabilities  as input to these codes.
For vibrational states, these properties can be computed using the TDHF approach in
the small amplitude response limit~\cite{simenel2003,stevenson2004,nakatsukasa2005,umar2005a,maruhn2005,stevenson2007,avez2008,simenel2009,stevenson2010,fracasso2012,scamps2013b,simenel2013a,simenel2013b,scamps2014a,vophuoc2016,bourgin2016} which is equivalent to the random-phase approximation (RPA). 

This makes use of the linear response theory where the evolution of the wave-function after a boost of the form
$|\Psi(t=0)\> = e^{-i\varepsilon \oQ } |\Psi_0\>,
$ applied on the ground state $|\Psi_0\>$,
leads to
\oeq
\Delta Q(t)=\<\Psi(t)|\oQ|\Psi(t)\>-\<\Psi_0|\oQ|\Psi_0\> = -2\varepsilon\sum_\nu |q_\nu|^2\sin \omega_\nu t +O(\varepsilon^2),
\label{eq:linresp}
\ceq
where  $q_\nu=  \<\Psi_\nu| \oQ |\Psi_0\>$ is the transition amplitude.
The strength function is then defined as
\oeq
R_{Q}(\omega) =\lim_{\varepsilon\rightarrow 0} \frac{-1 }{\pi \varepsilon}\,
\int_{0}^{\infty}  dt\, \Delta{Q}(t) \, \sin (\omega t) = \sum_\nu \, |q_\nu |^2 
 \delta (\omega - \omega_\nu). \label{eq:strengthfinal}
\ceq

\begin{figure}[!tb]
\centering
\includegraphics[width=7cm]{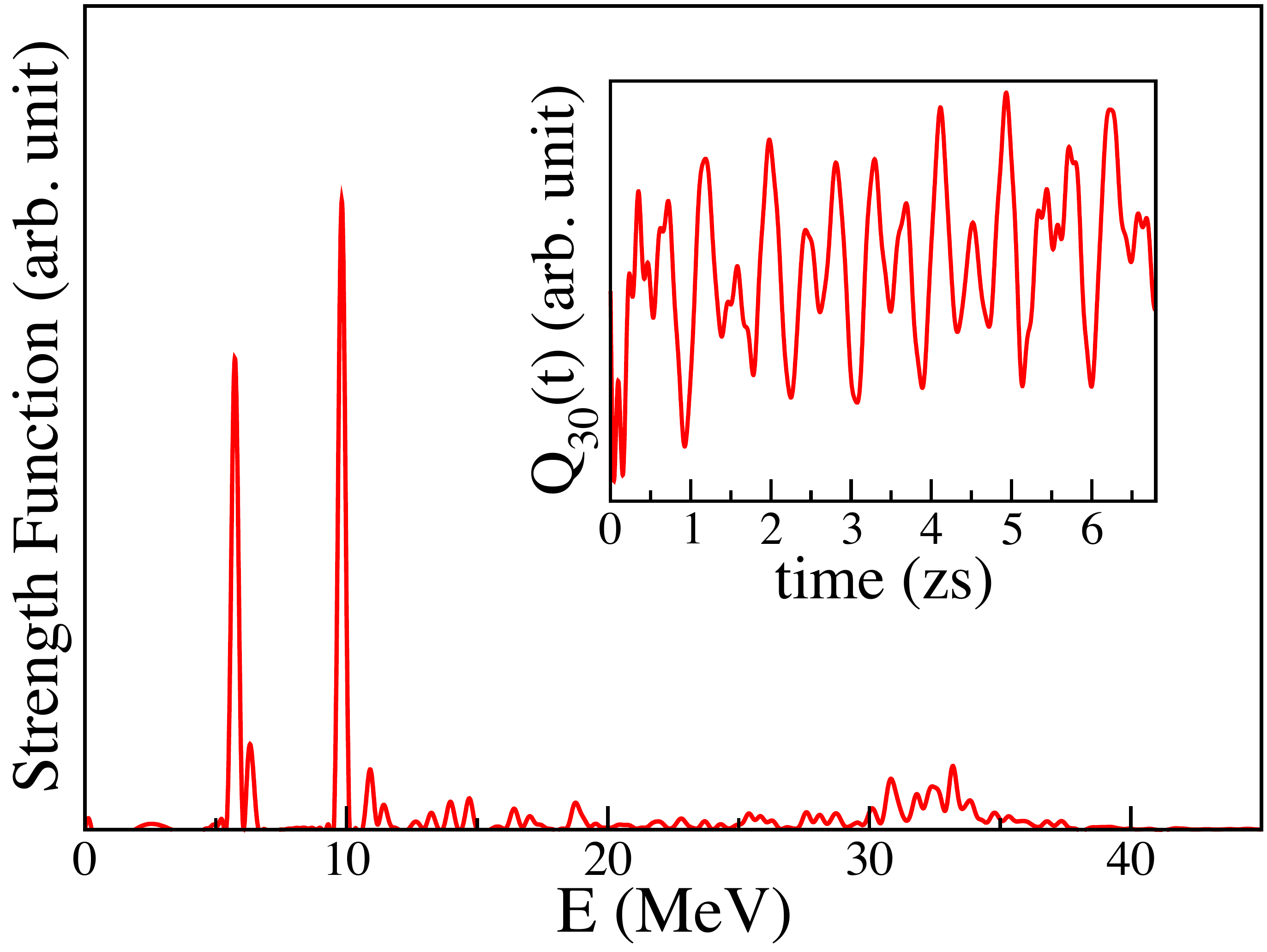}
\caption{Strength function of the octupole response in $^{48}$Ca obtained from the small amplitude limit of  TDHF with the SLy4 functional.
The time-evolution of the octupole moment following the octupole boost is shown in the inset. }
\label{fig:48Ca3-}   
\end{figure}
An example of RPA octupole strength function computed in the small amplitude limit of a TDHF evolution following an octupole excitation 
with the SLy4 Skyrme functional~\cite{chabanat1998a} is shown in Fig.~\ref{fig:48Ca3-}.
The time-evolution of the octupole moment used to compute the strength function from Eq.~(\ref{eq:strengthfinal}) is shown in the inset. 
Such strength function can be directly used to extract the energy of the low-lying collective vibrations (position of the peaks) as well as their transition amplitude (area of the peaks). 

\begin{figure}[!tb]
\centering
\includegraphics[width=12cm]{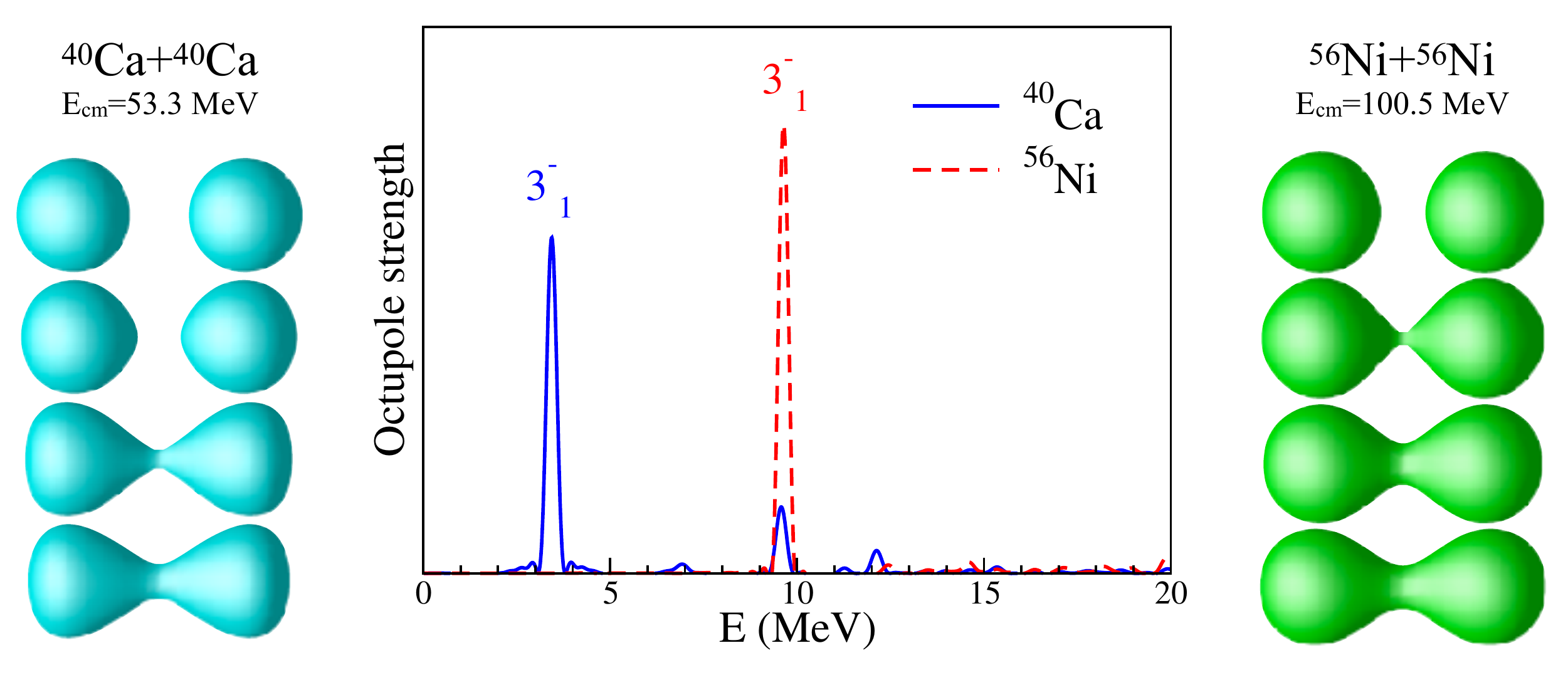}
\caption{Evolution of isodensities at half the saturation density $\rho_0/2=0.08$~fm$^{-3}$ in TDHF calculations of $^{40}$Ca$+^{40}$Ca (left) and $^{56}$Ni$+^{56}$Ni (right) central collisions at $E_{cm}=53.3$~MeV and $E_{cm}=100.3$~MeV, respectively, with the SLy4d Skyrme functional. 
The middle panel shows the RPA octupole strength function in $^{40}$Ca (solid line) and $^{56}$Ni (dashed line). The $3^-_1$ states are the low-lying collective octupole vibrations used in coupled-channels calculations. }
\label{fig:CaCa-NiNi}   
\end{figure}
Low-lying collective vibrational modes can get easily excited in heavy-ion collisions at near-barrier collisions.  
This is illustrated in Fig.~\ref{fig:CaCa-NiNi} comparing $^{40}$Ca$+^{40}$Ca and $^{56}$Ni$+^{56}$Ni near-barrier central collisions with TDHF~\cite{simenel2013b}. 
At the stage when the neck forms in $^{40}$Ca$+^{40}$Ca, the fragments have acquired a strong octupole shape 
due to the  coupling to the $3^-_1$ state which is predicted at $E_{3^-_1}^{TDHF}=3.44$~MeV in the strength function, in good agreement with the experimental energy $E_{3^-_1}^{exp}\simeq3.74$~MeV~\cite{kibedi2002}.
In the $^{56}$Ni$+^{56}$Ni reaction, however, such octupole deformation is not observed, which is consistent with the fact that the  $3^-_1$ state is predicted at much higher energy in $^{56}$Ni, and thus it is less excited in the collision. 

The impact of the coupling to octupole modes on the fusion barriers of $^A$Ca$+^{116}$Sn systems has been computed with the CCFULL code~\cite{hagino1999} in Ref.~\cite{vophuoc2016}.
The results are shown in Fig.~\ref{fig:CaSn} (triangles). 
It is observed that the couplings to these octupole modes systematically lower the average barrier. 
(This fact is well known from standard coupled-channels calculations, see, e.g.,~\cite{hagino2012}). 
Nevertheless, the effect of the neutron skin on the barrier is still present. 
It is interesting to see, however, that the effect of the neutron skin is completely washed out in direct TDHF calculations (squares) of the fusion thresholds 
(corresponding to the fusion barriers once all dynamics has been accounted for at the mean-field level).
Indeed, the change of slope at $A=48$ is  not observed in the TDHF calculations. 
This was interpreted as an effect of transfer channels in~\cite{vophuoc2016}.

Of course, the TDHF linear response (RPA) only provides an approximate description to basic vibrational modes in the harmonic limit. 
Anharmonicities can be studied by investigating, e.g.,  non-linear response in TDHF~\cite{simenel2003,reinhard2007,simenel2009} 
or boson mapping techniques~\cite{fallot2003}. 
An alternative approach based on the multireference covariant density functional theory (MCDFT) has also been used 
to compute the properties of low-lying collective excitations and their effect on fusion via  coupled-channel calculations~\cite{hagino2015,yao2016}.

\subsubsection{Density Constrained FHF \label{sec:DCFHF}}
While constructing the FHF potential, we have neglected all of the dynamical effects
as well as the Pauli principle between the nucleons belonging to different nuclei.
It is obvious that the Pauli exclusion principle will generate a repulsion between the two nuclei as they
begin to overlap. The use of Pauli orthogonalization results in altering the single-particle 
states in the overlay region to minimize their overlap and would have the effect of a static rearrangement.
The importance of Pauli orthogonalization was also recognized in the earlier work concerning $\alpha$-nucleus
scattering studies \,\cite{fliessbach1975},
where specialized normalization operators were introduced to reconstruct the states following a
Gram-Schmidt procedure. However, these techniques could only be applied using semi-analytic
methods.
In the same spirit, the Pauli repulsion should  be included in the nucleus-nucleus potentials used
to model reactions such as (in)elastic scattering, (multi)nucleon transfer, and fusion.

Pauli repulsion is ignored in double-folding and FHF potentials:
The argument has been that the outcome of a collision between nuclei is mostly determined
at a  distance where the nuclei do not overlap much, and thus the effects of the Pauli exclusion principle are negligible.
This is definitely the case in $^{16}$O$+^{16}$O as we see in Fig.~\ref{fig:Frozen_O+O} 
that the barrier is reached with very little overlap between the nuclei.  
This argument is based on the assumption that nuclei do not necessarily probe
the inner part of the fusion barrier.
However, at energies well above the barrier, the system could reach more compact shapes
where one cannot neglect the effect of the Pauli principle anymore,
as was shown by several authors in the 1970's\,\cite{fliessbach1971,brink1975,zint1975,beck1978,sinha1979}.
Similarly, for deep sub-barrier energies, the inner turning-point of the fusion barrier entails significant overlap
between the two nuclei\,\cite{dasso2003,umar2012a}.

The question then arises if one can construct a bare frozen-density nucleus-nucleus potential while incorporating
the Pauli exclusion principle at the same time.
An answer to this question can be found in the density-constrained method~\cite{cusson1985,umar1985}.
The density-constrained FHF (DCFHF) approach introduced in~\cite{simenel2017} facilitates the computation of the bare potential 
by using the self-consistent HF mean-field with exact frozen densities.
The Pauli exclusion principle is included exactly by allowing the single-particle states, 
constituting the combined nuclear density, to reorganize
to attain their minimum energy configuration and be properly antisymmetrized as the many-body
state is a Slater determinant of all the occupied single-particle wave-functions.
The HF minimization of the combined system is thus performed subject to the constraint that the
local proton ($p$) and neutron ($n$) densities do not change:
\begin{equation}
\delta \< \ H - \sum_{q=p,n}\int d\vr \ \lambda_q(\vr) \ [\rho_{1_q}(\vr)+\rho_{2_q}(\vr-\vR)] \ \> = 0\,,
\label{eq:var_dens}
\end{equation}
where the $\lambda_{n,p}(\vr)$ are Lagrange parameters at each point 
of space constraining the neutron and proton densities.
This equation determines the state vector (Slater determinant) $|\Phi(\vR)\>$.
The DCFHF potential, assumed to be central, is then defined as
\begin{equation}
V_{\mathrm{DCFHF}}(R)=\<\Phi(\vR) | H | \Phi(\vR) \>-E[\rho_1]-E[\rho_2]\,.
\label{eq:vr}
\end{equation}

\begin{figure}[!tb]
\centering
\includegraphics[width=7cm]{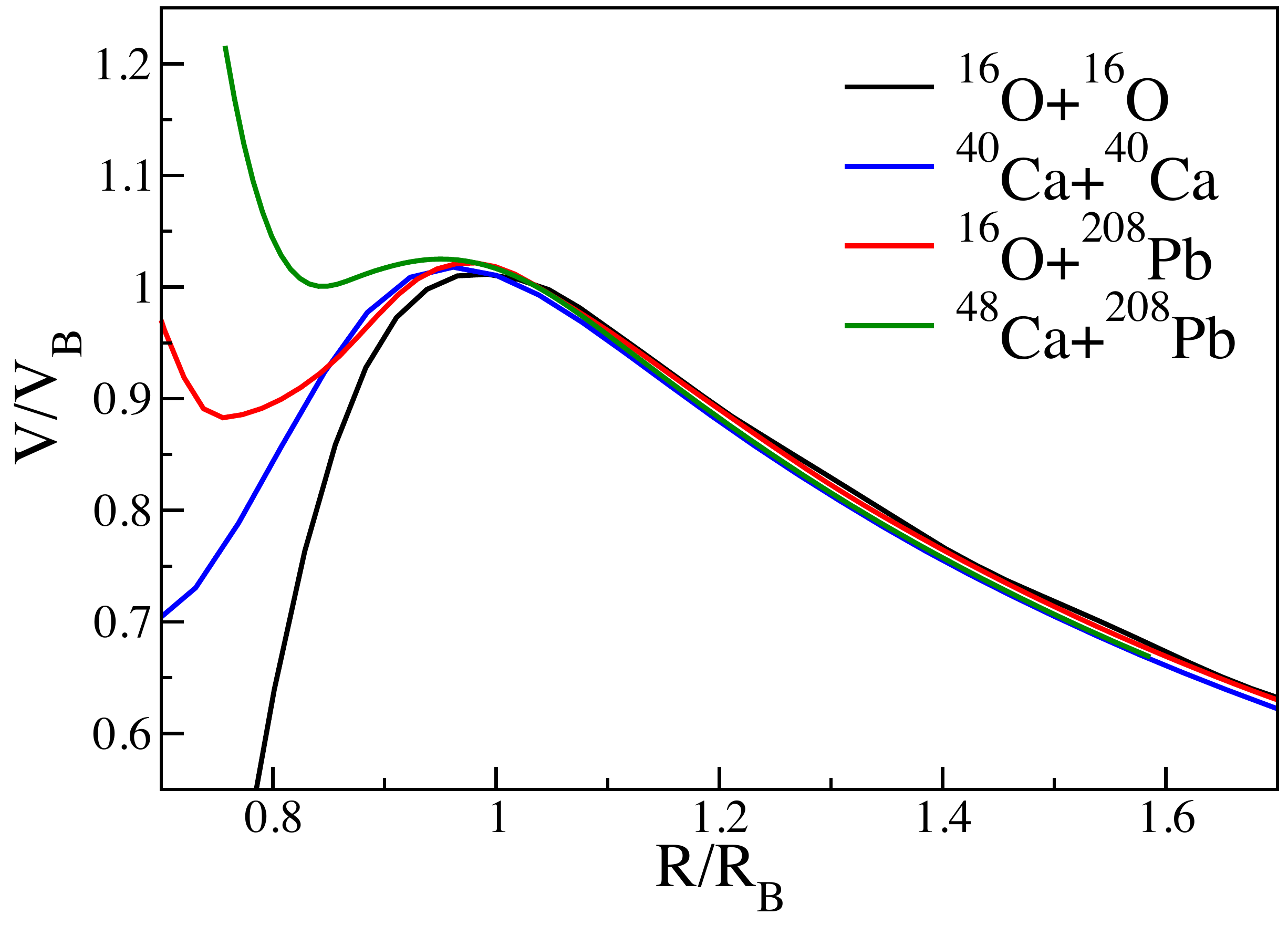}
\caption{DCFHF potentials in various systems. $R_B$ and $V_B$ denote the FHF barrier radius and height, respectively.}
\label{fig:Pauli}   
\end{figure}

An example of DCFHF potential in the $^{16}$O$+^{16}$O system  is shown in Fig.~\ref{fig:Frozen_O+O} (short dashed line).
We  see that the effect of the Pauli repulsion is to widen the barrier. 
In a light system such as $^{16}$O$+^{16}$O, the effect is relatively minor, 
though it is expected to induce a fusion hindrance at deep sub-barrier energies~\cite{simenel2017}.
In particular, Pauli repulsion  could have an important effect at energies relevant for astrophysical processes, i.e., in the Gamow window. 
In heavier systems, Pauli repulsion is expected to have a stronger effect at larger internuclear distance, 
and could even impact the barrier height.
This is illustrated in  Figure~\ref{fig:Pauli} where the nucleus-nucleus potentials for several systems are compared. 
We see that a potential pocket is produced inside the barrier, which becomes shallower in heavier systems. 

Of course, the proper inclusion of the Pauli
exclusion principle induces an apparent excitation of the system, e.g.,
increasing the local kinetic energy density in the overlap region.
Nevertheless, this type of excitation would have strictly no effect on
the shape as the density is frozen (coupled-channel calculations incorporate
effects of collective excitations via a change of density, so frozen
densities have no effect on the dynamics). The only place where the
excitation induced by the inclusion of the Pauli principle could in
principle affect the coupled-channel equations is in the increase of the excitation
energy which acts in the diagonal part of the coupled-channel Hamiltonian and thus
effectively corresponds to a modification of the potential. But this is
exactly what the Pauli repulsion potential does.

However, the fact that the nuclei cannot stay in their ground-states 
when they strongly overlap due to the Pauli exclusion principle raises another question
concerning the validity of the coupled-channel approach. In usual coupled-channel applications, 
a description of the internal states of the nuclei in terms of a ground-state 
and few low-lying collective excited states is assumed. When the overlap between the densities 
is significant, this assumption breaks down because of the change of internal 
structure induced by the Pauli exclusion principle and because of the transition 
toward an adiabatic potential for the compound system~\cite{ichikawa2009b}. 
Thus, we expect two effects from the Pauli exclusion principle: an increase of the potential (Pauli
repulsion) and a reduction of the coupling strength. 

To summarize, the DCFHF 
potential is a bare potential which can, in principle, be used in coupled-channel calculations if one neglects the modification of the internal states induced by the Pauli principle. 
The excitation induced by the inclusion of the Pauli principle has 
no other effect on the couplings as the density is frozen, and its excitation energy is already 
accounted for by the Pauli repulsion potential. The latter corresponds 
to the difference between the DCFHF and FHF potentials because $(i)$ they 
use exactly the same density and $(ii)$ FHF neglects the Pauli principle as the potential 
is obtained by adding the densities of the nuclei while in DCFHF the total state 
is described by a fully antisymmetrized Slater determinant. 

\subsubsection{Density Constrained TDHF \label{sec:DCTDHF}}
The TDHF method is probably the most established microscopic theory for studying the low-energy collisions of nuclei. 
In a direct TDHF calculation of a heavy-ion collision, 
two static many-body states calculated in the HF approximation 
are  boosted with a relative kinetic energy to initiate a nuclear collision.
 This evolution results in a self-organizing system which selects its evolutionary path by itself following the microscopic dynamics. 
 Some of the effects naturally included in the TDHF calculations are: neck formation (see, e.g., Fig.~\ref{fig:CaCa-NiNi}), 
 mass exchange, internal excitations, deformation effects to all orders, as well as the effect of nuclear alignment for deformed nuclei~\cite{simenel2004,umar2006a,umar2006d,umar2007}.
The main disadvantage of using TDHF directly to calculate fusion cross-sections, however,  is the fact that due to its semi-classical
character sub-barrier tunneling is not possible. This limits the fusion calculations to above barrier energies.
In direct TDHF calculations, the fusion probability can then be either 0 or 1.

The use of nucleus-nucleus potentials is currently the only way to calculate sub-barrier fusion cross-sections. 
As discussed above, the FHF and DCFHF potentials do not incorporate dynamical rearrangements of the density. 
One possibility mentioned earlier  is to include some dynamical effects via the coupled-channel method. 
Alternatively, one can compute potentials which incorporate the effects of the dynamics to some level. 
These ``polarization potentials'' are often energy dependent. 

A natural generalization of the FHF and DCFHF methods can be accomplished by using the dynamical nuclear densities
obtained from TDHF along with the density constraint
to extract ion-ion interaction potentials directly from the TDHF time-evolution of the nuclear system. 
This  density-constrained time-dependent Hartree-Fock (DCTDHF) method has been widely discussed and used in the literature~\cite{umar2006b,umar2006a,umar2012a,oberacker2013,jiang2014,umar2014a,umar2015a}.

The procedure for the DCTDHF method is as follows;  TDHF time-evolution takes place with no restrictions. At certain times,
corresponding to a nuclear distance $R(t)$ between the two nuclei during the evolution, the instantaneous density is used to perform a static Hartree-Fock minimization while holding the neutron and proton densities constrained to be the corresponding instantaneous TDHF densities,
\oeqn
E_{DC}(t)&=&\underset{\rho}{\min}\left\{E[\rho_n,\rho_p]+\int {\mbox d}\mathbf{r}\; \lambda_n(\mathbf{r})[\rho_n(\mathbf{r})-\rho_n^{TDHF}(\mathbf{r},t)]\right.\nonumber\\
&&\left.+\int {\mbox d}\mathbf{r}\; \lambda_p(\mathbf{r})[\rho_p(\mathbf{r})-\rho_p^{TDHF}(\mathbf{r},t)]\right\}.\nonumber
\ceqn
Here, $E[\rho_n,\rho_p]$ is the EDF used in the HF and TDHF calculations.
We refer to this minimized energy as the ``density constrained energy'',
$E_{\mathrm{DC}}$.
It is clear that the density constrained energy
plays the role of a collective potential, except for the fact that it contains the binding
energies of the two colliding nuclei. One can thus define the ion-ion
potential as~\cite{umar2006b}
\begin{equation}
V_{\mathrm{DCTDHF}}(R)=E_{DC}(R)-E[\rho_1]-E[\rho_2]\,.
\label{eq:vrdctdhf}
\end{equation}

This nucleus-nucleus potential is asymptotically correct
since at large initial separations it exactly reproduces $V_{Coulomb}(R_{max})$.
All of the dynamical features included in TDHF are naturally included in the DCTDHF potentials.
These effects include neck formation, particle exchange~\cite{umar2008a,simenel2010,sekizawa2013},
internal excitations, and deformation effects to all order, among others.

\begin{figure}[!tb]
\centering
\includegraphics[width=8cm]{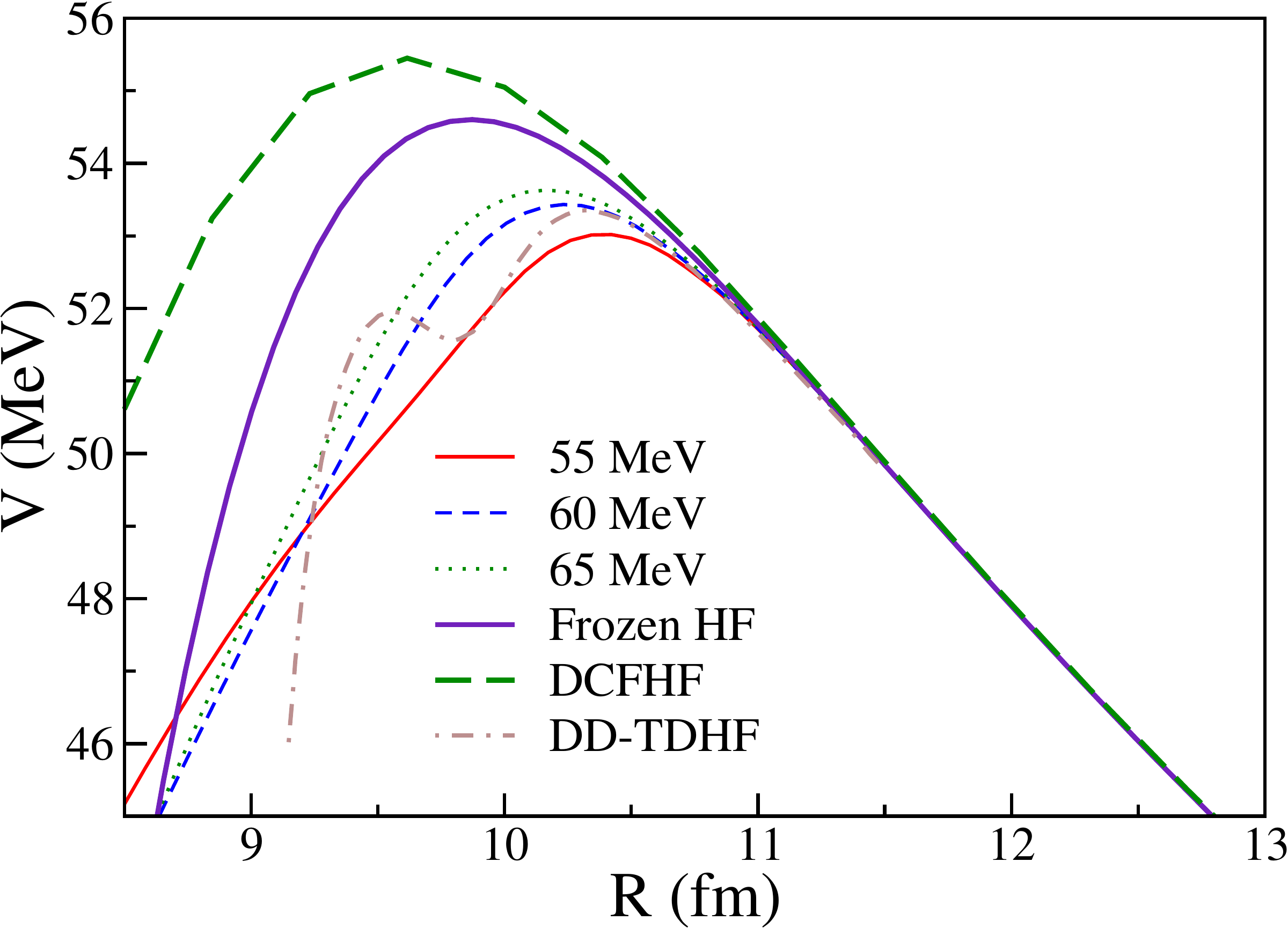}
\caption{\protect DCTDHF potentials in $^{40}$Ca$+^{40}$Ca obtained from TDHF central collisions at $E_{cm}=55$, 60 and 65~MeV, together with the FHF and DCFHF potentials. The dissipative dynamics TDHF (DD-TDHF) potential at $E_{cm}=55$~MeV from~\cite{washiyama2008} is also shown.}
\label{fig:Pot}   
\end{figure}

Examples of DCTDHF potentials are shown in Fig.~\ref{fig:Pot} for $^{40}$Ca$+^{40}$Ca at various TDHF energies~\cite{umar2014a}. 
The DCTDHF barriers are lower than the bare potential barriers obtained with the static FHF and DCFHF approaches. 
This is interpreted as a manifestation of the couplings to low-lying vibrational states which enhance fusion near the barrier.
Indeed, couplings between relative motion and internal structures are known to affect fusion
barriers by dynamically modifying the densities of the colliding nuclei (see Fig.~\ref{fig:CaCa-NiNi}).  

An energy dependence is also observed, with an increase of the barrier with collision energy. 
The lowering of the barrier due to the couplings is indeed expected to be stronger at energies near
the barrier top, where changes in density have longer time to develop than at higher energies.
A similar energy dependence is observed with the dissipative dynamics TDHF (DD-TDHF) method~\cite{washiyama2008}.
Note that at high energy the polarization potential should converge towards the bare potential. 
In fact, it is expected to converge towards the FHF potential 
and not the DC-FHF potential as nucleons of one fragment have very different momentum vectors 
than the nucleons in the other fragments, therefore reducing the Pauli repulsion~\cite{simenel2017}. 

There are different view points as to why DCTDHF works well. TDHF density
evolution does contain the relevant collective degrees of freedom as well as single-particle
dynamics during the early stages of the collision. Unlike static approaches,
the DCTDHF method incorporates  collision dynamics at the mean-field level, such as 
pre-equilibrium excitations, neck formation, and transfer. 
Naturally, the TDHF evolution takes place above the static
potential energy surface. However, despite this, single-particle friction can quickly absorb
this energy and lead to a configuration that may be considered a doorway state. 
As long as the average single-particle excitation energy per nucleon in this doorway state 
is less than the shell energy (about 4-8 MeV), the details of the ground-state potential-energy
surface are still felt, and shell-correction energies influence the TDHF dynamics. 
As a result, DCTDHF calculations reproduce ion-ion interaction
barriers for heavy-ion collisions.

\subsubsection{Isovector and isoscalar contributions to the potential}

It is sometimes convenient to separate isoscalar and isovector contributions to observables.
This is often particularly useful when one wants to investigate specific effects with exotic nuclei, 
such as the evolution of the fusion barrier with nuclei along isotopic chains. 

The isoscalar and isovector contributions in building
up the nucleus-nucleus interaction barrier have recently been isolated. 
This is possible because the Skyrme EDF can be decomposed as~\cite{dobaczewski1995};
\begin{equation}
\label{eq:edensity}
{\cal H}(\mathbf{r}) = \frac{\hbar^2}{2m}\tau_0
+ {\cal H}_0(\mathbf{r})
+ {\cal H}_1(\mathbf{r})
+ {\cal H}_{Coulomb}(\mathbf{r})\, .
\end{equation}
The isospin index $\mathrm{I}=0,1$ stands for isoscalar and isovector energy densities, respectively.
The isoscalar (isovector) energy density, ${\cal H}_0(\mathbf{r})$ (${\cal H}_1(\mathbf{r})$), depends on the isoscalar (isovector) particle
density, $\rho_0 = \rho_n + \rho_p$ ($\rho_1 = \rho_n - \rho_p$), with analogous expressions for other densities and
currents. The use of this decomposition in Eq.~(\ref{eq:vrdctdhf}) allows us to write
\begin{equation}
V(R) = \sum_{\mathrm{I}=0,1} v_\mathrm{I}(R) + V_C(R)\, ,
\end{equation}
where $v_\mathrm{I}(R)$ denotes the potential computed by using the isoscalar and isovector parts of
the Skyrme EDF~\cite{godbey2017}. 
\begin{figure}
\centering
\includegraphics[width=7cm]{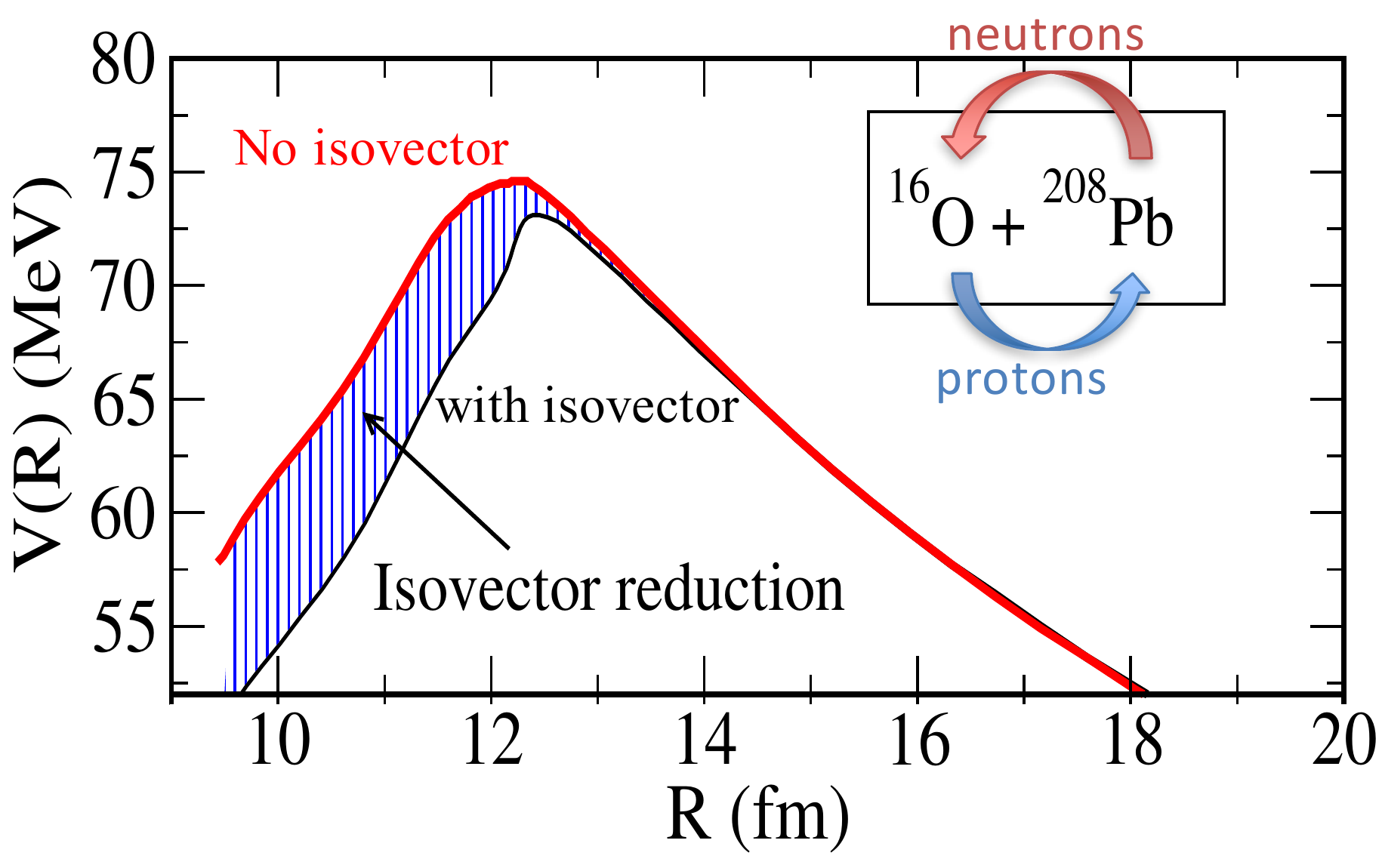}
\caption{DCTDHF potential in $^{16}$O$+^{208}$Pb with (black) and without (red) isovector contribution $v_1$ using TDHF densities from a central collision at $E_{cm}=75$~MeV.}
\label{fig:iso}       % Give a unique label
\end{figure}

In the static case the isovector potential $v_1$ is found to be small, and even vanishes in the FHF approximation. 
The isovector potential  is then mostly induced by dynamical effects. 
Figure~\ref{fig:iso} illustrates the impact of the isovector contribution to the potential in the $^{16}$O$+^{208}$Pb system~\cite{godbey2017}. 
An isovector reduction is observed in the inner part of the barrier. 
The origin of this reduction is interpreted as an effect of transfer of protons from $^{16}$O to $^{208}$Pb and of neutrons in the opposite direction. 
This is compatible with  recent experiments of multi-nucleon transfer at sub-barrier energies~\cite{evers2011,rafferty2016}.
The fact that protons and neutrons do not flow in the same direction induces a non-zero isovector density at the origin of the observed 
variation of the potential at short distances. 

\subsubsection{Interplay between transfer and fusion}

This approach has been employed to investigate the influence of transfer on fusion for a number of systems~\cite{godbey2017}, 
including on heavy systems such as $^{40,48}$Ca$+^{132}$Sn in which the experimental signatures are not so clear~\cite{back2014,liang2016}.
As shown in several works using  TDHF~\cite{simenel2001,simenel2007,iwata2010,oberacker2012,simenel2012b,umar2017}, 
systems with $N/Z$ asymmetries encounter a rapid charge equilibration (transfer of protons and neutrons in opposite directions). 
This charge equilibration has a strong impact on $v_1$~\cite{godbey2017}. 

\begin{figure}
    \centering
    \includegraphics[width=10cm]{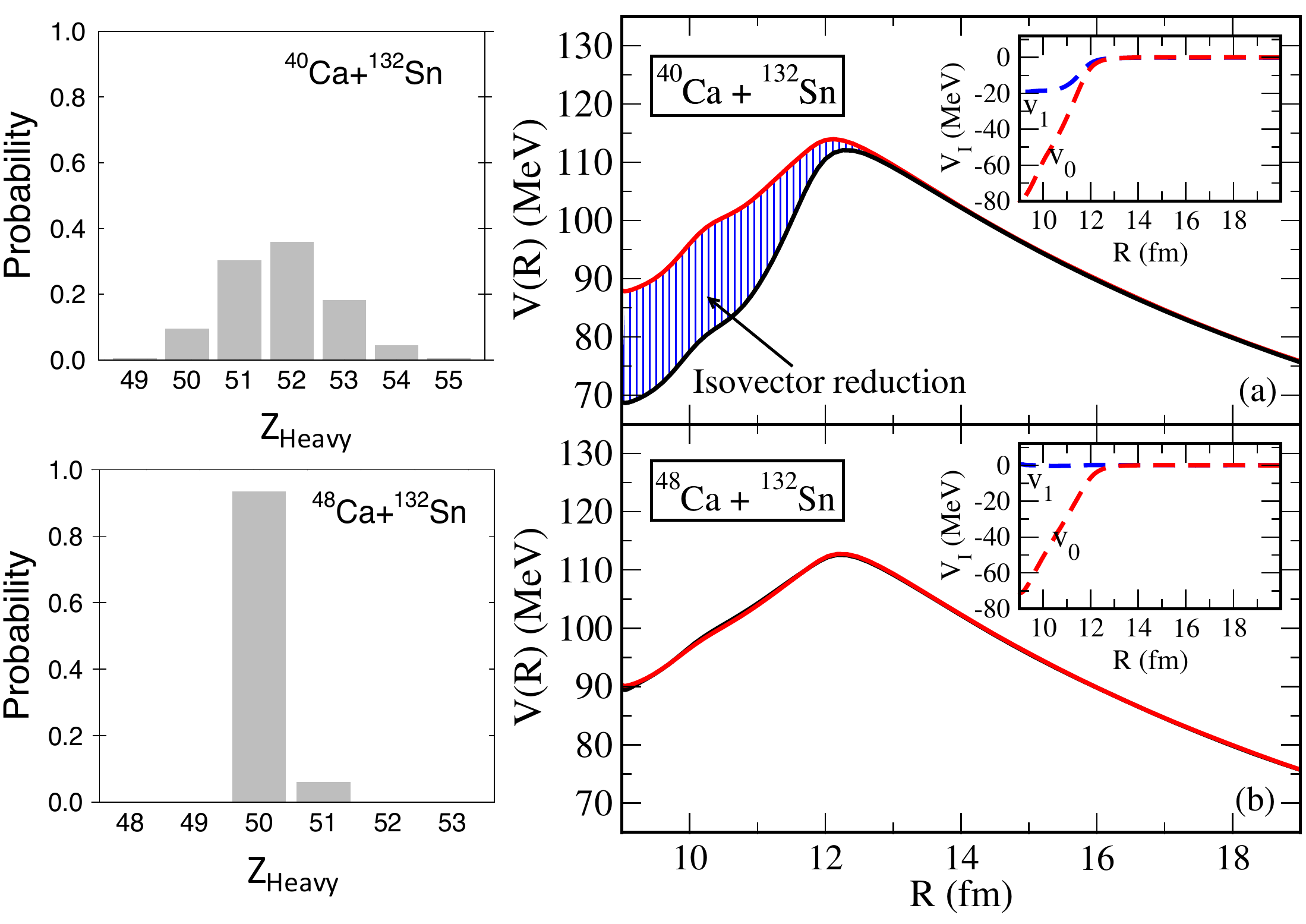}
    \caption{Charge distributions in the heavy fragments following a central collision with TDHF just below the barrier (left) 
        and DCTDHF potentials (right) without (solid red line) and with (solid black line) isovector contributions in $^{40}$Ca$+^{132}$Sn (top)  and in $^{48}$Ca$+^{132}$Sn (bottom). The insets show the isoscalar and isovector contributions to the nuclear part of the potential.}
    \label{fig:CaPb}       % Give a unique label
\end{figure}

The isovector reduction of the potential due to transfer  depends naturally on the presence of positive $Q-$value transfer channels~\cite{jiang2014a}.
As shown in Fig.~\ref{fig:CaPb}, an isovector reduction is observed in  $^{40}$Ca$+^{132}$Sn which has several positive $Q-$value transfer channels, but not in  $^{48}$Ca$+^{132}$Sn which has only negative $Q-$value transfer channel.
Another confirmation of this effect is obtained from the proton number distributions in the heavy fragments when the collision occurs just below the barrier 
(i.e., with two outgoing fragments in the  TDHF evolution).  
These distributions are obtained from a particle number projection technique~\cite{simenel2010} (see section~\ref{sec:projection}).
As a result, almost no charge transfer in observed in  $^{48}$Ca$+^{132}$Sn while significant transfer probabilities are found in $^{40}$Ca$+^{132}$Sn. 

Standard coupled-channels calculations without couplings to transfer channels then reproduce  
the $^{48}$Ca$+^{132}$Sn system relatively well~\cite{kolata2012}.
However, similar calculations underpredict the fusion cross sections in $^{40}$Ca$+^{132}$Sn. 
This is interpreted as a fusion enhancement at and below the barrier from couplings to transfer channels in the latter system. 
In DCTDHF, the origin of this enhancement is a reduction of the barrier width as seen in Fig.~\ref{fig:CaPb}-a).
As a result, sub-barrier fusion cross-sections calculated  from such potentials (see next section) 
are in relatively good agreement with experiment for these systems~\cite{oberacker2013} 
despite the fact that these calculations have no adjustable parameters.

\subsubsection{Fusion cross-sections \label{sec:cross-sections}}

The nucleus-nucleus potential is of course an essential ingredient to compute fusion cross-sections. 
As the DCTDHF potentials already account for effects of the dynamics (at the mean-field level), 
they should not be used in coupled-channels calculations as this would double count the effect of the couplings. 
Fusion probabilities can then be computed with DCTDHF potentials using a series of one-barrier penetration calculations.
In principle, one needs to do one calculation per incident energy as the DCTDHF potential is energy dependent. 
In practice, potentials computed at few energies in steps of few MeV near the barrier 
are usually sufficient for the calculation of fusion cross-sections~\cite{umar2008b,umar2009b,umar2014a}.
Note that one drawback of the approach is that it cannot provide barrier potentials from TDHF collisions at sub-barrier energies, 
as in this case the TDHF trajectory does not lead to fusion. 
Sub-barrier fusion cross-sections are then estimated with DCTDHF potentials computed with a near-barrier energy. 
%CS j'en suis la

\begin{figure}
\centering
\includegraphics[width=7cm]{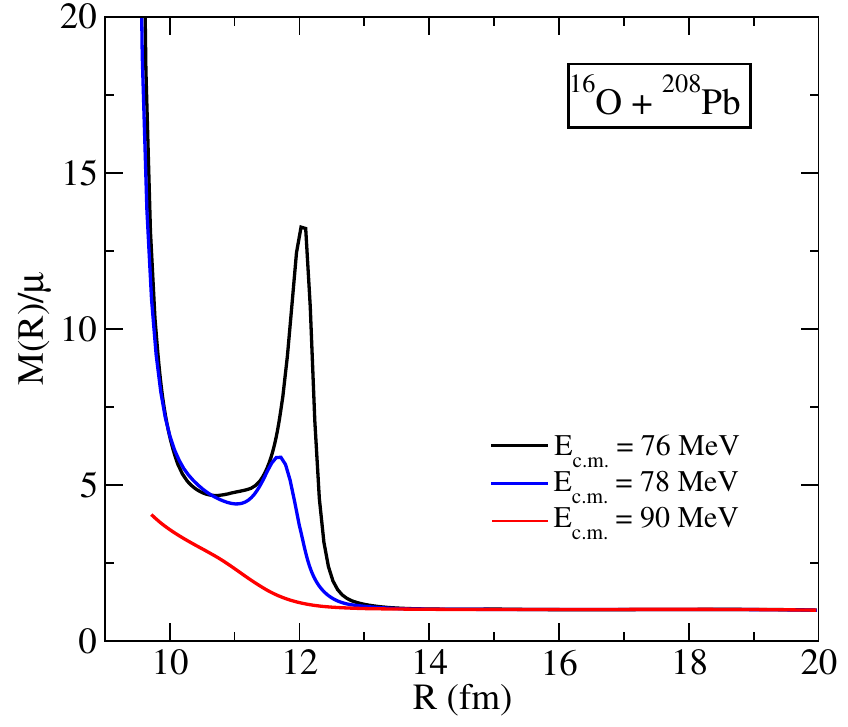}
\caption{Ratio of the coordinate-dependent mass $M(R)$ over the constant reduced mass obtained 
from TDHF $^{16}$O$+^{208}$Pb central collisions at various energies~\cite{umar2009b}.}
\label{fig:mass}       % Give a unique label
\end{figure}

Another important quantity entering the calculation of cross-sections is the coordinate-dependent mass $M(R)$.
The latter can be obtained from TDHF evolutions using energy conservation for a central collision
\begin{equation}
M(R)=\frac{2[E_{\mathrm{cm}}-V(R)]}{\dot{R}^{2}}\;,
\label{eq:mr}
\end{equation}
where the collective velocity $\dot{R}$ is directly obtained from the TDHF evolution.
Examples of evolutions of $M(R)$ at various energies are shown in Fig.~\ref{fig:mass} for the $^{16}$O$+^{208}$Pb reaction~\cite{umar2009b}.
The $R$-dependence of this mass at lower energies is
very similar to the one found in constrained Hartree-Fock calculations~\cite{goeke1983} with a constraint
on the quadrupole moment.
On the other hand, at higher energies the coordinate dependent mass essentially becomes flat,
which is again a sign that most dynamical effects are contained at lower energies.
The peak at small $R$ values is
due to the fact that the centre-of-mass energy is above the barrier and the
denominator of Eq.~(\ref{eq:mr}) becomes small due to the slowdown of the ions.

The fusion barrier penetrabilities $T_L(E_{\mathrm{cm}})$ can be
obtained by numerical integration of the Schr\"odinger equation
for the collective distance coordinate $R$, using the heavy-ion potential $V(R)$
with coordinate dependent mass parameter $M(R)$.
Alternatively, we can instead use the constant
reduced mass $\mu$ and transfer the coordinate-dependence of the mass to a scaled
potential using the well known exact point transformation~\cite{goeke1983,umar2009b}
\begin{equation}
dR\longrightarrow\left(\frac{M(R)}{\mu}\right)^{\frac{1}{2}}dR\;.
\label{eq:mrbar}
\end{equation}
The potential $V(R)$, which includes the coordinate-dependent mass effects differs from the
original only in the interior region of the barrier. Further details can
be found in Ref.~\cite{umar2009b}.
Using the transformed potential the
fusion barrier penetrabilities $T_L(E_{\mathrm{cm}})$
are obtained by numerical integration of the Schr\"odinger equation
\begin{equation}
\left[ \frac{-\hbar^2}{2\mu}\frac{d^2}{dR^2}+\frac{L(L+1)\hbar^2}{2\mu R^2}
+V(R)-E_{\mathrm{cm}}\right]\psi(R)=0\;,
\label{eq:xfus}
\end{equation}
using the incoming wave boundary condition (IWBC) method~\cite{rawitscher1964}.
IWBC assumes that once the minimum of the potential is reached fusion will
occur. In practice, the Schr\"odinger equation is integrated from the potential
minimum, $R_\mathrm{min}$, where only an incoming wave is assumed, to a large asymptotic distance,
where it is matched to incoming and outgoing Coulomb wavefunctions. The barrier
penetration factor, $T_L(E_{\mathrm{cm}})$, is the ratio of the
incoming flux at $R_\mathrm{min}$ to the incoming  flux at large distance.
Here, we implement the IWBC method exactly as it is
formulated for the coupled-channel code CCFULL described in Ref.~\cite{hagino1999}.
However, since the DCTDHF potential already includes excitations present at the mean-field
level it can no longer be considered as a bare nucleus-nucleus potential. Consequently,
it would not be appropriate to employ channel couplings using this potential and the potential
must be directly used.
This gives us a consistent way for calculating fusion cross-sections at energies below and above
the barrier via
\begin{equation}
\sigma_f(E_{\mathrm{cm}}) = \frac{\pi}{k^2} \sum_{L=0}^{\infty} (2L+1) T_L(E_{\mathrm{cm}})\;.
\label{eq:sigfus}
\end{equation}
At energies well above the barrier either the DC-TDHF method or direct TDHF
calculations [where $T_L(E_{\mathrm{cm}})=0$ or 1] can be used to determine the fusion cross-sections.
\begin{figure}
    \centering
    \includegraphics[width=7cm]{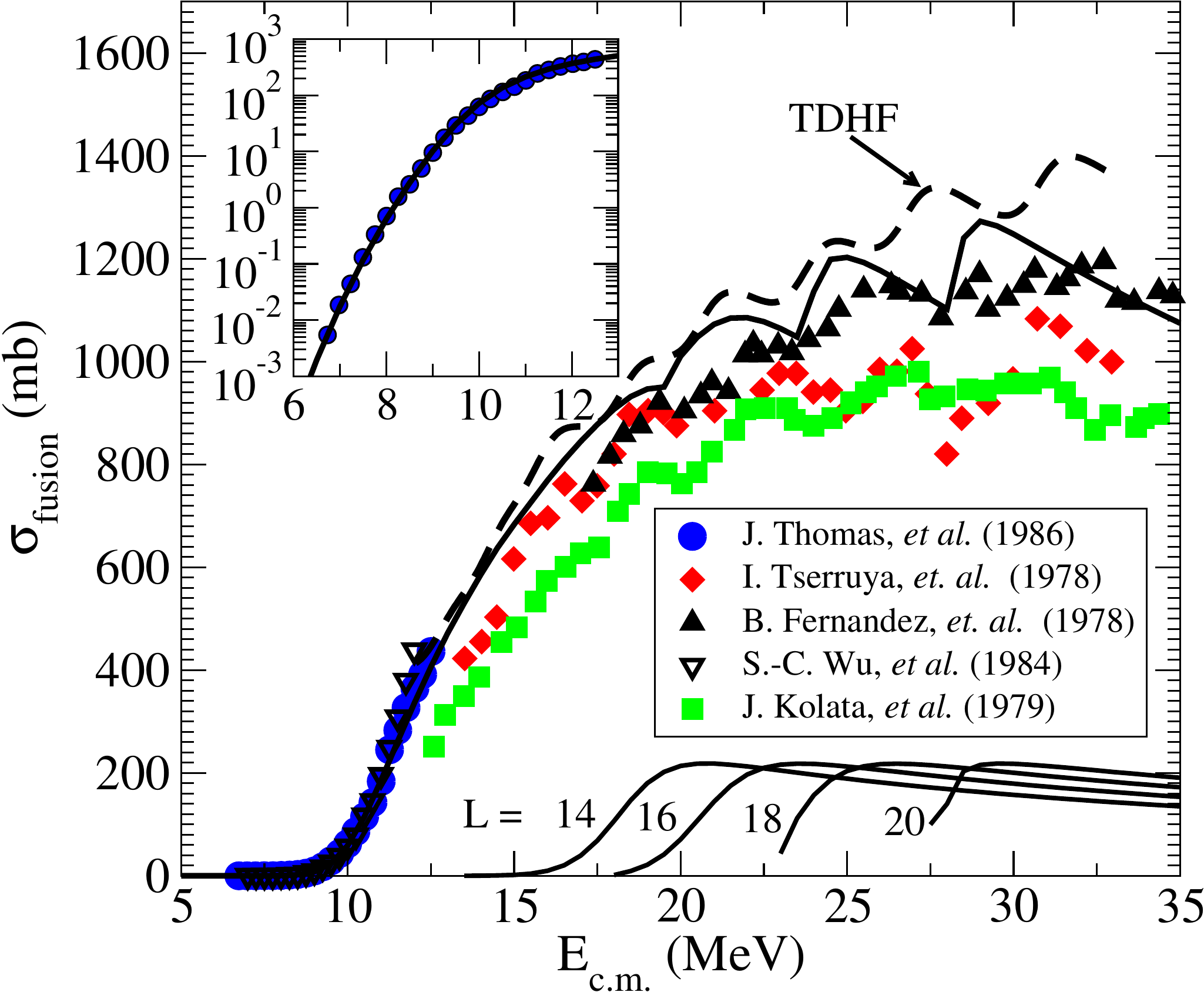}
    \caption{Fusion cross-sections in $^{16}$O$+^{16}$O in linear scale. 
        Logarithmic cross-sections are shown in the inset for the low energies. 
        Total fusion cross-sections computed from DCTDHF potential and the contributions from various angular momenta are shown with solid lines. 
        The dashed line gives the fusion cross-sections computed directly from TDHF.
        Experimental data are from~\cite{thomas1986,tserruya1978,fernandez1978,wu1984,kolata1979}.}
    \label{fig:fusOO}       % Give a unique label
\end{figure}

An example of calculations of fusion cross-sections in $^{16}$O$+^{16}$O is shown in Fig.~\ref{fig:fusOO}~\cite{simenel2013a}.
The oscillations observed in the cross-sections result from overcoming  angular momentum dependent fusion barriers. 
Cross-sections computed directly from TDHF agree relatively well with those computed from DCTDHF potential, 
except at the highest energies, which could possibly indicate a breakdown of the isocentrifugal approximation 
used in the one-barrier penetration approach. 
The calculations match well the experimental data near and below the barrier. 
Above the barrier, there is a strong dependence in the experimental results, with cross-sections found to be smaller than in the calculations. 
It would be interesting to perform new measurements on this system at above barrier energies. 
A similar comparison between TDHF predictions~\cite{simenel2008} and experimental data~\cite{morton1999} 
in the $^{16}$O$+^{208}$Pb system shows an overestimation of the fusion cross-sections of $\sim16\%$ above the barrier, 
while the position of the barrier is very well reproduced. 
The origin of this discrepancy remains to be investigated. 

\begin{figure}
\centering
\includegraphics[width=8cm]{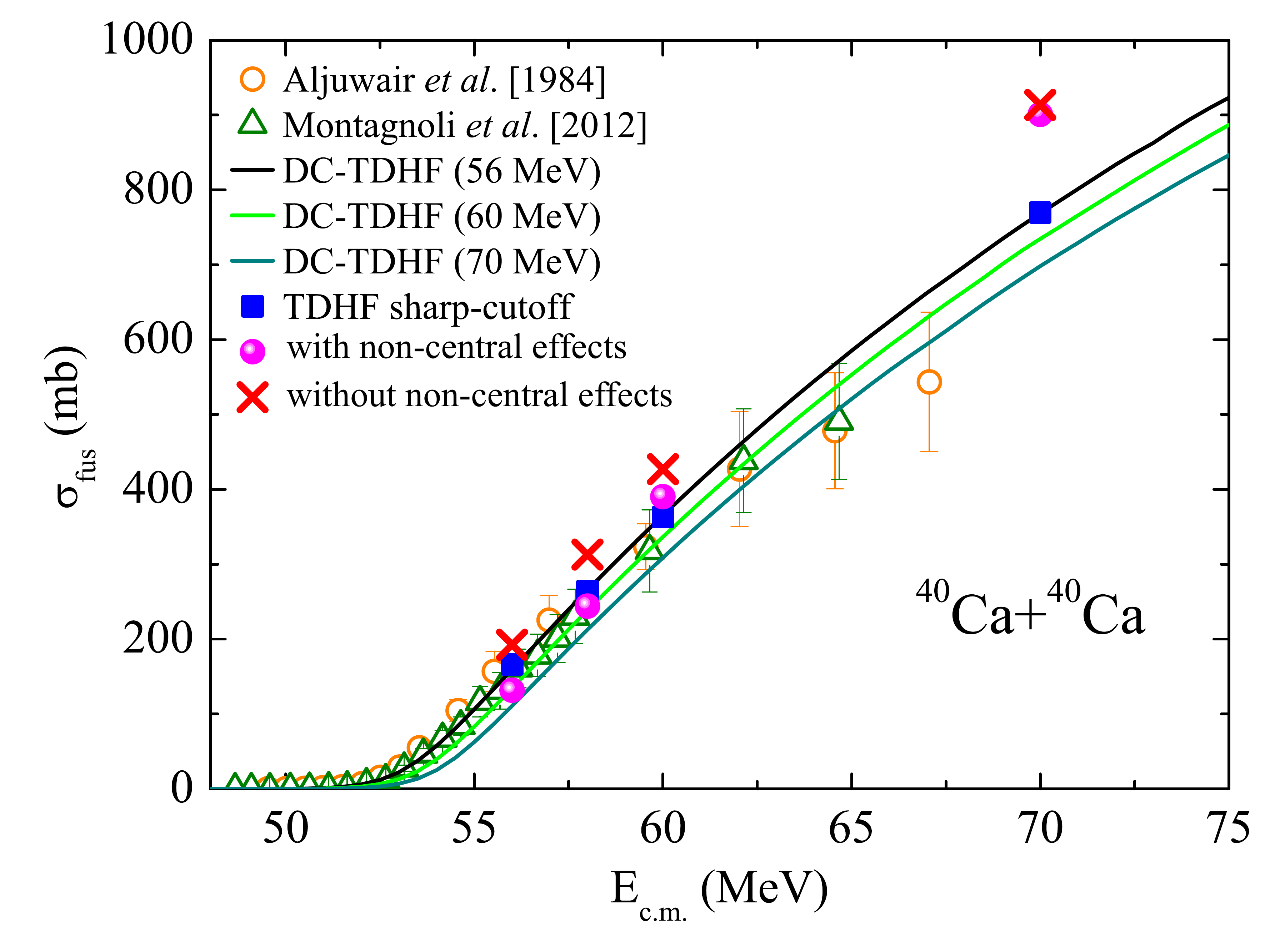}
\caption{Comparison of fusion cross-sections in $^{40}$Ca$+^{40}$Ca from experimental data~\cite{montagnoli2012,aljuwair1984} and from TDHF (squares) and DCTDHF predictions with (circles) and without (crosses) non-central effects. Adapted from~\cite{jiang2014}.}
\label{fig:jiang}       % Give a unique label
\end{figure}

The DCTDHF method has been recently implemented within the Sky3D solver~\cite{maruhn2014} by X.~Jiang and collaborators~\cite{jiang2014}.
In particular, they investigated non-central effects on the potential by computing the DCTDHF potentials at finite impact parameters~\cite{jiang2014,jiang2015a}.
They found that these non-central effects have a significant effect near the barrier, reducing the fusion cross-sections in this energy region 
and improving the agreement with experiment, as shown in Fig.~\ref{fig:jiang}.
At higher energies, however the non-central effects on the potential are shown to be negligible. 

\begin{figure}
\centering
\includegraphics[width=7cm]{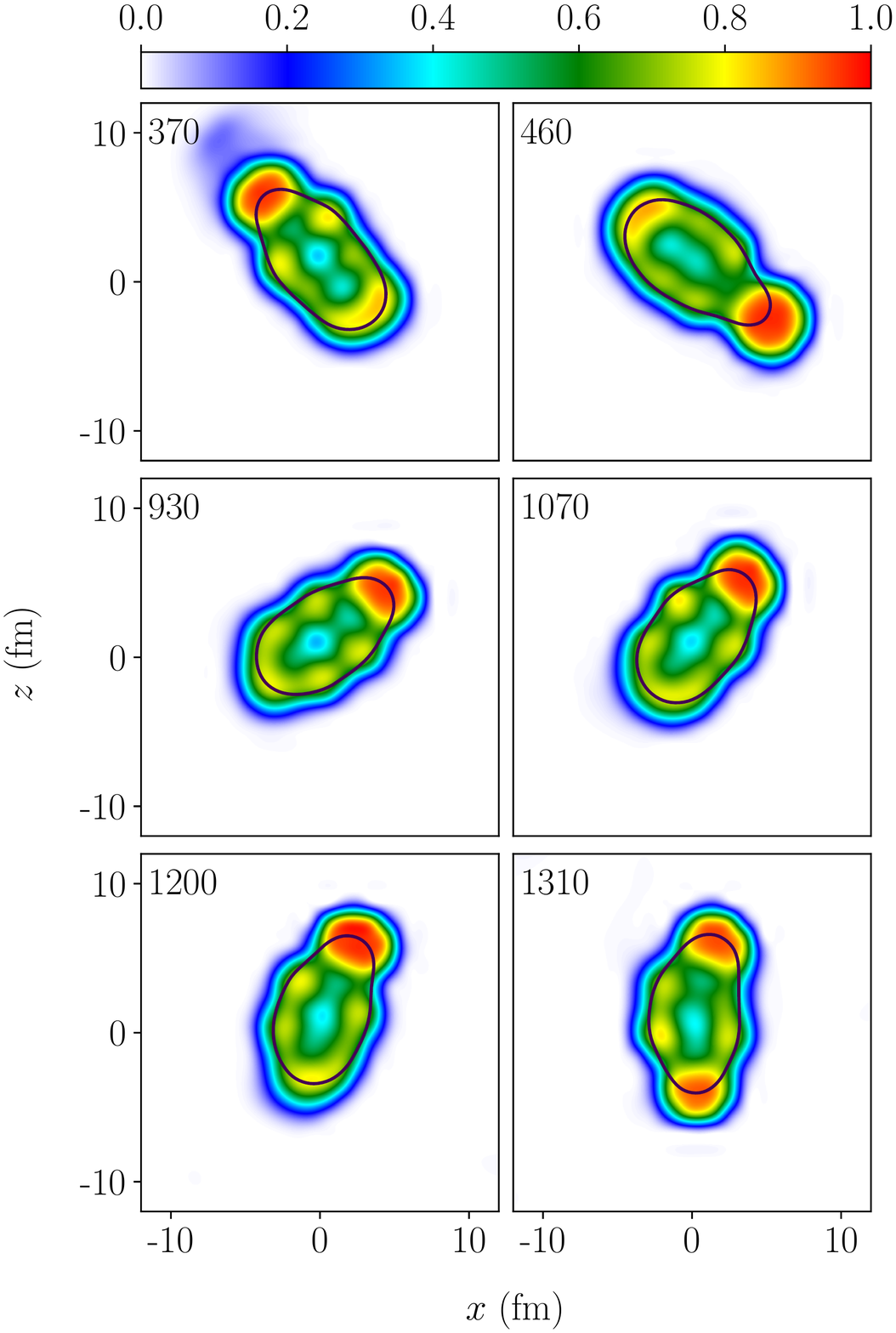}
\caption{$\alpha$-localization function in $^{18}$O$+^{12}$C at $E_{cm}=14$~MeV with an impact parameter of 2~fm~\cite{schuetrumpf2017}.
Hot colors indicate the likelihood for the presence of an $\al$-cluster. The numbers indicate time in fm$/c$. The solid line is an isodensity at 0.05~fm$^{-3}$.}
\label{fig:18O+12C}       % Give a unique label
\end{figure}

Other examples of comparisons between DCTDHF predictions and experimental fusion cross-sections 
include reactions with radioactive beams. 
DCTDHF calculations predict a large enhancement of  fusion  with neutron-rich light nuclei, 
such as $^{24}$O$+^{16}$O~\cite{umar2012a} which could have an impact on reaction rates 
in dense stellar matter such as neutron star crusts~\cite{gasques2007}.
In the same mass region, the reaction $^{20}$O$+^{12}$C has been measured~\cite{rudolph2012} and compared with DCTDHF calculations~\cite{desouza2013}.
While for this system the calculations underestimate the fusion cross-sections, 
it is interesting to note that a good agreement is obtained for the $^{19}$O$+^{12}$C system~\cite{singh2017}. 
A poor agreement is obtained for the $^{18}$O$+^{12}$C stable system~\cite{steinbach2014}. 
This could be due to the fact that $^{18}$O has a strong $^{14}$C$+\al$ cluster configuration, as shown by the strong probability for $\al-$transfer in $^{18}$O$+^{208}$Pb~\cite{astier2010,rafferty2016}.
These cluster configurations are difficult to account for in TDHF calculations, unless they are already present in the initial HF wave-function, as in the $^8$Be$+\alpha$ reaction~\cite{umar2010c} or with rod-shape configurations~\cite{iwata2013,iwata2015}. 
Note that Schuetrumpf and collaborators have recently developed a nucleon localization technique for nuclear systems~\cite{schuetrumpf2016}
which they applied to identify the presence of $\al$-clustering in heavy-ion collisions~\cite{schuetrumpf2017}. 
An example is shown in Fig.~\ref{fig:18O+12C}, where we see that an $\al$-cluster is likely to be found on one hedge of the rotating compound system. 
This technique could form the basis for a beyond TDHF approach to incorporate cluster effects in the dynamics. 

\subsubsection{Dissipation on the way to fusion \label{sec:dissipation}}

The ability for the nuclei to deform and get excited during a collision induces a dissipation mechanism 
where the kinetic energy for the relative motion is transformed into internal excitation energy. 
It is possible to use time-dependent microscopic calculations to compute  parameters associated with dissipative dynamics, e.g., 
friction coefficient entering the Langevin equation \cite{brink1981,washiyama2009a,wen2013}.
The dissipation mechanism has also been investigated using the DCTDHF technique in~\cite{umar2009a,umar2010a}.
\begin{figure}[!htb]
    \centering
    \includegraphics[width=7cm]{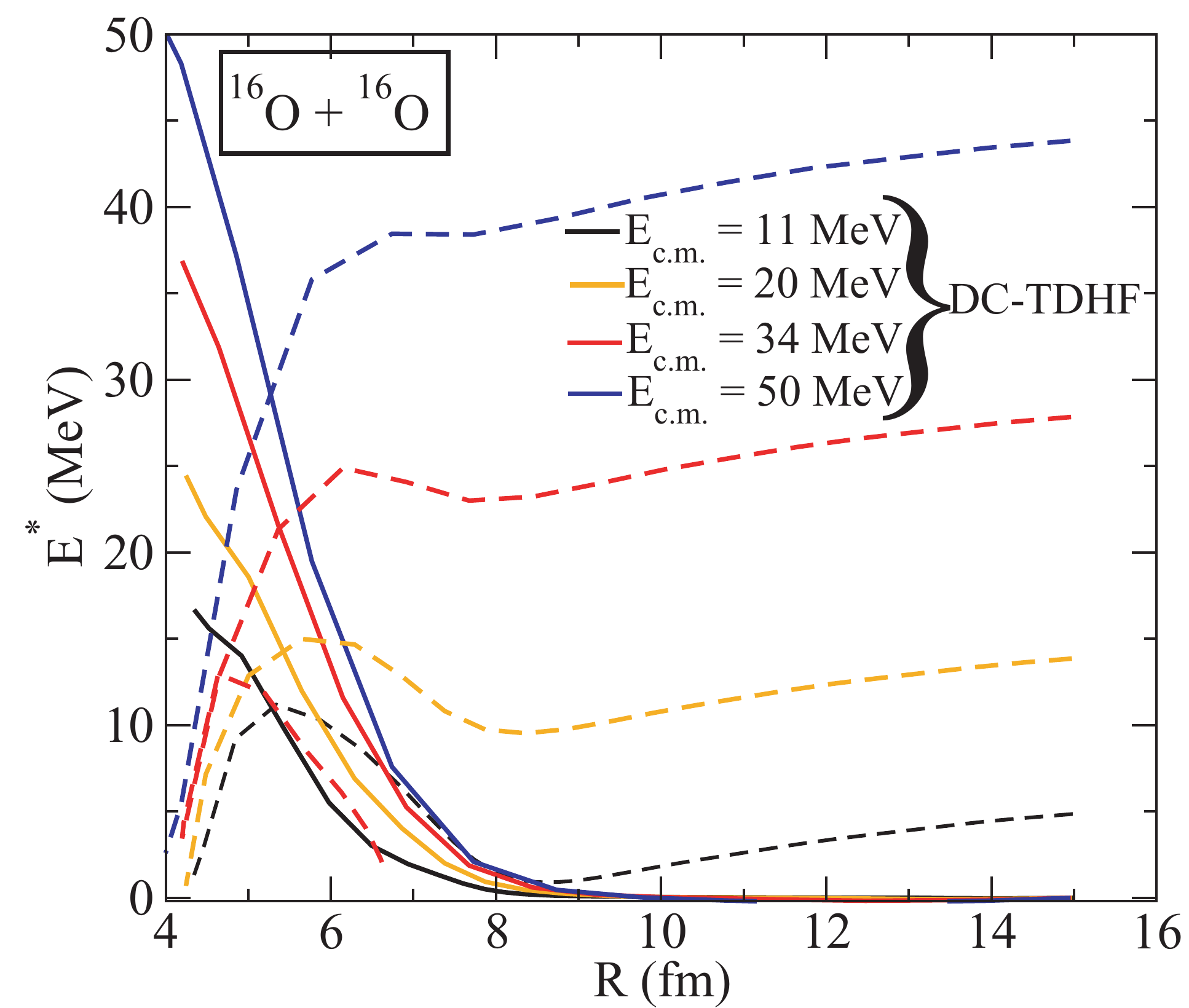}
    \caption{Excitation energy in $^{16}$O$+^{16}$O from DCTDHF (solid lines) and  corresponding collective kinetic energy (dashed lines) at various centre of mass  energies~\cite{umar2009a}.
    }
    \label{fig:OOE}       % Give a unique label
\end{figure}

Using the DCTDHF method, it is indeed possible to calculate dynamical excitation energies by making the
premise that the TDHF energy can be divided into a collective and an intrinsic part~\cite{umar2009a}
\begin{equation}
E^{*}(t)=E_{TDHF}-E_{coll}\left(\rho(t),\mathbf{j}(t)\right)\;,
\label{eq:es}
\end{equation}
where $E_\mathrm{TDHF}$ is the total energy of the dynamical system, which is a conserved quantity,
and $E_\mathrm{coll}$ represents the
collective energy of the system
\begin{equation}
E_{coll}\left(t\right)= E_{kin}\left(\rho(t),\mathbf{j}(t)\right) + E_{DC}\left(\rho(t)\right)\;.
\end{equation}
Here $E_\mathrm{kin}$ is the collective kinetic part and is given by
\begin{equation}
E_{kin}\left(\rho(t),\mathbf{j}(t)\right)=\frac{m}{2}\int {\mbox d}\mathbf{r}\;\;\mathbf{j}^2(\mathbf{r},t)/\rho(\mathbf{r},t)\;,
\end{equation}
which is asymptotically equivalent to the kinetic energy of the
relative motion, $\frac{1}{2}\mu\dot{R}^2$, where $\mu$ is the
reduced mass and $R(t)$ is the ion-ion separation distance.

An example of evolution of the excitation energy as a function of the relative distance between 
the nuclei is shown in Fig.~\ref{fig:OOE} for $^{16}$O$+^{16}$O central collisions~\cite{umar2009a}. 
The barrier radius is approximately 8.3~fm in this system (see Fig.~\ref{fig:Frozen_O+O}). 
We see in Fig.~\ref{fig:OOE} that, with such light collision partners, the excitation energy increases rapidly inside the barrier. 
The situation is different in heavier systems, e.g., in those used to form superheavy nuclei. 
Indeed, DCTDHF calculations show that, in such systems, the excitation energy increases before capture~\cite{umar2010a}, 
thus increasing the probability for the systems to encounter a quasi-fission process. 
These calculations also show that, in  the case of reactions with actinides, 
the evolution of the excitation energy depends  strongly on the orientation of the deformed target. 

\subsubsection{Adiabatic potentials for sub-barrier fusion}

As mentioned in section~\ref{sec:cross-sections}, one difficulty of the DCTDHF method is to determine  
the potential seen by the nuclei at sub-barrier energies when tunneling through the barrier. 
This is because DCTDHF is a method to extract the potential from TDHF trajectories, yet the latter do not lead to fusion at sub-barrier energies. 
This problem could be overcome with constrained Hartree-Fock potentials obtained by minimizing the HF energy under a collective constraint 
(e.g., a fixed value of the quadrupole and octupole moments) \cite{skalski2007}. 
This technique is bread and butter in fission studies \cite{schunck2016} as it leads to adiabatic potentials well suited for fission.
However, the choice of the constraint may sometime be problematic, 
in particular in asymmetric collisions where an octupole constraint 
may not be sufficient to ensure the correct asymmetry in the approach phase. 

The adiabatic self-consistent collective coordinate (ASCC) method provides an alternative approach to describe the fusion path and potential 
without relying on a specific choice of the collective coordinates~\cite{matsuo2000,nakatsukasa2016}. 
The method determines the optimal collective path which may differ from the constrained HF one~\cite{wen2016}.
 Another advantage of the ASCC method is to self-consistently determine the inertial mass to be used. 
 Wen and Nakatsukasa have applied the method recently to $\al,^{16}$O$+^{16}$O collisions~\cite{wen2017}.
 Although the calculations still use a simple functional (BKN), the method shows some promise in 
 automatically selecting the collective degrees of freedom which are relevant for the chosen reaction path (e.g., fusion).

\subsection{Superfluid dynamics for heavy-ion collisions and fission reactions \label{sec:pairing}}

It is well known that pairing plays a vital role in studies of structure of ground and excited states of nuclei~\cite{ring1980}.
The inclusion of pairing degrees of freedom in structure calculations is implemented via the 
Hartree-Fock-Bogoliubov (HFB) or the Hartree-Fock combined with BCS (HF+BCS)
approaches (see, e.g.,~\cite{dobaczewski1996,terasaki1996,stoitsov2000,bender2003,stoitsov2003,vretenar2005,blazkiewicz2005}).
Importance of pairing in nuclear fission has been the subject of a number of recent 
studies of multi-dimensional potential energy surfaces (PES)~\cite{staszczak2009,sadhukhan2013,sadhukhan2014,schunck2016,sadhukhan2016}.
It was also recognized as an essential ingredient to  investigations of the evolution from saddle to scission 
in fissioning nuclei with time-dependent microscopic approaches~\cite{negele1978}. 
One of the most important contributions of pairing to self-consistent mean-field calculations is the
ability of the system to allow for level crossings, which results in fragments establishing their
identity between the saddle and scission points~\cite{sadhukhan2013,sadhukhan2014,simenel2014a,scamps2015a,zhang2016,sadhukhan2017}.
Pairing is also expected to play a role as a residual interaction in dynamical calculations of heavy-ion collisions,
giving flexibility to the system to attain more compact shapes in fusion, influence transfer and breakup, or
lowering the effect of spherical magic shells and open other magic numbers for final fragment formation
in fission and quasifission studies~\cite{tao2017}.
While it is clear that pairing interaction plays an important role in multi-nucleon transfer reactions and for fission fragments acquiring their
identity after passing the saddle point, the influence of pairing in fusion, which involves
high excitation, is not so well established and is still under investigations~\cite{ebata2012,ebata2014a,scamps2015,hashimoto2016,magierski2017,scamps2018b}. 

\subsubsection{Time-dependent Hartree-Fock Bogoliubov calculations \label{sec:TDHFB}}

Formally, the time-dependent Hartree-Fock-Bogoliubov (TDHFB) theory should be the tool of choice for treating pairing
in dynamical calculations. 
To date, most applications of TDHFB have been confined to the small amplitude limit as
linear response calculations~\cite{matsuo2001,khan2002,avez2008,ebata2010,stetcu2011,hashimoto2012,nakatsukasa2016}.
The direct coordinate
space implementation of TDHFB equations for large amplitude reaction phenomena 
appears to be very complicated~\cite{jin2017}, requiring a vast computational effort, albeit recently significant progress
has been done in application to scission dynamics~\cite{bulgac2016}, neutron star crust vortex dynamics~\cite{wlazlowski2016}, and heavy-ion collisions~\cite{hashimoto2016,magierski2017,scamps2017b}.

In the traditional method to solve the HFB equations we express the many-body wavefunction in terms of the
Bogoliubov quasiparticle vacuum~\cite{dobaczewski1984,stoitsov2003,teran2003,blazkiewicz2005}, resulting in
the HFB supermatrix. 
As for the TDHF equation, the TDHFB equation can be derived starting from the action
between an initial and final time $t_i$ and $t_f$
\begin{equation}
S = \int_{t_i}^{t_f} dt \, \, \bra{\Psi(t)} i \hbar \frac{\partial}{\partial t} -\hat{H} \ket{\Psi(t)}
\label{PVfunc}
\end{equation}
 (see section~\ref{sec:Dirac}).
The difference with the independent particle case (TDHF) is that now the state $|\Psi(t)\>$ 
is constrained to be a  quasiparticle vacuum instead of a Slater determinant.
The basis of quasiparticle annihilators 
$\{\hat{\beta}\}$ is defined such that
 $ \hba_\mu \ket{\Psi} = 0$ for all $\mu$~\cite{ring1980} and can be related to 
 the particle creation and annihilation operators 
$\{\hat{a}^\dagger,\hat{a}\}$ through the Bogoliubov transformation~\cite{bogoliubov1958}
\begin{equation}
\hat{\beta}_\mu = \sum_\nu \left( U^*_{\nu \mu} \hat{a}_\nu 
+ V^*_{\nu \mu} \hat{a}^\dagger_\nu\right).
\label{eq:betadag}
\end{equation}

The variational principle $\delta S=0$ leads to the TDHFB equation~\cite{blaizot1986}
\begin{eqnarray}
i \hbar \frac{\partial}{\partial t} \Rbogo= \com{\Hbogo, \Rbogo},
\label{eq:TDHFB}
\end{eqnarray}
with the generalized one-body density matrix 
\begin{eqnarray}
\Rbogo(t)= \Smat{ \rho(t) }{ \kappa(t) }{ -\kappa^*(t) }{ 1-\rho^*(t) }
\end{eqnarray}
and the generalised Hamiltonian
 \begin{eqnarray}
\Hbogo=\Smat{ h }{ \Delta }{ -\Delta^* }{ -h^* }.
\end{eqnarray}
In addition to the one-body density matrix $\rho$ and the single-particle Hamiltonian $h$, 
we now have to account for the pairing tensor
 $\kappa_{\mu\nu}= \left< \psi|\anni_\nu \anni_\mu |\psi\right>$
as well as the pairing field 
\oeq
\Delta_{\mu\nu}=\frac{\delta \mathcal{E}[\rho,\kappa,\kappa^*]}{\delta
 \kappa^*_{\mu\nu}},
 \label{eq:field}
\ceq
where $\mathcal{E}[\rho,\kappa,\kappa^*]$ is the energy density functional accounting for pairing. 
In practice, the TDHFB equation is solved as a function of the 
 quasiparticle components $U$ and $V$ 
following
\begin{eqnarray}
i \hbar \frac{\partial}{\partial t} \Svec{U_{\nu \mu}}{V_{\nu \mu}} = 
\sum_{\eta} \Smat{h_{\nu \eta}}{\Delta_{\nu \eta}}{-\Delta_{\nu \eta}^*}{-h_{\nu \eta}^*}
\Svec{ U_{\eta \mu} }{ V_{\eta \mu} }.
\label{eq:TDHFB_QPcomp}
\end{eqnarray}

In calculations based on the Skyrme energy density functional, 
the  dependence  of the EDF on the pairing tensor is often written as 
a local pairing functional 
(see, e.g.,~\cite{bender2003} and references therein)
 \begin{eqnarray}
\mathcal{E}_{pair} &=& \int \!\!\! \mbox{d}\mathbf{r}
\,\,\,  \frac{g}{4} \, % \mbox{~g~} 
\left[ 1 - \left(\frac{\rho_0\ofR}{\rho_c}\right)^\gamma \right]% \times 
%\nonumber \\&&
 \sum_{q} \tilde{\rho}_q^*(\mathbf{r})\,  \tilde{\rho}_q(\mathbf{r}) 
\label{pairfunc}
\end{eqnarray}
where $\rho_q(\mathbf{r})=\sum_{s} \rho_q(\mathbf{r}s,\mathbf{r}s)$
and $\tilde{\rho}_q(\mathbf{r})=\sum_{s} \tilde{\rho}_q(\mathbf{r}s,\mathbf{r}s)$, and 
$\tilde{\rho}_q(\mathbf{r}s,\mathbf{r}'s') = -2s'\kappa_q(\mathbf{r}s,\mathbf{r}'-s')
$
is the anomalous density ($q$ and $s$ are the nucleon isospin and spin, respectively).
However, this choice of pairing functional leads to divergences~\cite{bulgac1999,bulgac2002b,dobaczewski1996} and 
it is necessary  to introduce a regularization scheme, e.g.,  by using a cutoff in the quasiparticle spectrum.
For instance, a quasiparticle
energy window of $80$~MeV (allowing for  two-quasiparticle excitations up to 160 MeV)
was necessary to ensure a good convergence in 
 a study of pairing vibrations with the TDHFB equation solved in spherical symmetry~\cite{avez2008}.
 Such high energy cutoff is often prohibitive in symmetry unrestricted calculations.
 
Because it is not an eigenstate of the particle number operator, 
a HFB ground state is defined up to a gauge angle. 
As a result, heavy-ion collisions computed in TDHFB exhibit a dependence 
with the relative gauge angle between the collision partners~\cite{hashimoto2016,magierski2017,scamps2017b}.
In particular, fusion thresholds are shown to  increase significantly when the gauge angles 
do not match in these calculations for symmetric head-on collisions. 
As shown by an analysis of frozen nucleus-nucleus potentials, this phenomenon is partly due to static effects~\cite{hashimoto2016}.
Nevertheless, dynamical effects such as solitonic excitations could contribute to the dissipation process~\cite{magierski2017}. 
Comparing fusion in superfluid and non-superfluid systems could lead to experimental evidence of these effects~\cite{scamps2018b}.
 
\subsubsection{Time-dependent BCS approximation}

 An alternate and much simplified approach can be obtained by expressing HFB solutions in the 
BCS form~\cite{blocki1976,reinhard1997,ebata2010,scamps2012}
\begin{equation}
|\Psi(t)\rangle = \prod_{k>0} \left( u_k(t) + v_k(t) a^{\dagger}_k(t) a^{\dagger}_{\overline k}(t) \right) | 0 \rangle, \label{eq:tdstate}
\end{equation} 
where $a^{\dagger}_k$ is the creation operator of a nucleon in the $k$-th HFB canonical or natural basis state.
The quantities $u_k(t)$ and $v_k(t)$ establish the link between the canonical states and the quasiparticle states.
This assumption results in a diagonal one-body density matrix as well as a diagonal pairing density.
The corresponding time-dependent equations can then be written in terms of the occupation numbers $n_k (t)= v^2_k(t)$ of single-particle states~\cite{ebata2010,scamps2012}
and anomalous density components $\kappa_k(t) = u_k^*(t) v_k(t)$
\begin{equation}
i\hbar\frac{d n_k}{dt} = \Delta^*_k \kappa_k - \Delta_k \kappa^*_k, \quad
i\hbar\frac{d \kappa_k}{dt} =  \kappa_i(\epsilon_k-\epsilon_{\overline{k}}) + \Delta_k(2n_k-1)\;,
\end{equation}
where $\Delta_k$ is the pairing field. 

There are numerous advantages of using the TDHF+BCS (also called TDBCS or canonical basis TDHFB in the literature)
approach instead of TDHFB.
First, the canonical states are localized and therefore instead of dealing with a very large number of states
proportional to the box volume we can deal with a much smaller number of states proportional to the nuclear
volume. Second, the evolution of the system closely resembles TDHF equations
\begin{equation}
i\hbar \frac{d}{dt}|\varphi_k\rangle = (h[\rho] - \epsilon_k(t)) |\varphi_k\rangle\;,
\end{equation}
where $h[\rho]$ is the self-consistent mean-field and
$\epsilon_k(t) = \langle \varphi_k |h[\rho]| \varphi_k \rangle $~\cite{scamps2012}.

Although the TDHF+BCS approach has been introduced a long time ago \cite{flocard1978},
it was applied with a simple ``constant G'' pairing interaction. 
Without an energy cut-off, this leads to only one degree of freedom for the pairing which is the pairing gap $\Delta$. 
In modern TDHF+BCS calculations with density-dependent interactions, the pairing field in the canonical basis 
$\Delta_{i\bar{i}}$ changes during the evolution and presents a non-integrable phase evolution different for each state $i$.

A word of caution when using the BCS approximation is necessary regarding the single-particle states
with positive energy. Use of these states should be avoided as they create an artificial low-density gas in the
numerical box and interfere with the time-evolution of the system by introducing an artificial damping due to
the interaction with box boundaries.
 It is also important to note that TDHF+BCS violates the one-body continuity equation.
This can lead to spurious transfer between fragments \cite{scamps2012}. 
In realistic calculations, one needs to monitor the effect. 
Although it can leads to uncertainties in transfer reactions \cite{scamps2013a},
it does not have a significant impact in fission studies \cite{scamps2015a,tanimura2015,tanimura2017}.
In addition, as discussed in section~\ref{sec:TDHFB}, in TDHF+BCS or 
TDHFB approaches the initial superfluid collision partners can 
each have different gauge angles. Realistic calculations of heavy ion
collisions would then require to sum these gauge angles coherently, 
thus calling for beyond mean-field techniques~\cite{scamps2017b}.

\begin{figure}
\centering
\includegraphics[width=9cm]{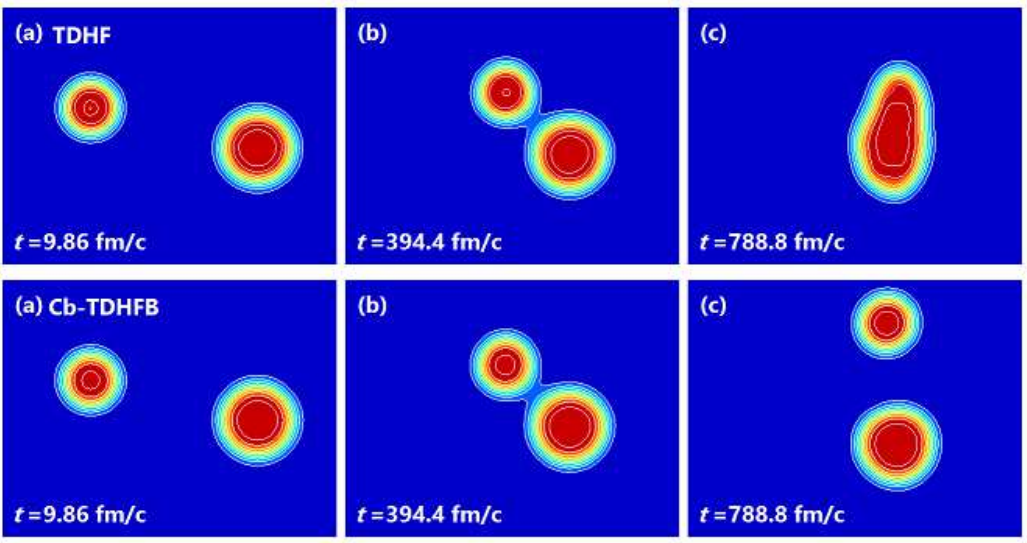}
\caption{TDHF (top) and TDHF+BCS (bottom) neutron density evolutions in $^{22}$O$+^{52}$Ca at $E_{cm}=25$~MeV and b=4.1~fm~\cite{ebata2014a}.
}
\label{fig:22O+52Ca}       % Give a unique label
\end{figure}

An interesting question is whether the TDHF+BCS approximation would capture the main features observed in TDHFB calculations. 
As an example, Fig.~\ref{fig:22O+52Ca} shows a comparison of density evolutions obtained from TDHF 
and TDHF+BCS in $^{22}$O$+^{52}$Ca at $E_{cm}=25$~MeV and for impact parameter $b=4.1$~fm~\cite{ebata2014a}.
It is observed that the inclusion of pairing dynamics at the BCS level prevents the system to fuse. 
Similar observations were also made by the full TDHFB calculations~\cite{hashimoto2016,magierski2017} discussed in section~\ref{sec:TDHFB}.
This is an indication that TDHF+BCS contains at least part of the dissipative mechanisms induced by the pairing correlations. 
Of course, the BCS approximation is much simpler than TDHFB, and some effects present in TDHFB could be missed in TDHF+BCS,
e.g., effects induced by the non-uniformity of the pairing field~\cite{magierski2017}.

\subsubsection{TDHF+BCS with the frozen occupation approximation}

When static pairing correlations are important for a proper description of the initial state
(e.g., to have a realistic deformation of the collision partners)
but are not expected to play a crucial role during the dynamics, 
then the so-called \textit{frozen occupation approximation}
 (FOA) is sometimes used.
 The FOA  further simplifies the BCS approach by  
freezing the occupation numbers for the time-evolution.
The FOA is expected to be a good approximation at the early stages of heavy-ion collisions, up to the capture of the collision partners. 
%, for example calculations of fusion and fusion barriers as evidenced by comparisons to experiments. 
For instance, TDHF+FOA calculations have been used to investigate above-barrier capture cross-sections with exotic nuclei~\cite{wakhle2018}. 

In Ref.~\cite{scamps2012} a comprehensive model study of TDHF+BCS and TDHFB approaches showed that the TDHF+BCS
approximation with time-dependent occupation numbers violates the continuity equation and results in
unphysical results for particle emission, while  the FOA approximation is found to be stable and
provide reasonable outcomes.
The parallel TDHFB study was found to give a good description of particle emission for short times but also
suffered from unphysical oscillations at later times. 

Applications of TDHF+BCS to study isovector dipole~\cite{ebata2014} 
and isoscalar and isovector  quadrupole vibrations~\cite{scamps2013b,scamps2014a}
were performed without the FOA approximation. 
However, in ~\cite{scamps2013b,scamps2014a} Scamps and Lacroix performed a comparison between
full TDHF+BCS results and  calculations using the FOA approximation.
An almost perfect agreement was found between the two levels of approximation.
Comparison to cases
without pairing was not straightforward since without pairing most of these nuclei are deformed. To remedy this
problem the filling approximation was used. It was found that the mean collective energy depends strongly on the
energy density functional, while other effects such as pairing resulted in local fluctuations about this mean.
However, the pairing was found to be essential when considering low-lying states. Note also that an excellent comparison
between the TDHF+BCS and QRPA (which is obtained by linearizing the full TDHFB equation) approaches was obtained.

%Several investigations have been done using the TDHF+BCS approach. The applications to study 
Another example of application of the FOA is to study transfer reactions (see also section \ref{sec:projection}). 
The effect of pairing on two-particle transfer has been investigated % with the FOA approximation 
by Scamps and Lacroix~\cite{scamps2013a,scamps2015}. It was found that
the pairing had a little effect on one-particle transfer but significant effect on two-particle transfer,
albeit still not enough transfer was obtained. The primary source of the increase in two-particle transfer was
traced to the partial occupations of the single-particle states.
This suggests that beyond mean-field approaches may be necessary for a
proper description of two-particle transfer. % but a full TDHFB study is also required to understand the effect
%of the FOA approximation. 

\subsubsection{Applications to fission  \label{sec:fission}}

Perhaps the most striking example of recent applications of TDHF and its extensions in the past few years 
has been to  fission~\cite{simenel2014a,goddard2015,scamps2015a,goddard2016,bulgac2016,tanimura2015,tanimura2017,scamps2018}  studies. 
These works have been performed with different level of approximation to the many-body dynamical problem, including standard TDHF, TDHF with BCS pairing correlations at the FOA level, and dynamical pairing with TDHF+BCS and TDHFB approaches.

\begin{figure}
\centering
\includegraphics[width=7cm]{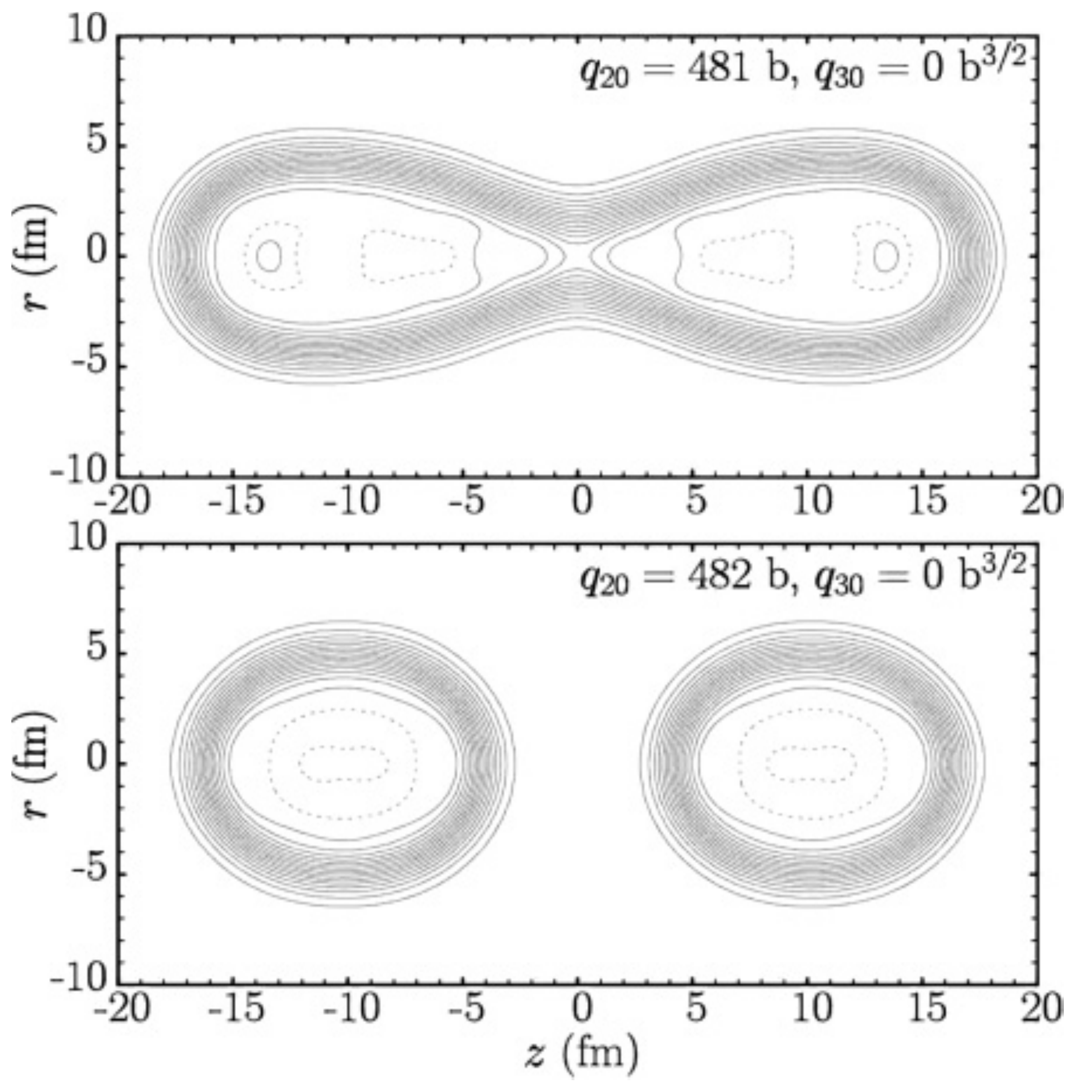}
\caption{Symmetric scission configurations of $^{226}$Th before (up) and after (down) scission~\cite{dubray2008}.
The quadrupole and octupole moments $q_{20}$ and $q_{30}$, respectively, are used as external constraints in the HFB calculations. 
}
\label{fig:scission}       % Give a unique label
\end{figure}

The description of fission with time-dependent microscopic theories has been a longstanding problem (see~\cite{schunck2016} for a recent review). 
One approach is to construct a fully microscopic potential energy surface (PES) using deformation constrained HFB states 
and to solve the time-dependent generator coordinate method (TDGCM) 
equation with the Gaussian overlap approximation (GOA)  to predict the dynamical evolution of the collective wave-function~\cite{goutte2005,regnier2016}. 
This approach is well suited to describe the overcoming of the barrier as well as fluctuations and repartition between fission valleys. 
However, the underlying adiabatic assumption used to construct the static constrained HFB states 
should break down close to scission where non-adiabatic effects are expected to play a role. 
As an illustration, Fig.~\ref{fig:scission}~\cite{dubray2008} shows the two consecutive HFB states obtained 
just before and just after scission (notice the very small difference in the quadrupole moment $q_{20}$). 
If pushed to the extreme,  the adiabatic approximation would lead to the spurious description that the systems evolves from the pre-scission 
to the post-scission state in one step, while what is expected is a rapid but continuous evolution at scission. 
The idea is then to use TDHF based approaches to improve the description of the microscopic states near scission. 

\begin{figure}
\centering
\includegraphics*[width=6cm]{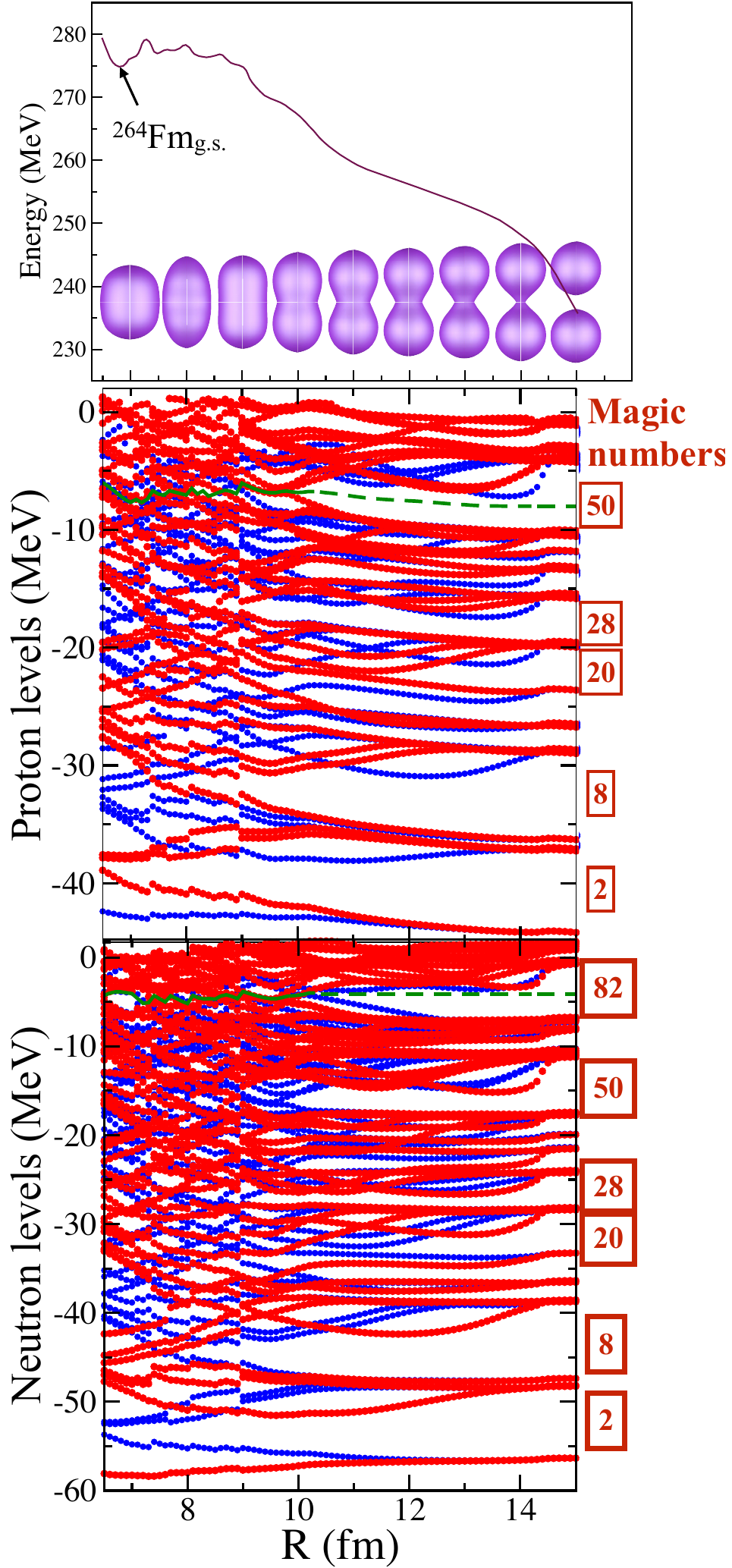}
\caption{Constrained HF+BCS calculations of $^{264}$Fm symmetric fission valley with the SLy4$d$ Skyrme functional and surface pairing as a function of the distance between the fragments $R$. (top) Adiabatic potential and isodensities at half the saturation density $\rho_0/2=0.08$~fm$^{-3}$. Single-particle levels are plotted for protons and neutrons in the middle and bottom panels, respectively. Positive and negative parity states are shown in red and blue, respectively. Adapted from~\cite{simenel2014a}.
}
\label{fig:264Fm}       % Give a unique label
\end{figure}

Fission dynamics along the mass symmetric valley of $^{264}$Fm has been investigated with TDHF calculations~\cite{simenel2014a}. 
As the $^{132}$Sn fragments are doubly magic, it is expected that pairing correlations disappear somewhere between the saddle and scission configurations,
with the formation of magic gaps in the single-particle energy spectra signing the pre-formation of the fragments, as shown in Fig.~\ref{fig:264Fm}.
Before the magic gap is formed, jumps in the single particle energies are observed due to single particle levels crossing the Fermi surface 
and inducing a  change of the shape of the system~\cite{dubray2012}. 
These jumps are often associated with barriers in the PES which cannot be overcome in mean-field based descriptions 
(even if one use the more elaborate TDHFB treatment) due to lack of quantum many-body tunneling.  
Time-dependent mean-field calculations then need to be started at configurations outside these barriers. 

The $^{264}$Fm symmetric fission is a special case as it can be studied at the TDHF level.
However, in the vast majority of systems, the inclusion of pairing correlations  at some level is mandatory. 
Goddard, Stevenson and Rios have investigated the dynamics of deformation and boost induced fission with the FOA~\cite{goddard2015,goddard2016}.
Although the systems could be made to fission in some cases, 
it was shown that the properties of the fragments (mass, charge, kinetic energies) 
were depending on the choice of the initial configuration used in the time-dependent calculation.
Nevertheless, a reasonably good agreement with experiment was achieved, showing promise for the technique. 
\begin{figure}[!htb]
    \centering
    \includegraphics[width=8cm]{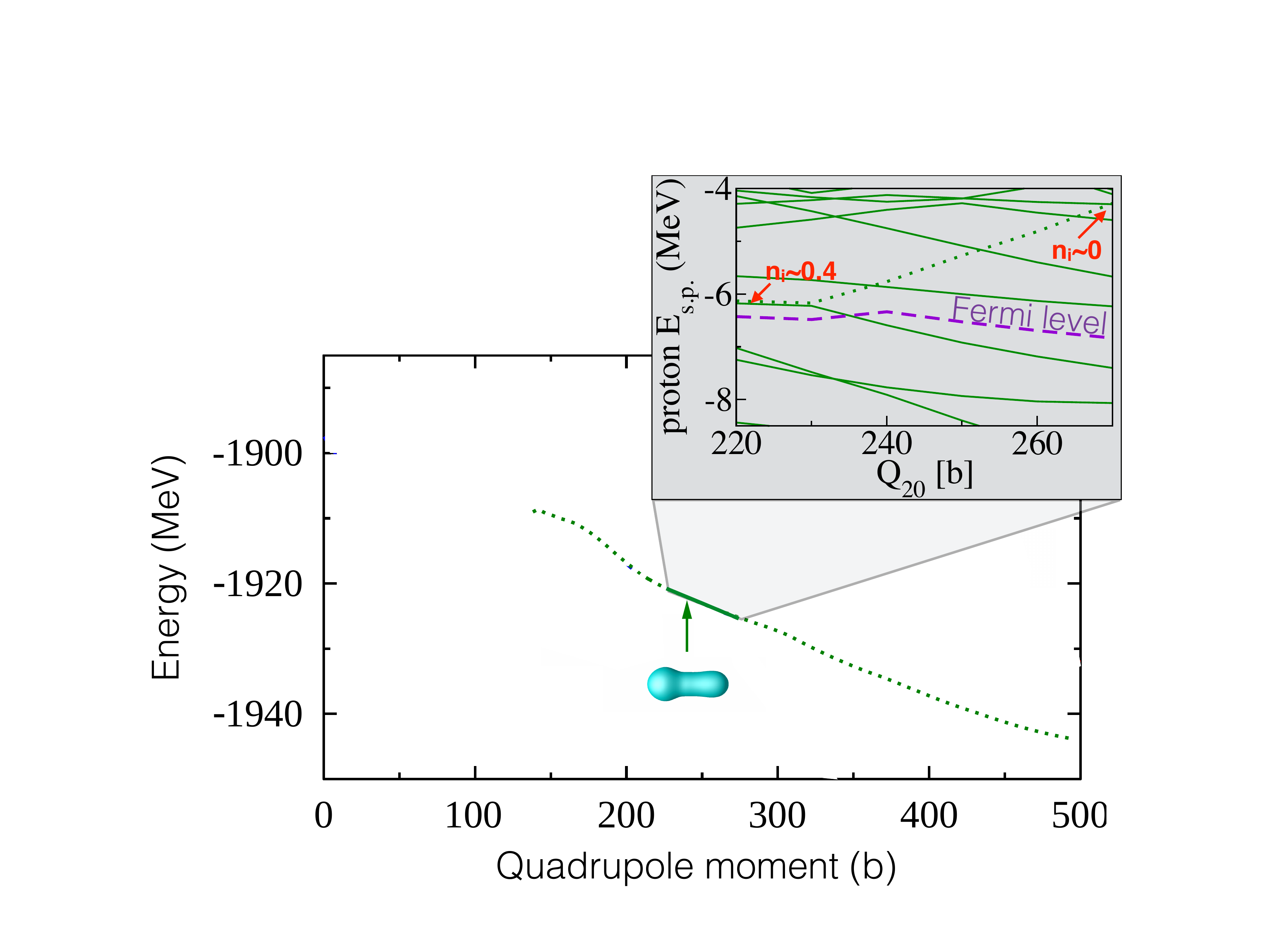}
    \caption{Potential energy in the asymmetric fission valley of $^{258}$Fm. 
        The inset shows the single-proton energies for a smaller range of deformations, with the Fermi energy shown by the dashed line and a specific level leaving the Fermi surface shown with a dotted line. Occupation numbers for this level are indicated.  Adapted from~\cite{scamps2015a}.
    }
    \label{fig:258Fm}       % Give a unique label
\end{figure}

\begin{figure}
    \centering
    \includegraphics[width=6cm]{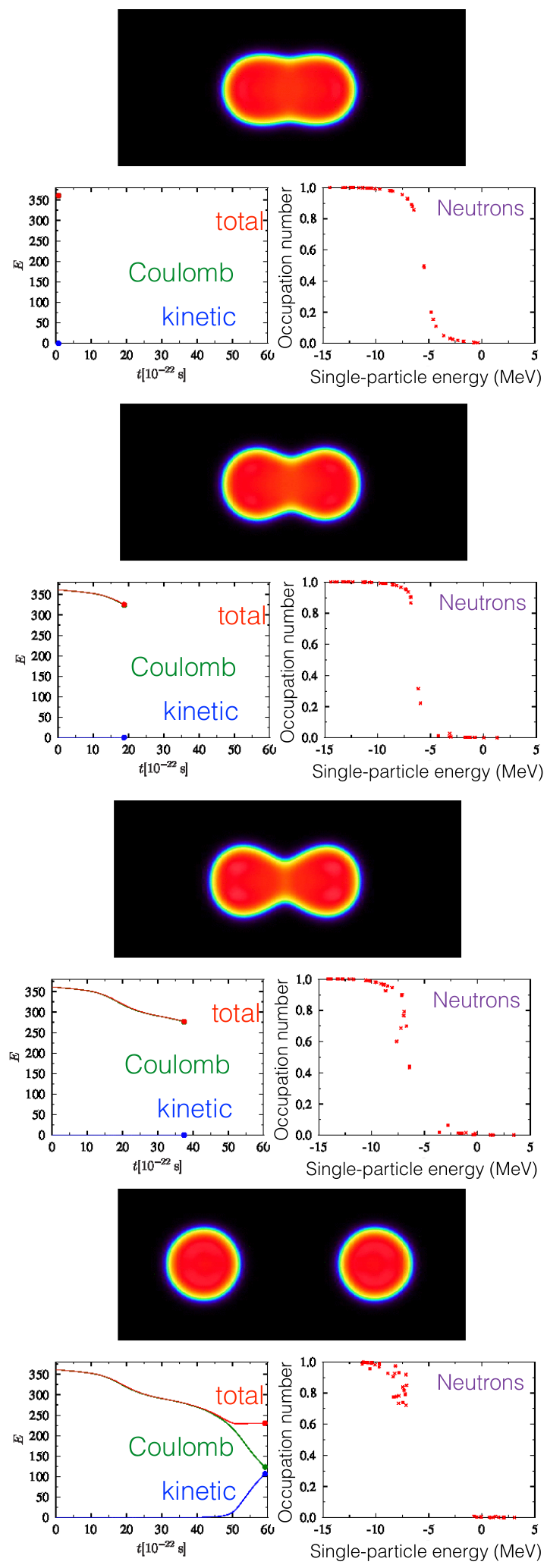}
    \caption{Density, energy repartition, and neutron occupation numbers in $^{258}$Fm symmetric fission with TDHF+BCS. Adapted from~\cite{scamps2015a}.
    }
    \label{fig:pairing}       % Give a unique label
\end{figure}

The dependence of the fission fragment properties with the initial configuration in the saddle to scission potential energy descent 
is to a large extent reduced when pairing is accounted for in a dynamical fashion. 
Scamps and collaborators have used the TDHF+BCS approach to study the fission dynamics of $^{258}$Fm~\cite{scamps2015a}. 
The time-dependent pairing correlations are able to describe the fact that when a single particle level leaves the Fermi surface, 
its occupation number goes to zero, as shown in Fig.~\ref{fig:258Fm}.  
The reorganisation of the neutron occupation numbers in the symmetric fission of $^{258}$Fm is also shown in Fig.~\ref{fig:pairing} 
together with the density evolution and the repartition of the energy in terms of kinetic and Coulomb potential energies. 
One particular advantage of these time-dependent descriptions of fission dynamics is that the final properties of the fission fragments
do not depend on how the scission configuration is defined.

\begin{figure}
\centering
\includegraphics[width=5cm]{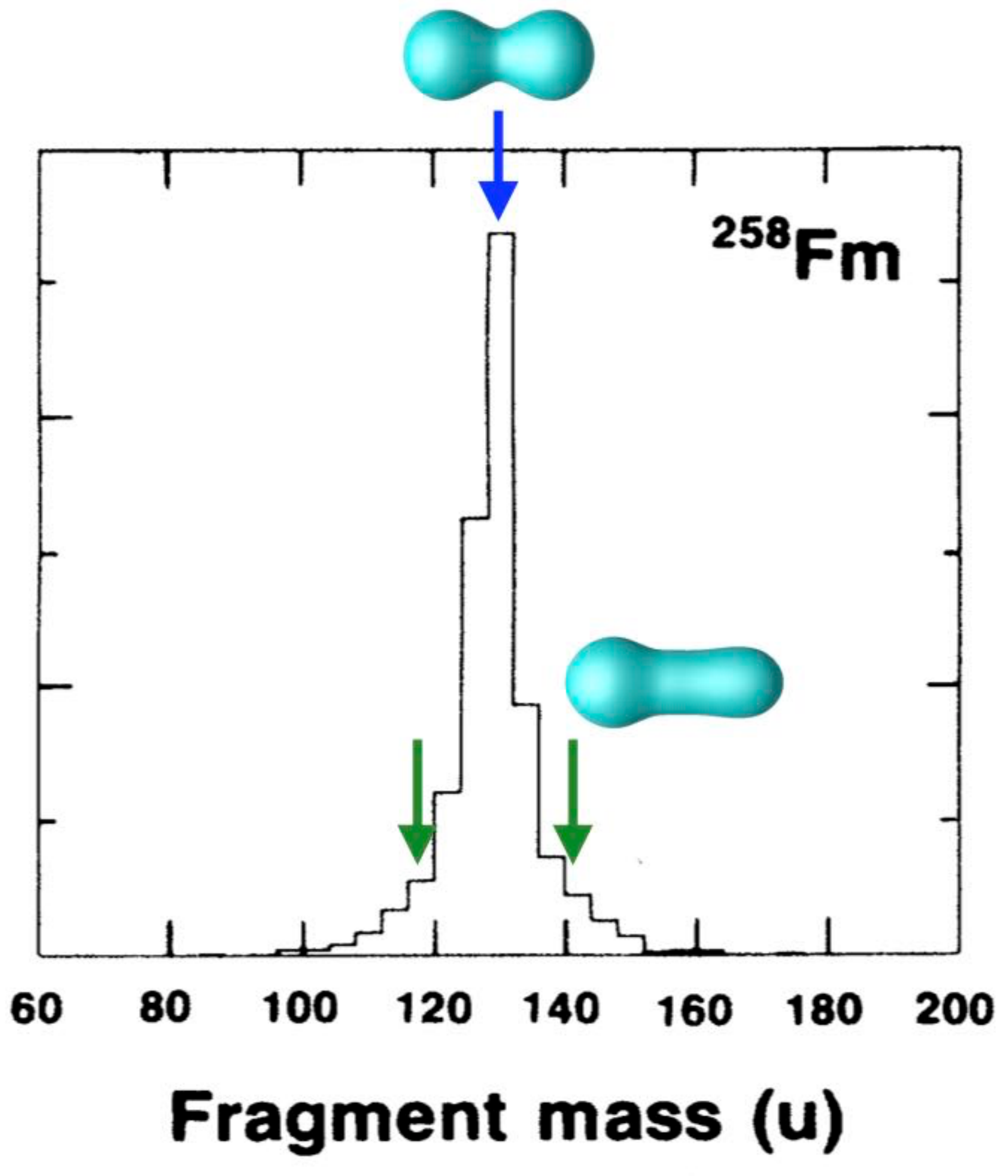}
\includegraphics[width=6cm]{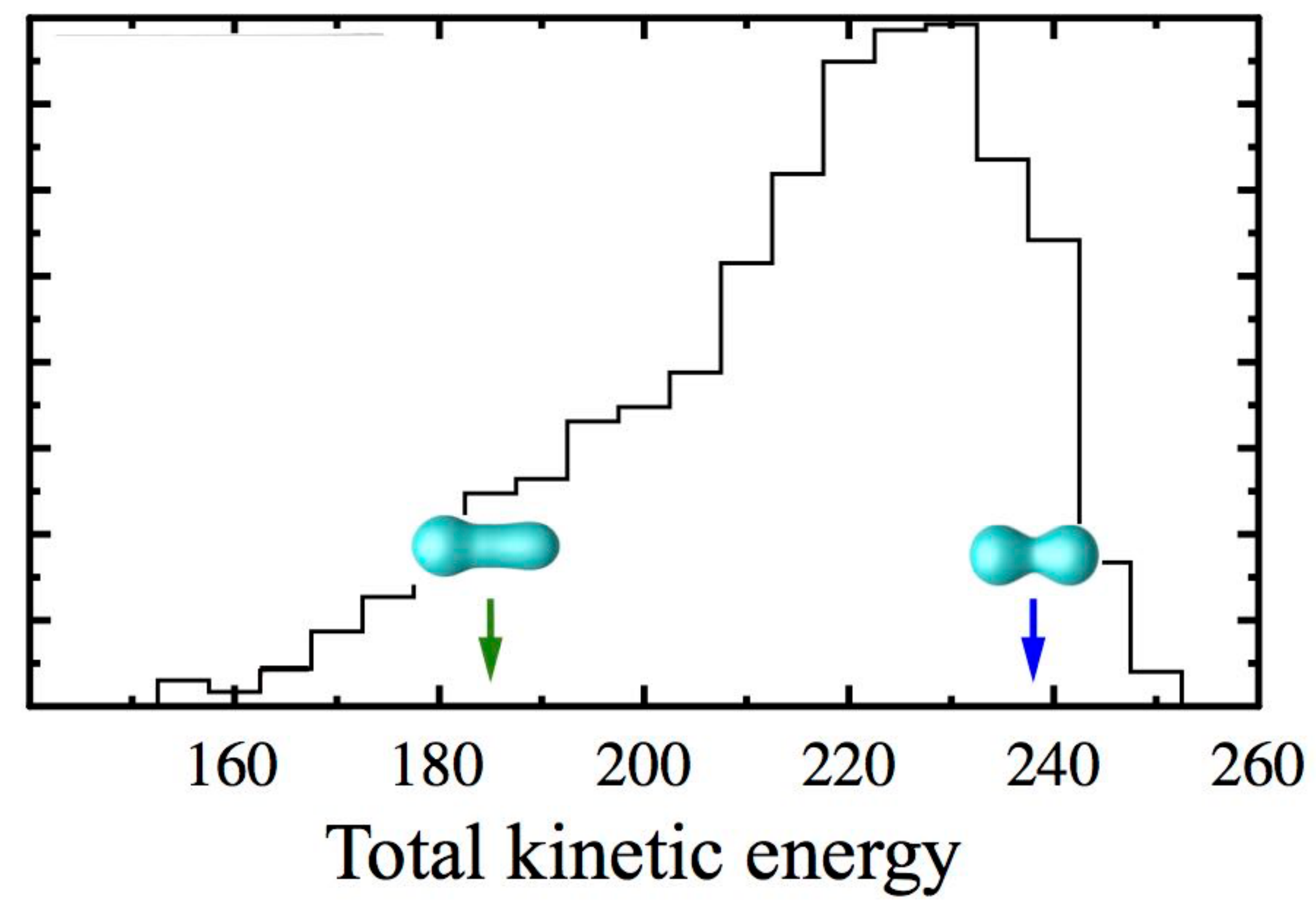}
\caption{Experimental mass (left) and TKE (right) distributions in $^{258}$Fm compared with TDHF+BCS predictions for the symmetric compact and asymmetric modes. Experimental data are from~\cite{hulet1986}. Adapted from~\cite{scamps2015a}.
}
\label{fig:TKE}       % Give a unique label
\end{figure}

In particular, this leads to unambiguous determination of the total kinetic energy (TKE) of the fragments. 
The TKE is obtained from the sum of the Coulomb and kinetic energies after scission.
Indeed, the nuclear interaction between the fragments has vanished when the fragments are well separated 
and the sum of the Coulomb and  kinetic energies remains constant. 
(Of course, asymptotically, all the Coulomb potential energy is transformed into kinetic energy.) 
This allows direct comparison of the predicted average mass and TKE of the fragments  with experimental distributions as illustrated in Fig.~\ref{fig:TKE}.

Recently, Bulgac and collaborators have used their TDHFB code to investigate fission dynamics of $^{240}$Pu~\cite{bulgac2016}.
An interesting finding is that the saddle to scission time seems to be longer than in the earlier studies described above. 
The explanation invoked in~\cite{bulgac2016} is that the system is allowed to explore more degrees of freedoms 
as the pairing field has  less spatial restrictions  than in TDHF+BCS. 
However, it has since been shown that TDHF+BCS could lead to similarly long saddle-to scission times \cite{tanimura2017,scamps2018}.

A drawback of the present mean-field approaches is that they do not account for the fluctuations of collective variables. 
If included, these fluctuations would allow a direct a comparison with the experimental distributions, i.e., not only the averages. 
The difficulty is that a large part of the fluctuations is expected to come from the early stage of the fission process,
when the barriers are overcome and where the evolution is essentially slow and adiabatic. 
Nevertheless, Tanimura, Lacroix and Ayik have recently applied their stochastic mean-field method (see section~\ref{sec:beyond}) 
to the symmetric fission of $^{258}$Fm and they showed that large fluctuations could be expected in the saddle to scission evolution~\cite{tanimura2017}.

\subsubsection{Applications to quasi-fission  \label{sec:QF}}

Applications of time-dependent microscopic methods to  quasi-fission has also attracted lots of efforts in the  past few years 
\cite{golabek2009,kedziora2010,wakhle2014,oberacker2014,umar2015a,hammerton2015,umar2016,sekizawa2016,yu2017,morjean2017,wakhle2018}.
Quasi-fission is a mechanism in which two fragments are emitted after capture 
and before the formation of a fully equilibrated compound nucleus could be achieved~\cite{toke1985}. 
It occurs in reactions with heavy nuclei (typically with a product of the charges $Z_1Z_2>1600$),
and it is the main obstacle for the formation of a compound superheavy nucleus.
One major difficulty in experimental studies of quasi-fission is that the outgoing fragments share many similarities with fragments formed in fusion-fission 
(i.e., where an equilibrated compound nucleus has actually been formed), making it difficult to separate both mechanisms.

\begin{figure}
\centering
\includegraphics[width=7cm]{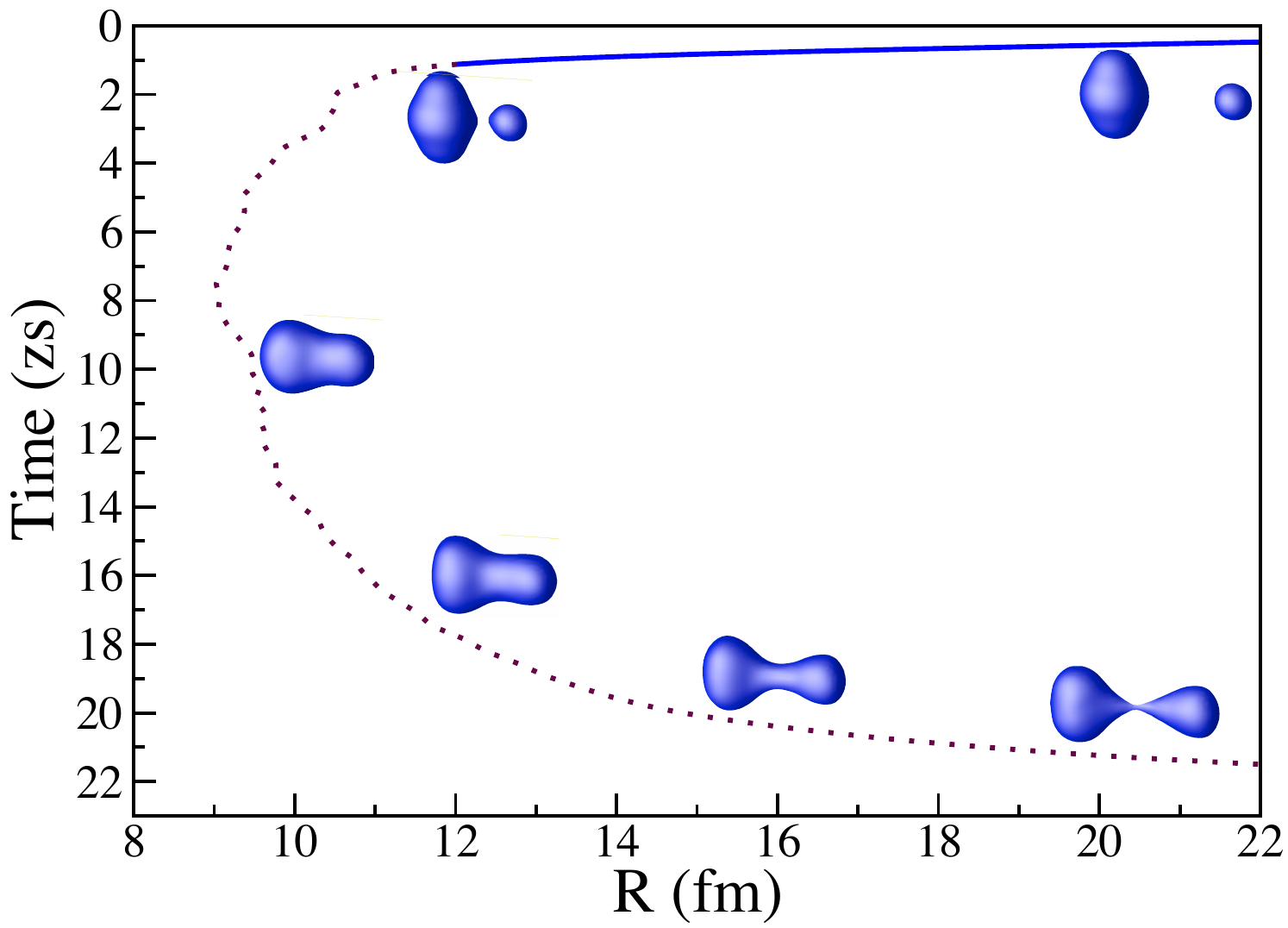}
\caption{ Time evolution of the distance between the centres of mass of the fragments in the $^{40}$Ca+$^{238}$U central collision  at $E_{cm}=205.2$~MeV~\cite{wakhle2014}. 
The snapshots show densities at half the saturation density. 
}
\label{fig:Ca+U}       % Give a unique label
\end{figure}

TDHF calculations have then been used  with the FOA to provide theoretical guidance and support in data analysis.
TDHF calculations can also be used to predict the energy threshold for fusion which is found to be higher than the static Coulomb barrier~\cite{guo2012,simenel2012,washiyama2015}.  
Figure~\ref{fig:Ca+U} shows a typical shape evolution observed in TDHF calculations of quasi-fission in the $^{40}$Ca$+^{238}$U system~\cite{wakhle2014}.
The contact time between the fragments of $\sim20$~zs is typical to quasi-fission.
Another characteristic of quasi-fission is the partial mass equilibration also observed in the final fragments ($^{103}$Mo$+^{175}$Yb) in Fig.~\ref{fig:Ca+U}.

An interesting finding of these TDHF studies of quasi-fission is the prediction of the role of shell effects
which favour the formation of magic fragments, in particular in the $Z=82$ region in reactions involving an actinide collision partner~\cite{wakhle2014}.
This prediction has been later confirmed experimentally by Morjean and collaborators~\cite{morjean2017}.
In addition, the calculations show that these shell effects strongly depend on the orientation of deformed actinide. 
Another finding is that only collisions with the side of the actinide can eventually lead to fusion.

%Similarly, the excellent agreement
%between the experiment and theory for mass-angle distributions and total kinetic energy (TKE) in 
%quasifission~\cite{wakhle2014,hammerton2015,umar2016} reactions also suggests that the major role
%played by pairing may be to allow for a wider range of fragment masses.

It would be interesting to apply the full TDHFB formalism to study quasi-fission reactions.
Indeed, TDHFB seems to predict long saddle to scission times \cite{bulgac2016}. 
The quasi-fission time distributions are relatively well known thanks to extensive experimental measurements of the correlations between 
the mass and the angle of the fragments~\cite{toke1985,durietz2013} and are thus ideal to benchmark microscopic theories. 

\subsection{Fragment particle number distributions in time-dependent mean-field theories\label{sec:projection}}

Predicting the properties of fragments formed in heavy-ion collisions or in fission, 
such as their mass, charge, angular momentum and kinetic energies, 
is particularly challenging when one wants to extract distributions for these observables. 
According to the Balian-V\'en\'eroni variational principle, the TDHF equation is optimized for the prediction of expectation values
of one-body observables, while their distributions (which involve quantum fluctuations) 
are in principle outside the range of application of the TDHF theory.
Nevertheless,  mean-field approaches based on the TDHF theory have been used to compute
 multi-nucleon transfer probabilities \cite{koonin1977,simenel2010,simenel2011,sekizawa2013,scamps2013a,sonika2015,
 sekizawa2016,scamps2017a,scamps2017b,sekizawa2017,sekizawa2017a,regnier2018} and to investigate particle number distributions in fission fragments \cite{scamps2015a,tanimura2017,williams2018}.

In TDHF calculations, the single-particle states are evolved according to Eq.~(\ref{eq:tdhf2}). 
Depending on the reaction, they can eventually spread in space once the fragments are formed.
While the single-particle wave functions are initially localized in a nuclear fragment at the beginning of a reaction,
they may partially transfer to the collision partner during the TDHF evolution~\cite{koonin1977,umar2008a,umar2009c}.
This process may result in a variation of the average particle number for each fragment, 
indicating that nucleon transfer has taken place during the reaction \cite{simenel2008}.

In quantum mechanics,  a system can be found in a coherent superposition of channels associated 
with different repartitions of particle numbers between its fragments. 
Formally, such a state is written as
\begin{equation}
|\Psi\>=\sum_N C_N |\Psi_N\>,
\end{equation}
where $|\Psi_N\>$ are eigenstates of a particle number operator $\oN$ counting the particles in a fragment:
\begin{equation}
\hat{N} |\Psi_N\> = N |\Psi_N\> .
\end{equation}
In the case of a nuclear reaction, the final fragments are entangled. 
For instance, measuring the number of particles in one fragment projects out the state of the other fragment. 

Note that, although the eigenvalue $N$ is an integer, $\<\oN\>\equiv\<\Psi|\oN|\Psi\>$ can be a real number, in particular when several coefficients $C_N$ are non-zero. 
Therefore, ``counting'' the particles involves the determination of  the distribution of probabilities $\mP(N)=|C_N|^2$. 
In particular, a non-integer value of $\<\oN\>=\overline{N}=\sum_N N|C_N|^2$ is a clear signature that the system is in a  superposition of different eigenstates of $\oN$.

\subsubsection{Evolution of single-particle states}

\begin{figure}[!htb] \begin{center}
\includegraphics[width=10.0cm]{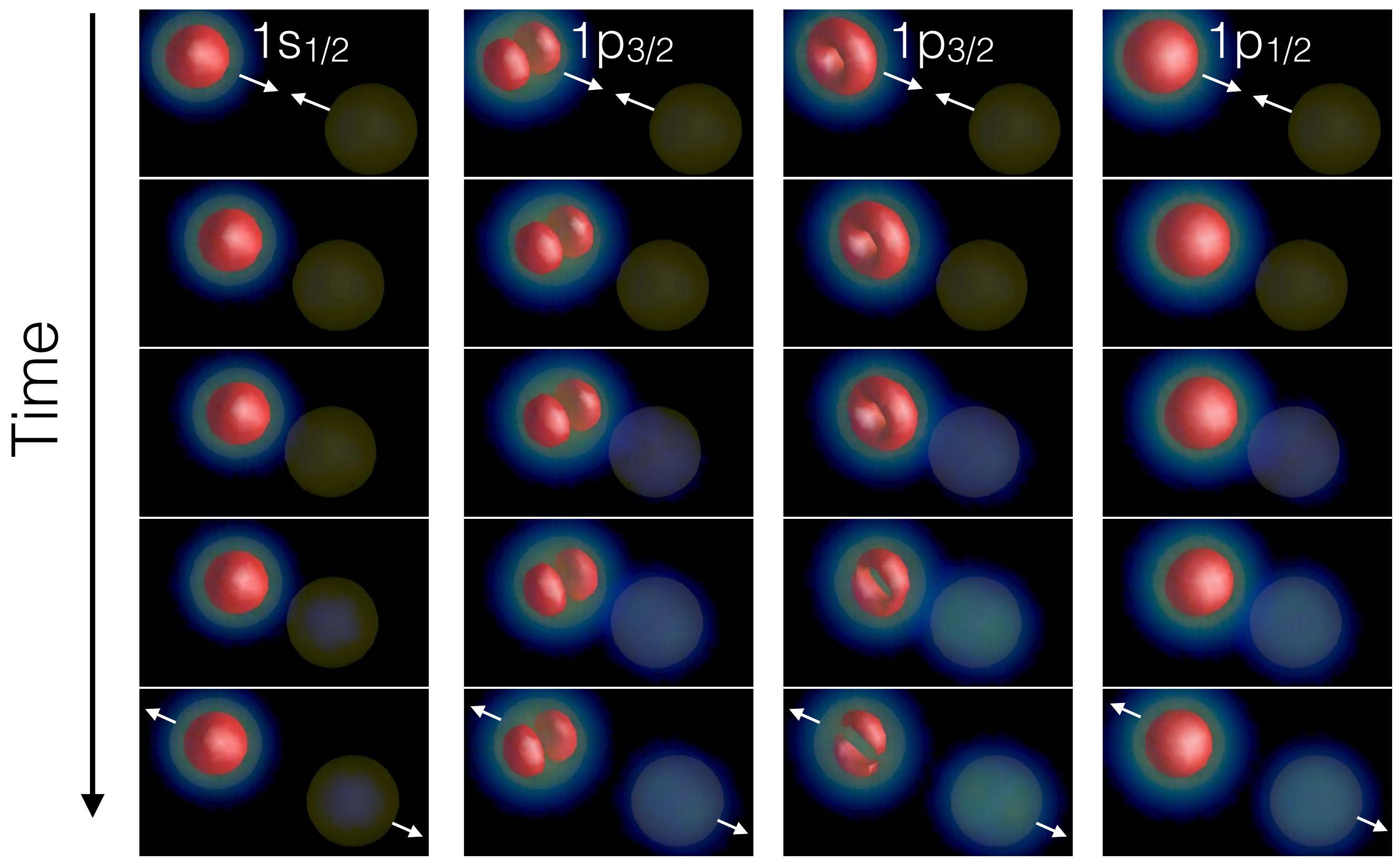}    
\caption{Time-evolution of single-particle probability spatial distributions associated with initial $^{16}$O neutron wave-functions in a near-barrier $^{16}$O+$^{16}$O central collision at $E_{cm}=10$~MeV.} 
\label{fig:sp_trans}
\end{center} \end{figure}

An example of  evolution of single-particle states is given in Fig.~\ref{fig:sp_trans} for a sub-barrier central collision of two $^{16}$O nuclei.
The figure shows a set of probability distributions $|\az_i(\vr,t)|^2$ at various times. 
Only neutron wave-functions initially located on the left fragment are shown. 
By symmetry, the same dynamics occurs with the wave functions initially in the right fragment. 
A partial transfer is observed to the collision partner for each wave-function. 
As expected, the transfer is more important for the least bound wave-functions ($p-$states in $^{16}$O)

In a semi-classical transport approach, we could approximate the variance $\sigma_N^2={\overline{N^2}-\overline{N}^2}$ of particle number in the fragments by the total number $N_{exc}$ of exchanged nucleons (in both directions) \cite{randrup1982}:
\oeq
\sigma_N^2\simeq N_{exc}.
\label{eq:var_class}
\ceq
Applying Eq.~(\ref{eq:var_class}) to the collision of two $^{16}$O shown in Fig.~\ref{fig:sp_trans}, we get $\sigma_N^2\simeq0.6$. 
In fact, this classical result is clearly unphysical as it does not account for the indistinguishability of identical particles. 
Consider a simple example where a single-particle state $|L\>$, initially in the left fragment, is partially transferred to the same state $|R\>$ in the right fragment:
$$|L\>\rightarrow \al|L\>+\be|R\>.$$
We can choose the phase such that $\alpha$ is real. If the collision is symmetric, we also have
$$|R\>\rightarrow \al'|R\>+\be'|L\>,$$
with $|\al'|=\al$ and $|\be'|=|\be|$.
Orthogonality of the states imposes $\al\be'+\be^*\al'=0$.
Thus, we have up to an irrelevant global phase
$$|R\>\rightarrow \al|R\>-\be^*|L\>.$$
For identical fermions, the state is antisymmetric. 
Up to a global phase and normalisation, we have
$$|L,R\>-|R,L\>\rightarrow  (\al^2+|\be|^2)\(|L,R\>-|R,L\>\).$$
Thus, the state is unchanged up to a global phase. 
As a consequence, we have $\sigma_N^2=0$ in this example. 
However, if we apply the semi-classical formula in Eq.~(\ref{eq:var_class}), we get $\sigma_N^2\simeq2|\be|^2$.
Therefore, Eq.~(\ref{eq:var_class}) is expected to overestimate the fluctuations of particle numbers in the fragments.

We see from the above example that we need a quantum approach in order to estimate the distribution of probability $\mP(N)$ to find $N$ nucleons in a fragment. 

\subsubsection{Variance}

Following Ref.~\cite{dasso1979}, we can calculate the variance $\sigma_N^2$ in TDHF using
anticommutation relations for fermions as well as the property
$$\<\oad_i\oad_j\oa_k\oa_l\>=n_in_j(\delta_{il}\delta_{jk}-\delta_{ik}\delta_{jl})$$
for a Slater determinant where $i$ and $j$ denote states in the canonical basis. 
The variance then reads
\oeq
\sigma_N^2=\<\oN_V\>-\sum_{i,j=1}^A |\<\az_i|\az_j\>_V|^2,
\label{eq:sig_TDHF}
\ceq
where 
\oeq
\oN_V = \int_V d^3r \, \oad(\vr) \oa(\vr)
\ceq
counts the number of particles in the volume $V$ and
\oeq
\<\az_i|\az_j\>_{V}=\int_{V} d^3r \, \az_i^*(\vr)  \az_j(\vr) \label{eq:sp_over}
\ceq
is the overlap of the single-particle wave-functions $\az_i$ and $\az_j$ over the volume $V$.
Spin and isospin degrees of freedom are omitted for simplicity. 
Note that the expectation value of $\oN_V$ can also be written with this notation as
\oeq
\<\oN_V\>=\sum_{i=1}^A \<\az_i|\az_i\>_V.
\ceq

Applying Eq.~(\ref{eq:sig_TDHF}) to compute the variance of the number of nucleons in the final fragments of the collision shown in Fig.~\ref{fig:sp_trans}, we get $\sigma_N^2$ of the order of $10^{-5}$, 
which is orders of magnitude smaller than the classical result obtained from Eq.~(\ref{eq:var_class}) as the indistinguishability of identical particles has been taken into account. 
Note that when the numerical value of the variance gets so small, it could easily be affected by an accumulation of numerical errors leading to small changes of the norm of the single-particle wave-functions during the evolution. 

\subsubsection{Combinatory approach}

We can go a step further and calculate the probability  $\mP(N)$ to find $N$ particles in a volume $V$ and the remaining $A-N$ particles in the complementary volume $\overline{V}$ using a simple combinatory approach \cite{koonin1977}:
\oeq
\mP(N) = \frac{A!}{N!(A-N)!}\int_V d^3r_1\cdots d^3r_N \int_{\overline{V}} d^3r_{N+1}\cdots d^3r_A |\Psi(\vr_1\cdots \vr_A)|^2.
\label{eq:combi}
\ceq
In the TDHF framework, the many-body wave-function is a Slater determinant of occupied single-particle wave-functions at all time. 
In this case, Eq.~(\ref{eq:combi}) leads to \cite{koonin1977,sekizawa2013}
\oeq
\mP(N) = \sum_{s(\{\tau_i\} : V^N\overline{V}^{A-N})} \det \{\<\az_i|\az_j\>_{\tau_i}\},
\label{eq:proj1}
\ceq
where $\tau_i=V$ or $\overline{V}$ [see Eq.~(\ref{eq:sp_over})]. 
The notation $s(\{\tau_i\} : V^N \overline{V}^{A-N})$ stands for all possible combinations of $N$ volumes $V$ and $A-N$ volumes $\overline{V}$.
Equation (\ref{eq:proj1}) becomes rapidly difficult to solve numerically when the number of particles $A$ increases due to the large number of determinants~\cite{sekizawa2013}.

\subsubsection{Particle number projection technique}

It is also possible to compute the probability to form a fragment with a given number of 
nucleons by employing the particle-number projection technique introduced in~\cite{simenel2010},
resulting in a distribution of multi-particle transfer probabilities.
The projection method has long been utilized in structure studies to resolve the particle-number symmetry violation~\cite{ring1980,valor2000,anguiano2001,samyn2004,dobaczewski2007,bender2008}.
To calculate transfer probabilities in heavy-ion collisions one can project the post-collision TDHF wave function, in a spatial volume $V$ containing the fragment, onto $N$ protons or neutrons that represent a certain transfer channel. 
The projection operator may be written as an integral over the gauge angle $\theta$ in the form~\cite{bender2003}
\begin{equation}
\hat{P}^q_V(N)=\frac{1}{2\pi}\int_{0}^{2\pi}d\theta\; e^{i\theta (\hat{N}^q_V - N)},
\end{equation}
where $\hat{N}^q_V$ counts the number of particles in the region $V$ with an isospin $q$. 
The probability of finding $N$ particles of isospin $q$ in the region $V$ 
%in the state $|\phi\rangle$ 
is~\cite{simenel2010}
\begin{equation}
\langle\Psi|\hat{P}^q_V(N)|\Psi\rangle=\frac{1}{2\pi}\int_{0}^{2\pi}d\theta\; e^{-i\theta N}\langle\Psi |e^{i\theta \hat{N}^q_V}|\Psi\rangle\;.
\label{eq:proj2}
\end{equation}
The expectation value in the integral of Eq.~(\ref{eq:proj2}) is the determinant of the overlap matrix
\begin{equation}
\langle\Psi|\Psi_V(\theta)\rangle=\text{det}(\mathbf{O}),
\end{equation}
with the state $|\Psi_V(\theta)\rangle = e^{i\theta \hat{N}^q_V}|\Psi\rangle$ representing a rotation in the 
gauge-space associated with the particle number degree of freedom.
The overlap matrix, $\mathbf{O}$, is given by
\begin{equation}
\left(\mathbf{O}\right)_{ij} = \sum_{\sigma}\int d\mathbf{r}\; \varphi^*_i\left(\mathbf{r}\sigma q\right)\varphi_j\left(\mathbf{r}\sigma q\right)e^{i\theta\Theta\left(\mathbf{r}\right)}=\delta_{ij}+\langle i|j\rangle_V\left(e^{i\theta}-1\right).
\end{equation}
The function $\Theta(\mathbf{r})$ takes the value of 1 in region $V$ and $0$ elsewhere.
The method allows one to obtain the particle number probability distributions for fragments by performing this procedure for differing values of $N$ in $V$.

Recently, the projection technique has been employed by Sekizawa and Yabana in the calculation of the expectation values of one-body and two-body operators using the particle-number projected states~\cite{sekizawa2014}. This extension allows for an analysis of other reaction observables associated with the target and projectile-like fragments that correspond to specific transfer channels, such as the excitation energy and intrinsic angular momenta of the number-projected
states~\cite{sekizawa2017}.
While the particle-number projection approach is a useful tool for obtaining
fragment mass and charge distributions, it often underestimates these in dissipative
collisions~\cite{dasso1979,simenel2011}. This suggests that other beyond mean-field correlations may be important.

\subsubsection{Numerical implementation}

\begin{figure}[!htb] \begin{center}
\includegraphics[width=7.0cm]{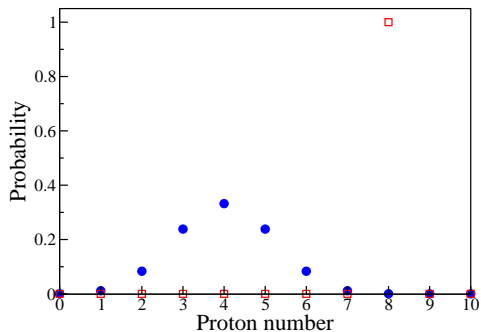}    
\caption{Proton number probability distribution with $V$ containing the entire $^{16}$O nucleus (open squares) and only half of it (full circles).} 
\label{fig:proba_16O}
\end{center} \end{figure}

Unlike expression (\ref{eq:proj1}), the number of determinants to compute in Eq.~(\ref{eq:proj2}) only depends on the number of discretised gauge angles $\theta$.
Numerical applications using Eqs.~(\ref{eq:proj1}) and~(\ref{eq:proj2}) lead to almost identical results \cite{sekizawa2013}.
In fact, the application of Eq.~(\ref{eq:proj2}) is rather straightforward. 

The shape and normalisation of the probability distribution $\mP(N)$ should be first tested in a simple case.
Let us consider the HF ground-state of a single $^{16}$O nucleus. 
Figure~\ref{fig:proba_16O} shows the proton probability distribution when $(i)$ $V$ contains the entire nucleus (open squares) and $(ii)$ $V$ includes only half of the nucleus (full circles).
We verify that both distributions have a norm one. 
As expected, when $V$ includes the entire nucleus, we find all the 8 protons with a probability one, while,
when only half the nucleus is in $V$, we find an average of 4 protons. 
Note that both  cases  give a probability zero (up to numerical noise) to find more than 8 protons. 

The distribution which is obtained for half the nucleus is particularly wide, with a variance of $\sim1.4$ protons. 
It would correspond to the distribution of fragments formed in a collision where the nucleus is suddenly cut in half. 
One would expect such a violent process to lead to the maximum possible variance. 
We see from Eq.~(\ref{eq:sig_TDHF}) that the variance  
is maximum when the overlap (restricted to the volume $V$) between 
different single-particle wave-functions is zero, {i.e.}, when  $\<\az_i|\az_j\>_V\propto \del_{ij}$.
Then, the maximum variance can be expressed as a binomial distribution~\cite{dasso1979}
\oeq
\si_{N}^2 \le \sum_{i=1}^A  \< \az_i|\az_i\>_V  \( 1- \< \az_i|\az_i\>_V \).
\ceq
This gives an upper limit for the variance 
\oeq
\si_N^2  \le  \<\oN_V\>  \(1-\frac{\<\oN_V\>}{A}\)  \le  A/4.
\label{eq:borne}
\ceq
This upper limit is an intrinsic limitation of independent particle systems.
Considering a binary partition of an oxygen nucleus, we see that the maximum variance of the proton number is 2, which is larger than, but of the same order of magnitude as the ``half-oxygen'' example (see Fig.~\ref{fig:proba_16O}).

\subsubsection{Multi-nucleon transfer probabilities with TDHF}

\begin{figure}[!htb] \begin{center}
\includegraphics[width=7.0cm]{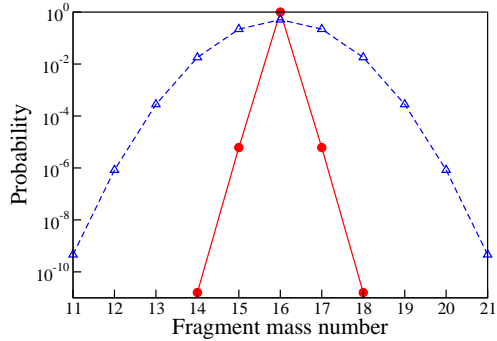}    
\caption{Fragment mass number probability distribution in $^{16}$O$+^{16}$O central collisions at $E_{cm}=10$~MeV. The TDHF result with the particle number projection technique is shown with solid line. The dashed line shows the probability distribution obtained from the semi-classical transport approach assuming a Gaussian distribution with $\sigma_N^2=N_{exc}=0.6$. } 
\label{fig:proba_coll}
\end{center} \end{figure}

We now consider the application of the particle number projection method to the case of sub-barrier transfer reactions studied with TDHF. 
Applying the method to the final state of the $^{16}$O$+^{16}$O central collision shown in Fig.~\ref{fig:sp_trans}, we obtain the distribution of probabilities given in Fig.~\ref{fig:proba_coll} with a solid line. 
As expected, this distribution is much narrower than the one obtained from the semi-classical transport approach of Ref.~\cite{randrup1982} (dashed line).
In fact, the transfer probability of one nucleon is smaller than $10^{-5}$, which would be at the limit of current experimental detection setups.

\begin{figure}[!htb] \begin{center}
\includegraphics[width=7.0cm]{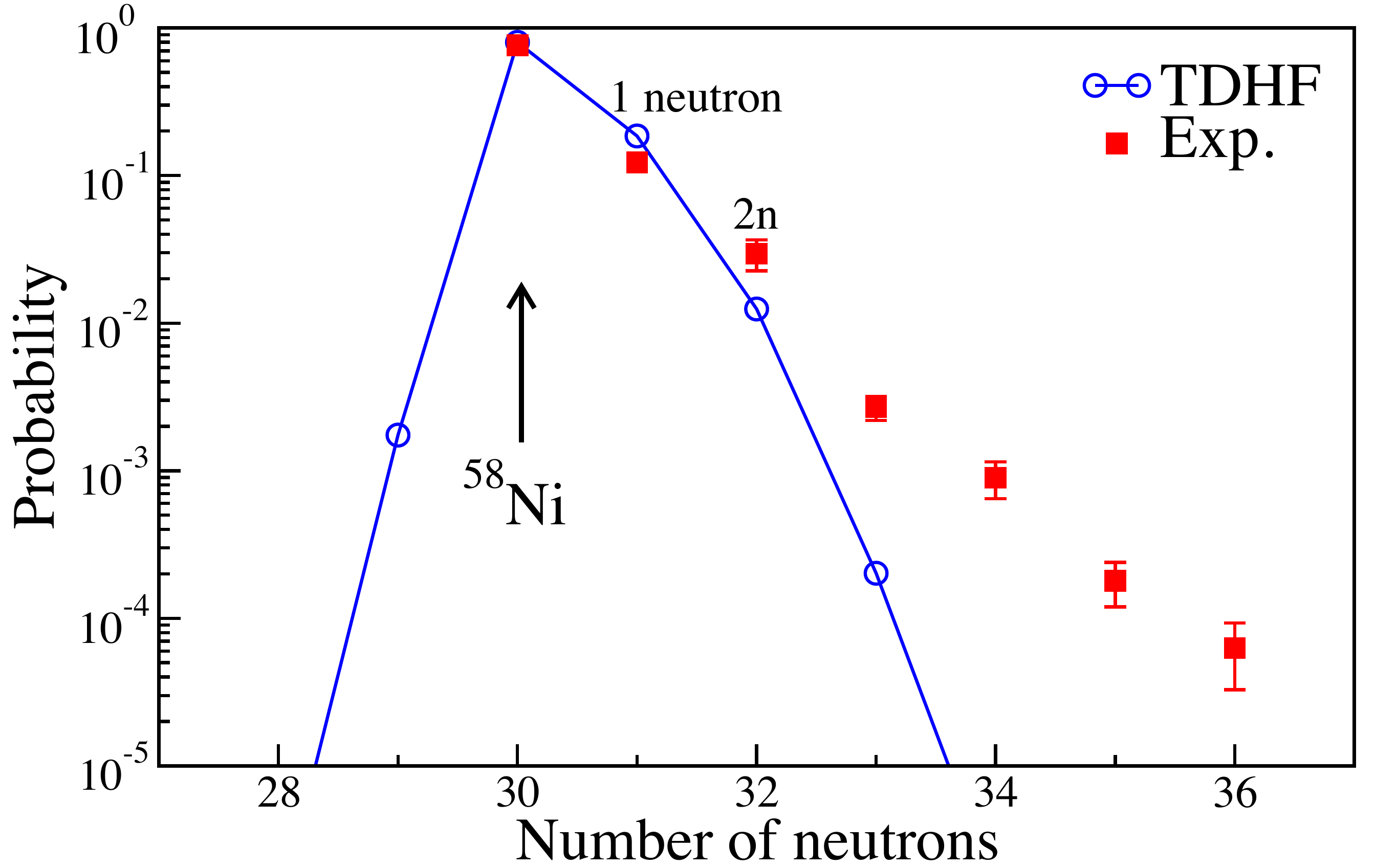}
\caption{Neutron number distribution of the light fragment in the exit channel of the $^{58}$Ni+$^{124}$Sn central collision at $E_{cm}=144.7$~MeV computed with TDHF (circles). 
The squares show experimental data extracted from Ref.~\cite{jiang1998} at $E_{cm}=153$~MeV$\simeq0.96V_B$ and $\theta_{cm}=127.5$~deg leading to the same distance of closest approach.
} 
\label{fig:proba_NiSn}   
\end{center} \end{figure}

The quantum mechanical effects responsible for the strong hindrance of transfer are expected to be less important in non-symmetric systems.
Figure~\ref{fig:proba_NiSn} shows an example of experimental near-barrier neutron transfer probabilities (filled squares) in $^{58}$Ni$+^{124}$Sn \cite{jiang1998}.
Positive $Q$-values for neutron stripping in this system allow for observation of up to 6 neutrons transferred from $^{58}$Ni to $^{124}$Sn.
In particular, a relatively large one-neutron transfer probability of $\sim0.1$ is observed.

The one-neutron stripping  probability is relatively well reproduced by TDHF calculations (open circles). 
However, these calculations strongly underpredict transfer of more than one neutron. 
This discrepancy comes from the mean-field approximation. 
Indeed, it is assumed that all transfer channels evolve in the same average mean-field. 
This approximation may be reasonable for the transfer of a very small number of particles (with respect to the total number of particles in the collision partners).
However, it is doomed to fail to describe reaction channels with very different mass and charge partitions than the average one. 

\begin{figure}[!htb] \begin{center}
\includegraphics[width=8.0cm]{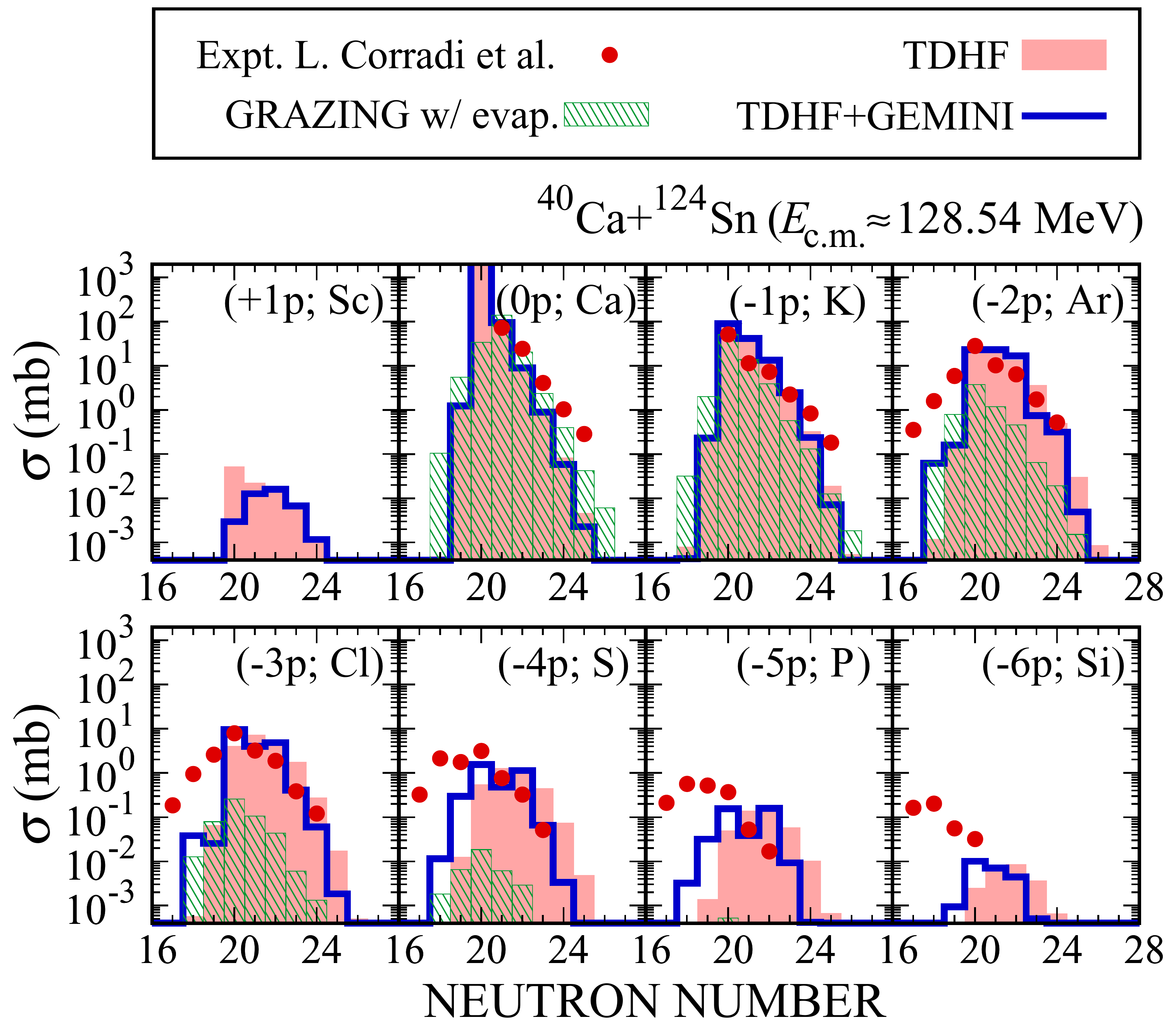}
\caption{Light fragment production cross sections in the $^{40}$Ca$+^{124}$Sn reaction at $E_{cm}=128.54$~MeV as a function of the neutron number and for various elements. Experimental data (circles) are from \cite{corradi1996}. TDHF results for primary reaction products are shown with filled areas. Cross sections for secondary reaction products obtained with TDHF+GEMINI are shown by blue solid lines. For comparison, GRAZING results including evaporation are also shown by green shaded histograms. From~\cite{sekizawa2017}.
} 
\label{fig:decay}   
\end{center} \end{figure}

Nevertheless, Sekizawa and Yabana have applied this method 
with some success to investigate multi-nucleon transfer reactions in various systems 
such as $^{40,48}$Ca+$^{124}$Sn, $^{40}$Ca$+^{208}$Pb, $^{58}$Ni$+^{208}$Pb \cite{sekizawa2013} and $^{64}$Ni$+^{238}$U \cite{sekizawa2016}.
In principle, the outgoing fragments are still excited at the end of a TDHF calculation.  
Therefore, they correspond to primary fragments before statistical  decay occurs. 
In order to compare with experimental data, it is then necessary to couple TDHF 
calculations with statistical decay calculations \cite{wang2016,umar2017,sekizawa2017}. 
An example performed by Sekizawa is shown in Fig.~\ref{fig:decay} for the $^{40}$Ca$+^{124}$Sn reaction \cite{sekizawa2017}.
Good agreement with experimental data is obtained except for large numbers of protons transferred 
due to the limitation of the mean-field approximation discussed above. 
It is also shown that this microscopic method is more reliable than the semi-classical approximation used in calculations with the GRAZING code. 

\subsubsection{Transfer reactions at sub-barrier energies}

Let us now investigate how the transfer of one nucleon evolves when going down in energy. 
At sub-barrier energies, the collision partners do not get close enough to form a neck through which nucleons can be transferred. 
In this case, particle transfer occurs only via quantum tunneling. 
It is therefore dominated by particles near the Fermi surface and it is expected to occur essentially at the distance of closest approach
\oeq
r_{min}\simeq \frac{Z_1Z_2e^2}{2E_{cm}} \(1+\frac{1}{\sin(\theta_{cm}/{2})}\),
\ceq
where $E_{cm}$ and $\theta_{cm}$ are the centre of mass energy and scattering angle, respectively. 
Due to the exponential decay of wave functions at the  surface, the transfer probability of one particle is expected to obey \cite{oertzen2001,corradi2009} 
\oeq
P\sim\exp(-2\kappa r_{min}).
\label{eq:exp_trans}
\ceq

\begin{figure}[!htb] \begin{center}
\includegraphics[width=7.0cm]{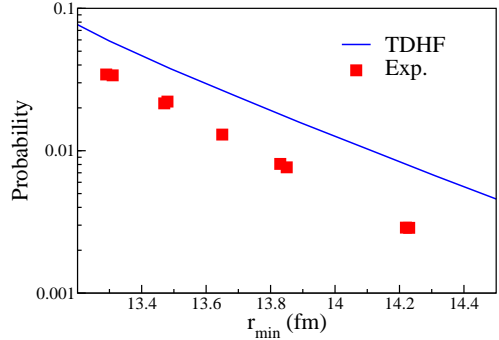}
\caption{One proton stripping probability in $^{16}$O+$^{208}$Pb deep sub-barrier central collisions computed with TDHF (solid line) \cite{simenel2010}. 
Filled squares show experimental data  from Ref.~\cite{evers2011}. The error bars are smaller than the size of the symbols. 
} 
\label{fig:proba_OPb}   
\end{center} \end{figure}

Deep sub-barrier transfer has been investigated experimentally, e.g., in the case of one-proton stripping in $^{16}$O$+^{208}$Pb \cite{evers2011}.
The resulting transfer probabilities are reported as a function of $r_{min}$ with filled squares in Fig.~\ref{fig:proba_OPb}.
We indeed observe an exponential decay of the transfer probability with increasing $r_{min}$.
A fit of the slope parameter for $r_{min}>13$~fm gives $\kappa\simeq1.3$~fm$^{-1}$. 
The same system has been investigated with TDHF calculations \cite{simenel2010}. 
The resulting one-proton transfer probabilities are shown in Fig.~\ref{fig:proba_OPb} with a solid line. 
Although they overestimate the experimental data, they qualitatively follow the experimental trend. 
In particular, they confirm the exponential decrease of transfer probabilities expected from Eq.~(\ref{eq:exp_trans}).
In addition, the value of the slope parameter $\kappa\simeq1.1$~fm$^{-1}$ is similar to the experimental one.

\subsubsection{Fragment distributions with pairing correlations using the double-projection technique}

Two nucleons could be transferred as a correlated Cooper pair in superfluid systems.
It is therefore necessary to include pairing correlations at some level in order to investigate transfer in superfluid systems.
Pairing correlations are indeed crucial in order to reproduce the transfer of two nucleons which is expected to increase in such systems.

As mentioned earlier, HFB and BCS states do not have a good number of particles. 
Therefore, the initial fragments of a collision are in a superposition of states with different particle numbers. 
Applying the particle number projection method to one initial fragment would give a distribution of probabilities even before transfer takes place. 
This is a  spurious effect which prevents the direct use of the particle number projection technique to compute transfer probabilities in superfluid systems. 

In models such as TDHFB and TDHF+BCS, where particle-number symmetry is violated, the same particle-number projection technique may be used to restore the symmetry by first projecting the system onto a space with $N_0 = N_{V} + N_{\bar{V}}$ total particles~\cite{scamps2013a,scamps2015}.
Once the initial projection onto $N_0$ total nucleons has been performed, the above prescription may be followed for the projection of each fragment onto a particle-number $N$.

Using this idea, Scamps and Lacroix have proposed a double projection technique \cite{scamps2013a}. 
Instead of applying Eq.~(\ref{eq:proj2}) with the BCS state $|\Psi\>$, they consider the normalised state with good number of particles $N_0$:
\oeq
|N_0\>=\frac{\oP'(N_0)|\Psi\>}{\sqrt{\<\Psi|\oP'(N_0)|\Psi\>}},
\ceq
where $\oP'$ is a particle number projector on the entire system, i.e., on the total volume $V+\bar{V}$. 
Then, they apply the same projection technique as before in order to compute the particle number  distribution in the fragments using
\oeq
\mP(N)=\<N_0|\oP(N)|N_0\>.
\ceq

The double projection technique has been successfully applied to transfer reactions involving one superfluid nucleus~\cite{scamps2013a}. 
As a result, the pair transfer is indeed enhanced as compared to calculations without pairing correlations. 

\subsubsection{Application to superfluid fission fragments}

The double projection technique has also been recently applied to investigate fission fragment properties~\cite{scamps2015a}. 
As discussed in section~\ref{sec:fission}, the inclusion of dynamics \cite{simenel2014a,goddard2015} 
and in particular dynamical pairing correlations~\cite{scamps2015a,bulgac2016} is crucial in order to describe fission. 

In the TDHF+BCS application to describe the formation and dynamics of fragment produced in $^{258}$Fm fission~\cite{scamps2015a}, 
three different valleys \cite{bonneau2006} were studied: one symmetric valley 
with compact (quasi-spherical) fragments, one with elongated fragments, and one asymmetric valley. 
It has been argued that the symmetric compact valley is strongly affected by shell effects due to the $Z=50$ magic number \cite{hulet1986}.
This is supported by the quasi-spherical shape of the fragments in this valley.

\begin{figure}[!htb] \begin{center}
\includegraphics[width=6.0cm]{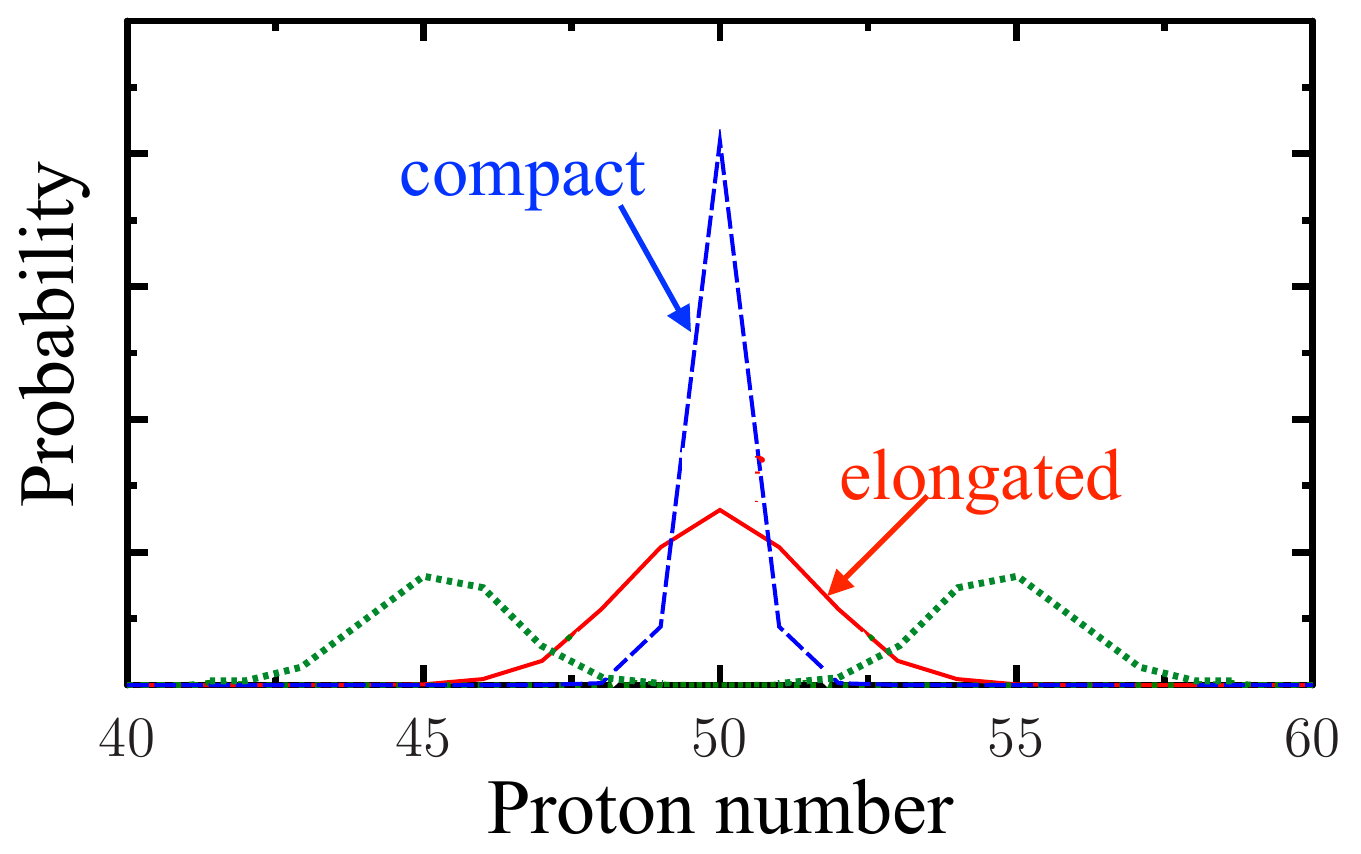}
\includegraphics[width=6.0cm]{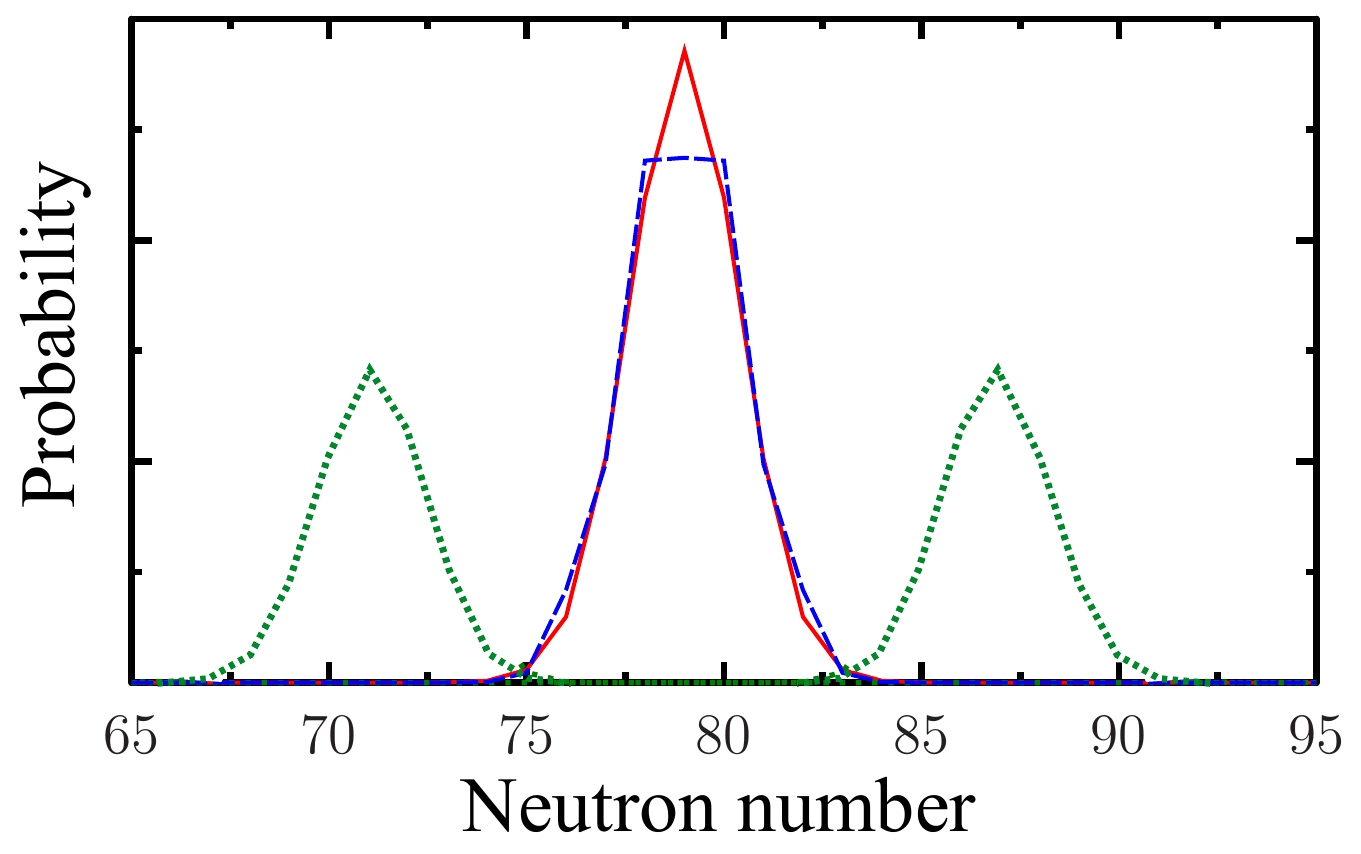}
\caption{TDHF+BCS prediction of the proton (left) and (neutron) number probability distributions in fission fragments of $^{258}$Fm for different fission valleys. Adapted from~\cite{scamps2015a}.} 
\label{fig:fission}   
\end{center} \end{figure}

The TDHF+BCS calculations of Ref.~\cite{scamps2015a} cannot predict what would be the repartition between these valleys. 
However,  the role of shell effects in the particle number distribution can be investigated at the mean-field level thanks to the double projection method. 
The resulting probability distributions are shown in Fig.~\ref{fig:fission}  for protons and neutrons.
We observe that all distributions exhibit  similar widths, except for the proton distribution in the symmetric compact mode which is much narrower. 
This is interpreted as an effect of the spherical magic proton shell in tin isotopes ($Z=50$). 
It is interesting to see that the mean-field dynamics, combined with projection techniques, is indeed able to exhibit signatures of magic shells in the fragment properties. 

\subsection{Beyond Mean-Field Methods \label{sec:beyond}}

One limitation of TDHF lies in the semi-classical nature of the theory. 
According to the Balian-V\'en\'eroni variational principle~\cite{balian1981}, only average quantities may in principle be obtained from the study of a reaction with the TDHF theory.
In recent years,
much effort has been done to improve the mean-field approximation by incorporating the fluctuation mechanism
into the description. At low energies, the one-body fluctuations make the dominant contribution to the fluctuation
mechanism of the collective motion.
Various extensions have been developed to study the correlation of one-body observables.
These include the time-dependent random phase approximation (TDRPA) approach of Balian and V\'en\'eroni~\cite{balian1992}, the time-dependent generator coordinate method
(TDGCM)~\cite{goutte2005}, and the stochastic mean-field (SMF) method~\cite{ayik2008}.
The effects of two-body dissipation on reactions of heavy systems using the 
time-dependent density-matrix (TDDM)~\cite{tohyama1985,tohyama2002a}
approach have also been recently reported~\cite{assie2009,tohyama2016,wen2018}.

Among the most important observables that are affected by these fluctuations are the widths of nucleon transfer in
heavy-ion collisions and fission fragment yields.
The mean-field description provides the mean values of the proton and neutron numbers transferred in heavy-ion collisions, sometimes
referred to as the drift, as well as the mean-value of the fragment masses and charges in fission reactions.

\subsubsection{Time-dependent Random Phase Approximation\label{sec:TDRPA}}

In Ref.~\cite{balian1981}, the Balian-V\'en\'eroni (BV) variational principle is used to show that the TDHF
mean-field approach is specially suited for the prediction of one-body observable averages.
However, as illustrated in Ref.~\cite{dasso1979}, the limitation of a single-determinant TDHF formalism 
results in the underestimation of fluctuations and correlations of one-body observables.
Indeed, comparisons of TDHF with deep-inelastic collision experiments show that TDHF consistently underestimates experimental mass distributions~\cite{koonin1977,davies1978a}.
The BV approach can be used to better estimate these missing correlations~\cite{balian1984}.

The general formula for the correlation of two one-body operators, $\hat{X}$ and $\hat{Y}$, is given by
\begin{equation}
\sigma_{XY}=\sqrt{\langle\hat{X}\hat{Y}\rangle - \langle\hat{X}\rangle\langle\hat{Y}\rangle},
\end{equation}
with the same formula providing the fluctuations in the case of $\hat{X}=\hat{Y}$.
%can include this or not, not sure
In TDHF, the expression for the correlation at a final time, $t_f$, is
\begin{equation}
{\sigma_{XY}^2}^{(TDHF)}(t_f)=\text{Tr}\left\lbrace Y\rho(t_f) X[I-\rho(t_f)]\right\rbrace,
\end{equation}
where $I$ is the identity matrix, $X$ and $Y$ are the matrix representation of the operators 
$\hat{X}$ and $\hat{Y}$, and $\rho(t)$ is the one-body density matrix at time $t$.
However, the derivation of the TDHF equation from the Balian-V\'en\'eroni variational principle demonstrates that TDHF is optimized to the expectation values of one-body observables only.
Fluctuations of one-body observables involve the expectation value of the square of a one-body operator, for which there is no guarantee that TDHF is predictive. 
To better recover correlations and fluctuations of one-body operators, the variational space of the observable must be expanded from one-body observables to exponential of one-body operators as detailed in Ref.~\cite{balian1984}. In this expanded space, one obtains the following result for the correlation at time $t_f$~(see, e.g., \cite{simenel2011} for a detailed derivation)
\begin{equation}
\label{eq:corr}
\sigma_{XY}^2(t_f)=\lim\limits_{\epsilon\rightarrow 0}\frac{\text{Tr}\left\lbrace\left[\rho(t_0)-\rho_X(t_0,\epsilon)\right]\left[\rho(t_0)-\rho_Y(t_0,\epsilon)\right]\right\rbrace}{2\epsilon^2},
\end{equation}
%cite tdhf here
where the one-body density matrices $\rho_{X,Y}(t,\epsilon)$ are solutions to the TDHF equation with the boundary condition
\begin{equation}
\label{eq:transform}
\rho_{X,Y}(t_f,\epsilon)=e^{i\epsilon X,Y} \rho(t_f) e^{-i\epsilon X,Y}.
\end{equation}
The result in Eq.~(\ref{eq:corr}) considers the fluctuations of the observable about the mean-field evolution at the RPA level~\cite{balian1984,balian1992,ayik2008,simenel2011}. To calculate the correlations, the state at $t_f$ is propagated backwards in time to the initial time $t_0$, explaining why the correlations at the time of interest, $t_f$, depends on the density matrices at the initial time, $t_0$.

One useful application of the BV approach is in the investigation of particle number fluctuations for reaction products in deep-inelastic collisions, as in Ref.~\cite{simenel2011}.
In this case, the one-body operator employed in Eq.~(\ref{eq:corr}) is the number operator, $\hat{N}^q_V$
\begin{equation}
\hat{X}=\hat{N}^q_V = \sum_{s}\int d\mathbf{r}~\hat{a}^\dagger(\mathbf{r}sq)\hat{a}(\mathbf{r}sq)~\Theta(\mathbf{r}),
\end{equation}
with $\Theta(\mathbf{r})$ being 1 in the fragment region $V$ and $0$ elsewhere.
The purpose of this operator is to count all particles with isospin $q$ in the region $V$. 
The total number operator is then $\hat{N}_V = \hat{N}^{(n)}_V + \hat{N}^{(p)}_V$.
%We may simplify $\hat{N}^{(q)}_V$ to the form
%\begin{equation}
%\hat{N}^q_V = \xi_X \Theta(\mathbf{r}),
%\end{equation}
We transform the single-particle states at the final time $t_f$ in accordance with Eq.~(\ref{eq:transform}) as follows~\cite{marston1985}
\begin{equation}
\psi^X_i(\mathbf{r}\sigma q, t_1;\epsilon) = e^{-i \epsilon \xi_X \Theta_V(\mathbf{r})} \varphi_i(\mathbf{r}\sigma q, t_1),
\end{equation}
where $\xi_X=1$ for the total number operator, $\hat{N}_V$, and $\xi_X=\delta_{qq'}$ for the isospin dependent number operator, $\hat{N}^{q'}_V$.
The small parameter $\epsilon$ is  varied to check convergence~\cite{marston1985,simenel2011}.
These transformed states are then propagated backwards in time to the initial time $t_0$ as described above.
To obtain the correlations and fluctuations, one must evaluate Eq.~(\ref{eq:corr}), which can be reduced to
\begin{equation}
\sigma_{XY}(t_0)=\sqrt{\frac{\eta_{00}(t_0)+\eta_{XY}(t_0)-\eta_{0X}(t_0)-\eta_{0Y}(t_0)}{2\epsilon^2}},
\end{equation}
where $\eta_{XX'}$ is
\begin{equation}
\eta_{XX'}=\sum_{ij}\left|\left\langle\psi^X_i \big| \psi^{X'}_j \right\rangle\right|^2,
\end{equation}
with the sum over $i$ and $j$ being over occupied states and 
$X=0$ and $Y=0$ denoting the use of untransformed states, $\varphi_i$.
In practice, the final untransformed states are also propagated backwards in time to reduce numerical inaccuracies introduced by the TDHF evolution, as indicated in Ref.~\cite{bonche1985}.

The BV approach has  been used in the past to study mass dispersion from giant resonance decay \cite{troudet1985,broomfield2008,broomfield2009} 
as well as to compute the fluctuation of the particle number in fragments produced in deep-inelastic collisions \cite{marston1985,bonche1985,zielinska1988,broomfield2009,simenel2011,williams2018}.
This approach was also used to study hot Fermi gas~\cite{martin1991}, $\phi^4$ theory~\cite{martin1995}, and Boson systems~\cite{benarous1999,boudjemaa2010,boudjemaa2015}.

\begin{figure}
\begin{center}
\includegraphics[width=7cm]{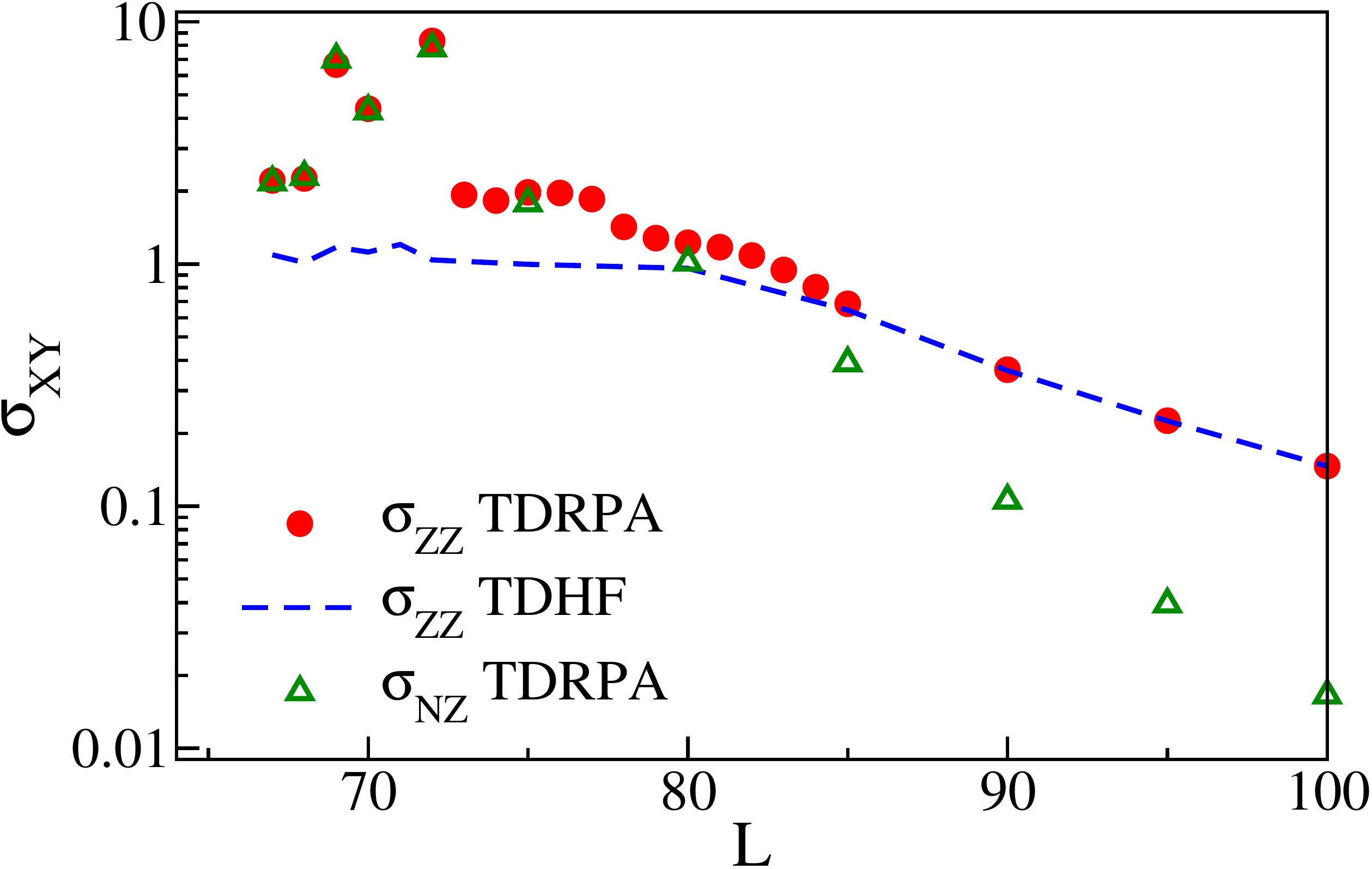}
\end{center}
\caption{Variance of the fragment charge distribution from TDHF (dashed line) and TDRPA (filled circles) calculations as a function of the angular momentum $L$ in units of~$\hb$ for $^{40}$Ca$+^{40}$Ca collisions at $E_{cm}=128$~MeV. Correlations $\sigma_{NZ}$ between proton and neutron numbers obtained from TDRPA calculations are also shown (open triangles). Adapted from Ref.~\cite{simenel2011}.  
\label{fig:CaCasigma}}
\end{figure}

An example of application of the TDRPA to fragment charge distributions in $^{40}$Ca+$^{40}$Ca deep-inelastic collisions is shown in Fig.~\ref{fig:CaCasigma} \cite{simenel2011}. 
Deep-inelastic collisions are characterised by a strong orbiting of the di-nuclear system, a significant damping of the initial relative kinetic energy, and a large width of the fragment mass and charge distributions which are usually underestimated in TDHF.
Figure~\ref{fig:CaCasigma} shows the standard deviation of the fragment charge distribution obtained from TDHF (dashed line) and TDRPA (filled circles)  as a function of the angular momentum $L$ at $E_{cm}=128$~MeV (approximately $2.5$ times the Coulomb barrier).
Both calculations agree at large $L$, dominated by quasi-elastic scattering.
However, TDHF calculations strongly underestimate the fluctuations for more violent collisions at smaller $L$. 
The fluctuations predicted by the TDRPA are in better agreement with experimental data of Ref.~\cite{roynette1977}.

\begin{figure}
\begin{center}
\includegraphics[width=6cm]{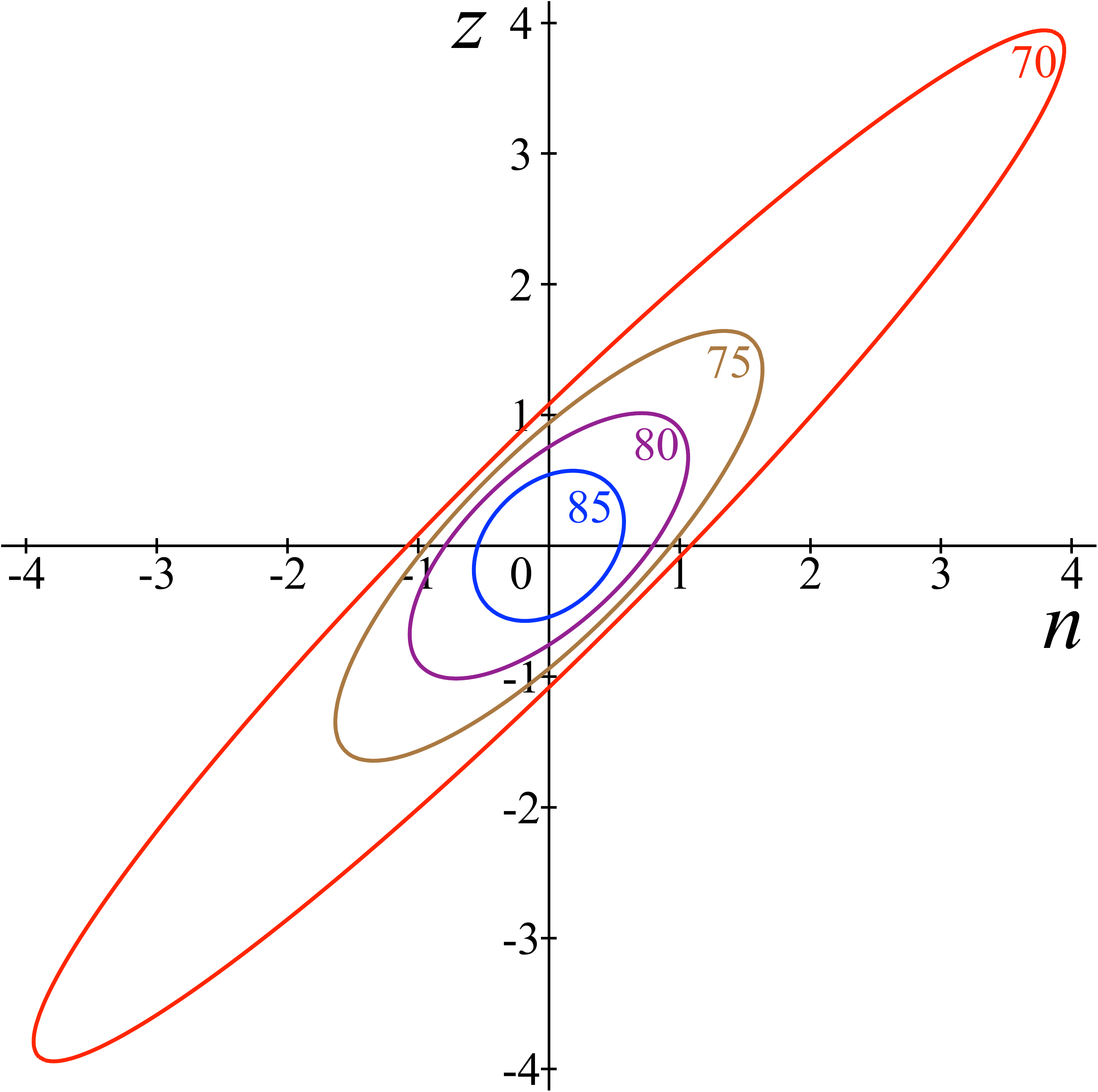}
\end{center}
\caption{Iso-probabilities $\mP(n,z)=\mP(0,0)/2$ for the transfer of $n$ neutrons and $z$ protons in $^{40}$Ca+$^{40}$Ca at $E_{cm}=128$~MeV using fluctuations and correlations computed with the TDRPA. The numbers on the curves indicate the angular momentum in units of~$\hb$.  \label{fig:CaCacorrel}}
\end{figure}

Correlations between proton and neutron numbers obtained from TDRPA calculations are also reported in Fig.~\ref{fig:CaCasigma} (open triangles).
These correlations are zero in standard TDHF calculations.
Although they are negligible for quasi-elastic scattering, they are of similar order as the charge fluctuations for deep-inelastic events. 
This can be interpreted as an effect of the symmetry energy which favours a transfer of both protons and neutrons in the same direction.
This is illustrated in Fig.~\ref{fig:CaCacorrel} which shows the expected distribution of fragments in the $N-Z$ plane for various angular momenta assuming Gaussian probability distribution of the form
\oeq
\mathcal{P}(n,z) = \mP(0,0)\exp\[ -\frac{1}{1-\rho^2} \( \frac{n^2}{\sigma_N^2}+\frac{z^2}{\sigma_Z^2} - \frac{2\rho nz}{\sigma_N\sigma_Z}\) \], 
\label{eq:Pnz}
\ceq
where $n$ and $z$ are the number of transferred neutrons and protons, respectively.  
The correlations between $N$ and $Z$ are quantified by the parameter
\oeq
\rho = \mbox{sign}(\sigma_{NZ})\frac{\sigma_{NZ}^2}{\sigma_N\sigma_Z}=\frac{\<nz\>}{\sqrt{\<n^2\>\<z^2\>}}.
\ceq
We see that the correlations have a noticeable impact in the $N-Z$ distributions at smaller angular momenta associated with deep-inelastic collisions.

\subsubsection{Stochastic Mean Field Approach}

As mentioned above the standard TDHF approach provides a deterministic description of a collision process, i.e. the system
evolves from a specified initial condition to a single final state.
In the stochastic mean-field (SMF) approach, the fluctuations in the initial
state are incorporated in a stochastic manner by introducing a proper distribution of the initial single-particle density
matrices~\cite{ayik2008,lacroix2014}. This results in an ensemble of single-particle density matrices generated by evolving each
density in its own mean-field Hamiltonian. 

In a single event labeled by $\lambda $, the single-particle density matrix is determined by evolving the
single-particle wave functions $\phi _{j}^{\lambda } (\vec{r},t)$ according to the self-consistent
Hamiltonian in that event. Consequently, in a given event, nucleon density is
given by
\begin{equation} \label{eq1} 
\rho ^{\lambda } (\vec{r},t)=\sum _{ij}\phi _{j}^{*\lambda } (\vec{r},t)\rho _{ji}^{\lambda}  \phi _{i}^{\lambda } (\vec{r},t), 
\end{equation} 
where labels $(i,j)$ indicate a complete set of quantum numbers for specifying single-particle wave
functions.
The mean values of the elements of density matrices are given
by $\overline{\rho }_{ji}^{\lambda}=\delta _{ji} n_{j}$ and the second moments of fluctuating parts
are determined by
\begin{align} \label{eq5}
\overline{\delta \rho _{ji}^{\lambda } \delta \rho _{i'j'}^{\lambda } }=\frac{1}{2} \delta _{ii'}
\delta _{jj'} \left[n_{i} (1-n_{j} )+n_{j} (1-n_{i} )\right],
\end{align}
where $n_{j}$ are the average occupation numbers of the single-particle states.
Equation~(\ref{eq5}) lies at the root of the SMF approach as it can be employed in the study of various reaction
phenomena.

By calculating the expectation values of an observable in each event, it is possible to determine probability distributions
of observables. In a number of studies
 \cite{ayik2009,washiyama2009b,yilmaz2011,ayik2008a,ayik2009a,ayik2011,yilmaz2011a,yilmaz2013} 
 (see also \cite{lacroix2014} for a recent review), including studies of mass dispersions in transfer reactions~\cite{washiyama2009b,yilmaz2011,ayik2017}, 
it has been shown that the SMF approach provides a good approximation for
the description of quantal fluctuations of the collective motion. 
In particular, in small amplitude limit, the approach gives
rise to the same expression for the dispersion of one-body observables familiar from the variational approach of  Balian
and V\'en\'eroni~\cite{balian1992}.

Recently, the SMF approach was generalized to include pairing degrees of freedom~\cite{lacroix2013} and subsequently
applied to study fission fragment mass and TKE distributions~\cite{tanimura2017}. This study assumed that the
fluctuations on the way to scission, at some point after the saddle configuration, arise from fluctuation of
states in the vicinity of the Fermi energy. Thus, by considering an ensemble of such states generated by
taking $\rho _{ji}^{\lambda}$
to be a Gaussian distribution of random numbers,
the authors were able to obtain a set of initial conditions, which were then propagated using the TDHF+BCS method
mentioned earlier. However, in this case density matrices used in TDHF+BCS equations are no longer diagonal in
the canonical basis, due to the fluctuations introduced in Eq.~(\ref{eq1})~\cite{tanimura2017}. This requires
modification of codes and a higher computational cost.
It has been found  that the dynamical pairing effects wash out due to the introduction of quantum fluctuations via
the initial conditions. In particular,  very similar outcomes between the cases of dynamical occupations
and the FOA approach were obtained.

In the di-nuclear regime, the SMF approach gives rise to Langevin description for nucleon exchange between
projectile-like and target-like nuclei characterized by diffusion and drift coefficients~\cite{ayik2009}. It is well
known that the Langevin description is equivalent to the evolution of the distribution function of the collective
variable according to the Fokker-Planck equation. As a result of this equivalence, instead of carrying out stochastic
simulations, it is more convenient to calculate the transport coefficients and employ the Fokker-Planck approach. When
the drift coefficients are linear functions of mass and charge asymmetry variables, the primary fragment charge and mass
distribution $P \left( N,Z,t \right)$ is given by Eq.~(\ref{eq:Pnz}). 
%a correlated Gaussian in the neutron-proton plane
%\begin{align} \label{eq12}
%P(N,Z,t)=\frac{1}{2\pi \sigma _{NN} \sigma _{ZZ} \sqrt{1-\rho ^{2} } } \exp \left(-C\right).
%\end{align}
%Here, the exponent $C$ is
%\begin{align} \label{eq13}
%C=\frac{1}{2\left(1-\rho ^{2} \right)} &\left[\left(\frac{Z-\overline{Z}}{\sigma _{ZZ} }\right)^{2}+\left(\frac{N-\overline{N}}{\sigma _{NN}}\right)^{2}
%- 2\rho \left(\frac{Z-\overline{Z}}{\sigma_{ZZ} } \right)\left(\frac{N-\overline{N}}{\sigma _{NN} } \right) \right],
%\end{align}
%where $\rho=\sigma_{NZ}^{2}/\sigma _{ZZ}\sigma_{NN}$ is the correlation coefficient. 
%This distribution
%function is specified by the mean neutron, $\overline{N}$, and mean proton, $\overline{Z}$, numbers of target-like or
%project-like fragments, and
The co-variances $\sigma_{NN}$, $\sigma_{ZZ}$, and $\sigma_{NZ}$ are
determined by a set of coupled differential equations in which the inputs are provided by the neutron and proton
diffusion and  drift coefficients.  

In the SMF approach these transport coefficients are calculated in terms of the
solutions of the TDHF equations. Calculations take into account the full collision geometry and do not involve any
fitting parameter other than the standard parameters of the Skyrme interaction. In the earlier investigations, transport
coefficients were calculated in the semi-classical approximation in Markovian limit~\cite{yilmaz2014,ayik2015a}.
Recently, it has become possible to calculate these transport coefficients in the quantal framework by 
including the shell
structure and the Pauli blocking in an exact manner~\cite{ayik2015,ayik2016}.

\subsection{Nuclear Forces}

For all the reaction theories it is desirable to be based on the same footing as the underlying structure theory.
The TDHF theory employs the same energy density functional (EDF) developed by the nuclear structure community
with no additional parameters. The determination of the parameters of these EDFs contains no reaction information.
Almost all TDHF calculations done to date have used the Skyrme energy density functional~\cite{skyrme1956}, 
although recent TDHFB calculations with the Gogny force have been performed~\cite{hashimoto2012,scamps2017b}.

The Skyrme EDF was developed as a zero-range low-momentum expansion of the effective two-body nucleon-nucleon interaction
by Skyrme~\cite{skyrme1956,skyrme1958}.
In recent years, the quality of effective interactions has been substantially
improved~\cite{chabanat1998a,kluepfel2009,kortelainen2010,kortelainen2012}.
The Skyrme EDF contains terms
which depend on the nuclear density, $\rho$, kinetic-energy density, $\tau$,  vector current density, $\mathbf{j}$,
 pseudovector spin density, $\mathbf{s}$,  pseudovector spin kinetic density, $\mathbf{T}$, 
pseudotensor spin-current density,
$\mathbf{J}$, and pseudovector tensor kinetic-density, $\mathbf{F}$,
\begin{eqnarray}
E &=& \int d^3r\sum_{t=0,1}\Bigg\{C_t^\rho[\rho_0]\rho_t^2+C_t^s[\rho_0]\mathbf{s}_t^2+C_t^{\Delta\rho}\rho_t\mathbf{\nabla}^2\rho_t\nonumber \\
&+&C_t^{\nabla s}(\mathbf{\nabla}\cdot\mathbf{s})^2+C_t^{\Delta s}\mathbf{s}_t\cdot\mathbf{\nabla}^2\mathbf{s}_t+C_t^{\tau}(\rho_t\tau_t-\mathbf{j}_t^2) \nonumber \\
&+&C_t^T\left(\mathbf{s}_t\cdot\mathbf{T}_t-\sum_{\mu,\nu=x}^zJ_{t,\mu\nu}J_{t,\mu\nu}\right) \nonumber\\
&+&C_t^F\Bigg[\mathbf{s}_t\cdot\mathbf{F}_t-\frac{1}{2}\left(\sum_{\mu=x}^zJ_{t,\mu\mu}\right)^2-\frac{1}{2}\sum_{\mu,\nu=x}^zJ_{t,\mu\nu}J_{t,\nu\mu}\Bigg]\nonumber\\
&+&C_t^{\nabla\cdot J}\left(\rho_t\mathbf{\nabla}\cdot\mathbf{J}_t+\mathbf{s}_t\cdot\mathbf{\nabla}\times\mathbf{j}_t\right)\Bigg\},
\label{eq:edens}
\end{eqnarray}
where the detailed definitions of densities and currents can be found in Refs.~\cite{engel1975,lesinski2007,dobaczewski2000}.
%\begin{equation}
%E=\int d^{3}r\;{\cal H}(\rho ,\tau,\mathbf{j},\mathbf{{s}},\mathbf{T},\mathbf{J},\mathbf{F};\mathbf{{r}})\;.
%\end{equation}

The time-odd densities ($\mathbf{j}$, $\mathbf{s}$, $\mathbf{T}$, $\mathbf{F}$) vanish
for static calculations of even-even nuclei, while they are present for odd mass nuclei, in cranking calculations, as well
as in TDHF. The spin-current pseudotensor, $\mathbf{J}$, is time-even and does not vanish for
static calculations of even-even nuclei.
It has been shown~\cite{umar1986a,reinhard1988,umar1989,maruhn2006b,umar2006c,suckling2010,fracasso2012,dai2014} that the presence of these
extra time-odd terms are necessary for preserving the Galilean
invariance and make an appreciable contribution to the dissipative properties of the collision.
Although this  one-body dissipation plays an important role in the collisions, 
it is however not sufficient to lead to a fully equilibrated compound nucleus.
This lack of equilibration mechanism has been illustrated by Loebl and collaborators using the Wigner distribution function \cite{loebl2011,loebl2012}.

The TDHF calculations performed within the last decade have generally incorporated the complete Skyrme interaction with
the exception of the tensor part. A comprehensive discussion of the history of the Skyrme tensor force is given 
in Ref.~\cite{lesinski2007}. 
Recently, the tensor force was independently incorporated in TDHF calculations by two
groups~\cite{fracasso2012,dai2014a} into the Sky3D TDHF code~\cite{maruhn2014}. 
The effect of the tensor force on vibration has been studied \cite{fracasso2012,barton2017} and shown to affect low-energy fusion~\cite{guo2018}. 
In addition, investigation of the high energy limit to fusion in the $^{16}$O + $^{16}$O collision found an effect on the fusion threshold energy of the
order of several MeV \cite{stevenson2016}. 
This suggests that the tensor force can
play a non-negligible role in dynamic processes in nuclei.
It is also suggested that the tensor force plays a more important role in heavier nuclei \cite{iwata2011}.
 However, there seems to be relatively significant variations
among the Skyrme parametrizations including the tensor force~\cite{dai2014a}, which suggests that further studies will be needed
before the large scale use of these terms.

\subsection{Boundary Conditions}

Generally, the numerical solutions of TDHF equation are performed in a box with periodic or static boundary conditions.
In few cases, such as $0^+$ pairing vibration~\cite{avez2008} and giant monopole resonance (GMR)~\cite{stevenson2004,almehed2005,reinhard2006,stevenson2007,avez2013} studies in spherical nuclei, 
the system exhibits a spherical symmetry at all time and the TDHF equation can be solved in one dimension (the radial coordinate), 
allowing for very large boxes without spurious reflection of the wave functions at the boundaries. 
In Ref.~\cite{avez2013}, large spherical grids were then used to perform spectral analyses 
(via spatial Fourier transforms) of the  nucleon wave-functions emitted from unbound GMR direct decay.
However, most of TDHF applications (e.g., to heavy-ion collisions), involve Cartesian grids with sizes which are limited by the available computational power.

In TDHF calculations of nuclear reactions, the single-particle states undergo rapid changes which may lead to the development of low-density tails reaching
the box boundaries. For hard boundary conditions, this may lead to 
spurious effects such as a non-conservation of the momentum. For periodic boundary conditions the reflecting single-particle states may cause further
damping of the collective motion. Such dependence on the boundary conditions can of course be tested by changing the size of the
numerical box. 
This is not a serious problem in most cases except at higher energies, where
the emission of low-density material from
the nuclei can interfere with the dynamics in the neighboring box and
cause problems in the conservation of energy and angular momentum~\cite{guo2008}.

Various numerical solutions to this issue have been proposed, including a multigrid approach~\cite{degiovannini2012} to
extend the boundaries, and methods to remove the emitted particles via radiating or exact 
boundary conditions~\cite{boucke1997,mangin-brinet1998,pardi2013,pardi2014}. However, these methods
remain numerically very costly for fully three-dimensional calculations. On the other hand the approach of
using absorbing boundary conditions (ABC)~\cite{nakatsukasa2005} offers a feasible alternative, as well as
applying a mask function during the TDHF evolution~\cite{reinhard2006}.

Recently, the twist-averaged boundary conditions (TABC) have been employed to remove the spurious finite-volume effects for
periodic boundary conditions \cite{schuetrumpf2015a,schuetrumpf2016}. 
Periodic boundary conditions (PBC) reflect the boundary condition at one edge of the 
numerical box to the other edge. 
However, the Floquet-Bloch theorem states that a wave function in a periodic potential is periodic up
to a complex phase shift (twist) between adjacent numerical cells.
Bloch boundary conditions do the same but when reflecting also apply a phase correction, which can be averaged
to correct for the phase. The TABC Bloch boundary condition for single-particle states is
\begin{equation}
\phi_{\lambda\theta}(\mathbf{r}+\mathbf{L}_i)=e^{i\theta_i}\phi_{\lambda\theta}(\mathbf{r})\;,
\label{tabc}
\end{equation}
where $\mathbf{L}_i$ are the box lattice vectors ($i=x,y,z$) and $\phi_{\lambda\theta}(\mathbf{r})$ are the
solution of HF singe-particle equations subject to the above boundary condition
\begin{equation}
\hat{h}_{\theta}\phi_{\lambda\theta}(\mathbf{r})=\epsilon_{\lambda\theta}\phi_{\lambda\theta}(\mathbf{r})\;.
\end{equation}
This boundary condition can be implemented by multiplying the original PBC state with a phase,
$\phi_{\lambda\theta}(\mathbf{r})=e^{-i\theta_ix_i/L}\phi_{\lambda}(\mathbf{r})$. With this substitution, density still
remains periodic but the first and second derivatives must be computed with the above form in constructing currents
\textit{etc}.
In TABC, observables are computed by averaging over different Bloch phases~\cite{schuetrumpf2015a}:
\begin{equation}
<\hat{O}>=\int \; \frac{{\bf d}\theta}{\pi^3}<\Phi_{\theta}|\hat{O}|\Phi_{\theta}>\;,
\end{equation}
where $\Phi_{\theta}$ is the corresponding HF many-body wavefunction.
The angles $\theta_i$ change between zero (PBC) and $\pi$ (anti-PBC),
as the time-reversal symmetry is assumed. The integral can be done with Gauss-Legendre quadrature as it was done for the application of calculating nuclear pasta
phases found in the structure of the neutron star crust~\cite{schuetrumpf2015a}.

The treatment of boundary conditions can also be utilized in the study of widths of collective
excitations using the time-dependent linear response
approach~\cite{nakatsukasa2016}. The total width of giant resonances is comprised of the spreading width,
escape width, and the width coming from the Landau
damping. The Landau damping is due to the coupling of the collective states incoherently with $1p1h$ excitations, while the
escape width is due to continuum excitations or particle emission. The spreading width is due to higher order correlations and
requires beyond mean-field methods, such as the TDDM approach~\cite{tohyama2001,tohyama2002a}.
Applications of ABC~~\cite{nakatsukasa2005}, masking~\cite{reinhard2006}, and TABC~\cite{schuetrumpf2016}
have all shown that they can effectively reduce the effects of the finite volume discretization in calculating
the escape width when using TDHF codes to perform time-dependent linear response calculations.

\section{Conclusions and perspectives}

Quantum microscopic approaches to time-dependent many-body problems have been discussed
within the framework of nuclear dynamics. 
The time-dependent Hartree-Fock theory, which is obtained via  different approaches (variational or based on perturbation schemes), 
has become the tool of choice to investigate low-energy heavy-ion collisions and fission dynamics. 
The increase of computational power, the better performance of modern algorithms, 
as well as the improvement in the nuclear energy density functional (now including spin-orbit and sometimes tensor terms) have made TDHF codes a great tool 
to help analyze experimental data and guide future experimental programs, e.g., with exotic beams. 

Fusion near the barrier is ideal to investigate the interplay between structure and reaction mechanisms. 
In particular, TDHF calculations, which incorporate all sorts of vibrational modes as well as transfer channels (at the mean-field level), 
are used to study how these coherent couplings, which are built up during the approach of two nuclei, impact fusion. 
Microscopic approaches like TDHF are not competitors to coupled-channels models, 
but instead bring complementary information on the mechanisms at play. 
It is then natural to attempt to merge both approaches, a program which has been initiated only recently.

By construction, microscopic approaches incorporate the dynamics of single-particles. 
This allowed studies of nucleon transfer in heavy-ion collisions and the development of new tools 
such as the particle number projection technique to compute transfer probabilities. 
The range of applications has been recently extended to include multi-nucleon transfer and quasi-fission reactions. 
In particular, the ability of mean-field dynamics to capture the influence of magic shell effects 
in the formation of quasi-fission fragments has been a surprise. 

Some extensions of TDHF to incorporate part of the residual interaction have reached the level of realistic applications. 
For instance, the inclusion of pairing correlations allows to study the dynamics of superfluid systems. 
The latter is crucial in fission and enabled a series of recent breakthroughs in the description of the saddle to scission 
evolution in fissioning nuclei including non-adiabatic effects. 
A major outcome of these studies has been the ability to predict the total kinetic energy of the fission fragments 
without relying on any prescription for the definition of the scission point. 
Of course, there are still open problems. 
For instance, the treatment of the relative phase between colliding superfluid nuclei, which has been shown to impact the reaction outcome, 
may require beyond mean-field techniques. 

Nowadays, various beyond mean-field approaches are also used to incorporate quantum fluctuations 
and to study their effect on the distributions of fragments formed in nuclear processes. 
This is done at the TDRPA level, or with a stochastic mean-field method which, in the limit of small fluctuations,
leads to the TDRPA. 
Inclusion of these fluctuations is important for the description of the fragment properties in deep-inelastic collisions as well as  in fission. 

Some major problems remain. 
The main challenge is probably to find a way to treat quantum tunneling with a fully microscopic approach. 
Indeed, the current method to describe sub-barrier fusion using  a potential extracted microscopically from mean-field approaches
is only a temporary solution. 
The ultimate goal is to describe the fusion of heavy-nuclei via tunneling without invoking a nucleus-nucleus potential at all. 
Currently, this is only possible at above barrier energies. 
The problem relies on the fact that more than one mean-fields are necessary to describe the transmitted and reflected fluxes simultaneously. 
A possible solution is to use the time-dependent generator coordinate method (TDGCM) where the collective wave-function is developed on 
TDHF mean-field trajectories. 
However, how to fix the collective coordinate?
Using the path integral approach with complex time and invoking  the stationary phase approximation might offer an alternative solution. 
Nevertheless many technical problems remain to be solved before one can hope to achieve realistic applications to fusion with this approach. 

Another challenge is to develop an approach which accounts for each reaction channel properly, for instance in multi-nucleon transfer reactions. 
The difficulty with current mean-field approaches is that, despite the fact that the total state is a coherent superposition of different transfer channels, 
all these channels are assumed to evolve with the same mean-field. 
This might be a reasonable approximation for the transfer of few nucleons. 
However, this approach is doomed to fail for the description of transfer  channels leading to fragments which are very different than the average ones. 
Here, a TDGCM description with a difference of chemical potentials between the fragments  as collective coordinate might be a solution \cite{simenel2014b}.
Another difficulty in describing transfer channels is to include the effect of clustering. 
The recent technique developed in \cite{schuetrumpf2017} to identify the presence of clusters in the wave-function may help for this purpose. 

Finally, we are still far from a consistent microscopic and non-adiabatic treatment of fission 
from the compound system to the post-scission configuration.
The progress discussed in this review concern the later stage of fission, namely the evolution from saddle to scission. 
Time-dependent approaches are currently unable to describe the entire evolution from the compound nucleus. 
No need to say that, for spontaneous fission, the problem of tunneling in complex system is also present. 
A standard technique to describe fission is based on the TDGCM, where a collective wave-function is evolved on an adiabatic potential energy surface. 
A possible improvement will be to couple this approach 
with TDHF trajectories for the latter stage in order to account for the dynamical effects near scission.

\section*{Acknowledgments}
The authors thank S. Ayik, M. Dasgupta, K. Godbey, L. Guo, K. Hagino, D.J. Hinde, D. Lacroix, J. Maruhn, T. Nakatsukasa, W. Nazarewicz, 
V.E. Oberacker, G. Scamps, B. Schuetrumpf, K. Sekizawa, E.C. Simpson, P.D. Stevenson, K. Vo-Phuoc, and E. Williams for useful discussions. 
N. Dubray, S. Ebata, T. Nakatsukasa, B. Schuetrumpf, K. Sekizawa, K. Vo-Phuoc, P.W. Wein, and S. Yan are also thanked 
for authorising the use of their figures. 
This work has been supported by the
Australian Research Council Grants No. FT120100760 and DP160101254, 
and by the U.S. Department of Energy under grant No.
DE-SC0013847 with Vanderbilt University.
Part of this research was undertaken with the assistance of resources 
from the National Computational Infrastructure (NCI), which is supported by the Australian Government.
\setcounter{secnumdepth}{0}
\section{Appendix: Feynman rules}
%\color{magenta}
The Feynman rules for the $n$-th order contribution to the single-particle Green's function $G(x,y)$ in quantum many-body perturbation theory 
depends on the choice of the interaction. For an interaction of the form $\mathcal{V}(x,y)=V(\vx,\vy)\delta(t_x-t_y)$, and omitting spin, these rules are
 (see, e.g., \cite{fetter2003}):
\begin{enumerate}
\item Draw all topologically distinct connected diagrams with $n$ interaction (wavy) lines  and $2n+1$ non-interacting propagators $G^0(x,y)$, running from $y$ to $x$. 
Each vertex connects one interaction line with two propagators. 
\item Label each vertex with a 4-dimensional space-time point $z=(\vz,t_z)$. 
\item Each wavy line represents an interaction $\mathcal{V}(x,y)=V(\vx,\vy)\delta(t_x-t_y)$ with extremity at $x$ and $y$.
\item Integrate all internal variables over space and time.
\item Each closed fermion loop gives a minus sign.
\item Assign a global factor $(i/\hb)^n$ to each $n$-th order diagram
\item Green's functions with equal time variables must be interpreted as $G(\vx t,\vx t^+)$
\end{enumerate}

%\section*{References}
\bibliography{VU_bibtex_master}

\begin{thebibliography}{100}
\expandafter\ifx\csname url\endcsname\relax
  \def\url#1{\texttt{#1}}\fi
\expandafter\ifx\csname urlprefix\endcsname\relax\def\urlprefix{URL }\fi
\expandafter\ifx\csname href\endcsname\relax
  \def\href#1#2{#2} \def\path#1{#1}\fi

\bibitem{bardeen1957}
J.~Bardeen, L.~N. Cooper, J.~R. Schrieffer, Theory of {S}uperconductivity,
  Phys. Rev. 108 (1957) 1175.
\newblock \href {http://dx.doi.org/10.1103/PhysRev.108.1175}
  {\path{doi:10.1103/PhysRev.108.1175}}.

\bibitem{dasgupta1998}
M.~Dasgupta, D.~J. Hinde, N.~Rowley, A.~M. Stefanini, {Measuring Barriers to
  Fusion}, Annu. Rev. Nucl. Part. Sci. 48 (1998) 401--461.
\newblock \href {http://dx.doi.org/10.1146/annurev.nucl.48.1.401}
  {\path{doi:10.1146/annurev.nucl.48.1.401}}.

\bibitem{hagino2012}
K.~Hagino, N.~Takigawa, {S}ubbarrier {F}usion {R}eactions and {M}any-{P}article
  {Q}uantum {T}unneling, Prog. Theor. Phys. 128 (2012) 1001--1060.
\newblock \href {http://dx.doi.org/10.1143/PTP.128.1061}
  {\path{doi:10.1143/PTP.128.1061}}.

\bibitem{zurek1991}
W.~H. Zurek, Decoherence and the {T}ransition from {Q}uantum to {C}lassical,
  Phys. Today 44 (1991) 36.
\newblock \href {http://dx.doi.org/10.1063/1.881293}
  {\path{doi:10.1063/1.881293}}.

\bibitem{joos2003}
E.~Joos, H.~D. Zeh, C.~Kiefer, D.~Giulini, J.~Kupsch, I.~Stamatescu,
  Decoherence and the appearance of a classical world in quantum theory,
  Springer, Berlin, 2003, second {E}d.

\bibitem{bohr1975}
A.~Bohr, B.~Mottelson, {Nuclear Structure, Vol. 2}, W.A. Benjamin, Inc., 1975.

\bibitem{caldeira1981}
A.~O. Caldeira, A.~J. Leggett, Influence of {D}issipation on {Q}uantum
  {T}unneling in {M}acroscopic {S}ystems, Phys. Rev. Lett. 46 (1981) 211--214.
\newblock \href {http://dx.doi.org/10.1103/PhysRevLett.46.211}
  {\path{doi:10.1103/PhysRevLett.46.211}}.

\bibitem{lamehi-rachti1976}
M.~{Lamehi--Rachti}, W.~Mittig, Quantum mechanics and hidden variables: {A}
  test of {B}ell's inequality by the measurement of the spin correlation in
  lowxi---energy proton--proton scattering, Phys. Rev. D 14 (1976) 2543--2555.
\newblock \href {http://dx.doi.org/10.1103/PhysRevD.14.2543}
  {\path{doi:10.1103/PhysRevD.14.2543}}.

\bibitem{dirac1930}
P.~A.~M. Dirac, {N}ote on {E}xchange {P}henomena in the {T}homas {A}tom, Math.
  Proc. Camb. Phil. Soc. 26 (1930) 376.
\newblock \href {http://dx.doi.org/10.1017/S0305004100016108}
  {\path{doi:10.1017/S0305004100016108}}.

\bibitem{bonche1976}
P.~Bonche, S.~Koonin, J.~W. Negele, {O}ne-dimensional nuclear dynamics in
  time-dependent {H}artree-{F}ock approximation, Phys. Rev. C 13 (1976)
  1226--1258.
\newblock \href {http://dx.doi.org/10.1103/PhysRevC.13.1226}
  {\path{doi:10.1103/PhysRevC.13.1226}}.

\bibitem{simenel2011}
C.~Simenel, {P}article-{N}umber {F}luctuations and {C}orrelations in {T}ransfer
  {R}eactions {O}btained {U}sing the {B}alian-{V}\'en\'eroni {V}ariational
  {P}rinciple, Phys. Rev. Lett. 106 (2011) 112502.
\newblock \href {http://dx.doi.org/10.1103/PhysRevLett.106.112502}
  {\path{doi:10.1103/PhysRevLett.106.112502}}.

\bibitem{balian1981}
R.~Balian, M.~V\'en\'eroni, {T}ime-{D}ependent {V}ariational {P}rinciple for
  {P}redicting the {E}xpectation {V}alue of an {O}bservable, Phys. Rev. Lett.
  47 (1981) 1353.
\newblock \href {http://dx.doi.org/10.1103/PhysRevLett.47.1353}
  {\path{doi:10.1103/PhysRevLett.47.1353}}.

\bibitem{iwata2010}
Y.~Iwata, T.~Otsuka, J.~A. Maruhn, N.~Itagaki, {Suppression of Charge
  Equilibration Leading to the Synthesis of Exotic Nuclei}, Phys. Rev. Lett.
  104 (2010) 252501.
\newblock \href {http://dx.doi.org/10.1103/PhysRevLett.104.252501}
  {\path{doi:10.1103/PhysRevLett.104.252501}}.

\bibitem{stone2017}
J.~R. Stone, P.~Danielewicz, Y.~Iwata, Proton and neutron density distributions
  at supranormal density in low- and medium-energy heavy-ion collisions, Phys.
  Rev. C 96 (2017) 014612.
\newblock \href {http://dx.doi.org/10.1103/PhysRevC.96.014612}
  {\path{doi:10.1103/PhysRevC.96.014612}}.

\bibitem{negele1982}
J.~W. Negele, {T}he mean-field theory of nuclear-structure and dynamics, Rev.
  Mod. Phys. 54 (1982) 913--1015.
\newblock \href {http://dx.doi.org/10.1103/RevModPhys.54.913}
  {\path{doi:10.1103/RevModPhys.54.913}}.

\bibitem{simenel2012}
C.~Simenel, {N}uclear quantum many-body dynamics, Eur. Phys. J. A 48 (2012)
  152.
\newblock \href {http://dx.doi.org/10.1140/epja/i2012-12152-0}
  {\path{doi:10.1140/epja/i2012-12152-0}}.

\bibitem{nakatsukasa2016}
T.~Nakatsukasa, K.~Matsuyanagi, M.~Matsuo, K.~Yabana, Time-dependent
  density-functional description of nuclear dynamics, Rev. Mod. Phys. 88 (2016)
  045004.
\newblock \href {http://dx.doi.org/10.1103/RevModPhys.88.045004}
  {\path{doi:10.1103/RevModPhys.88.045004}}.

\bibitem{thouless1960}
D.~J. Thouless, Stability conditions and nuclear rotations in the
  {H}artree--{F}ock theory, Nucl. Phys. 21 (1960) 225--232.
\newblock \href {http://dx.doi.org/10.1016/0029-5582(60)90048-1}
  {\path{doi:10.1016/0029-5582(60)90048-1}}.

\bibitem{lacroix2014}
D.~Lacroix, S.~Ayik, {S}tochastic quantum dynamics beyond mean field, Eur.
  Phys. J. A 50 (2014) 95.
\newblock \href {http://dx.doi.org/10.1140/epja/i2014-14095-8}
  {\path{doi:10.1140/epja/i2014-14095-8}}.

\bibitem{blaizot1981}
J.~P. Blaizot, G.~Ripka, A variational principle for the calculation of
  transition amplitudes, Phys. Lett. B 105 (1981) 1.
\newblock \href {http://dx.doi.org/10.1016/0370-2693(81)90026-5}
  {\path{doi:10.1016/0370-2693(81)90026-5}}.

\bibitem{fetter2003}
A.~L. Fetter, J.~D. Walecka,
  \href{https://books.google.com.au/books?id=0wekf1s83b0C}{Quantum {T}heory of
  {M}any-particle {S}ystems}, Dover Books on Physics, Dover Publications, 2003.
\newline\urlprefix\url{https://books.google.com.au/books?id=0wekf1s83b0C}

\bibitem{bogoliubov1946}
N.~N. Bogoliubov, Kinetic {E}quations, J. Phys. (USSR) 10 (1946) 256.

\bibitem{born1946}
H.~Born, H.~S. Green, A {G}eneral {K}inetic {T}heory of {L}iquids, Proc. Roy.
  Soc. A188 (1946) 10.
\newblock \href {http://dx.doi.org/10.1098/rspa.1946.0093}
  {\path{doi:10.1098/rspa.1946.0093}}.

\bibitem{kirkwood1946}
J.~G. Kirkwood, {The Statistical Mechanical Theory of Transport Processes I.
  General Theory}, J. Chem. Phys. 14 (1946) 180.
\newblock \href {http://dx.doi.org/10.1063/1.1724117}
  {\path{doi:10.1063/1.1724117}}.

\bibitem{avez2008}
B.~Avez, C.~Simenel, P.~Chomaz, {P}airing vibrations study with the
  time-dependent {H}artree-{F}ock-{B}ogoliubov theory, Phys. Rev. C 78 (2008)
  044318.
\newblock \href {http://dx.doi.org/10.1103/PhysRevC.78.044318}
  {\path{doi:10.1103/PhysRevC.78.044318}}.

\bibitem{ebata2010}
S.~Ebata, T.~Nakatsukasa, T.~Inakura, K.~Yoshida, Y.~Hashimoto, K.~Yabana,
  {C}anonical--basis time-dependent {H}artree-{F}ock-{B}ogoliubov theory and
  linear--response calculations, Phys. Rev. C 82 (2010) 034306.
\newblock \href {http://dx.doi.org/10.1103/PhysRevC.82.034306}
  {\path{doi:10.1103/PhysRevC.82.034306}}.

\bibitem{scamps2013a}
G.~Scamps, D.~Lacroix, {E}ffect of pairing on one- and two-nucleon transfer
  below the {C}oulomb barrier: {A} time-dependent microscopic description,
  Phys. Rev. C 87 (2013) 014605.
\newblock \href {http://dx.doi.org/10.1103/PhysRevC.87.014605}
  {\path{doi:10.1103/PhysRevC.87.014605}}.

\bibitem{stetcu2011}
I.~Stetcu, A.~Bulgac, P.~Magierski, K.~J. Roche, {I}sovector giant dipole
  resonance from the 3{D} time--dependent density functional theory for
  superfluid nuclei, Phys. Rev. C 84 (2011) 051309.
\newblock \href {http://dx.doi.org/10.1103/PhysRevC.84.051309}
  {\path{doi:10.1103/PhysRevC.84.051309}}.

\bibitem{dasso1979}
C.~H. Dasso, T.~Dossing, H.~C. Pauli, {O}n the mass distribution in
  {T}ime-{D}ependent {H}artree-{F}ock calculations of heavy-ion collisions, Z.
  Phys. A 289 (1979) 395--398.
\newblock \href {http://dx.doi.org/10.1007/BF01409391}
  {\path{doi:10.1007/BF01409391}}.

\bibitem{balian1984}
R.~Balian, M.~V\'en\'eroni, {F}luctuations in a time-dependent mean-field
  approach, Phys. Lett. B 136 (1984) 301--306.
\newblock \href {http://dx.doi.org/10.1016/0370-2693(84)92008-2}
  {\path{doi:10.1016/0370-2693(84)92008-2}}.

\bibitem{ayik2008}
S.~Ayik, A stochastic mean-field approach for nuclear dynamics, Phys. Lett. B
  658 (2008) 174.
\newblock \href {http://dx.doi.org/10.1016/j.physletb.2007.09.072}
  {\path{doi:10.1016/j.physletb.2007.09.072}}.

\bibitem{kim1997}
K.-H. Kim, T.~Otsuka, P.~Bonche, {T}hree-dimensional {TDHF} calculations for
  reactions of unstable nuclei, J. Phys. G 23 (1997) 1267.
\newblock \href {http://dx.doi.org/10.1088/0954-3899/23/10/014}
  {\path{doi:10.1088/0954-3899/23/10/014}}.

\bibitem{umar2005a}
A.~S. Umar, V.~E. Oberacker, {T}ime-dependent response calculations of nuclear
  resonances, Phys. Rev. C 71 (2005) 034314.
\newblock \href {http://dx.doi.org/10.1103/PhysRevC.71.034314}
  {\path{doi:10.1103/PhysRevC.71.034314}}.

\bibitem{nakatsukasa2005}
T.~Nakatsukasa, K.~Yabana, {L}inear response theory in the continuum for
  deformed nuclei: {Green}'s function vs time-dependent {H}artree-{F}ock with
  the absorbing boundary condition, Phys. Rev. C 71 (2005) 024301.
\newblock \href {http://dx.doi.org/10.1103/PhysRevC.71.024301}
  {\path{doi:10.1103/PhysRevC.71.024301}}.

\bibitem{maruhn2014}
J.~A. Maruhn, P.-G. Reinhard, P.~D. Stevenson, A.~S. Umar, {T}he {TDHF C}ode
  {S}ky3{D}, Comput. Phys. Commun. 185 (2014) 2195--2216.
\newblock \href {http://dx.doi.org/10.1016/j.cpc.2014.04.008}
  {\path{doi:10.1016/j.cpc.2014.04.008}}.

\bibitem{umar2008a}
A.~S. Umar, V.~E. Oberacker, J.~A. Maruhn, {N}eutron transfer dynamics and
  doorway to fusion in time-dependent {H}artree-{F}ock theory, Eur. Phys. J. A
  37 (2008) 245--250.
\newblock \href {http://dx.doi.org/10.1140/epja/i2008-10614-6}
  {\path{doi:10.1140/epja/i2008-10614-6}}.

\bibitem{feynman1948}
R.~P. Feynman, Space-{T}ime {A}pproach to {N}on-{R}elativistic {Q}uantum
  {M}echanics, Rev. Mod. Phys. 20 (1948) 367--387.
\newblock \href {http://dx.doi.org/10.1103/RevModPhys.20.367}
  {\path{doi:10.1103/RevModPhys.20.367}}.

\bibitem{levit1980a}
S.~Levit, Time-dependent mean-field approximation for nuclear dynamical
  problems, Phys. Rev. C 21 (1980) 1594--1602.
\newblock \href {http://dx.doi.org/10.1103/PhysRevC.21.1594}
  {\path{doi:10.1103/PhysRevC.21.1594}}.

\bibitem{levit1980b}
S.~Levit, J.~W. Negele, Z.~Paltiel, Time-dependent mean-field theory and
  quantized bound states, Phys. Rev. C 21 (1980) 1603--1625.
\newblock \href {http://dx.doi.org/10.1103/PhysRevC.21.1603}
  {\path{doi:10.1103/PhysRevC.21.1603}}.

\bibitem{bass1974}
R.~Bass, {F}usion of heavy nuclei in a classical model, Nucl. Phys. A 231
  (1974) 45--63.
\newblock \href {http://dx.doi.org/10.1016/0375-9474(74)90292-9}
  {\path{doi:10.1016/0375-9474(74)90292-9}}.

\bibitem{bass1980}
R.~Bass, Nuclear {R}eactions with {H}eavy {I}ons, Springer--{V}erlag, New York,
  1980.

\bibitem{blocki1977}
J.~B{\l}ocki, J.~Randrup, W.~J. Swiatecki, C.~F. Tsang, Proximity forces, Ann.
  Phys. 105 (1977) 427.
\newblock \href {http://dx.doi.org/10.1016/0003-4916(77)90249-4}
  {\path{doi:10.1016/0003-4916(77)90249-4}}.

\bibitem{randrup1978a}
J.~Randrup, J.~S. Vaagen, On the proximity treatment of the interaction between
  deformed nuclei, Phys. Lett. B 77 (1978) 170--173.
\newblock \href {http://dx.doi.org/10.1016/0370-2693(78)90613-5}
  {\path{doi:10.1016/0370-2693(78)90613-5}}.

\bibitem{seiwert1984}
M.~Seiwert, W.~Greiner, V.~Oberacker, M.~J. Rhoades-Brown, Test of the
  proximity theorem for deformed nuclei, Phys. Rev. C 29 (1984) 477--485.
\newblock \href {http://dx.doi.org/10.1103/PhysRevC.29.477}
  {\path{doi:10.1103/PhysRevC.29.477}}.

\bibitem{birkelund1983}
J.~R. Birkelund, J.~R. Huizenga, {F}usion {R}eactions {B}etween {H}eavy
  {N}uclei, Annu. Rev. Nucl. Part. Sci. 33 (1983) 265--322.
\newblock \href {http://dx.doi.org/10.1146/annurev.ns.33.120183.001405}
  {\path{doi:10.1146/annurev.ns.33.120183.001405}}.

\bibitem{satchler1979}
G.~R. Satchler, W.~G. Love, {F}olding model potentials from realistic
  interactions for heavy-ion scattering, Phys. Rep. 55 (1979) 183--254.
\newblock \href {http://dx.doi.org/10.1016/0370-1573(79)90081-4}
  {\path{doi:10.1016/0370-1573(79)90081-4}}.

\bibitem{bertsch1977}
G.~Bertsch, J.~Borysowicz, H.~Mc{M}anus, W.~G. Love, {I}nteractions for
  inelastic-scattering derived from realistic potentials, Nucl. Phys. A 284
  (1977) 399--419.
\newblock \href {http://dx.doi.org/10.1016/0375-9474(77)90392-X}
  {\path{doi:10.1016/0375-9474(77)90392-X}}.

\bibitem{takigawa1984}
N.~Takigawa, G.~F. Bertsch, Semiclassical theory of quantum tunneling in
  multidimensional systems, Phys. Rev. C 29 (1984) 2358--2361.
\newblock \href {http://dx.doi.org/10.1103/PhysRevC.29.2358}
  {\path{doi:10.1103/PhysRevC.29.2358}}.

\bibitem{balantekin1998}
A.~B. Balantekin, N.~Takigawa, Quantum tunneling in nuclear fusion, Rev. Mod.
  Phys. 70 (1998) 77--100.
\newblock \href {http://dx.doi.org/10.1103/RevModPhys.70.77}
  {\path{doi:10.1103/RevModPhys.70.77}}.

\bibitem{landowne1984}
S.~Landowne, S.~C. Pieper, {C}oupled-channels fusion calculations for
  $^{58}\mathrm{Ni}$+$^{58}\mathrm{Ni}$, Phys. Rev. C 29 (1984) 1352--1357.
\newblock \href {http://dx.doi.org/10.1103/PhysRevC.29.1352}
  {\path{doi:10.1103/PhysRevC.29.1352}}.

\bibitem{hagino1999}
K.~Hagino, N.~Rowley, A.~Kruppa, A program for coupled-channel calculations
  with all order couplings for heavy-ion fusion reactions, Comput. Phys.
  Commun. 123 (1999) 143--152.
\newblock \href {http://dx.doi.org/10.1016/s0010-4655(99)00243-x}
  {\path{doi:10.1016/s0010-4655(99)00243-x}}.

\bibitem{esbensen2004}
H.~Esbensen, Challenges in {C}oupled--{C}hannels {C}alculations of
  {H}eavy--{I}on {F}usion {R}eactions, Prog. Theor. Phys. Suppl. 154 (2004)
  11--20.
\newblock \href {http://dx.doi.org/10.1143/PTPS.154.11}
  {\path{doi:10.1143/PTPS.154.11}}.

\bibitem{karpov2015}
A.~V. Karpov, V.~A. Rachkov, V.~V. Samarin, Quantum coupled-channels model of
  nuclear fusion with a semiclassical consideration of neutron rearrangement,
  Phys. Rev. C 92 (2015) 064603.
\newblock \href {http://dx.doi.org/10.1103/PhysRevC.92.064603}
  {\path{doi:10.1103/PhysRevC.92.064603}}.

\bibitem{gasques2004}
L.~R. Gasques, L.~C. Chamon, D.~Pereira, M.~A.~G. Alvarez, E.~S. Rossi, C.~P.
  Silva, B.~V. Carlson, {G}lobal and consistent analysis of the heavy-ion
  elastic scattering and fusion processes, Phys. Rev. C 69 (2004) 034603.
\newblock \href {http://dx.doi.org/10.1103/PhysRevC.69.034603}
  {\path{doi:10.1103/PhysRevC.69.034603}}.

\bibitem{chamon2002}
L.~C. Chamon, B.~V. Carlson, L.~R. Gasques, D.~Pereira, C.~{De Conti}, M.~A.~G.
  Alvarez, M.~S. Hussein, M.~A.~C. Ribeiro, E.~S. Rossi, C.~P. Silva, {T}oward
  a global description of the nucleus-nucleus interaction, Phys. Rev. C 66
  (2002) 014610.
\newblock \href {http://dx.doi.org/10.1103/PhysRevC.66.014610}
  {\path{doi:10.1103/PhysRevC.66.014610}}.

\bibitem{ichikawa2015b}
T.~Ichikawa, K.~Matsuyanagi, Universal damping mechanism of quantum vibrations
  in deep sub-barrier fusion reactions, Phys. Rev. C 92 (2015) 021602.
\newblock \href {http://dx.doi.org/10.1103/physrevc.92.021602}
  {\path{doi:10.1103/physrevc.92.021602}}.

\bibitem{misicu2006}
\c{S}. Mi\c{s}icu, H.~Esbensen, {H}indrance of {H}eavy-{I}on {F}usion due to
  {N}uclear {I}ncompressibility, Phys. Rev. Lett. 96 (2006) 112701.
\newblock \href {http://dx.doi.org/10.1103/PhysRevLett.96.112701}
  {\path{doi:10.1103/PhysRevLett.96.112701}}.

\bibitem{ichikawa2007}
T.~Ichikawa, K.~Hagino, A.~Iwamoto, {E}xistence of a one-body barrier revealed
  in deep subbarrier fusion, Phys. Rev. C 75 (2007) 057603.
\newblock \href {http://dx.doi.org/10.1103/physrevc.75.057603}
  {\path{doi:10.1103/physrevc.75.057603}}.

\bibitem{weizsacker1935}
C.~F. {von Weizs\"acker}, {Z}ur {T}heorie der {K}ernmassen, Z. Phys. A 96
  (1935) 431--458.
\newblock \href {http://dx.doi.org/10.1007/BF01337700}
  {\path{doi:10.1007/BF01337700}}.

\bibitem{bethe1936}
H.~A. Bethe, R.~F. Bacher, {N}uclear {P}hysics {A}. {S}tationary {S}tates of
  {N}uclei, Rev. Mod. Phys. 8 (1936) 82--229.
\newblock \href {http://dx.doi.org/10.1103/RevModPhys.8.82}
  {\path{doi:10.1103/RevModPhys.8.82}}.

\bibitem{brueckner1968}
K.~A. Brueckner, J.~R. Buchler, M.~M. Kelly, {N}ew {T}heoretical {A}pproach to
  {N}uclear {H}eavy-{I}on {S}cattering, Phys. Rev. 173 (1968) 944--949.
\newblock \href {http://dx.doi.org/10.1103/physrev.173.944}
  {\path{doi:10.1103/physrev.173.944}}.

\bibitem{skyrme1956}
T.~H.~R. Skyrme, {CVII}. {T}he nuclear surface, Phil. Mag. 1 (1956) 1043--1054.
\newblock \href {http://dx.doi.org/10.1080/14786435608238186}
  {\path{doi:10.1080/14786435608238186}}.

\bibitem{reinhard2016a}
P.-G. Reinhard, A.~S. Umar, P.~D. Stevenson, J.~Piekarewicz, V.~E. Oberacker,
  J.~A. Maruhn, {S}ensitivity of the fusion cross section to the density
  dependence of the symmetry energy, Phys. Rev. C 93 (2016) 044618.
\newblock \href {http://dx.doi.org/10.1103/PhysRevC.93.044618}
  {\path{doi:10.1103/PhysRevC.93.044618}}.

\bibitem{denisov2002}
V.~Y. Denisov, W.~N\"orenberg, {E}ntrance channel potentials in the synthesis
  of the heaviest nuclei, Eur. Phys. J. A 15 (2002) 375--388.
\newblock \href {http://dx.doi.org/10.1140/epja/i2002-10039-3}
  {\path{doi:10.1140/epja/i2002-10039-3}}.

\bibitem{washiyama2008}
K.~Washiyama, D.~Lacroix, {E}nergy dependence of the nucleus-nucleus potential
  close to the {C}oulomb barrier, Phys. Rev. C 78 (2008) 024610.
\newblock \href {http://dx.doi.org/10.1103/PhysRevC.78.024610}
  {\path{doi:10.1103/PhysRevC.78.024610}}.

\bibitem{simenel2008}
C.~Simenel, B.~Avez, Time--dependent {H}artree--{F}ock description of heavy
  {io}ns fusion, Intl. J. Mod. Phys. E 17 (2008) 31--40.
\newblock \href {http://dx.doi.org/10.1142/S0218301308009525}
  {\path{doi:10.1142/S0218301308009525}}.

\bibitem{vophuoc2016}
K.~Vo-Phuoc, C.~Simenel, E.~C. Simpson, Dynamical effects in fusion with exotic
  nuclei, Phys. Rev. C 94 (2016) 024612.
\newblock \href {http://dx.doi.org/10.1103/physrevc.94.024612}
  {\path{doi:10.1103/physrevc.94.024612}}.

\bibitem{akyuz1981}
O.~Aky\"uz, A.~Winther, Nuclear surface-surface interaction in the folding
  model, in: R.~A. Broglia, R.~A. Ricci (Eds.), Nuclear Structure and Heavy-Ion
  Physics, Proc. Int. School of Physics "Enrico Fermi", Varenna, North Holland,
  Amsterdam, 1982, p. 492.

\bibitem{simenel2013b}
C.~Simenel, M.~Dasgupta, D.~J. Hinde, E.~Williams, {M}icroscopic approach to
  coupled-channels effects on fusion, Phys. Rev. C 88 (2013) 064604.
\newblock \href {http://dx.doi.org/10.1103/PhysRevC.88.064604}
  {\path{doi:10.1103/PhysRevC.88.064604}}.

\bibitem{bourgin2016}
D.~Bourgin, C.~Simenel, S.~Courtin, F.~Haas, {M}icroscopic study of $^{40}${C}a
  $+$ $^{58,64}${N}i fusion reactions, Phys. Rev. C 93 (2016) 034604.
\newblock \href {http://dx.doi.org/10.1103/PhysRevC.93.034604}
  {\path{doi:10.1103/PhysRevC.93.034604}}.

\bibitem{simenel2003}
C.~Simenel, P.~Chomaz, {N}onlinear vibrations in nuclei, Phys. Rev. C 68 (2003)
  024302.
\newblock \href {http://dx.doi.org/10.1103/PhysRevC.68.024302}
  {\path{doi:10.1103/PhysRevC.68.024302}}.

\bibitem{stevenson2004}
P.~D. Stevenson, M.~R. Strayer, J.~R. Stone, W.~G. Newton, {G}iant resonances
  from {TDHF}, Intl. J. Mod. Phys. E 13 (2004) 181--185.
\newblock \href {http://dx.doi.org/10.1142/S0218301304001928}
  {\path{doi:10.1142/S0218301304001928}}.

\bibitem{maruhn2005}
J.~A. Maruhn, P.~G. Reinhard, P.~D. Stevenson, J.~{Rikovska Stone}, M.~R.
  Strayer, {D}ipole giant resonances in deformed heavy nuclei, Phys. Rev. C 71
  (2005) 064328.
\newblock \href {http://dx.doi.org/10.1103/PhysRevC.71.064328}
  {\path{doi:10.1103/PhysRevC.71.064328}}.

\bibitem{stevenson2007}
P.~D. Stevenson, D.~Almehed, P.-G. Reinhard, J.~A. Maruhn, {M}onopole giant
  resonances and {TDHF} boundary conditions, Nucl. Phys. A 788 (2007)
  343C--348C.
\newblock \href {http://dx.doi.org/10.1016/j.nuclphysa.2007.01.091}
  {\path{doi:10.1016/j.nuclphysa.2007.01.091}}.

\bibitem{simenel2009}
C.~Simenel, P.~Chomaz, {C}ouplings between dipole and quadrupole vibrations in
  tin isotopes, Phys. Rev. C 80 (2009) 064309.
\newblock \href {http://dx.doi.org/10.1103/PhysRevC.80.064309}
  {\path{doi:10.1103/PhysRevC.80.064309}}.

\bibitem{stevenson2010}
P.~D. Stevenson, S.~Fracasso, {E}xtracting structure information from {TDHF},
  J. Phys. G 37 (2010) 064030.
\newblock \href {http://dx.doi.org/10.1088/0954-3899/37/6/064030}
  {\path{doi:10.1088/0954-3899/37/6/064030}}.

\bibitem{fracasso2012}
S.~Fracasso, E.~B. Suckling, P.~D. Stevenson, {U}nrestricted {S}kyrme-tensor
  time-dependent {H}artree-{F}ock model and its application to the nuclear
  response from spherical to triaxial nuclei, Phys. Rev. C 86 (2012) 044303.
\newblock \href {http://dx.doi.org/10.1103/PhysRevC.86.044303}
  {\path{doi:10.1103/PhysRevC.86.044303}}.

\bibitem{scamps2013b}
G.~Scamps, D.~Lacroix, {S}ystematics of isovector and isoscalar giant
  quadrupole resonances in normal and superfluid spherical nuclei, Phys. Rev. C
  88 (2013) 044310.
\newblock \href {http://dx.doi.org/10.1103/PhysRevC.88.044310}
  {\path{doi:10.1103/PhysRevC.88.044310}}.

\bibitem{simenel2013a}
C.~Simenel, R.~Keser, A.~S. Umar, V.~E. Oberacker, Microscopic study of
  ${}^{16}\mathrm{O}+{}^{16}\mathrm{O}$ fusion, Phys. Rev. C 88 (2013) 024617.
\newblock \href {http://dx.doi.org/10.1103/PhysRevC.88.024617}
  {\path{doi:10.1103/PhysRevC.88.024617}}.

\bibitem{scamps2014a}
G.~Scamps, D.~Lacroix, {S}ystematic study of isovector and isoscalar giant
  quadrupole resonances in normal and superfluid deformed nuclei, Phys. Rev. C
  89 (2014) 034314.
\newblock \href {http://dx.doi.org/10.1103/PhysRevC.89.034314}
  {\path{doi:10.1103/PhysRevC.89.034314}}.

\bibitem{chabanat1998a}
E.~Chabanat, P.~Bonche, P.~Haensel, J.~Meyer, R.~Schaeffer, {A S}kyrme
  parametrization from subnuclear to neutron star densities {P}art {II}.
  {N}uclei far from stabilities, Nucl. Phys. A 635 (1998) 231--256.
\newblock \href {http://dx.doi.org/10.1016/S0375-9474(98)00180-8}
  {\path{doi:10.1016/S0375-9474(98)00180-8}}.

\bibitem{kibedi2002}
T.~Kib\'edi, R.~H. Spear, {Reduced electric-octupole transition probabilities,
  $B(E3;0_1^+\rightarrow3_1^-)$--AN UPDATE}, At. Data Nucl. Data Tables 80
  (2002) 35.
\newblock \href {http://dx.doi.org/10.1006/adnd.2001.0871}
  {\path{doi:10.1006/adnd.2001.0871}}.

\bibitem{reinhard2007}
P.-G. Reinhard, {Lu Guo}, J.~Maruhn, {N}uclear giant resonances and linear
  response, Eur. Phys. J. A 32 (2007) 19--23.
\newblock \href {http://dx.doi.org/10.1140/epja/i2007-10366-9}
  {\path{doi:10.1140/epja/i2007-10366-9}}.

\bibitem{fallot2003}
M.~Fallot, P.~Chomaz, M.~V. Andr\'es, F.~Catara, E.~G. Lanza, J.~A. Scarpaci,
  Anharmonic vibrations in nuclei, Nucl. Phys. A 729 (2003) 699.
\newblock \href {http://dx.doi.org/10.1016/j.nuclphysa.2003.10.001}
  {\path{doi:10.1016/j.nuclphysa.2003.10.001}}.

\bibitem{hagino2015}
K.~Hagino, J.~M. Yao, Semimicroscopic modeling of heavy-ion fusion reactions
  with multireference covariant density functional theory, Phys. Rev. C 91
  (2015) 064606.
\newblock \href {http://dx.doi.org/10.1103/PhysRevC.91.064606}
  {\path{doi:10.1103/PhysRevC.91.064606}}.

\bibitem{yao2016}
J.~M. Yao, K.~Hagino, Anharmonicity of multi--octupole-phonon excitations in
  $^{208}\mathrm{Pb}$: {A}nalysis with multireference covariant density
  functional theory and subbarrier fusion of
  $^{16}\mathrm{O}+^{208}\mathrm{Pb}$, Phys. Rev. C 94 (2016) 011303.
\newblock \href {http://dx.doi.org/10.1103/PhysRevC.94.011303}
  {\path{doi:10.1103/PhysRevC.94.011303}}.

\bibitem{fliessbach1975}
T.~Fliessbach, The reduced width amplitude in the reaction theory for composite
  particles, Z. Phys. A 272 (1975) 39--46.
\newblock \href {http://dx.doi.org/10.1007/bf01408426}
  {\path{doi:10.1007/bf01408426}}.

\bibitem{fliessbach1971}
T.~Fliessbach, The optical potential for the elastic heavy ion scattering, Z.
  Phys. 247 (1971) 117--126.
\newblock \href {http://dx.doi.org/10.1007/bf01395288}
  {\path{doi:10.1007/bf01395288}}.

\bibitem{brink1975}
D.~M. Brink, F.~Stancu, Interaction potential between two $^{16}${O} nuclei
  derived from the {S}kyrme interaction, Nucl. Phys. A 243 (1975) 175--188.
\newblock \href {http://dx.doi.org/10.1016/0375-9474(75)90027-5}
  {\path{doi:10.1016/0375-9474(75)90027-5}}.

\bibitem{zint1975}
P.~G. Zint, U.~Mosel, Kinetic energy contributions to heavy ion potentials,
  Phys. Lett. B 56 (1975) 424--426.
\newblock \href {http://dx.doi.org/10.1016/0370-2693(75)90402-5}
  {\path{doi:10.1016/0370-2693(75)90402-5}}.

\bibitem{beck1978}
F.~Beck, K.-H. M\"uller, H.~S. K\"ohler, {M}omentum {D}ependence of the
  {I}on-{I}on {P}otential in a {M}icroscopic {T}heory, Phys. Rev. Lett. 40
  (1978) 837--840.
\newblock \href {http://dx.doi.org/10.1103/physrevlett.40.837}
  {\path{doi:10.1103/physrevlett.40.837}}.

\bibitem{sinha1979}
B.~Sinha, S.~A. Moszkowski, The nucleus-nucleus interaction potential using
  density-dependent delta interaction, Phys. Lett. B 81 (1979) 289--294.
\newblock \href {http://dx.doi.org/10.1016/0370-2693(79)90337-X}
  {\path{doi:10.1016/0370-2693(79)90337-X}}.

\bibitem{dasso2003}
C.~H. Dasso, G.~Pollarolo, Investigating the nucleus-nucleus potential at very
  short distances, Phys. Rev. C 68 (2003) 054604.
\newblock \href {http://dx.doi.org/10.1103/physrevc.68.054604}
  {\path{doi:10.1103/physrevc.68.054604}}.

\bibitem{umar2012a}
A.~S. Umar, V.~E. Oberacker, C.~J. Horowitz, {M}icroscopic sub-barrier fusion
  calculations for the neutron star crust, Phys. Rev. C 85 (2012) 055801.
\newblock \href {http://dx.doi.org/10.1103/PhysRevC.85.055801}
  {\path{doi:10.1103/PhysRevC.85.055801}}.

\bibitem{cusson1985}
R.~Y. Cusson, P.-G. Reinhard, M.~R. Strayer, J.~A. Maruhn, W.~Greiner, Density
  as a constraint and the separation of internal excitation energy in {TDHF},
  Z. Phys. A 320 (1985) 475--482.
\newblock \href {http://dx.doi.org/10.1007/BF01415725}
  {\path{doi:10.1007/BF01415725}}.

\bibitem{umar1985}
A.~S. Umar, M.~R. Strayer, R.~Y. Cusson, P.-G. Reinhard, D.~A. Bromley,
  {T}ime-dependent {H}artree-{F}ock calculations of
  $^{4}${H}e${+\mathrm{}}^{14}${C},
  $^{12}${C}${+\mathrm{}}^{12}${C}${(0}^{+})$, and
  $^{4}${H}e${+\mathrm{}}^{20}${N}e molecular formations, Phys. Rev. C 32
  (1985) 172--183.
\newblock \href {http://dx.doi.org/10.1103/PhysRevC.32.172}
  {\path{doi:10.1103/PhysRevC.32.172}}.

\bibitem{simenel2017}
C.~Simenel, A.~S. Umar, K.~Godbey, M.~Dasgupta, D.~J. Hinde, How the {P}auli
  exclusion principle affects fusion of atomic nuclei, Phys. Rev. C 95 (2017)
  031601.
\newblock \href {http://dx.doi.org/10.1103/physrevc.95.031601}
  {\path{doi:10.1103/physrevc.95.031601}}.

\bibitem{ichikawa2009b}
T.~Ichikawa, K.~Hagino, A.~Iwamoto, {S}ignature of {S}mooth {T}ransition from
  {S}udden to {A}diabatic {S}tates in {H}eavy-{I}on {F}usion {R}eactions at
  {D}eep {S}ub-{B}arrier {E}nergies, Phys. Rev. Lett. 103 (2009) 202701.
\newblock \href {http://dx.doi.org/10.1103/PhysRevLett.103.202701}
  {\path{doi:10.1103/PhysRevLett.103.202701}}.

\bibitem{simenel2004}
C.~Simenel, P.~Chomaz, G.~{de France}, {Q}uantum {C}alculations of {C}oulomb
  {R}eorientation for {S}ub-{B}arrier {F}usion, Phys. Rev. Lett. 93 (2004)
  102701.
\newblock \href {http://dx.doi.org/10.1103/PhysRevLett.93.102701}
  {\path{doi:10.1103/PhysRevLett.93.102701}}.

\bibitem{umar2006a}
A.~S. Umar, V.~E. Oberacker, {D}ynamical deformation effects in subbarrier
  fusion of $^{64}${N}i+$^{132}${S}n, Phys. Rev. C 74 (2006) 061601.
\newblock \href {http://dx.doi.org/10.1103/PhysRevC.74.061601}
  {\path{doi:10.1103/PhysRevC.74.061601}}.

\bibitem{umar2006d}
A.~S. Umar, V.~E. Oberacker, {T}ime dependent {H}artree-{F}ock fusion
  calculations for spherical, deformed systems, Phys. Rev. C 74 (2006) 024606.
\newblock \href {http://dx.doi.org/10.1103/PhysRevC.74.024606}
  {\path{doi:10.1103/PhysRevC.74.024606}}.

\bibitem{umar2007}
A.~S. Umar, V.~E. Oberacker, $^{64}\mathrm{Ni}+^{132}\mathrm{Sn}$ fusion within
  the density-constrained time-dependent {H}artree-{F}ock formalism, Phys. Rev.
  C 76 (2007) 014614.
\newblock \href {http://dx.doi.org/10.1103/PhysRevC.76.014614}
  {\path{doi:10.1103/PhysRevC.76.014614}}.

\bibitem{umar2006b}
A.~S. Umar, V.~E. Oberacker, {H}eavy-ion interaction potential deduced from
  density-constrained time-dependent {H}artree-{F}ock calculation, Phys. Rev. C
  74 (2006) 021601.
\newblock \href {http://dx.doi.org/10.1103/PhysRevC.74.021601}
  {\path{doi:10.1103/PhysRevC.74.021601}}.

\bibitem{oberacker2013}
V.~E. Oberacker, A.~S. Umar, {M}icroscopic analysis of sub-barrier fusion
  enhancement in ${}^{132}\mathrm{Sn}+{}^{40}${C}a versus
  ${}^{132}\mathrm{Sn}+{}^{48}${C}a, Phys. Rev. C 87 (2013) 034611.
\newblock \href {http://dx.doi.org/10.1103/PhysRevC.87.034611}
  {\path{doi:10.1103/PhysRevC.87.034611}}.

\bibitem{jiang2014}
X.~Jiang, J.~A. Maruhn, S.~Yan, {M}icroscopic study of noncentral effects in
  heavy-ion fusion reactions with spherical nuclei, Phys. Rev. C 90 (2014)
  064618.
\newblock \href {http://dx.doi.org/10.1103/PhysRevC.90.064618}
  {\path{doi:10.1103/PhysRevC.90.064618}}.

\bibitem{umar2014a}
A.~S. Umar, C.~Simenel, V.~E. Oberacker, {E}nergy dependence of potential
  barriers and its effect on fusion cross sections, Phys. Rev. C 89 (2014)
  034611.
\newblock \href {http://dx.doi.org/10.1103/PhysRevC.89.034611}
  {\path{doi:10.1103/PhysRevC.89.034611}}.

\bibitem{umar2015a}
A.~S. Umar, V.~E. Oberacker, C.~Simenel, {S}hape evolution and collective
  dynamics of quasifission in the time-dependent {H}artree-{F}ock approach,
  Phys. Rev. C 92 (2015) 024621.
\newblock \href {http://dx.doi.org/10.1103/PhysRevC.92.024621}
  {\path{doi:10.1103/PhysRevC.92.024621}}.

\bibitem{simenel2010}
C.~Simenel, {P}article {T}ransfer {R}eactions with the {T}ime-{D}ependent
  {H}artree-{F}ock {T}heory {U}sing a {P}article {N}umber {P}rojection
  {T}echnique, Phys. Rev. Lett. 105 (2010) 192701.
\newblock \href {http://dx.doi.org/10.1103/PhysRevLett.105.192701}
  {\path{doi:10.1103/PhysRevLett.105.192701}}.

\bibitem{sekizawa2013}
K.~Sekizawa, K.~Yabana, {T}ime-dependent {H}artree-{F}ock calculations for
  multinucleon transfer processes in $^{40,48}${C}a+$^{124}${S}n,
  $^{40}${C}a+$^{208}${P}b, and $^{58}${N}i+$^{208}${P}b reactions, Phys. Rev.
  C 88 (2013) 014614.
\newblock \href {http://dx.doi.org/10.1103/PhysRevC.88.014614}
  {\path{doi:10.1103/PhysRevC.88.014614}}.

\bibitem{dobaczewski1995}
J.~Dobaczewski, J.~Dudek, {T}ime-odd components in the mean field of rotating
  superdeformed nuclei, Phys. Rev. C 52 (1995) 1827--1839.
\newblock \href {http://dx.doi.org/10.1103/PhysRevC.52.1827}
  {\path{doi:10.1103/PhysRevC.52.1827}}.

\bibitem{godbey2017}
K.~Godbey, A.~S. Umar, C.~Simenel, Dependence of fusion on isospin dynamics,
  Phys. Rev. C 95 (2017) 011601(R).
\newblock \href {http://dx.doi.org/10.1103/PhysRevC.95.011601}
  {\path{doi:10.1103/PhysRevC.95.011601}}.

\bibitem{evers2011}
M.~Evers, M.~Dasgupta, D.~J. Hinde, D.~H. Luong, R.~Rafiei, R.~du~Rietz,
  C.~Simenel, {C}luster transfer in the reaction $^{16}${O}+$^{208}${P}b at
  energies well below the fusion barrier: {A} possible doorway to energy
  dissipation, Phys. Rev. C 84 (2011) 054614.
\newblock \href {http://dx.doi.org/10.1103/PhysRevC.84.054614}
  {\path{doi:10.1103/PhysRevC.84.054614}}.

\bibitem{rafferty2016}
D.~C. Rafferty, M.~Dasgupta, D.~J. Hinde, C.~Simenel, E.~C. Simpson,
  E.~Williams, I.~P. Carter, K.~J. Cook, D.~H. Luong, S.~D. {McNeil},
  K.~Ramachandran, K.~Vo-Phuoc, A.~Wakhle, Multinucleon transfer in
  $^{16,18}${O}, $^{19}${F}+$^{208}${P}b reactions at energies near the fusion
  barrier, Phys. Rev. C 94 (2016) 024607.
\newblock \href {http://dx.doi.org/10.1103/physrevc.94.024607}
  {\path{doi:10.1103/physrevc.94.024607}}.

\bibitem{back2014}
B.~B. Back, H.~Esbensen, C.~L. Jiang, K.~E. Rehm, {R}ecent developments in
  heavy-ion fusion reactions, Rev. Mod. Phys. 86 (2014) 317--360.
\newblock \href {http://dx.doi.org/10.1103/RevModPhys.86.317}
  {\path{doi:10.1103/RevModPhys.86.317}}.

\bibitem{liang2016}
J.~F. Liang, J.~M. Allmond, C.~J. Gross, P.~E. Mueller, D.~Shapira, R.~L.
  Varner, M.~Dasgupta, D.~J. Hinde, C.~Simenel, E.~Williams, K.~{Vo--Phuoc},
  M.~L. Brown, I.~P. Carter, M.~Evers, D.~H. Luong, T.~Ebadi, A.~Wakhle,
  Examining the role of transfer coupling in sub-barrier fusion of
  $^{46,50}${T}i+$^{124}${S}n, Phys. Rev. C 94 (2016) 024616.
\newblock \href {http://dx.doi.org/10.1103/physrevc.94.024616}
  {\path{doi:10.1103/physrevc.94.024616}}.

\bibitem{simenel2001}
C.~Simenel, P.~Chomaz, G.~{de France}, {Q}uantum {C}alculation of the {D}ipole
  {E}xcitation in {F}usion {R}eactions, Phys. Rev. Lett. 86 (2001) 2971--2974.
\newblock \href {http://dx.doi.org/10.1103/PhysRevLett.86.2971}
  {\path{doi:10.1103/PhysRevLett.86.2971}}.

\bibitem{simenel2007}
C.~Simenel, P.~Chomaz, G.~{de France}, {F}usion process studied with a
  preequilibrium giant dipole resonance in time-dependent {H}artree-{F}ock
  theory, Phys. Rev. C 76 (2007) 024609.
\newblock \href {http://dx.doi.org/10.1103/PhysRevC.76.024609}
  {\path{doi:10.1103/PhysRevC.76.024609}}.

\bibitem{oberacker2012}
V.~E. Oberacker, A.~S. Umar, J.~A. Maruhn, P.-G. Reinhard, {D}ynamic
  microscopic study of pre-equilibrium giant resonance excitation and fusion in
  the reactions ${}^{132}${S}n $+$ ${}^{48}${C}a and ${}^{124}${S}n $+$
  ${}^{40}${C}a, Phys. Rev. C 85 (2012) 034609.
\newblock \href {http://dx.doi.org/10.1103/PhysRevC.85.034609}
  {\path{doi:10.1103/PhysRevC.85.034609}}.

\bibitem{simenel2012b}
C.~Simenel, D.~J. Hinde, R.~{du Rietz}, M.~Dasgupta, M.~Evers, C.~J. Lin, D.~H.
  Luong, A.~Wakhle, {I}nfluence of entrance-channel magicity and isospin on
  quasi-fission, Phys. Lett. B 710 (2012) 607--611.
\newblock \href {http://dx.doi.org/10.1016/j.physletb.2012.03.063}
  {\path{doi:10.1016/j.physletb.2012.03.063}}.

\bibitem{umar2017}
A.~S. Umar, C.~Simenel, W.~Ye, Transport properties of isospin asymmetric
  nuclear matter using the time-dependent {H}artree--{F}ock method, Phys. Rev.
  C 96 (2017) 024625.
\newblock \href {http://dx.doi.org/10.1103/PhysRevC.96.024625}
  {\path{doi:10.1103/PhysRevC.96.024625}}.

\bibitem{jiang2014a}
C.~L. Jiang, K.~E. Rehm, B.~B. Back, H.~Esbensen, R.~V.~F. Janssens, A.~M.
  Stefanini, G.~Montagnoli, Influence of heavy-ion transfer on fusion
  reactions, Phys. Rev. C 89 (2014) 051603(R).
\newblock \href {http://dx.doi.org/10.1103/physrevc.89.051603}
  {\path{doi:10.1103/physrevc.89.051603}}.

\bibitem{kolata2012}
J.~J. Kolata, A.~Roberts, A.~M. Howard, D.~Shapira, J.~F. Liang, C.~J. Gross,
  R.~L. Varner, Z.~Kohley, A.~N. Villano, H.~Amro, W.~Loveland, E.~Chavez,
  {F}usion of ${}^{124,132}${S}n with ${}^{40,48}${C}a, Phys. Rev. C 85 (2012)
  054603.
\newblock \href {http://dx.doi.org/10.1103/PhysRevC.85.054603}
  {\path{doi:10.1103/PhysRevC.85.054603}}.

\bibitem{umar2008b}
A.~S. Umar, V.~E. Oberacker, $^{64}${N}i+$^{64}${N}i fusion reaction calculated
  with the density-constrained time-dependent {H}artree-{F}ock formalism, Phys.
  Rev. C 77 (2008) 064605.
\newblock \href {http://dx.doi.org/10.1103/PhysRevC.77.064605}
  {\path{doi:10.1103/PhysRevC.77.064605}}.

\bibitem{umar2009b}
A.~S. Umar, V.~E. Oberacker, {D}ensity-constrained time-dependent
  {H}artree-{F}ock calculation of $^{16}${O}+$^{208}${P}b fusion
  cross-sections, Eur. Phys. J. A 39 (2009) 243--247.
\newblock \href {http://dx.doi.org/10.1140/epja/i2008-10712-5}
  {\path{doi:10.1140/epja/i2008-10712-5}}.

\bibitem{goeke1983}
K.~Goeke, F.~Gr\"ummer, P.-G. Reinhard, {T}hree-dimensional nuclear dynamics in
  the quantized {ATDHF} approach, Ann. Phys. 150 (1983) 504--551.
\newblock \href {http://dx.doi.org/10.1016/0003-4916(83)90025-8}
  {\path{doi:10.1016/0003-4916(83)90025-8}}.

\bibitem{rawitscher1964}
G.~H. Rawitscher, {A}pproximate {I}ndependence of {O}ptical-{M}odel {E}lastic
  {S}cattering {C}alculations on the {P}otential at {S}mall {D}istances, Phys.
  Rev. 135 (1964) B605--B612.
\newblock \href {http://dx.doi.org/10.1103/PhysRev.135.B605}
  {\path{doi:10.1103/PhysRev.135.B605}}.

\bibitem{thomas1986}
J.~Thomas, Y.~T. Chen, S.~Hinds, D.~Meredith, M.~Olson, {S}ub-barrier fusion of
  the oxygen isotopes: {A} more complete picture, Phys. Rev. C 33 (1986)
  1679--1689.
\newblock \href {http://dx.doi.org/10.1103/PhysRevC.33.1679}
  {\path{doi:10.1103/PhysRevC.33.1679}}.

\bibitem{tserruya1978}
I.~Tserruya, Y.~Eisen, D.~Pelte, A.~Gavron, H.~Oeschler, D.~Berndt, H.~L.
  Harney, {T}otal fusion cross section for the $^{16}${O}+$^{16}${O} system,
  Phys. Rev. C 18 (1978) 1688--1699.
\newblock \href {http://dx.doi.org/10.1103/PhysRevC.18.1688}
  {\path{doi:10.1103/PhysRevC.18.1688}}.

\bibitem{fernandez1978}
B.~Fernandez, C.~Gaarde, J.~S. Larsen, S.~Pontoppidan, F.~Videbaek, {F}usion
  cross sections for the $^{16}${O} + $^{16}${O} reaction, Nucl. Phys. A 306
  (1978) 259--284.
\newblock \href {http://dx.doi.org/10.1016/0375-9474(78)90327-5}
  {\path{doi:10.1016/0375-9474(78)90327-5}}.

\bibitem{wu1984}
S.-C. Wu, C.~A. Barnes, {F}usion and elastic scattering cross sections for the
  $^{16}${O} + $^{16}${O} reactions near the {C}oulomb barrier, Nucl. Phys. A
  422 (1984) 373--396.
\newblock \href {http://dx.doi.org/10.1016/0375-9474(84)90523-2}
  {\path{doi:10.1016/0375-9474(84)90523-2}}.

\bibitem{kolata1979}
J.~J. Kolata, R.~M. Freeman, F.~Haas, B.~Heusch, A.~Gallmann, {G}ross and
  intermediate-width structure in the interaction of $^{16}${O} with
  $^{16}${O}, Phys. Rev. C 19 (1979) 2237--2245.
\newblock \href {http://dx.doi.org/10.1103/PhysRevC.19.2237}
  {\path{doi:10.1103/PhysRevC.19.2237}}.

\bibitem{morton1999}
C.~R. Morton, A.~C. Berriman, M.~Dasgupta, D.~J. Hinde, J.~O. Newton,
  K.~Hagino, I.~J. Thompson, {C}oupled-channels analysis of the
  $^{16}${O}+$^{208}${P}b fusion barrier distribution, Phys. Rev. C 60 (1999)
  044608.
\newblock \href {http://dx.doi.org/10.1103/PhysRevC.60.044608}
  {\path{doi:10.1103/PhysRevC.60.044608}}.

\bibitem{montagnoli2012}
G.~Montagnoli, A.~M. Stefanini, C.~L. Jiang, H.~Esbensen, L.~Corradi,
  S.~Courtin, E.~Fioretto, A.~Goasduff, F.~Haas, A.~F. Kifle, C.~Michelagnoli,
  D.~Montanari, T.~Mijatovi\c{c}, K.~E. Rehm, R.~Silvestri, P.~P. Singh,
  F.~Scarlassara, S.~Szilner, X.~D. Tang, C.~A. Ur, {F}usion of
  ${}^{\text{40}}\text{Ca}+{}^{\text{40}}\text{Ca}$ and other
  $\text{Ca}+\text{Ca}$ systems near and below the barrier, Phys. Rev. C 85
  (2012) 024607.
\newblock \href {http://dx.doi.org/10.1103/PhysRevC.85.024607}
  {\path{doi:10.1103/PhysRevC.85.024607}}.

\bibitem{aljuwair1984}
H.~A. Aljuwair, R.~J. Ledoux, M.~Beckerman, S.~B. Gazes, J.~Wiggins, E.~R.
  Cosman, R.~R. Betts, S.~Saini, O.~Hansen, {Isotopic effects in the fusion of
  $^{40}\mathrm{Ca}$ with $^{40,44,48}\mathrm{Ca}$},, Phys. Rev. C 30 (1984)
  1223--1227.
\newblock \href {http://dx.doi.org/10.1103/PhysRevC.30.1223}
  {\path{doi:10.1103/PhysRevC.30.1223}}.

\bibitem{jiang2015a}
X.~Jiang, J.~A. Maruhn, S.~W. Yan, Configuration transition effect in heavy-ion
  fusion reactions with deformed nuclei, EPL 112 (2015) 12001.
\newblock \href {http://dx.doi.org/10.1209/0295-5075/112/12001}
  {\path{doi:10.1209/0295-5075/112/12001}}.

\bibitem{schuetrumpf2017}
B.~Schuetrumpf, W.~Nazarewicz, Cluster formation in precompound nuclei in the
  time--dependent framework, Phys. Rev. C 96 (2017) 064608.
\newblock \href {http://dx.doi.org/10.1103/PhysRevC.96.064608}
  {\path{doi:10.1103/PhysRevC.96.064608}}.

\bibitem{gasques2007}
L.~R. Gasques, A.~V. Afanasjev, M.~Beard, J.~Lubian, T.~Neff, M.~Wiescher,
  D.~G. Yakovlev, {S}\~ao {P}aulo potential as a tool for calculating ${S}$
  factors of fusion reactions in dense stellar matter, Phys. Rev. C 76 (2007)
  045802.
\newblock \href {http://dx.doi.org/10.1103/PhysRevC.76.045802}
  {\path{doi:10.1103/PhysRevC.76.045802}}.

\bibitem{rudolph2012}
M.~J. Rudolph, Z.~Q. Gosser, K.~Brown, S.~Hudan, R.~T. {de Souza}, A.~Chbihi,
  B.~Jacquot, M.~Famiano, J.~F. Liang, D.~Shapira, D.~Mercier, {N}ear- and
  sub-barrier fusion of $^{20}${O} incident ions with $^{12}${C} target nuclei,
  Phys. Rev. C 85 (2012) 024605.
\newblock \href {http://dx.doi.org/10.1103/PhysRevC.85.024605}
  {\path{doi:10.1103/PhysRevC.85.024605}}.

\bibitem{desouza2013}
R.~T. deSouza, S.~Hudan, V.~E. Oberacker, A.~S. Umar, {C}onfronting measured
  near- and sub-barrier fusion cross sections for $^{20}${O}+$^{12}${C} with a
  microscopic method, Phys. Rev. C 88 (2013) 014602.
\newblock \href {http://dx.doi.org/10.1103/PhysRevC.88.014602}
  {\path{doi:10.1103/PhysRevC.88.014602}}.

\bibitem{singh2017}
V.~Singh, J.~Vadas, T.~K. Steinbach, B.~B. Wiggins, S.~Hudan, R.~T. {deSouza},
  {Zidu Lin}, C.~J. Horowitz, L.~T. Baby, S.~A. Kuvin, {Vandana Tripathi},
  I.~Wiedenh\"over, A.~S. Umar, Fusion enhancement at near and sub-barrier
  energies in $^{19}${O}+$^{12}${C}, Phys. Lett. B 765 (2017) 99--103.
\newblock \href {http://dx.doi.org/10.1016/j.physletb.2016.12.017}
  {\path{doi:10.1016/j.physletb.2016.12.017}}.

\bibitem{steinbach2014}
T.~K. Steinbach, J.~Vadas, J.~Schmidt, C.~Haycraft, S.~Hudan, R.~T. deSouza,
  L.~T. Baby, S.~A. Kuvin, I.~Wiedenh\"over, A.~S. Umar, V.~E. Oberacker,
  Sub-barrier enhancement of fusion as compared to a microscopic method in
  $^{18}\mathrm{O}+^{12}\mathrm{C}$, Phys. Rev. C 90 (2014) 041603.
\newblock \href {http://dx.doi.org/10.1103/PhysRevC.90.041603}
  {\path{doi:10.1103/PhysRevC.90.041603}}.

\bibitem{astier2010}
A.~Astier, P.~Petkov, M.-G. Porquet, D.~S. Delion, P.~Schuck, {Novel
  Manifestation of $\alpha$-Clustering Structures: New
  $\alpha+^{208}\mathrm{Pb}$ States in $^{212}\mathrm{Po}$ Revealed by Their
  Enhanced $E1$ Decays}, Phys. Rev. Lett. 104 (2010) 042701.
\newblock \href {http://dx.doi.org/10.1103/PhysRevLett.104.042701}
  {\path{doi:10.1103/PhysRevLett.104.042701}}.

\bibitem{umar2010c}
A.~S. Umar, J.~A. Maruhn, N.~Itagaki, V.~E. Oberacker, {M}icroscopic {S}tudy of
  the {T}riple$-\alpha$ {R}eaction, Phys. Rev. Lett. 104 (2010) 212503.
\newblock \href {http://dx.doi.org/10.1103/PhysRevLett.104.212503}
  {\path{doi:10.1103/PhysRevLett.104.212503}}.

\bibitem{iwata2013}
Y.~Iwata, K.~Iida, N.~Itagaki, Synthesis of thin, long heavy nuclei in ternary
  collisions, Phys. Rev. C 87 (2013) 014609.
\newblock \href {http://dx.doi.org/10.1103/PhysRevC.87.014609}
  {\path{doi:10.1103/PhysRevC.87.014609}}.

\bibitem{iwata2015}
Y.~Iwata, T.~Ichikawa, N.~Itagaki, J.~A. Maruhn, T.~Otsuka, Examination of the
  stability of a rod-shaped structure in $^{24}\mathrm{Mg}$, Phys. Rev. C 92
  (2015) 011303.
\newblock \href {http://dx.doi.org/10.1103/PhysRevC.92.011303}
  {\path{doi:10.1103/PhysRevC.92.011303}}.

\bibitem{schuetrumpf2016}
B.~Schuetrumpf, W.~Nazarewicz, P.-G. Reinhard, Time-dependent density
  functional theory with twist--averaged boundary conditions, Phys. Rev. C 93
  (2016) 054304.
\newblock \href {http://dx.doi.org/10.1103/PhysRevC.93.054304}
  {\path{doi:10.1103/PhysRevC.93.054304}}.

\bibitem{brink1981}
D.~M. Brink, F.~Stancu, {T}ime-dependent {H}artree-{F}ock and the one-body
  dissipation for head-on collisions, Phys. Rev. C 24 (1981) 144--147.
\newblock \href {http://dx.doi.org/10.1103/physrevc.24.144}
  {\path{doi:10.1103/physrevc.24.144}}.

\bibitem{washiyama2009a}
K.~Washiyama, D.~Lacroix, S.~Ayik, {O}ne-body energy dissipation in fusion
  reactions from mean-field theory, Phys. Rev. C 79 (2009) 024609.
\newblock \href {http://dx.doi.org/10.1103/PhysRevC.79.024609}
  {\path{doi:10.1103/PhysRevC.79.024609}}.

\bibitem{wen2013}
K.~Wen, F.~Sakata, Z.-X. Li, X.-Z. Wu, Y.-X. Zhang, S.-G. Zhou,
  {N}on-{G}aussian {F}luctuations and {N}on-{M}arkovian {E}ffects in the
  {N}uclear {F}usion {P}rocess: {L}angevin {D}ynamics {E}merging from {Q}uantum
  {M}olecular {D}ynamics {S}imulations, Phys. Rev. Lett. 111 (2013) 012501.
\newblock \href {http://dx.doi.org/10.1103/PhysRevLett.111.012501}
  {\path{doi:10.1103/PhysRevLett.111.012501}}.

\bibitem{umar2009a}
A.~S. Umar, V.~E. Oberacker, J.~A. Maruhn, P.-G. Reinhard, {M}icroscopic
  calculation of precompound excitation energies for heavy-ion collisions,
  Phys. Rev. C 80 (2009) 041601.
\newblock \href {http://dx.doi.org/10.1103/PhysRevC.80.041601}
  {\path{doi:10.1103/PhysRevC.80.041601}}.

\bibitem{umar2010a}
A.~S. Umar, V.~E. Oberacker, J.~A. Maruhn, P.-G. Reinhard, {E}ntrance channel
  dynamics of hot and cold fusion reactions leading to superheavy elements,
  Phys. Rev. C 81 (2010) 064607.
\newblock \href {http://dx.doi.org/10.1103/PhysRevC.81.064607}
  {\path{doi:10.1103/PhysRevC.81.064607}}.

\bibitem{skalski2007}
J.~Skalski, Adiabatic fusion barriers from self-consistent calculations, Phys.
  Rev. C 76 (2007) 044603.
\newblock \href {http://dx.doi.org/10.1103/PhysRevC.76.044603}
  {\path{doi:10.1103/PhysRevC.76.044603}}.

\bibitem{schunck2016}
N.~Schunck, L.~M. Robledo, Microscopic theory of nuclear fission: a review,
  Rep. Prog. Phys. 79 (2016) 116301.
\newblock \href {http://dx.doi.org/10.1088/0034-4885/79/11/116301}
  {\path{doi:10.1088/0034-4885/79/11/116301}}.

\bibitem{matsuo2000}
M.~Matsuo, T.~Nakatsukasa, K.~Matsuyanagi, {A}diabatic {S}elfconsistent
  {C}ollective {C}oordinate {M}ethod for {L}arge {A}mplitude {C}ollective
  {M}otion in {N}uclei with {P}airing {C}orrelations, Prog. Theor. Phys. 103
  (2000) 959--979.
\newblock \href {http://dx.doi.org/10.1143/ptp.103.959}
  {\path{doi:10.1143/ptp.103.959}}.

\bibitem{wen2016}
K.~Wen, T.~Nakatsukasa, Self-consistent collective coordinate for reaction path
  and inertial mass, Phys. Rev. C 94 (2016) 054618.
\newblock \href {http://dx.doi.org/10.1103/PhysRevC.94.054618}
  {\path{doi:10.1103/PhysRevC.94.054618}}.

\bibitem{wen2017}
K.~Wen, T.~Nakatsukasa, Adiabatic self-consistent collective path in nuclear
  fusion reactions, Phys. Rev. C 96 (2017) 014610.
\newblock \href {http://dx.doi.org/10.1103/PhysRevC.96.014610}
  {\path{doi:10.1103/PhysRevC.96.014610}}.

\bibitem{ring1980}
P.~Ring, P.~Schuck, {T}he {N}uclear {M}any--{B}ody {P}roblem, Springer--Verlag,
  New York, 1980.
\newblock \href {http://dx.doi.org/10.1007/978-3-642-61852-9}
  {\path{doi:10.1007/978-3-642-61852-9}}.

\bibitem{dobaczewski1996}
J.~Dobaczewski, W.~Nazarewicz, T.~R. Werner, J.~F. Berger, C.~R. Chinn,
  J.~Decharge, {M}ean-field description of ground-state properties of drip-line
  nuclei: {P}airing and continuum effects, Phys. Rev. C 53 (1996) 2809--2840.
\newblock \href {http://dx.doi.org/10.1103/PhysRevC.53.2809}
  {\path{doi:10.1103/PhysRevC.53.2809}}.

\bibitem{terasaki1996}
J.~Terasaki, P.-H. Heenen, H.~Flocard, P.~Bonche, 3{D} solution of
  {H}artree-{F}ock-{B}ogoliubov equations for drip-line nuclei, Nucl. Phys. A
  600 (1996) 371--386.
\newblock \href {http://dx.doi.org/10.1016/0375-9474(96)00036-X}
  {\path{doi:10.1016/0375-9474(96)00036-X}}.

\bibitem{stoitsov2000}
M.~V. Stoitsov, J.~Dobaczewski, P.~Ring, S.~Pittel, {Q}uadrupole deformations
  of neutron-drip-line nuclei studied within the {S}kyrme
  {H}artree-{F}ock-{B}ogoliubov approach, Phys. Rev. C 61 (2000) 034311.
\newblock \href {http://dx.doi.org/10.1103/PhysRevC.61.034311}
  {\path{doi:10.1103/PhysRevC.61.034311}}.

\bibitem{bender2003}
M.~Bender, P.-H. Heenen, P.-G. Reinhard, {S}elf-consistent mean-field models
  for nuclear structure, Rev. Mod. Phys. 75 (2003) 121--180.
\newblock \href {http://dx.doi.org/10.1103/RevModPhys.75.121}
  {\path{doi:10.1103/RevModPhys.75.121}}.

\bibitem{stoitsov2003}
M.~V. Stoitsov, J.~Dobaczewski, W.~Nazarewicz, S.~Pittel, D.~J. Dean,
  {S}ystematic study of deformed nuclei at the drip lines and beyond, Phys.
  Rev. C 68 (2003) 054312.
\newblock \href {http://dx.doi.org/10.1103/PhysRevC.68.054312}
  {\path{doi:10.1103/PhysRevC.68.054312}}.

\bibitem{vretenar2005}
D.~Vretenar, A.~Afanasjev, G.~Lalazissis, P.~Ring, {R}elativistic
  {H}artree--{B}ogoliubov theory: static and dynamic aspects of exotic nuclear
  structure, Phys. Rep. 409 (2005) 101--259.
\newblock \href {http://dx.doi.org/10.1016/j.physrep.2004.10.001}
  {\path{doi:10.1016/j.physrep.2004.10.001}}.

\bibitem{blazkiewicz2005}
A.~Blazkiewicz, V.~E. Oberacker, A.~S. Umar, M.~Stoitsov, {C}oordinate space
  {H}artree-{F}ock-{B}ogoliubov calculations for the zirconium isotope chain up
  to the two-neutron drip line, Phys. Rev. C 71 (2005) 054321.
\newblock \href {http://dx.doi.org/10.1103/PhysRevC.71.054321}
  {\path{doi:10.1103/PhysRevC.71.054321}}.

\bibitem{staszczak2009}
A.~Staszczak, A.~Baran, J.~Dobaczewski, W.~Nazarewicz, {M}icroscopic
  description of complex nuclear decay: {M}ultimodal fission, Phys. Rev. C 80
  (2009) 014309.
\newblock \href {http://dx.doi.org/10.1103/PhysRevC.80.014309}
  {\path{doi:10.1103/PhysRevC.80.014309}}.

\bibitem{sadhukhan2013}
J.~Sadhukhan, K.~Mazurek, A.~Baran, J.~Dobaczewski, W.~Nazarewicz, J.~A.
  Sheikh, {S}pontaneous fission lifetimes from the minimization of
  self-consistent collective action, Phys. Rev. C 88 (2013) 064314.
\newblock \href {http://dx.doi.org/10.1103/PhysRevC.88.064314}
  {\path{doi:10.1103/PhysRevC.88.064314}}.

\bibitem{sadhukhan2014}
J.~Sadhukhan, J.~Dobaczewski, W.~Nazarewicz, J.~A. Sheikh, A.~Baran,
  Pairing-induced speedup of nuclear spontaneous fission, Phys. Rev. C 90
  (2014) 061304.
\newblock \href {http://dx.doi.org/10.1103/PhysRevC.90.061304}
  {\path{doi:10.1103/PhysRevC.90.061304}}.

\bibitem{sadhukhan2016}
J.~Sadhukhan, W.~Nazarewicz, N.~Schunck, Microscopic modeling of mass and
  charge distributions in the spontaneous fission of $^{240}\mathrm{Pu}$, Phys.
  Rev. C 93 (2016) 011304.
\newblock \href {http://dx.doi.org/10.1103/PhysRevC.93.011304}
  {\path{doi:10.1103/PhysRevC.93.011304}}.

\bibitem{negele1978}
J.~W. Negele, S.~E. Koonin, P.~M\"oller, J.~R. Nix, A.~J. Sierk, {D}ynamics of
  induced fission, Phys. Rev. C 17 (1978) 1098--1115.
\newblock \href {http://dx.doi.org/10.1103/PhysRevC.17.1098}
  {\path{doi:10.1103/PhysRevC.17.1098}}.

\bibitem{simenel2014a}
C.~Simenel, A.~S. Umar, {F}ormation and dynamics of fission fragments, Phys.
  Rev. C 89 (2014) 031601(R).
\newblock \href {http://dx.doi.org/10.1103/PhysRevC.89.031601}
  {\path{doi:10.1103/PhysRevC.89.031601}}.

\bibitem{scamps2015a}
G.~Scamps, C.~Simenel, D.~Lacroix, {S}uperfluid dynamics of $^{258}\mathrm{Fm}$
  fission, Phys. Rev. C 92 (2015) 011602(R).
\newblock \href {http://dx.doi.org/10.1103/PhysRevC.92.011602}
  {\path{doi:10.1103/PhysRevC.92.011602}}.

\bibitem{zhang2016}
C.~L. Zhang, B.~Schuetrumpf, W.~Nazarewicz, Nucleon localization and fragment
  formation in nuclear fission, Phys. Rev. C 94 (2016) 064323.
\newblock \href {http://dx.doi.org/10.1103/PhysRevC.94.064323}
  {\path{doi:10.1103/PhysRevC.94.064323}}.

\bibitem{sadhukhan2017}
J.~Sadhukhan, C.~Zhang, W.~Nazarewicz, N.~Schunck, Formation and distribution
  of fragments in the spontaneous fission of ${}^{\mathrm{240}}\mathrm{Pu}$,
  Phys. Rev. C 96 (2017) 061301.
\newblock \href {http://dx.doi.org/10.1103/PhysRevC.96.061301}
  {\path{doi:10.1103/PhysRevC.96.061301}}.

\bibitem{tao2017}
H.~Tao, J.~Zhao, Z.~P. Li, T.~Nik{\v{s}}i{\'{c}}, D.~Vretenar, Microscopic
  study of induced fission dynamics of $^{226}\mathrm{Th}$ with covariant
  energy density functionals, Phys. Rev. C 96 (2017) 024319.
\newblock \href {http://dx.doi.org/10.1103/PhysRevC.96.024319}
  {\path{doi:10.1103/PhysRevC.96.024319}}.

\bibitem{ebata2012}
S.~Ebata, T.~Nakatsukasa, Pairing {E}ffects in {N}uclear {F}usion {R}eaction,
  JPS Conf. Proc. 1 (2012) 013038.
\newblock \href {http://dx.doi.org/10.7566/JPSCP.1.013038}
  {\path{doi:10.7566/JPSCP.1.013038}}.

\bibitem{ebata2014a}
S.~Ebata, T.~Nakatsukasa, Repulsive aspects of pairing correlation in nuclear
  fusion reaction, JPS Conf. Proc. 6 (2014) 020056.
\newblock \href {http://dx.doi.org/10.7566/JPSCP.1.013038}
  {\path{doi:10.7566/JPSCP.1.013038}}.

\bibitem{scamps2015}
G.~Scamps, D.~Lacroix, Effect of pairing on transfer and fusion reactions, EPJ
  Web Conf. 86 (2015) 00042.
\newblock \href {http://dx.doi.org/10.1051/epjconf/20158600042}
  {\path{doi:10.1051/epjconf/20158600042}}.

\bibitem{hashimoto2016}
Y.~Hashimoto, G.~Scamps, Gauge angle dependence in time-dependent
  {H}artree-{F}ock-{B}ogoliubov calculations of
  $^{20}\mathrm{O}+{}^{20}\mathrm{O}$ head-on collisions with the {G}ogny
  interaction, Phys. Rev. C 94 (2016) 014610.
\newblock \href {http://dx.doi.org/10.1103/PhysRevC.94.014610}
  {\path{doi:10.1103/PhysRevC.94.014610}}.

\bibitem{magierski2017}
P.~Magierski, K.~Sekizawa, G.~Wlaz\l{}owski, Novel {R}ole of {S}uperfluidity in
  {L}ow--{E}nergy {N}uclear {R}eactions, Phys. Rev. Lett. 119 (2017) 042501.
\newblock \href {http://dx.doi.org/10.1103/PhysRevLett.119.042501}
  {\path{doi:10.1103/PhysRevLett.119.042501}}.

\bibitem{scamps2018b}
G.~Scamps, Examining empirical evidence of the effect of superfluidity on the
  fusion barrier, Phys. Rev. C 97 (2018) 044611.
\newblock \href {http://dx.doi.org/10.1103/PhysRevC.97.044611}
  {\path{doi:10.1103/PhysRevC.97.044611}}.

\bibitem{matsuo2001}
M.~Matsuo, Continuum linear response in coordinate space
  \text{Hartree-Fock-Bogoliubov} formalism for collective excitations in
  drip-line nuclei, Nucl. Phys. A 696 (2001) 371.
\newblock \href {http://dx.doi.org/10.1016/S0375-9474(01)01133-2}
  {\path{doi:10.1016/S0375-9474(01)01133-2}}.

\bibitem{khan2002}
E.~Khan, N.~Sandulescu, M.~Grasso, N.~Van~Giai, Continuum quasiparticle random
  phase approximation and the time-dependent \text{Hartree-Fock-Bogoliubov}
  approach, Phys. Rev. C 66 (2002) 024309.
\newblock \href {http://dx.doi.org/10.1103/PhysRevC.66.024309}
  {\path{doi:10.1103/PhysRevC.66.024309}}.

\bibitem{hashimoto2012}
Y.~Hashimoto, {Linear responses in time-dependent Hartree-Fock-Bogoliubov
  method with {G}ogny interaction}, Eur. Phys. J. A 48 (2012) 55.
\newblock \href {http://dx.doi.org/10.1140/epja/i2012-12055-0}
  {\path{doi:10.1140/epja/i2012-12055-0}}.

\bibitem{jin2017}
S.~Jin, A.~Bulgac, K.~Roche, G.~Wlaz\l{}owski, Coordinate-space solver for
  superfluid many-fermion systems with the shifted conjugate-orthogonal
  conjugate-gradient method, Phys. Rev. C 95 (2017) 044302.
\newblock \href {http://dx.doi.org/10.1103/PhysRevC.95.044302}
  {\path{doi:10.1103/PhysRevC.95.044302}}.

\bibitem{bulgac2016}
A.~Bulgac, P.~Magierski, K.~J. Roche, I.~Stetcu, {I}nduced {F}ission of
  $^{240}${P}u within a {R}eal-{T}ime {M}icroscopic {F}ramework, Phys. Rev.
  Lett. 116 (2016) 122504.
\newblock \href {http://dx.doi.org/10.1103/physrevlett.116.122504}
  {\path{doi:10.1103/physrevlett.116.122504}}.

\bibitem{wlazlowski2016}
G.~Wlaz\l{}owski, K.~Sekizawa, P.~Magierski, A.~Bulgac, M.~M. Forbes, Vortex
  {P}inning and {D}ynamics in the {N}eutron {S}tar {C}rust, Phys. Rev. Lett.
  117 (2016) 232701.
\newblock \href {http://dx.doi.org/10.1103/PhysRevLett.117.232701}
  {\path{doi:10.1103/PhysRevLett.117.232701}}.

\bibitem{scamps2017b}
G.~Scamps, Y.~Hashimoto, Transfer probabilities for the reactions
  $^{14,20}\mathrm{O}+^{20}\mathrm{O}$ in terms of multiple time-dependent
  {H}artree-{F}ock-{B}ogoliubov trajectories, Phys. Rev. C 96 (2017) 031602.
\newblock \href {http://dx.doi.org/10.1103/PhysRevC.96.031602}
  {\path{doi:10.1103/PhysRevC.96.031602}}.

\bibitem{dobaczewski1984}
J.~Dobaczewski, H.~Flocard, J.~Treiner, Hartree--{F}ock--{B}ogolyubov
  description of nuclei near the neutron--drip line, Nucl. Phys. A 422 (1984)
  103--139.
\newblock \href {http://dx.doi.org/10.1016/0375-9474(84)90433-0}
  {\path{doi:10.1016/0375-9474(84)90433-0}}.

\bibitem{teran2003}
E.~Ter\'an, V.~E. Oberacker, A.~S. Umar, {A}xially symmetric
  {H}artree-{F}ock-{B}ogoliubov calculations for nuclei near the drip lines,
  Phys. Rev. C 67 (2003) 064314.
\newblock \href {http://dx.doi.org/10.1103/PhysRevC.67.064314}
  {\path{doi:10.1103/PhysRevC.67.064314}}.

\bibitem{bogoliubov1958}
N.~N. Bogoliubov, A {N}ew method in the theory of superconductivity. {I}, Sov.
  Phys. JETP 7 (1958) 41.

\bibitem{blaizot1986}
J.~Blaizot, G.~Ripka, Quantum {T}heory of {F}inite {S}ystems, {MIT Press},
  Cambridge, MA, 1985.

\bibitem{bulgac1999}
A.~Bulgac, \href{http://arXiv:nucl-th/9907088}{{H}artree-{F}ock-{B}ogoliubov
  {A}pproximation for {F}inite {S}ystems}, arXiv:nucl-th/9907088.
\newline\urlprefix\url{http://arXiv:nucl-th/9907088}

\bibitem{bulgac2002b}
A.~Bulgac, Y.~Yu, {R}enormalization of the {H}artree-{F}ock-{B}ogoliubov
  {E}quations in the {C}ase of a {Z}ero {R}ange {P}airing {I}nteraction, Phys.
  Rev. Lett. 88 (2002) 042504.
\newblock \href {http://dx.doi.org/10.1103/PhysRevLett.88.042504}
  {\path{doi:10.1103/PhysRevLett.88.042504}}.

\bibitem{blocki1976}
J.~Blocki, H.~Flocard, {S}imple dynamical models including pairing residual
  interaction, Nucl. Phys. A 273 (1976) 45--60.
\newblock \href {http://dx.doi.org/10.1016/0375-9474(76)90299-2}
  {\path{doi:10.1016/0375-9474(76)90299-2}}.

\bibitem{reinhard1997}
P.-G. Reinhard, M.~Bender, K.~Rutz, J.~A. Maruhn, An {HFB} scheme in natural
  orbitals, Z. Phys. A 358 (1997) 277--278.
\newblock \href {http://dx.doi.org/10.1007/s002180050328}
  {\path{doi:10.1007/s002180050328}}.

\bibitem{scamps2012}
G.~Scamps, D.~Lacroix, G.~F. Bertsch, K.~Washiyama, Pairing dynamics in
  particle transport, Phys. Rev. C 85 (2012) 034328.
\newblock \href {http://dx.doi.org/10.1103/PhysRevC.85.034328}
  {\path{doi:10.1103/PhysRevC.85.034328}}.

\bibitem{flocard1978}
H.~Flocard, S.~E. Koonin, M.~S. Weiss, {T}hree-dimensional time-dependent
  {H}artree-{F}ock calculations: {A}pplication to $^{16}${O}+$^{16}${O}
  collisions, Phys. Rev. C 17 (1978) 1682--1699.
\newblock \href {http://dx.doi.org/10.1103/PhysRevC.17.1682}
  {\path{doi:10.1103/PhysRevC.17.1682}}.

\bibitem{tanimura2015}
Y.~Tanimura, D.~Lacroix, G.~Scamps, Collective aspects deduced from
  time-dependent microscopic mean-field with pairing: {A}pplication to the
  fission process, Phys. Rev. C 92 (2015) 034601.
\newblock \href {http://dx.doi.org/10.1103/PhysRevC.92.034601}
  {\path{doi:10.1103/PhysRevC.92.034601}}.

\bibitem{tanimura2017}
Y.~Tanimura, D.~Lacroix, S.~Ayik, Microscopic {P}hase--{S}pace {E}xploration
  {M}odeling of $^{258}\mathrm{Fm}$ {S}pontaneous {F}ission, Phys. Rev. Lett.
  118 (2017) 152501.
\newblock \href {http://dx.doi.org/10.1103/PhysRevLett.118.152501}
  {\path{doi:10.1103/PhysRevLett.118.152501}}.

\bibitem{wakhle2018}
A.~Wakhle, K.~Hammerton, Z.~Kohley, D.~J. Morrissey, K.~Stiefel, J.~Yurkon,
  J.~Walshe, K.~J. Cook, M.~Dasgupta, D.~J. Hinde, D.~J. Jeung, E.~Prasad,
  D.~C. Rafferty, C.~Simenel, E.~C. Simpson, K.~Vo-Phuoc, J.~King, W.~Loveland,
  R.~Yanez, Capture cross sections for the synthesis of new heavy nuclei using
  radioactive beams, Phys. Rev. C 97 (2018) 021602.
\newblock \href {http://dx.doi.org/10.1103/PhysRevC.97.021602}
  {\path{doi:10.1103/PhysRevC.97.021602}}.

\bibitem{ebata2014}
S.~Ebata, T.~Nakatsukasa, T.~Inakura, Systematic investigation of low-lying
  dipole modes using the canonical-basis time-dependent
  {H}artree-{F}ock-{B}ogoliubov theory, Phys. Rev. C 90 (2014) 024303.
\newblock \href {http://dx.doi.org/10.1103/PhysRevC.90.024303}
  {\path{doi:10.1103/PhysRevC.90.024303}}.

\bibitem{goddard2015}
P.~M. Goddard, P.~D. Stevenson, A.~Rios, {F}ission dynamics within
  time-dependent {H}artree-{F}ock: deformation-induced fission, Phys. Rev. C 92
  (2015) 054610.
\newblock \href {http://dx.doi.org/10.1103/PhysRevC.92.054610}
  {\path{doi:10.1103/PhysRevC.92.054610}}.

\bibitem{goddard2016}
P.~M. Goddard, P.~D. Stevenson, A.~Rios, Fission dynamics within
  time--dependent {H}artree--{F}ock. {II}. {B}oost-induced fission, Phys. Rev.
  C 93 (2016) 014620.
\newblock \href {http://dx.doi.org/10.1103/PhysRevC.93.014620}
  {\path{doi:10.1103/PhysRevC.93.014620}}.

\bibitem{scamps2018}
G.~Scamps, C.~Simenel, \href{https://arxiv.org/abs/1804.03337}{{O}n the origin
  of asymmetric fission of actinides}, arXiv:1804.03337.
\newline\urlprefix\url{https://arxiv.org/abs/1804.03337}

\bibitem{dubray2008}
N.~Dubray, H.~Goutte, J.-P. Delaroche, {S}tructure properties of $^{226}${T}h
  and $^{256,258,260}${F}m fission fragments: {M}ean-field analysis with the
  {G}ogny force, Phys. Rev. C 77 (2008) 014310.
\newblock \href {http://dx.doi.org/10.1103/PhysRevC.77.014310}
  {\path{doi:10.1103/PhysRevC.77.014310}}.

\bibitem{goutte2005}
H.~Goutte, J.~F. Berger, P.~Casoli, D.~Gogny, {M}icroscopic approach of fission
  dynamics applied to fragment kinetic energy and mass distributions in
  $^{238}${U}, Phys. Rev. C 71 (2005) 024316.
\newblock \href {http://dx.doi.org/10.1103/PhysRevC.71.024316}
  {\path{doi:10.1103/PhysRevC.71.024316}}.

\bibitem{regnier2016}
D.~Regnier, N.~Dubray, N.~Schunck, M.~Verri\`ere, Fission fragment charge and
  mass distributions in $^{239}\mathrm{Pu}(n,f)$ in the adiabatic nuclear
  energy density functional theory, Phys. Rev. C 93 (2016) 054611.
\newblock \href {http://dx.doi.org/10.1103/PhysRevC.93.054611}
  {\path{doi:10.1103/PhysRevC.93.054611}}.

\bibitem{dubray2012}
N.~Dubray, D.~Regnier, Numerical search of discontinuities in self-consistent
  potential energy surfaces, Comput. Phys. Commun. 183 (2012) 2035--2041.
\newblock \href {http://dx.doi.org/10.1016/j.cpc.2012.05.001}
  {\path{doi:10.1016/j.cpc.2012.05.001}}.

\bibitem{hulet1986}
E.~K. Hulet, J.~F. Wild, R.~J. Dougan, R.~W. Lougheed, J.~H. Landrum, A.~D.
  Dougan, M.~Schadel, R.~L. Hahn, P.~A. Baisden, C.~M. Henderson, R.~J. Dupzyk,
  K.~S\"ummerer, G.~R. Bethune, {B}iomodal symmetrical fission observed in the
  heaviest elements, Phys. Rev. Lett. 56 (1986) 313--316.
\newblock \href {http://dx.doi.org/10.1103/PhysRevLett.56.313}
  {\path{doi:10.1103/PhysRevLett.56.313}}.

\bibitem{golabek2009}
C.~Golabek, C.~Simenel, {C}ollision {D}ynamics of {T}wo $^{238}${U A}tomic
  {N}uclei, Phys. Rev. Lett. 103 (2009) 042701.
\newblock \href {http://dx.doi.org/10.1103/PhysRevLett.103.042701}
  {\path{doi:10.1103/PhysRevLett.103.042701}}.

\bibitem{kedziora2010}
D.~J. Kedziora, C.~Simenel, {N}ew inverse quasifission mechanism to produce
  neutron-rich transfermium nuclei, Phys. Rev. C 81 (2010) 044613.
\newblock \href {http://dx.doi.org/10.1103/PhysRevC.81.044613}
  {\path{doi:10.1103/PhysRevC.81.044613}}.

\bibitem{wakhle2014}
A.~Wakhle, C.~Simenel, D.~J. Hinde, M.~Dasgupta, M.~Evers, D.~H. Luong,
  R.~du~Rietz, E.~Williams, {I}nterplay between {Q}uantum {S}hells and
  {O}rientation in {Q}uasifission, Phys. Rev. Lett. 113 (2014) 182502.
\newblock \href {http://dx.doi.org/10.1103/PhysRevLett.113.182502}
  {\path{doi:10.1103/PhysRevLett.113.182502}}.

\bibitem{oberacker2014}
V.~E. Oberacker, A.~S. Umar, C.~Simenel, {D}issipative dynamics in
  quasifission, Phys. Rev. C 90 (2014) 054605.
\newblock \href {http://dx.doi.org/10.1103/PhysRevC.90.054605}
  {\path{doi:10.1103/PhysRevC.90.054605}}.

\bibitem{hammerton2015}
K.~Hammerton, Z.~Kohley, D.~J. Hinde, M.~Dasgupta, A.~Wakhle, E.~Williams,
  V.~E. Oberacker, A.~S. Umar, I.~P. Carter, K.~J. Cook, J.~Greene, D.~Y.
  Jeung, D.~H. Luong, S.~D. {McNeil}, C.~S. Palshetkar, D.~C. Rafferty,
  C.~Simenel, K.~Stiefel, {R}educed quasifission competition in fusion
  reactions forming neutron-rich heavy elements, Phys. Rev. C 91 (2015)
  041602(R).
\newblock \href {http://dx.doi.org/10.1103/PhysRevC.91.041602}
  {\path{doi:10.1103/PhysRevC.91.041602}}.

\bibitem{umar2016}
A.~S. Umar, V.~E. Oberacker, C.~Simenel, Fusion and quasifission dynamics in
  the reactions $^{48}\mathrm{Ca}+^{249}\mathrm{Bk}$ and
  $^{50}\mathrm{Ti}+^{249}\mathrm{Bk}$ using a time-dependent {H}artree-{F}ock
  approach, Phys. Rev. C 94 (2016) 024605.
\newblock \href {http://dx.doi.org/10.1103/PhysRevC.94.024605}
  {\path{doi:10.1103/PhysRevC.94.024605}}.

\bibitem{sekizawa2016}
K.~Sekizawa, K.~Yabana, {T}ime-dependent {H}artree-{F}ock calculations for
  multinucleon transfer and quasifission processes in the
  $^{64}\text{Ni}+^{238}\text{U}$ reaction, Phys. Rev. C 93 (2016) 054616.
\newblock \href {http://dx.doi.org/10.1103/PhysRevC.93.054616}
  {\path{doi:10.1103/PhysRevC.93.054616}}.

\bibitem{yu2017}
C.~Yu, L.~Guo, Angular momentum dependence of quasifission dynamics in the
  reaction $^{48}${C}a+$^{244}${P}u, Sci. China Phys. 60 (2017) 092011.
\newblock \href {http://dx.doi.org/10.1007/s11433-017-9063-3}
  {\path{doi:10.1007/s11433-017-9063-3}}.

\bibitem{morjean2017}
M.~Morjean, D.~J. Hinde, C.~Simenel, D.~Y. Jeung, M.~Airiau, K.~J. Cook,
  M.~Dasgupta, A.~Drouart, D.~Jacquet, S.~Kalkal, C.~S. Palshetkar, E.~Prasad,
  D.~Rafferty, E.~C. Simpson, L.~Tassan-Got, K.~Vo-Phuoc, E.~Williams, Evidence
  for the {R}ole of {P}roton {S}hell {C}losure in {Q}uasifission {R}eactions
  from {X--Ray} {F}luorescence of {M}ass--{I}dentified {F}ragments, Phys. Rev.
  Lett. 119 (2017) 222502.
\newblock \href {http://dx.doi.org/10.1103/PhysRevLett.119.222502}
  {\path{doi:10.1103/PhysRevLett.119.222502}}.

\bibitem{toke1985}
J.~T{\~{o}}ke, R.~Bock, G.~X. Dai, A.~Gobbi, S.~Gralla, K.~D. Hildenbrand,
  J.~Kuzminski, W.~F.~J. M\"uller, A.~Olmi, H.~Stelzer, B.~B. Back,
  S.~Bj\o{}rnholm, {Q}uasi-fission: {T}he mass-drift mode in heavy-ion
  reactions, Nucl. Phys. A 440 (1985) 327--365.
\newblock \href {http://dx.doi.org/10.1016/0375-9474(85)90344-6}
  {\path{doi:10.1016/0375-9474(85)90344-6}}.

\bibitem{guo2012}
L.~Guo, T.~Nakatsukasa, {T}ime-dependent {H}artree-{F}ock studies of the
  dynamical fusion threshold, EPJ Web Conf. 38 (2012) 09003.
\newblock \href {http://dx.doi.org/10.1051/epjconf/20123809003}
  {\path{doi:10.1051/epjconf/20123809003}}.

\bibitem{washiyama2015}
K.~Washiyama, {M}icroscopic analysis of fusion hindrance in heavy nuclear
  systems, Phys. Rev. C 91 (2015) 064607.
\newblock \href {http://dx.doi.org/10.1103/PhysRevC.91.064607}
  {\path{doi:10.1103/PhysRevC.91.064607}}.

\bibitem{durietz2013}
R.~du~Rietz, E.~Williams, D.~J. Hinde, M.~Dasgupta, M.~Evers, C.~J. Lin, D.~H.
  Luong, C.~Simenel, A.~Wakhle, Mapping quasifission characteristics and
  timescales in heavy element formation reactions, Phys. Rev. C 88 (2013)
  054618.
\newblock \href {http://dx.doi.org/10.1103/PhysRevC.88.054618}
  {\path{doi:10.1103/PhysRevC.88.054618}}.

\bibitem{koonin1977}
S.~E. Koonin, K.~T.~R. Davies, V.~Maruhn-Rezwani, H.~Feldmeier, S.~J. Krieger,
  J.~W. Negele, {T}ime-dependent {H}artree-{F}ock calculations for $^{16}${O}
  $+$ $^{16}${O} and $^{40}${C}a $+$ $^{40}${C}a reactions, Phys. Rev. C 15
  (1977) 1359--1374.
\newblock \href {http://dx.doi.org/10.1103/PhysRevC.15.1359}
  {\path{doi:10.1103/PhysRevC.15.1359}}.

\bibitem{sonika2015}
Sonika, B.~J. Roy, A.~Parmar, U.~K. Pal, H.~Kumawat, V.~Jha, S.~K. Pandit,
  V.~V. Parkar, K.~Ramachandran, K.~Mahata, A.~Pal, S.~Santra, A.~K. Mohanty,
  K.~Sekizawa, Multinucleon transfer study in
  $^{206}${P}b$\left(^{18}\mathrm{O},x\right)$ at energies above the {C}oulomb
  barrier, Phys. Rev. C 92 (2015) 024603.
\newblock \href {http://dx.doi.org/10.1103/physrevc.92.024603}
  {\path{doi:10.1103/physrevc.92.024603}}.

\bibitem{scamps2017a}
G.~Scamps, C.~Rodr\'{\i}guez-Tajes, D.~Lacroix, F.~Farget, Time-dependent
  mean-field determination of the excitation energy in transfer reactions:
  {A}pplication to the reaction $^{238}\mathrm{U}$ on $^{12}\mathrm{C}$ at 6.14
  {MeV}/nucleon, Phys. Rev. C 95 (2017) 024613.
\newblock \href {http://dx.doi.org/10.1103/PhysRevC.95.024613}
  {\path{doi:10.1103/PhysRevC.95.024613}}.

\bibitem{sekizawa2017}
K.~Sekizawa, Microscopic description of production cross sections including
  deexcitation effects, Phys. Rev. C 96 (2017) 014615.
\newblock \href {http://dx.doi.org/10.1103/physrevc.96.014615}
  {\path{doi:10.1103/physrevc.96.014615}}.

\bibitem{sekizawa2017a}
K.~Sekizawa, Enhanced nucleon transfer in tip collisions of
  $^{238}\mathrm{U}+^{124}\mathrm{Sn}$, Phys. Rev. C 96 (2017) 041601(R).
\newblock \href {http://dx.doi.org/10.1103/PhysRevC.96.041601}
  {\path{doi:10.1103/PhysRevC.96.041601}}.

\bibitem{regnier2018}
D.~Regnier, D.~Lacroix, G.~Scamps, Y.~Hashimoto, Microscopic description of
  pair transfer between two superfluid {F}ermi systems: {C}ombining phase-space
  averaging and combinatorial techniques, Phys. Rev. C 97 (2018) 034627.
\newblock \href {http://dx.doi.org/10.1103/PhysRevC.97.034627}
  {\path{doi:10.1103/PhysRevC.97.034627}}.

\bibitem{williams2018}
E.~Williams, K.~Sekizawa, D.~J. Hinde, C.~Simenel, M.~Dasgupta, I.~P. Carter,
  K.~J. Cook, D.~Y. Jeung, S.~D. McNeil, C.~S. Palshetkar, D.~C. Rafferty,
  K.~Ramachandran, A.~Wakhle, Exploring {Z}eptosecond {Q}uantum {E}quilibration
  {D}ynamics: {F}rom {D}eep-{I}nelastic to {F}usion-{F}ission {O}utcomes in
  $^{58}\mathrm{Ni}+^{60}\mathrm{Ni}$ {R}eactions, Phys. Rev. Lett. 120 (2018)
  022501.
\newblock \href {http://dx.doi.org/10.1103/PhysRevLett.120.022501}
  {\path{doi:10.1103/PhysRevLett.120.022501}}.

\bibitem{umar2009c}
A.~S. Umar, V.~E. Oberacker, {C}enter-of-mass motion and cross-channel coupling
  in the time-dependent {H}artree-{F}ock theory, J. Phys. G 36 (2009) 025101.
\newblock \href {http://dx.doi.org/10.1088/0954-3899/36/2/025101}
  {\path{doi:10.1088/0954-3899/36/2/025101}}.

\bibitem{randrup1982}
J.~Randrup, Transport of angular momentum in damped nuclear reactions, Nucl.
  Phys. A 383 (1982) 468--508.
\newblock \href {http://dx.doi.org/10.1016/0375-9474(82)90087-2}
  {\path{doi:10.1016/0375-9474(82)90087-2}}.

\bibitem{valor2000}
A.~Valor, P.-H. Heenen, P.~Bonche, Configuration mixing of mean-field wave
  functions projected on angular momentum and particle number: {A}pplication to
  $^{24}\mathrm{Mg}$, Nucl. Phys. A 671 (2000) 145.
\newblock \href {http://dx.doi.org/10.1016/S0375-9474(99)00830-1}
  {\path{doi:10.1016/S0375-9474(99)00830-1}}.

\bibitem{anguiano2001}
M.~Anguiano, J.~L. Egido, L.~M. Robledo, {P}article number projection with
  effective forces, Nucl. Phys. A 696 (2001) 467--493.
\newblock \href {http://dx.doi.org/10.1016/s0375-9474(01)01219-2}
  {\path{doi:10.1016/s0375-9474(01)01219-2}}.

\bibitem{samyn2004}
M.~Samyn, S.~Goriely, M.~Bender, J.~M. Pearson, {F}urther explorations of
  {S}kyrme-{H}artree-{F}ock-{B}ogoliubov mass formulas. {III}. {R}ole of
  particle-number projection, Phys. Rev. C 70 (2004) 044309.
\newblock \href {http://dx.doi.org/10.1103/PhysRevC.70.044309}
  {\path{doi:10.1103/PhysRevC.70.044309}}.

\bibitem{dobaczewski2007}
J.~Dobaczewski, M.~V. Stoitsov, W.~Nazarewicz, P.-G. Reinhard,
  {P}article-number projection and the density functional theory, Phys. Rev. C
  76 (2007) 054315.
\newblock \href {http://dx.doi.org/10.1103/physrevc.76.054315}
  {\path{doi:10.1103/physrevc.76.054315}}.

\bibitem{bender2008}
M.~Bender, P.-H. Heenen, {C}onfiguration mixing of angular-momentum and
  particle-number projected triaxial {H}artree-{F}ock-{B}ogoliubov states using
  the {S}kyrme energy density functional, Phys. Rev. C 78 (2008) 024309.
\newblock \href {http://dx.doi.org/10.1103/PhysRevC.78.024309}
  {\path{doi:10.1103/PhysRevC.78.024309}}.

\bibitem{sekizawa2014}
K.~Sekizawa, K.~Yabana, {P}article-number projection method in time-dependent
  {H}artree-{F}ock theory: {P}roperties of reaction products, Phys. Rev. C 90
  (2014) 064614.
\newblock \href {http://dx.doi.org/10.1103/PhysRevC.90.064614}
  {\path{doi:10.1103/PhysRevC.90.064614}}.

\bibitem{jiang1998}
C.~L. Jiang, K.~E. Rehm, H.~Esbensen, D.~J. Blumenthal, B.~Crowell, J.~Gehring,
  B.~Glagola, J.~P. Schiffer, A.~H. Wuosmaa, {M}ultineutron transfer in
  $^{58}${N}i + $^{124}${S}n collisions at sub-barrier energies, Phys. Rev. C
  57 (1998) 2393--2400.
\newblock \href {http://dx.doi.org/10.1103/physrevc.57.2393}
  {\path{doi:10.1103/physrevc.57.2393}}.

\bibitem{corradi1996}
L.~Corradi, J.~H. He, D.~Ackermann, A.~M. Stefanini, A.~Pisent, S.~Beghini,
  G.~Montagnoli, F.~Scarlassara, G.~F. Segato, G.~Pollarolo, C.~H. Dasso,
  A.~Winther, Multinucleon transfer reactions in $^{40}${C}a+$^{124}${S}n,
  Phys. Rev. C 54 (1996) 201--205.
\newblock \href {http://dx.doi.org/10.1103/physrevc.54.201}
  {\path{doi:10.1103/physrevc.54.201}}.

\bibitem{wang2016}
N.~Wang, L.~Guo, New neutron-rich isotope production in
  $^{154}${S}m+$^{160}${G}d, Phys. Lett. B 760 (2016) 236--241.
\newblock \href {http://dx.doi.org/10.1016/j.physletb.2016.06.073}
  {\path{doi:10.1016/j.physletb.2016.06.073}}.

\bibitem{oertzen2001}
W.~von Oertzen, A.~Vitturi, Pairing correlations of nucleons and multi-nucleon
  transfer between heavy nuclei, Rep. Prog. Phys. 64 (2001) 1247.
\newblock \href {http://dx.doi.org/10.1088/0034-4885/64/10/202}
  {\path{doi:10.1088/0034-4885/64/10/202}}.

\bibitem{corradi2009}
L.~Corradi, G.~Pollarolo, S.~Szilner, Multinucleon transfer processes in
  heavy-ion reactions, J. Phys. G 36 (2009) 113101.
\newblock \href {http://dx.doi.org/10.1088/0954-3899/36/11/113101}
  {\path{doi:10.1088/0954-3899/36/11/113101}}.

\bibitem{bonneau2006}
L.~Bonneau, {F}ission modes of $^{256}${F}m and $^{258}${F}m in a microscopic
  approach, Phys. Rev. C 74 (2006) 014301.
\newblock \href {http://dx.doi.org/10.1103/PhysRevC.74.014301}
  {\path{doi:10.1103/PhysRevC.74.014301}}.

\bibitem{balian1992}
R.~Balian, M.~V\'en\'eroni, Correlations and fluctuations in static and dynamic
  mean-field approaches, Ann. Phys. 216 (1992) 351.
\newblock \href {http://dx.doi.org/10.1016/0003-4916(92)90181-K}
  {\path{doi:10.1016/0003-4916(92)90181-K}}.

\bibitem{tohyama1985}
M.~Tohyama, {T}wo-body collision effects on the low-{L} fusion window in
  $^{16}${O}+$^{16}${O} reactions, Phys. Lett. B 160 (1985) 235--238.
\newblock \href {http://dx.doi.org/10.1016/0370-2693(85)91317-6}
  {\path{doi:10.1016/0370-2693(85)91317-6}}.

\bibitem{tohyama2002a}
M.~Tohyama, A.~S. Umar, {Q}uadrupole resonances in unstable oxygen isotopes in
  time-dependent density-matrix formalism, Phys. Lett. B 549 (2002) 72--78.
\newblock \href {http://dx.doi.org/10.1016/S0370-2693(02)02885-X}
  {\path{doi:10.1016/S0370-2693(02)02885-X}}.

\bibitem{assie2009}
M.~Assi\'e, D.~Lacroix, {P}robing {N}eutron {C}orrelations through {N}uclear
  {B}reakup, Phys. Rev. Lett. 102 (2009) 202501.
\newblock \href {http://dx.doi.org/10.1103/PhysRevLett.102.202501}
  {\path{doi:10.1103/PhysRevLett.102.202501}}.

\bibitem{tohyama2016}
M.~Tohyama, A.~S. Umar, {T}wo-body dissipation effects on the synthesis of
  superheavy elements, Phys. Rev. C 93 (2016) 034607.
\newblock \href {http://dx.doi.org/10.1103/PhysRevC.93.034607}
  {\path{doi:10.1103/PhysRevC.93.034607}}.

\bibitem{wen2018}
K.~Wen, M.~C. Barton, A.~Rios, P.~D. Stevenson,
  \href{https://link.aps.org/doi/10.1103/PhysRevC.98.014603}{Two-body
  dissipation effect in nuclear fusion reactions}, Phys. Rev. C 98 (2018)
  014603.
\newblock \href {http://dx.doi.org/10.1103/PhysRevC.98.014603}
  {\path{doi:10.1103/PhysRevC.98.014603}}.
\newline\urlprefix\url{https://link.aps.org/doi/10.1103/PhysRevC.98.014603}

\bibitem{davies1978a}
K.~T.~R. Davies, V.~Maruhn-Rezwani, S.~E. Koonin, J.~W. Negele, {Test of the
  Time-Dependent Mean-Field Theory in Kr-Induced Strongly Damped Collisions},
  Phys. Rev. Lett. 41 (1978) 632.
\newblock \href {http://dx.doi.org/10.1103/PhysRevLett.41.632}
  {\path{doi:10.1103/PhysRevLett.41.632}}.

\bibitem{marston1985}
J.~B. Marston, S.~E. Koonin, {M}ean-{F}ield {C}alculations of {F}luctuations in
  {N}uclear {C}ollisions, Phys. Rev. Lett. 54 (1985) 1139--1141.
\newblock \href {http://dx.doi.org/10.1103/PhysRevLett.54.1139}
  {\path{doi:10.1103/PhysRevLett.54.1139}}.

\bibitem{bonche1985}
P.~Bonche, H.~Flocard, {D}ispersion of one-body operators with the
  {B}alian-{V}eneroni variational principle, Nucl. Phys. A 437 (1985) 189--207.
\newblock \href {http://dx.doi.org/10.1016/0375-9474(85)90232-5}
  {\path{doi:10.1016/0375-9474(85)90232-5}}.

\bibitem{troudet1985}
T.~Troudet, D.~Vautherin, Mass dispersion through particle emission from a
  vibrating nucleus, Phys. Rev. C 31 (1985) 278.
\newblock \href {http://dx.doi.org/10.1103/PhysRevC.31.278}
  {\path{doi:10.1103/PhysRevC.31.278}}.

\bibitem{broomfield2008}
J.~M.~A. Broomfield, P.~D. Stevenson, {M}ass dispersions from giant dipole
  resonances using the {B}alian-{V}\'en\'eroni variational approach, J. Phys. G
  35 (2008) 095102.
\newblock \href {http://dx.doi.org/10.1088/0954-3899/35/9/095102}
  {\path{doi:10.1088/0954-3899/35/9/095102}}.

\bibitem{broomfield2009}
J.~M.~A. Broomfield, Calculations of {M}ass {D}istributions using the
  {B}alian-{V}\'en\'eroni {V}ariational {A}pproach, Ph.D. thesis, University of
  Surrey (2009).

\bibitem{zielinska1988}
M.~Zielinska-Pfab\'e, C.~Gr\'egoire, Dispersions in semiclassical dynamics,
  Phys. Rev. C 37 (1988) 2594.
\newblock \href {http://dx.doi.org/10.1103/PhysRevC.37.2594}
  {\path{doi:10.1103/PhysRevC.37.2594}}.

\bibitem{martin1991}
C.~{M}artin, D.~{V}autherin, A variational derivation of semiclassical
  mean-field evolution equations for a hot \text{Fermi} gas, Phys. Lett. B 260
  (1991) 1.
\newblock \href {http://dx.doi.org/10.1016/0370-2693(91)90959-T}
  {\path{doi:10.1016/0370-2693(91)90959-T}}.

\bibitem{martin1995}
C.~Martin, {Variational approximation for two-time correlation functions in
  $\Phi^{4}$ theory: {O}ptimization of the dynamics}, Phys. Rev. D 52 (1995)
  7121.
\newblock \href {http://dx.doi.org/10.1103/PhysRevD.52.7121}
  {\path{doi:10.1103/PhysRevD.52.7121}}.

\bibitem{benarous1999}
M.~Benarous, H.~Flocard, {Time-Dependent Variational Approach for Boson
  Systems}, Ann. Phys. 273 (1999) 242.
\newblock \href {http://dx.doi.org/10.1006/aphy.1998.5901}
  {\path{doi:10.1006/aphy.1998.5901}}.

\bibitem{boudjemaa2010}
A.~Boudjem{\^a}a, M.~Benarous, Variational self-consistent theory for trapped
  {Bose} gases at finite temperature, Eur. Phys. J. D 59 (2010) 427.
\newblock \href {http://dx.doi.org/10.1140/epjd/e2010-00177-5}
  {\path{doi:10.1140/epjd/e2010-00177-5}}.

\bibitem{boudjemaa2015}
A.~Boudjem{\^a}a,
  \href{http://stacks.iop.org/1751-8121/48/i=4/a=045002}{Self-consistent theory
  of a bose-einstein condensate with impurity at finite temperature}, J. Phys.
  A 48~(4) (2015) 045002.
\newline\urlprefix\url{http://stacks.iop.org/1751-8121/48/i=4/a=045002}

\bibitem{roynette1977}
J.~C. Roynette, H.~Doubre, N.~Frascaria, J.~C. Jacmart, N.~Poff\'e, M.~Riou, On
  the time scale of $^{40}\mathrm{Ca}+{}^{40}\mathrm{Ca}$ strongly damped
  reactions, Phys. Lett. B 67 (1977) 395.
\newblock \href {http://dx.doi.org/10.1016/0370-2693(77)90428-2}
  {\path{doi:10.1016/0370-2693(77)90428-2}}.

\bibitem{ayik2009}
S.~{Ayik}, K.~{Washiyama}, D.~{Lacroix}, Fluctuation and dissipation dynamics
  in fusion reactions from a stochastic mean-field approach, Phys. Rev. C 79
  (2009) 054606.
\newblock \href {http://dx.doi.org/10.1103/PhysRevC.79.054606}
  {\path{doi:10.1103/PhysRevC.79.054606}}.

\bibitem{washiyama2009b}
K.~Washiyama, S.~Ayik, D.~Lacroix, {M}ass dispersion in transfer reactions with
  a stochastic mean-field theory, Phys. Rev. C 80 (2009) 031602.
\newblock \href {http://dx.doi.org/10.1103/PhysRevC.80.031602}
  {\path{doi:10.1103/PhysRevC.80.031602}}.

\bibitem{yilmaz2011}
B.~Yilmaz, S.~Ayik, D.~Lacroix, K.~Washiyama, {N}ucleon exchange mechanism in
  heavy-ion collisions at near-barrier energies, Phys. Rev. C 83 (2011) 064615.
\newblock \href {http://dx.doi.org/10.1103/PhysRevC.83.064615}
  {\path{doi:10.1103/PhysRevC.83.064615}}.

\bibitem{ayik2008a}
S.~Ayik, N.~Er, O.~Yilmaz, A.~Gokalp, Quantal effects on spinodal instabilities
  in charge asymmetric nuclear matter, Nucl. Phys. A 812 (2008) 44--57.
\newblock \href {http://dx.doi.org/10.1016/j.nuclphysa.2008.08.007}
  {\path{doi:10.1016/j.nuclphysa.2008.08.007}}.

\bibitem{ayik2009a}
S.~Ayik, O.~Yilmaz, N.~Er, A.~Gokalp, P.~Ring, Spinodal instabilities in
  nuclear matter in a stochastic relativistic mean-field approach, Phys. Rev. C
  80 (2009) 034613.
\newblock \href {http://dx.doi.org/10.1103/PhysRevC.80.034613}
  {\path{doi:10.1103/PhysRevC.80.034613}}.

\bibitem{ayik2011}
S.~Ayik, O.~Yilmaz, F.~Acar, B.~Danisman, N.~Er, A.~Gokalp, Investigations of
  instabilities in nuclear matter in stochastic relativistic models, Nucl.
  Phys. A 859 (2011) 73--86.
\newblock \href {http://dx.doi.org/10.1016/j.nuclphysa.2011.04.004}
  {\path{doi:10.1016/j.nuclphysa.2011.04.004}}.

\bibitem{yilmaz2011a}
O.~Yilmaz, S.~Ayik, A.~Gokalp, Quantal description of instabilities in nuclear
  matter in a stochastic relativistic model, Eur. Phys. J. A 47 (2011) 123.
\newblock \href {http://dx.doi.org/10.1140/epja/i2011-11123-3}
  {\path{doi:10.1140/epja/i2011-11123-3}}.

\bibitem{yilmaz2013}
O.~Yilmaz, S.~Ayik, F.~Acar, S.~Saatci, A.~Gokalp, Investigations of spinodal
  dynamics in asymmetric nuclear matter within a stochastic relativistic model,
  Eur. Phys. J. A 49 (2013) 33.
\newblock \href {http://dx.doi.org/10.1140/epja/i2013-13033-8}
  {\path{doi:10.1140/epja/i2013-13033-8}}.

\bibitem{ayik2017}
S.~Ayik, B.~Yilmaz, O.~Yilmaz, A.~S. Umar, G.~Turan, Multinucleon transfer in
  central collisions of $^{238}\mathrm{U}+^{238}\mathrm{U}$, Phys. Rev. C 96
  (2017) 024611.
\newblock \href {http://dx.doi.org/10.1103/PhysRevC.96.024611}
  {\path{doi:10.1103/PhysRevC.96.024611}}.

\bibitem{lacroix2013}
D.~Lacroix, D.~Gambacurta, S.~Ayik, Quantal corrections to mean--field dynamics
  including pairing, Phys. Rev. C 87 (2013) 061302.
\newblock \href {http://dx.doi.org/10.1103/PhysRevC.87.061302}
  {\path{doi:10.1103/PhysRevC.87.061302}}.

\bibitem{yilmaz2014}
B.~Yilmaz, S.~Ayik, D.~Lacroix, O.~Yilmaz, Nucleon exchange in heavy-ion
  collisions within a stochastic mean-field approach, Phys. Rev. C 90 (2014)
  024613.
\newblock \href {http://dx.doi.org/10.1103/physrevc.90.024613}
  {\path{doi:10.1103/physrevc.90.024613}}.

\bibitem{ayik2015a}
S.~Ayik, B.~Yilmaz, O.~Yilmaz, Multinucleon exchange in quasifission reactions,
  Phys. Rev. C 92 (2015) 064615.
\newblock \href {http://dx.doi.org/10.1103/physrevc.92.064615}
  {\path{doi:10.1103/physrevc.92.064615}}.

\bibitem{ayik2015}
S.~Ayik, O.~Yilmaz, B.~Yilmaz, A.~S. Umar, A.~Gokalp, G.~Turan, D.~Lacroix,
  {Q}uantal description of nucleon exchange in a stochastic mean-field
  approach, Phys. Rev. C 91 (2015) 054601.
\newblock \href {http://dx.doi.org/10.1103/PhysRevC.91.054601}
  {\path{doi:10.1103/PhysRevC.91.054601}}.

\bibitem{ayik2016}
S.~Ayik, O.~Yilmaz, B.~Yilmaz, A.~S. Umar, Quantal nucleon diffusion: {C}entral
  collisions of symmetric nuclei, Phys. Rev. C 94 (2016) 044624.
\newblock \href {http://dx.doi.org/10.1103/PhysRevC.94.044624}
  {\path{doi:10.1103/PhysRevC.94.044624}}.

\bibitem{skyrme1958}
T.~Skyrme, The effective nuclear potential, Nucl. Phys. 9 (1958) 615--634.
\newblock \href {http://dx.doi.org/10.1016/0029-5582(58)90345-6}
  {\path{doi:10.1016/0029-5582(58)90345-6}}.

\bibitem{kluepfel2009}
P.~Kl\"uepfel, P.-G. Reinhard, T.~J. B\"urvenich, J.~A. Maruhn, {V}ariations on
  a theme by {S}kyrme: {A} systematic study of adjustments of model parameters,
  Phys. Rev. C 79 (2009) 034310.
\newblock \href {http://dx.doi.org/10.1103/PhysRevC.79.034310}
  {\path{doi:10.1103/PhysRevC.79.034310}}.

\bibitem{kortelainen2010}
M.~Kortelainen, T.~Lesinski, J.~More, W.~Nazarewicz, J.~Sarich, N.~Schunck,
  M.~V. Stoitsov, S.~Wild, {N}uclear energy density optimization, Phys. Rev. C
  82 (2010) 024313.
\newblock \href {http://dx.doi.org/10.1103/PhysRevC.82.024313}
  {\path{doi:10.1103/PhysRevC.82.024313}}.

\bibitem{kortelainen2012}
M.~Kortelainen, J.~{McDonnell}, W.~Nazarewicz, P.-G. Reinhard, J.~Sarich,
  N.~Schunck, M.~V. Stoitsov, S.~M. Wild, {N}uclear energy density
  optimization: {L}arge deformations, Phys. Rev. C 85 (2012) 024304.
\newblock \href {http://dx.doi.org/10.1103/PhysRevC.85.024304}
  {\path{doi:10.1103/PhysRevC.85.024304}}.

\bibitem{engel1975}
Y.~M. Engel, D.~M. Brink, K.~Goeke, S.~J. Krieger, D.~Vautherin,
  {T}ime-dependent {H}artree-{F}ock theory with {S}kyrme's interaction, Nucl.
  Phys. A 249 (1975) 215--238.
\newblock \href {http://dx.doi.org/10.1016/0375-9474(75)90184-0}
  {\path{doi:10.1016/0375-9474(75)90184-0}}.

\bibitem{lesinski2007}
T.~Lesinski, M.~Bender, K.~Bennaceur, T.~Duguet, J.~Meyer, Tensor part of the
  {S}kyrme energy density functional: {S}pherical nuclei, Phys. Rev. C 76
  (2007) 014312.
\newblock \href {http://dx.doi.org/10.1103/PhysRevC.76.014312}
  {\path{doi:10.1103/PhysRevC.76.014312}}.

\bibitem{dobaczewski2000}
J.~Dobaczewski, J.~Dudek, S.~G. Rohozi\ifmmode~\acute{n}\else \'{n}\fi{}ski,
  T.~R. Werner, Point symmetries in the {H}artree-{F}ock approach. {I}.
  {D}ensities, shapes, and currents, Phys. Rev. C 62 (2000) 014310.
\newblock \href {http://dx.doi.org/10.1103/PhysRevC.62.014310}
  {\path{doi:10.1103/PhysRevC.62.014310}}.

\bibitem{umar1986a}
A.~S. Umar, M.~R. Strayer, P.-G. Reinhard, {R}esolution of the {F}usion
  {W}indow {A}nomaly in {H}eavy-{I}on {C}ollisions, Phys. Rev. Lett. 56 (1986)
  2793--2796.
\newblock \href {http://dx.doi.org/10.1103/PhysRevLett.56.2793}
  {\path{doi:10.1103/PhysRevLett.56.2793}}.

\bibitem{reinhard1988}
P.-G. Reinhard, A.~S. Umar, K.~T.~R. Davies, M.~R. Strayer, S.-J. Lee,
  {D}issipation and forces in time-dependent {H}artree-{F}ock calculations,
  Phys. Rev. C 37 (1988) 1026--1035.
\newblock \href {http://dx.doi.org/10.1103/PhysRevC.37.1026}
  {\path{doi:10.1103/PhysRevC.37.1026}}.

\bibitem{umar1989}
A.~S. Umar, M.~R. Strayer, P.-G. Reinhard, K.~T.~R. Davies, S.-J. Lee,
  {S}pin-orbit force in time-dependent {H}artree-{F}ock calculations of
  heavy-ion collisions, Phys. Rev. C 40 (1989) 706--714.
\newblock \href {http://dx.doi.org/10.1103/PhysRevC.40.706}
  {\path{doi:10.1103/PhysRevC.40.706}}.

\bibitem{maruhn2006b}
J.~A. Maruhn, P.-G. Reinhard, P.~D. Stevenson, M.~R. Strayer, {S}pin-excitation
  mechanisms in {S}kyrme-force time-dependent {H}artree-{F}ock calculations,
  Phys. Rev. C 74 (2006) 027601.
\newblock \href {http://dx.doi.org/10.1103/PhysRevC.74.027601}
  {\path{doi:10.1103/PhysRevC.74.027601}}.

\bibitem{umar2006c}
A.~S. Umar, V.~E. Oberacker, {T}hree-dimensional unrestricted time-dependent
  {H}artree-{F}ock fusion calculations using the full {S}kyrme interaction,
  Phys. Rev. C 73 (2006) 054607.
\newblock \href {http://dx.doi.org/10.1103/PhysRevC.73.054607}
  {\path{doi:10.1103/PhysRevC.73.054607}}.

\bibitem{suckling2010}
E.~B. Suckling, P.~D. Stevenson, {T}he effect of the tensor force on the
  predicted stability of superheavy nuclei, EPL 90 (2010) 12001.
\newblock \href {http://dx.doi.org/10.1209/0295-5075/90/12001}
  {\path{doi:10.1209/0295-5075/90/12001}}.

\bibitem{dai2014}
G.-F. Dai, L.~Guo, E.-G. Zhao, S.-G. Zhou, Dissipation dynamics and spin--orbit
  force in time--dependent {H}artree--{F}ock theory, Phys. Rev. C 90 (2014)
  044609.
\newblock \href {http://dx.doi.org/10.1103/PhysRevC.90.044609}
  {\path{doi:10.1103/PhysRevC.90.044609}}.

\bibitem{loebl2011}
N.~Loebl, J.~A. Maruhn, P.-G. Reinhard, Equilibration in the time-dependent
  \text{Hartree-Fock} approach probed with the \text{Wigner} distribution
  function, Phys. Rev. C 84 (2011) 034608.
\newblock \href {http://dx.doi.org/10.1103/PhysRevC.84.034608}
  {\path{doi:10.1103/PhysRevC.84.034608}}.

\bibitem{loebl2012}
N.~Loebl, A.~S. Umar, J.~A. Maruhn, P.-G. Reinhard, P.~D. Stevenson, V.~E.
  Oberacker, {S}ingle-particle dissipation in a time-dependent {H}artree-{F}ock
  approach studied from a phase-space perspective, Phys. Rev. C 86 (2012)
  024608.
\newblock \href {http://dx.doi.org/10.1103/PhysRevC.86.024608}
  {\path{doi:10.1103/PhysRevC.86.024608}}.

\bibitem{dai2014a}
G.~Dai, L.~Guo, E.~Zhao, S.~Zhou, Effect of tensor force on dissipation
  dynamics in time-dependent {H}artree--{F}ock theory, Sci. China Phys. 57
  (2014) 1618--1622.
\newblock \href {http://dx.doi.org/10.1007/s11433-014-5536-8}
  {\path{doi:10.1007/s11433-014-5536-8}}.

\bibitem{barton2017}
M.~C. Barton, P.~D. Stevenson,
  \href{https://arxiv.org/abs/1709.07823}{{M}agnetic {D}ipole {T}ransitions
  with the {F}ull {S}kyrme {I}nteraction}, arXiv:1709.07823.
\newline\urlprefix\url{https://arxiv.org/abs/1709.07823}

\bibitem{guo2018}
L.~Guo, C.~Simenel, L.~Shi, C.~Yu,
  \href{http://www.sciencedirect.com/science/article/pii/S0370269318304301}{The
  role of tensor force in heavy-ion fusion dynamics}, Phys. Lett. B 782 (2018)
  401 -- 405.
\newblock \href
  {http://dx.doi.org/https://doi.org/10.1016/j.physletb.2018.05.066}
  {\path{doi:https://doi.org/10.1016/j.physletb.2018.05.066}}.
\newline\urlprefix\url{http://www.sciencedirect.com/science/article/pii/S0370269318304301}

\bibitem{stevenson2016}
P.~D. Stevenson, E.~B. Suckling, S.~Fracasso, M.~C. Barton, A.~S. Umar, Skyrme
  tensor force in heavy ion collisions, Phys. Rev. C 93 (2016) 054617.
\newblock \href {http://dx.doi.org/10.1103/physrevc.93.054617}
  {\path{doi:10.1103/physrevc.93.054617}}.

\bibitem{iwata2011}
Y.~Iwata, J.~A. Maruhn, Enhanced spin-current tensor contribution in collision
  dynamics, Phys. Rev. C 84 (2011) 014616.
\newblock \href {http://dx.doi.org/10.1103/PhysRevC.84.014616}
  {\path{doi:10.1103/PhysRevC.84.014616}}.

\bibitem{almehed2005}
D.~Almehed, P.~D. Stevenson, Isovector giant monopole resonances in spherical
  nuclei, J. Phys. G 31 (2005) S1819.
\newblock \href {http://dx.doi.org/10.1088/0954-3899/31/10/079}
  {\path{doi:10.1088/0954-3899/31/10/079}}.

\bibitem{reinhard2006}
P.-G. Reinhard, P.~D. Stevenson, D.~Almehed, J.~A. Maruhn, M.~R. Strayer,
  {R}ole of boundary conditions in dynamic studies of nuclear giant resonances
  and collisions, Phys. Rev. E 73 (2006) 036709.
\newblock \href {http://dx.doi.org/10.1103/PhysRevE.73.036709}
  {\path{doi:10.1103/PhysRevE.73.036709}}.

\bibitem{avez2013}
B.~Avez, C.~Simenel, {S}tructure and direct decay of {G}iant {M}onopole
  {R}esonances, Eur. Phys. J. A 49 (2013) 76.
\newblock \href {http://dx.doi.org/10.1140/epja/i2013-13076-9}
  {\path{doi:10.1140/epja/i2013-13076-9}}.

\bibitem{guo2008}
L.~Guo, J.~A. Maruhn, P.-G. Reinhard, Y.~Hashimoto, {C}onservation properties
  in the time-dependent {H}artree {F}ock theory, Phys. Rev. C 77 (2008) 041301.
\newblock \href {http://dx.doi.org/10.1103/PhysRevC.77.041301}
  {\path{doi:10.1103/PhysRevC.77.041301}}.

\bibitem{degiovannini2012}
U.~{De Giovannini}, D.~Varsano, M.~A.~L. Marques, H.~Appel, E.~K.~U. Gross,
  A.~Rubio, \textit{Ab initio} angle-- and energy--resolved photoelectron
  spectroscopy with time--dependent density--functional theory, Phys. Rev. A 85
  (2012) 062515.
\newblock \href {http://dx.doi.org/10.1103/PhysRevA.85.062515}
  {\path{doi:10.1103/PhysRevA.85.062515}}.

\bibitem{boucke1997}
K.~Boucke, H.~Schmitz, H.-J. Kull, Radiation conditions for the time--dependent
  {S}chr\"odinger equation: {A}pplication to strong--field photoionization,
  Phys. Rev. A 56 (1997) 763--771.
\newblock \href {http://dx.doi.org/10.1103/PhysRevA.56.763}
  {\path{doi:10.1103/PhysRevA.56.763}}.

\bibitem{mangin-brinet1998}
M.~Mangin-Brinet, J.~Carbonell, C.~Gignoux, Exact boundary conditions at finite
  distance for the time-dependent {S}chr\"odinger equation, Phys. Rev. A 57
  (1998) 3245--3255.
\newblock \href {http://dx.doi.org/10.1103/PhysRevA.57.3245}
  {\path{doi:10.1103/PhysRevA.57.3245}}.

\bibitem{pardi2013}
C.~I. Pardi, P.~D. Stevenson, {C}ontinuum time-dependent {H}artree-{F}ock
  method for giant resonances in spherical nuclei, Phys. Rev. C 87 (2013)
  014330.
\newblock \href {http://dx.doi.org/10.1103/PhysRevC.87.014330}
  {\path{doi:10.1103/PhysRevC.87.014330}}.

\bibitem{pardi2014}
C.~I. Pardi, P.~D. Stevenson, K.~Xu, {E}xtension of the continuum
  time-dependent {H}artree-{F}ock method to proton states, Phys. Rev. E 89
  (2014) 033312.
\newblock \href {http://dx.doi.org/10.1103/PhysRevE.89.033312}
  {\path{doi:10.1103/PhysRevE.89.033312}}.

\bibitem{schuetrumpf2015a}
B.~Schuetrumpf, W.~Nazarewicz, {T}wist-averaged boundary conditions for nuclear
  pasta {H}artree-{F}ock calculations, Phys. Rev. C 92 (2015) 045806.
\newblock \href {http://dx.doi.org/10.1103/PhysRevC.92.045806}
  {\path{doi:10.1103/PhysRevC.92.045806}}.

\bibitem{tohyama2001}
M.~Tohyama, A.~S. Umar, {D}ipole resonances in oxygen isotopes in
  time-dependent density-matrix theory, Phys. Lett. B 516 (2001) 415--420.
\newblock \href {http://dx.doi.org/10.1016/S0370-2693(01)00925-X}
  {\path{doi:10.1016/S0370-2693(01)00925-X}}.

\bibitem{simenel2014b}
C.~{S}imenel, Challenges in description of heavy-ion collisions with
  microscopic time-dependent approaches, J. Phys. G 41 (2014) 094007.
\newblock \href {http://dx.doi.org/10.1088/0954-3899/41/9/094007}
  {\path{doi:10.1088/0954-3899/41/9/094007}}.

\end{thebibliography}

\end{document}